\documentclass[longauth]{aa}
\pdfoutput=1
\usepackage[varg]{txfonts}

\usepackage[T1]{fontenc}
\usepackage{ae,aecompl}
\usepackage{multirow}

\usepackage{grffile}
\usepackage{float}
\usepackage{graphicx}	% Including figure files
\usepackage{hyperref}
\usepackage{subcaption}
\usepackage{booktabs}

\usepackage{bm}	% Advanced maths commands

\begin{document}

\title{Unveiling the rarest morphologies of the LOFAR Two-metre Sky Survey radio source population with self-organised maps} 

\subtitle{} 
\authorrunning{Mostert et al.}
\titlerunning{Unveiling radio morphology using self-organising maps}

\author{Rafa\"el I.J. Mostert\thanks{E-mail: mostert@strw.leidenuniv.nl}\inst{1,2}\and
Kenneth J. Duncan\thanks{E-mail: kdun@roe.ac.uk}\inst{1,3}\and
Huub J.A. R\"{o}ttgering\inst{1}\and
Kai L. Polsterer\inst{4}\and
Philip N. Best\inst{3}\and
 Marisa Brienza\inst{5,6}\and
 Marcus Brüggen\inst{7}\and
 Martin J. Hardcastle\inst{8}\and
 Nika Jurlin\inst{2,9}\and
  Beatriz Mingo\inst{10}\and
 Raffaella Morganti\inst{2,9}\and
 Tim Shimwell\inst{2,1}\and
 Dan Smith\inst{8}\and
 Wendy L. Williams\inst{1}\\}
 
\institute{Leiden Observatory, Leiden University, PO Box 9513, NL-2300 RA Leiden, The Netherlands \and
ASTRON, the Netherlands Institute for Radio Astronomy, Postbus 2, 7990 AA, Dwingeloo, The Netherlands\and
SUPA, Institute for Astronomy, Royal Observatory, Blackford Hill, Edinburgh, EH9 3HJ, UK \and
HITS gGmbH (Heidelberg Institute for Theoretical Studies), AstroinformaticsSchloss-Wolfsbrunnenweg 35, 69118 Heidelberg, Germany \and
Dipartimento di Fisica e Astronomia, Università di Bologna, via P. Gobetti 93/2, 40129, Bologna, Italy\and
INAF - Istituto di Radioastronomia, Via P. Gobetti 101, 40129, Bologna, Italy\and
University of Hamburg, Gojenbergsweg 112, 21029 Hamburg, Germany\and
Centre for Astrophysics Research, University of Hertfordshire, College Lane, Hatfield AL10 9AB, UK\and
Kapteyn Astronomical Institute, University of Groningen, P.O. Box 800, 9700 AV Groningen, The Netherlands\and
School of Physical Sciences, The Open University, Walton Hall, Milton Keynes, MK7 6AA, UK 
 }

\date{Received 26 May 2020 / Accepted 29 October 2020}

\abstract
%context
{The Low Frequency Array (LOFAR) Two-metre Sky Survey (LoTSS) is a low-frequency radio continuum survey of the Northern sky at an unparalleled resolution and sensitivity.}
%aims
{In order to fully exploit this huge dataset and those produced by the Square Kilometre Array in the next decade, automated methods in machine learning and data-mining will be increasingly essential both for morphological classifications and for identifying optical counterparts to the radio sources.}
%method
{Using self-organising maps (SOMs), a form of unsupervised machine learning, we created a dimensionality reduction of the radio morphologies for the $\sim$25k extended radio continuum sources in the LoTSS first data release, which is only $\sim$2 percent of the final LoTSS survey. We made use of \textsc{PINK}, a code which extends the SOM algorithm with rotation and flipping invariance, increasing its suitability and effectiveness for training on astronomical sources.} 
%results
{After training, the SOMs can be used for a wide range of science exploitation and we present an illustration of their potential by finding an arbitrary number of morphologically rare sources in our training data (424 square degrees) and subsequently in an area of the sky ($\sim$5300 square degrees) outside the training data. 
Objects found in this way span a wide range of morphological and physical categories: extended jets of radio active galactic nuclei, diffuse cluster haloes and relics, and nearby spiral galaxies. 
Finally, to enable accessible, interactive, and intuitive data exploration, we showcase the LOFAR-PyBDSF Visualisation Tool, which allows users to explore the LoTSS dataset through the trained SOMs.} 
{}

\maketitle 

\section{Introduction}
The morphology of a radio source is an important tool for studying the nature of the source emitting the radio waves and the environment or medium around the radio-emitting source \citep[e.g.][]{Miley1980, kempner2004conference}.
In radio astronomy, the most time-resistant morphological classification scheme for radio galaxies is the one presented by \citet{Fanaroff1974}, which classifies radio galaxies based on their extended radio jets.
The binary classification scheme is based on the location of the brightest hot-spots within the lobes of an extended source. 
\citet{Fanaroff1974} used their scheme to classify a number of sources from the revised third Cambridge catalogue of radio sources \citep[3CR;][]{mackay1971observations} and found a distinct separation in the luminosities of these sources at $2 \times 10^{25}\,\textup{W Hz}^{-1} \textup{sr}^{-1}$ at 178MHz, the Fanaroff-Riley class I (FRI) sources being below this separation and the Fanaroff-Riley class II (FRII) sources being systematically above it. 
 
A new generation of radio surveys, such as the Low Frequency Array \citep[LOFAR;][]{vanHaarlem2013} Two-metre Sky Survey \citep[LoTSS;][]{Shimwell2017}, the Evolutionary Map of the Universe \citep[EMU;][]{Norris2011}, and the MeerKAT international GHz tiered extragalactic exploration \citep[MIGHTEE;][]{mightee} are producing ever larger samples of highly resolved radio sources. Furthermore, LoTSS combines high angular resolution with a high sensitivity to diffuse radio emission, which has never been achieved before over large area surveys, and thus it probes a different range of morphologies \citep[e.g.][]{hardcastle2019ngc, mandal2020revived}. 
The extent to which these new radio surveys may redefine our understanding of radio morphology is illustrated by the recent results of \citet{mingo2019revisiting}, who find that the dichotomy between FRI and FRII luminosities does not seem to hold (supporting  \citealt{best2009radio}) and that the radio active galactic nuclei (AGN) population might be more heterogeneous than previously assumed.

Radio morphological information is also used when performing optical or infrared cross-identifications, providing us with valuable information on the nature of the radio sources.
For unresolved radio sources, if present, a host galaxy should be located at the same location as the radio emission. For resolved radio AGN, we expect a host galaxy to be at the origin of its jets.
This origin is not necessarily close to the flux-weighted centre of the radio emission.
In these cases, we need to use morphological information (the orientation of the jets projected onto the sky plane) to find the potential host galaxy.

A classical classification of radio morphology consists of automated source detection (using a simple signal-to-noise criterion), followed by a manual label process. 
With the LOFAR surveys and the future surveys of the Square Kilometer Array \citep[SKA;][]{schilizzi2004square} and its pathfinders, it is essential to explore methods in machine learning and data-mining to deal, in a more automated and therefore inherently statistical way, with the large sample coming our way. 

Conveniently, in the past few years, in the field of computer science, significant improvements have been made in the fields of data mining and machine learning in general and computer vision specifically.
These improvements are generally recognised to be enabled by the availability of larger datasets, increasingly powerful graphics processing unit (GPU) accelerated compute power and the development and refinement of machine learning algorithms \citep[e.g.][]{Halevy2009unreasonable, Goodfellow2016, Sun2017revisiting}.

We can roughly divide machine learning approaches into two categories: supervised and unsupervised learning \citep[e.g.][]{Goodfellow2016}.
Supervised approaches are fundamentally limited by the requirement for a labelled training sample, which can be limited in size, or by the fact that such a sample is only available for certain surveys.
Human-created labels or human-annotated data are valuable but costly as the annotation process scales linearly with the number of samples in a dataset.
Unsupervised learning does not require labels for its training dataset 
and can be used for density estimations or to cluster data in groups according to patterns in the data \citep[e.g.][]{Goodfellow2016}. 
The lack of labels means these techniques are not biased by preconceived human-created categories; however, they do not necessarily relate to intrinsic physical source properties.

Unsupervised learning approaches have been applied to astronomical datasets before. \cite{Baron2016} took an unsupervised approach to study galaxy evolution by using random forests to find spectroscopic outliers in the Sloan Digital Sky Survey (SDSS).
\cite{Segal2018} used `apparent complexity' as a metric to describe radio morphology. In this approach, images of radio sources are compressed using the gzip compression algorithm after which the size of the resulting file is a course-grained measure for the morphological complexity of the source. 
The one-dimensional nature of the result makes it a potentially useful addition to a catalogue where it can serve as input to supervised learning methods. 
On its own, the metric is not so valuable, as the degeneracies are numerous and no information about the shape, size or number of components of the source is retained.

Our ultimate aim is to robustly classify the radio sources in new generations of wide and deep radio surveys according to their morphology.
Given the new parameter space being probed by these surveys and the potential new morphological regime they reveal \citep{mingo2019revisiting}, unsupervised clustering offers an approach that minimises our assumptions and any biases inherent within them. 

In this paper we use a rotation and flipping invariant implementation of the self-organised maps \citep[SOM;][]{Kohonen1989,Kohonen2001} dimensionality reduction algorithm  \citep{Polsterer2015} to explore the morphologies of radio sources in the LOFAR Two-metre Sky Survey First Data Release \citep{Shimwell2017}.
The dimensionality reduction will be a model that represents the most frequently occurring shapes in our data, regardless of whether they conform to any pre-existing morphological classification scheme. 
In addition to providing a data-driven model of the representative radio morphologies within LoTSS, an SOM can be used to select the radio objects that most diverge from this model -- these will be morphologically rare or outlier sources that can automatically be identified within potentially unexplored parameter space. This method does not limit our search to many forms of AGN (bent, assymetric, remnant and restarted), it also leads us to nearby spiral galaxies and cluster emission many of which may be previously undiscovered in radio observations. We do not tackle the FRI or FRII classification of LoTSS radio sources in this paper as this requires additional completeness simulations. 

This paper is set out as follows: Section 2 presents the LOFAR radio continuum dataset used in this work. 
Section 3 introduces the rotation and flipping invariant SOM technique and outlines its application to the LoTSS sample of radio continuum sources.
Section 4 presents the resulting trained SOMs, including the range and distribution of morphological representative images within the LoTSS extended radio source population. We illustrate how the trained map can be used to automatically identify morphologically unique sources in new datasets.
In Section 5 we discuss our research and its place within the wider picture of large survey science.
Finally, Section 6 presents the summary and conclusions of the paper.

\section{Data}
The primary data used for training and optimising the SOMs in our study is taken from the first data release of the LOFAR Two-metre Sky Survey \citep[LoTSS-DR1;][]{Shimwell2019} and consists of 58 pointings that make up a mosaic that covers 424 square degrees in the HETDEX region of the sky (right ascension 10h45m00s to 15h30m00s and declination $45^{\circ}$ to $57^{\circ}$).  The survey, observed at 120--168MHz with a median rms sensitivity of $S_{144MHz} = 71\mu$Jy/beam and a resolution of $6\arcsec$, will eventually cover the entire Northern sky.

The data are accompanied by a corresponding catalogue containing $352,694$ source-entries, generated by the Python Blob Detection and Source Finder (PyBDSF)\footnote{\url{github.com/lofar-astron/PyBDSF}} application \citep{Mohan2015}; see \citet{Shimwell2019} for the parameters used. 
PyBDSF builds a catalogue from emission islands with peak intensity values that exceed the surrounding noise in the image by, in this case, five standard deviations.
One or more Gaussians are fit to these islands. If these Gaussians overlap\footnote{Grouping of Gaussians into sources: \url{http://www.astron.nl/citt/pybdsf/algorithms.html\#grouping}}, they will enter the catalogue as a single entry. The RA and DEC of the entry is set to the centroid of its Gaussians, which is determined using moment analysis.
$93\%$ of the catalogue entries correspond to unresolved sources. As these contain no morphological information beyond an upper limit on their angular size, we only train on the $24,601$ objects composed of multiple Gaussians.\footnote{That is, entries with PyBDSF $S\_Code = `M'$ or $`C'$. 
See \url{astron.nl/citt/pybdsf/write_catalog.html\#write-catalog} and \url{astron.nl/citt/pybdsf/algorithms.html\#gaussian-fitting}.}

In Section \ref{discover}, we make use of additional LoTSS data 
consisting of 841 pointings that make up a mosaic that covers $5,720$ square degrees, fully overlapping with the $424$ square degrees of the HETDEX region of the sky.
This dataset, constitutes the second LoTSS data release (LoTSS-DR2; Tasse et al. 2020, Shimwell et al. in prep) and contains $4,395,448$ PyBDSF-generated source-entries. We use the part of this larger dataset that is outside our initial dataset to show that an SOM trained with sources on a small patch of the sky can be used to cluster and find outliers in sources from a different part of the sky without retraining.

\section{Method: Rotation invariant self-organised maps}
\label{sec:som}
A self-organising map \citep[SOM;][]{Kohonen1989,Kohonen2001}, also known as a Kohonen self-organising map, Kohonen map or Kohonen network, is an unsupervised artificial neural network used to reduce high-dimensional data to a low-dimensional (usually two or three) representation (known as the `map' or the `lattice'). 
An SOM belongs to the family of dimensionality reduction techniques and is an especially useful starting point for visualisation and clustering and hence data-exploration.
An SOM aims to capture the properties of the elements of a dataset by creating a small number of representative elements on a fixed lattice. 
One key property of SOMs that makes them particularly useful for morphological studies is that they are coherent; similar representative images should be close to each other on the lattice while dissimilar representative images should be further apart on the lattice.

Our dataset consists of cutouts from Stokes-I LOFAR images, with each cutout centred on a radio source that has been detected in the associated PyBDSF source catalogue (see Section \ref{prepro} for details). 
Thus our SOM should capture the overall properties of the images in our dataset by creating a small number of representative images on a fixed lattice. 
Formally, an SOM consists of a lattice of `neurons' and each neuron has weights (pixels of the representative image). 
By iteratively training these weights (the pixels of the representative images will be iteratively adjusted), the aim is to maximise the similarity between the neurons (images) and the training dataset.

The metric we use for determining the similarity between a representative image and an image from the training dataset is the Euclidean norm, meaning that we subtract the representative image from the image in the training dataset and take the square root of the squared sums of the pixel values in the residual image. 
By minimising the value of the norm of each image in the training dataset to its corresponding most similar representative image, we ensure that the representative images of the SOM are a good representation of the images in the training set.

For determining the coherence we count the number of images in our training dataset for which the second best matching representative image is not located directly next to the best matching representative image on the lattice. By minimising this number we ensure that the SOM is coherent.

In machine learning, the distinction is made between model parameters and model hyper-parameters. Model parameters are initialised by the user and updated by the training algorithm, while the hyper-parameters are set (and may be updated) by the user. The pixel-values of the representative images are model parameters, the dimensions of the SOM lattice and the dimensions of the cutouts are hyper-parameters.

For a full outline of the SOM algorithm and its basic implementation, we refer the reader to \citet{Kohonen2001}.
In the next section we describe the rotation and flipping-invariant SOM algorithm employed in this work and subsequently its hyper-parameters.

\subsection{Rotation invariant SOM}
The classification of radio source morphologies should not depend on the orientation of the source on the sky. Hence, the classification should be invariant to rotation and flipping. 
Even so, most algorithms, supervised or unsupervised, are not fully rotation and flipping invariant. 
For supervised convolutional neural networks this problem is often handled by simplified approximation by inserting (many) rotated copies of each source in the training dataset \citep[e.g.][]{Dieleman2015, Aniyan2017, alhassan2018first, Dai2018, Lukic2018, Lukic2019}.

\citet{Polsterer2015} proposed a rotation and flipping invariant SOM algorithm:

\begin{enumerate}
\item Initialise the pixel-values of the images in the adopted lattice to some arbitrary value. In our case we initialise with zeros.
\item For each image in the training dataset:
\begin{enumerate}
\item  Create rotated and flipped copies.
\item For each representative image $\bm{Rep}$, calculate the Euclidean norm to each of the copied images $\bm{Im_{copy}}$ in the SOM, where the Euclidean norm is defined as:
\begin{equation}
\lVert \bm{Im_{copy}} - \bm{Rep} \rVert_2 = \sqrt{\sum_i(\bm{Im_{copy_i}} - \bm{Rep}_i)^2},
\end{equation}
and the summation is over all the pixels $i$ of the images.
\item For each representative image, find the image copy $\bm{Im_{copy,best}}$ to which it has the smallest Euclidean norm. If multiple image copies produce the same norm, randomly select one of those image copies.
\item  From the set of image copies  to each representative image found in 2c, find the single combination of representative image and copied image $\bm{Im_{copy,best}}$ that have the smallest Euclidean norm to each other. We refer to this representative image as $\bm{Rep_{best}}$. If multiple combinations produce the same norm, randomly select one of those combinations. $\bm{ Rep_{best,loc}}$ denotes the $(x,y)$ location of $\bm{Rep_{best}}$ on the SOM lattice.
\item  Update the pixels of each representative image $\bm{Rep}$ such that they are more like their respective $\bm{Im_{copy,best}}$: 
\begin{equation}
\begin{split}
\bm{ Rep_{new}} = \bm{ Rep} + \\
\alpha \cdot \theta \left(\lVert  \bm{ Rep_{best}} - \bm{ Rep_{loc}} \rVert_2 \right) \cdot(\bm{ Im_{copy,best}} - \bm{ Rep}),
\label{update_BMU}
\end{split}
\end{equation}
where $\bm{ Rep_{new}}$ is the updated representative image, $\alpha$ is a scalar known as the learning constraint and $\theta \left(\lVert  \bm{ Rep_{best}} - \bm{ Rep_{loc}} \rVert_2 \right)$ is a function known as the neighbourhood-function. We note that $\alpha$ regulates to what degree all representative images should adapt to $\bm{ Im_{copy,best}}$; $\theta$ is a function of the Euclidean norm between the location of the considered representative image on the lattice $\bm{ Rep_{loc}}$ and $\bm{ Rep_{best,loc}}$. It causes representative images that are farther apart from $\bm{ Rep_{best,loc}}$ on the lattice to adapt less to their respective $\bm{ Im_{copy,best}}$. We chose the boundaries of the SOM to be periodic.\footnote{The representative images at the right border are connected to those at the left and the representative images at the bottom are connected to the representative images at the top.} 
\end{enumerate}

\item Repeat step 2 (except for 2a) a fixed number of times (or `epochs') or until a user-defined stopping condition is met. 
\item Now that the SOM is trained, optionally repeat step 2b and 2c once. Then for each image, return the smallest Euclidean norm of each representative image to their $\bm{ Im_{copy,best}}$.
\end{enumerate}

\citet{Polsterer2015} developed \textsc{PINK}\footnote{Parallelised rotation and flipping INvariant Kohonen map. See github: \url{https://github.com/HITS-AIN/PINK}}, a GPU-optimised code for step 1, 2 and 4 of this algorithm. 
In this study we build on the core \textsc{PINK} algorithm to appropriately preprocess LOFAR data for \textsc{PINK} version 0.23 and enable automated and flexible implementation of step 3 such that $\theta$ and $\alpha$ are scalar functions that also depend on the number of completed epochs. 

For SOMs, tuning hyper-parameters comes down to a trade-off between compute-time, how similar the images in our dataset are to the representative images in the (trained) SOM and the coherency across the SOM. 
Good coherence implies that the similarity between the representative images decreases gradually as a function of distance on the lattice between them.

To quantify the coherency of the SOM and how similar the images of our dataset are to the representative images in the (trained) SOM, we used two metrics.
The Average Quantisation Error (AQE), as described by \citet{Kohonen2001}, is defined as the average summed Euclidean norm from each image in our dataset to its corresponding best matching representative image:
\begin{equation}
AQE = \left(\sum^{|D|} \lVert \bm{ Im_{copy,best}} - \bm{ Rep_{best}} \rVert_2\right) / |D|,
\end{equation}
where the summation iterates over all images in our dataset $D$ and $|D|$ is the number of images in our dataset.
A lower AQE equals a better representation of the data.
The Topological Error (TE) as described by \citet{Villmann1994}, is a measure for the coherence of the SOM. It is defined as the percentage of images for which the second best matching representative image is \emph{not} a direct neighbour of the best matching representative image, where direct neighbour is defined as all eight neighbours for representative images on a rectangular lattice. 
A lower TE equals better coherency. 
We note that AQE and TE are relative measures and can only be used to monitor progress of an SOM during training (for example after every completed epoch) or to make a comparison between different SOMs that have been trained with the same dataset and the same image and lattice dimensions.

A dimensionality reduction technique can produce a closer approximation of a dataset when given more model parameters to model this dataset. For an SOM, having more representative images is equivalent to more model parameters.
Therefore, an SOM with more representative images leads to a better representation of the images in the training dataset (lower AQE and TE). 
The size of the SOM lattice is arbitrary and we can change it in accordance with the purpose of the dimensionality reduction. Using hundreds of representative images for the visual inspection of the most common morphologies in the data is impractical, and a small trained lattice (< $10\times10$ representative images) might suffice. In general, smaller lattices lead to a smaller number of discernible morphological groups. In the case of a $4\times4$ SOM, elongated singles and compact doubles were the only discernible groups. For a $20\times20$ SOM we find additional neurons representing slightly bent extended doubles. 
For this paper we adopt a $10\times 10$ lattice.
SOMs can be trained on a lattice with more than two dimensions, but in this study we make use of only two-dimensional SOMs, as higher dimensional SOMs are harder to visualise on a 2D surface.

For the neighbourhood-function \textsc{PINK} adopts a commonly used 2D-symmetric Gaussian of the form \begin{equation}\begin{split}
\theta(\sigma, \bm{ Rep_{best}}, \bm{ Rep_{loc}}) = \frac{1}{\sigma\sqrt{2 \pi}} \\ \cdot \exp\left(-\frac{1}{2} \left(\frac{\lVert  \bm{ Rep_{best}} - \bm{ Rep_{loc}} \rVert_2}{\sigma}\right)^2\right)
\end{split},
\end{equation} where $\sigma$ is known as the neighbourhood radius.
Generally a larger neighbourhood radius will result in a lower TE and higher AQE. We can intuitively understand this, as a larger neighbourhood radius will cause each training image to leave its imprint on a larger part of the SOM lattice and as a result, representative images will be more similar across the lattice: creating better coherence but with the representative image set encompassing less well the variety of images in the dataset.
As stated by \citet{Kohonen2001}, by decreasing the neighbourhood radius with each training epoch, updates to the lattice will at first be global (ensuring coherence) and then become ever more local (ensuring a good representation of the individual images in our dataset).
Therefore, we adopt and implement a decrease in the neighbourhood radius such that 
$\sigma(t) = \sigma_0 \times \sigma_d ^t$
with $t$ the epoch number, $\sigma_0$ the starting radius and $\sigma_{d}$ the radius decrease rate.
For the value of $\sigma_0$, we adopted the rule of thumb from \citet{Kohonen2001}: we start with a neighbourhood radius half the size of the largest dimension of the SOM lattice.

The size of the learning rate $\alpha$ determines the size of the step we take in the model-landscape. Small steps will generally slowly take us in the right direction but can get us stuck in a local optimum, while big steps give faster results but might overshoot the (local) optimum.
The best results within a given compute-time is achieved by starting out with a large value for the learning rate and gradually decreasing its size with every epoch.
We adopt and implement a decrease in the learning rate such that 
$\alpha(t) = \alpha_0 \times \alpha_d^t \times 	\sigma(t) \sqrt{2 \pi}$
with $t$ the epoch number, $\alpha_0$ the starting learning rate and $\alpha_d$ the learning rate decrease.
We also use the learning rate to undo the normalisation of the neighbourhood-function: we keep its peak constant at the value of 1. 
If we were to keep the 2D-Gaussian normalised, for small neighbourhood radii, the impact of a single training image on the representative images of the SOM would be too great. For each subsequent image, the best matching representative image would be updated to be very much like this image, thereby erasing the similarity to previous images.

To determine reasonable values for the SOM lattice dimensions, $\sigma_{d}$, $\alpha_0$ and $\alpha_d$, we tested the SOM algorithm and its hyperparameters on the well known MNIST handwritten digits dataset.\footnote{The dataset is available at \url{http://yann.lecun.com/exdb/mnist/}}
Subsequently, we tuned the parameters on our own dataset based on the AQE and TE metrics. We stop training once the AQE improved (declined) by less than 1\% over the last epoch.
To prevent over-fitting, we evaluated the quality of the SOM based on the AQE and TE metrics in a holdout-part of the dataset: a randomly drawn subset of the dataset that had not been used for training. Once we had settled on reasonable parameters (see Table \ref{Run_parameters}), we adopted those for the full dataset.

At the start of the training process, we initialise the pixel-values of our representative images with zeros.
Different initialisation, for example using random numbers, changes the initial place on which the images leave their inprint on the SOM. As we train for more than 20 epochs, the initialisation -- given it is of similar magnitude to the training images -- only leads to changes in the final location of the groups we find. It does not expose new morphological groups.

\begin{table*}[]
\centering
\caption{Parameters used for the first SOM training run}
\label{Run_parameters}
\begin{tabular}{l|l}
SOM lattice dimensions (w x h x d)                      & $10\times10\times1$ \\
Number of channels or layers	& 1 \\ 
Representative image dimensions	& $67 \times 67$ pixels\textsuperscript{2}, equivalent to $100\times 100$ arcsec\textsuperscript{2} \\
Neighbourhood radius start $\sigma_0$ | decrease $\sigma_d$ & 5 | 0.9       \\
Learning rate start $\alpha_0$ | decrease $\alpha_d$ & 1 | 0.7 \\
Periodic boundary conditions	& True \\
Stopping condition & AQE improvement per epoch < $1\%$ \\
Resulting number of training epochs & 23 \\
Initialisation	 & Zeros \\
\end{tabular}
\end{table*}

\subsection{Preprocessing: Creating a training dataset from LoTSS images}
\label{prepro}
Our training set only contains images -- no labels, or catalogue information, just pixel-information.
We created images for our training dataset by making a square cutout from the LOFAR intensity maps for each catalogue entry. Each cut-out has a fixed angular (or on-sky) size, centred on the source right ascension and declination. We proceeded by applying a circular mask, removed all flux below $1.5$ times the local noise and rescaled the remaining flux to the continuous [0,1] range. Below we elaborate on these steps.

Given the varying intrinsic physical sizes and redshifts of the radio source population, the choice for the on-sky size is thus a cause of degeneracy in our trained SOM. 
Sources with similar morphology but different apparent size will best match different neurons on the SOM. 
Using different on-sky sizes for each source and then rescaling the dimensions of each image such that the extent of the radio emission for all resolved sources spans a fixed number of pixels would avoid this issue. 
However, this is not a trivial task for several key reasons. 
Firstly, a radio source can consist of two lobes of emission that are spatially separated, such that automatic source extraction software is not able to recognise that the two islands of emission belong to a single radio source. The source size reported by source extraction software may thus refer to a small emission structure that is part of a larger structure. 
Secondly, the precise size or extent of a radio source has no fixed definition and will also depend on the morphology itself.
We refer the reader to Section \ref{linear-invariance} for more on this topic.

In the remainder of this paper, we proceed to use fixed on-sky size images to train our SOM. 
Using fixed on-sky size images, the SOM will need more neurons to represent the images in the training dataset compared to the ideal case where the extent of the emission is normalised. 
Nevertheless, after training, the SOM will still be a good representation of the images in the training dataset. 
Compared to regular sources, sources with outlying morphologies will still have a larger Euclidean norm with respect to the weights of their best matching neurons.

To capture most of the source morphology, the fixed on-sky size should be chosen big enough that most sources fit inside and small enough to minimise the amount of nearby unrelated radio emission. 
Ideally, we would remove the emission from neighbouring sources that spuriously entered our image. 
This is not practical due to the difficulty of correctly associating emission with a single radio source. 
Our source might be part of a larger structure which we risk incorrectly removing (i.e. our catalogue entry might be centred on a single lobe of a double-lobed radio source in which case we would remove the second lobe).
Limitations to the fixed image size are discussed in section \ref{linear-invariance}.

\begin{figure}\begin{center}
\includegraphics[width=0.49\textwidth]{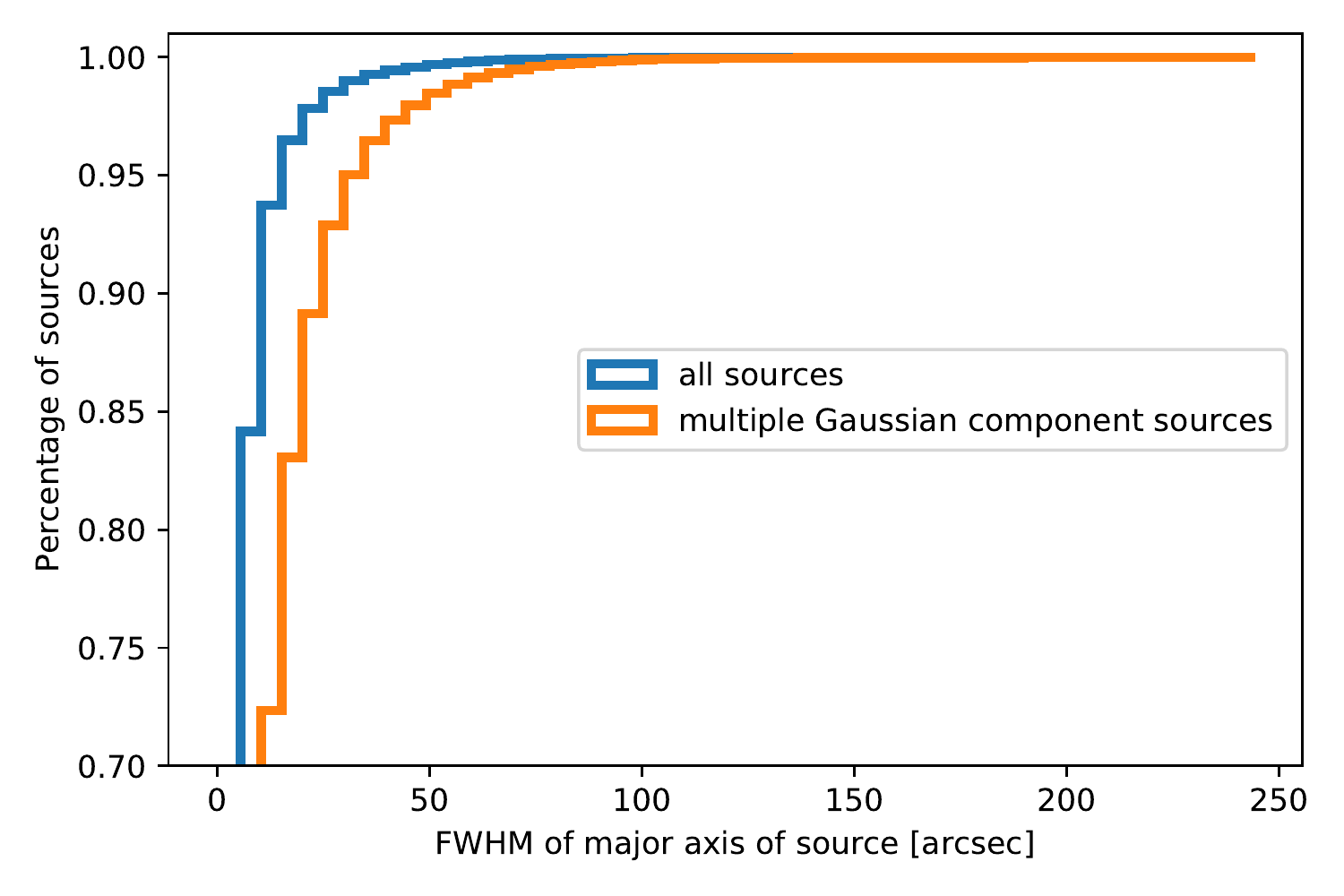}
\caption{Cumulative distribution function (CDF) of the full-width half maximum (FWHM) of the $325,694$ radio-sources in the catalogue and the subset of $24,601$ sources that are composed of multiple Gaussians by PyBDSF. The reported FWHMs of the sources are on the low side as PyBDSF breaks up large apparent objects into multiple catalogue entries with smaller sized objects.}
\label{fig:size_cdf}
\end{center}\end{figure}

\begin{figure}\begin{center}
\includegraphics[width=0.49\textwidth]{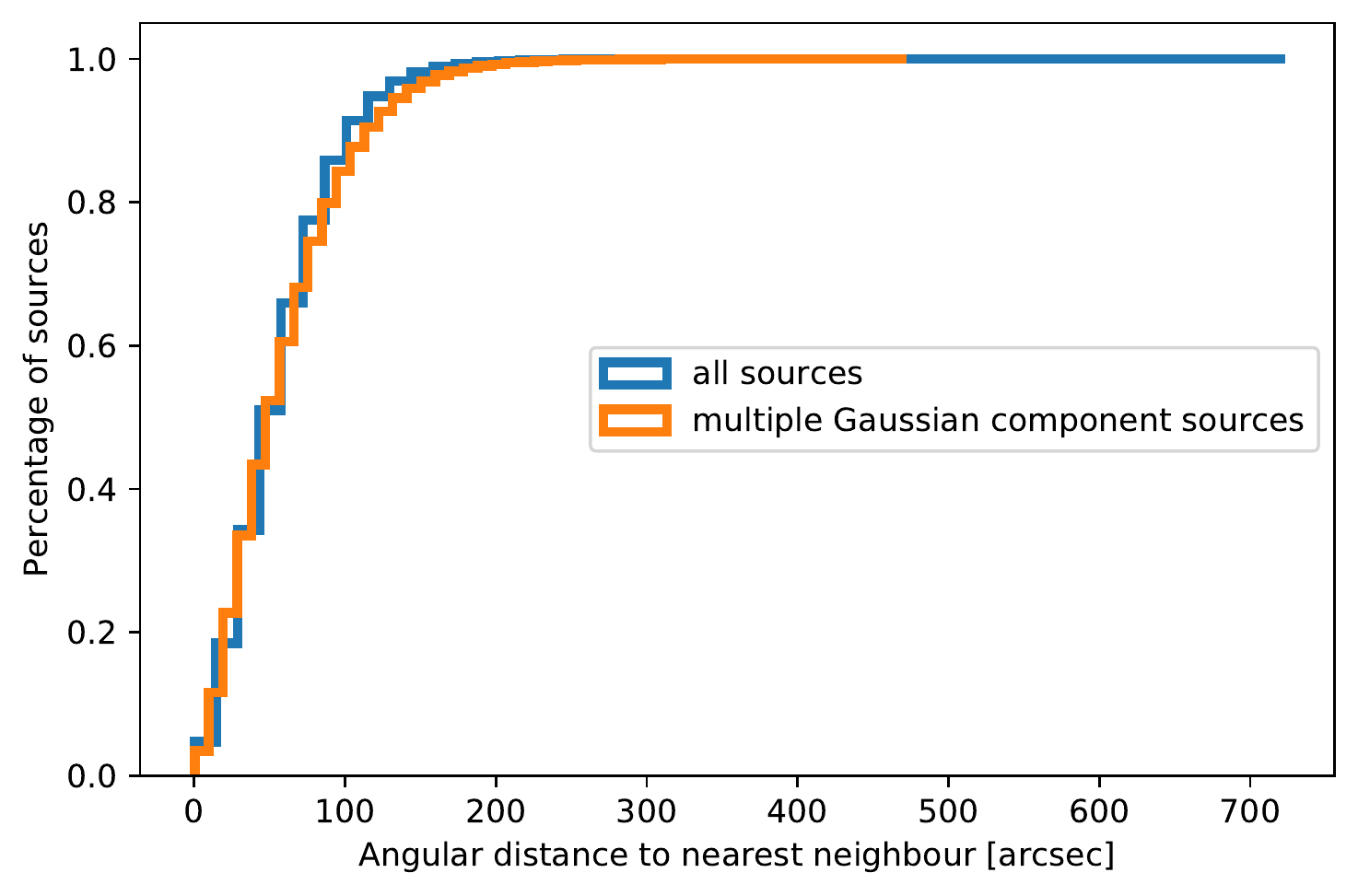}
\caption{CDF of the angular distance to the nearest neighbour for all $325,694$ sources and for the subset of $24,601$ multiple Gaussian-component sources to all $325,694$ sources.}
\label{fig:nn_distance_cdf}
\end{center}\end{figure}

We informed our decision process by gathering information on the nearest-neighbour distances of all $325,694$ radio-sources in the catalogue and the different sizes of all $24,601$ multiple-component objects that we used for training.
Figure  \ref{fig:size_cdf} indicates that a cutout size between roughly 50 and 150 arcsec is able to encompass most extended sources as reported by the catalogue. Figure \ref{fig:nn_distance_cdf} shows that avoiding any contamination from unrelated sources is impossible; we should at least stay well below 200 arcsec to avoid contamination in virtually all cutouts.
We adopt a fixed on-sky size of $100\times 100$ arcsec\textsuperscript{2}, which translates into $67 \times 67$ pixel\textsuperscript{2} images as the pixel scale of our FITS-files is $1.5\times 1.5$ arcsec\textsuperscript{2}. 


In the rotation procedure of the training algorithm (step 2a) we create image copies that are flipped and rotated by increments of 1 degree using bilinear interpolation. Including flipping we end up with 720 rotated and or flipped image copies for each image in our training dataset.
To ensure that these images do not have empty corners, initial images of $95\times 95$ pixels\textsuperscript{2} ($142.5\times 142.5$ arcsec\textsuperscript{2}) were extracted, which were then cropped to $67\times 67$ pixels\textsuperscript{2} after rotation.

We do not want our SOM to learn the correlated noise-patterns around our sources during training.
Therefore, we tested preprocessing the images in the form of clipping the data above or below a certain brightness threshold and in the form of non-linear rescaling of the intensity. 
In a supervised convolution neural network approach to classify FRIs and FRIIs, \citet{Aniyan2017} report best results with sigma-clipping, removing all values below 3 times the local noise.
We tested a range of sigma-clip values based on the local mean of the PyBDSF-generated noise map \citep[see][]{Shimwell2019} 
and find that a $1.5$ sigma-clip threshold results in a good balance between noise-reduction and retaining diffuse parts of the emission. 

As the similarity measure in our algorithm is the Euclidean norm, the intensity of the images in our training dataset affects the outcome of the trained SOM.
Without rescaling the intensities, the SOM clustering will be dominated by the apparent brightness of the sources instead of by morphology. 
The goal of this study is to find rare morphologies across the full dynamic range of LoTSS, we therefore normalise the intensity of all sources by linearly scaling the intensity values of each image in our training dataset to between 0 and 1 (after sigma-clipping).
During training, because of the rotation and subsequent cropping, only the pixels within the circle with a diameter equal to the width of our image are used to compare the image to the neuron weights. 
Therefore, we only consider these pixels during the intensity rescaling and mask all pixels outside of this circle.

In future research simultaneous training on multiple layers will be considered: multiple instances of the same image with different scaling or clipping applied to highlight different features of each source. See section \ref{sec:future} for a description of a multi-layer SOM.

\section{Results}
In this section we present and inspect the trained SOMs and the outlying sources that we can find using these SOMs.
\subsection{Initial SOM training}

\begin{figure}\begin{center}
\includegraphics[width=0.5\textwidth]{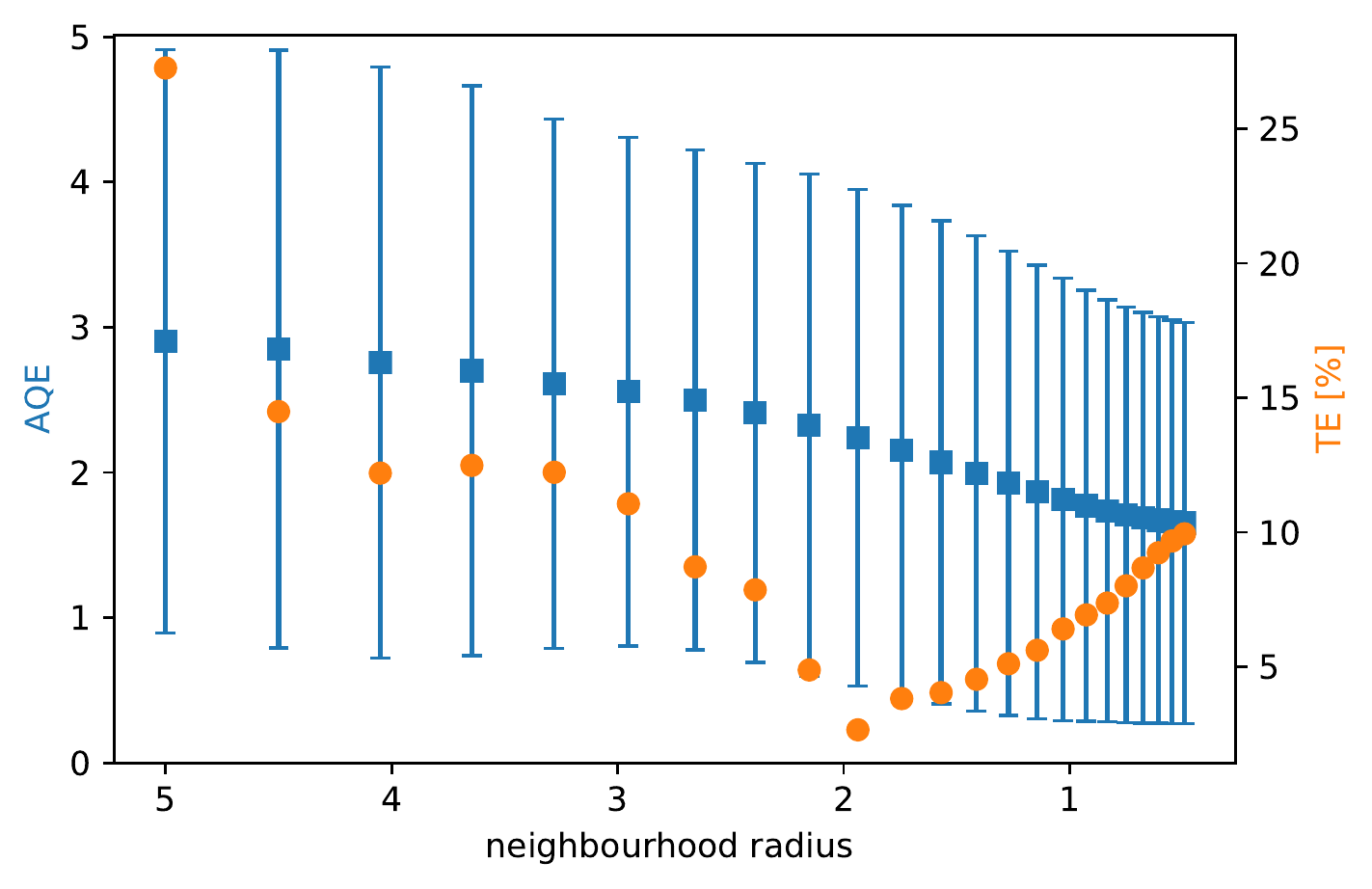}
\caption{Training process of the final $10\times10$ cyclic SOM. Neighbourhood radius is the radius on the SOM lattice. With each epoch we decrease the neighbourhood radius which results in an increasingly accurate SOM (lower AQE) and higher coherency (lower TE). Eventually we increase the accuracy at the cost of the coherency (higher TE). As the AQE is the average of the Euclidean norm between each cutout and its best matching neuron, the errorbars indicate the standard deviation of these values for our data set. }
\label{fig:neighbourhood_vs_aqe}
\end{center}\end{figure}

We first trained the SOM with the parameters reported in Table \ref{Run_parameters}, starting with our initial sample of $24,601$ multiple Gaussian-component sources.
Visually inspecting the resulting trained map revealed that a large part of the SOM contains morphologically similar looking neurons that represent a large number of unresolved or barely resolved sources still present in our training set (see Appendix~\ref{app:firstrun} for more details).

Therefore, to increase the diversity of the different neurons of the SOM, images from our training set that best matched one of the 10 least unique neurons (10\% of the total number of neurons) were removed from our training sample.\footnote{These representative images were selected based on the U-matrix of the SOM. The U-matrix \citep{Ultsch1990} is a metric that shows how similar each representative image in a trained SOM is to its neighbouring representative images.} 
In this way we removed sources that added the least extra value to our exploration of the morphologies in LoTSS. 
Future versions of \textsc{PINK} might make this step obsolete by introducing a learning rate that is adaptable per representative image. This will enable us to lower the learning rate specifically for neurons that are often selected as best matching neuron. As a result, often occurring shapes (of which unresolved or barely resolved sources are the most frequent) will not be so dominant in the trained SOM.

After removal of the unresolved (or marginally resolved) sources, the SOM training procedure was repeated using the same hyper-parameters (Table \ref{Run_parameters}), on the reduced sample of $19,544$ images -- $5,057$ fewer than the training set used in the first run.
Figure \ref{fig:neighbourhood_vs_aqe} shows the SOM training progress using the values of our two performance metrics, we expect both values to gradually decrease. We see that training with a neighbourhood radius that decreases with every epoch results in an ever more accurate SOM (reflected in the lower AQE and the smaller standard deviation on the AQE) but eventually this comes at the cost of the overall coherency (higher TE).
We stopped the training process at the point where the AQE declined less than $1\%$ from the last epoch. 
However, if coherency is deemed more important than training set representation one can decide to stop when TE is at its local minimum.
After SOM training was complete, in the mapping phase (step 4 of the algorithm as described in Section \ref{sec:som}), the $19,544$ sources used for training were compared to the trained SOM to find the best-matching neuron for each image and to calculate the smallest Euclidean norm between each image and each neuron.

\begin{figure*}\begin{center}
\includegraphics[width=0.9\textwidth]{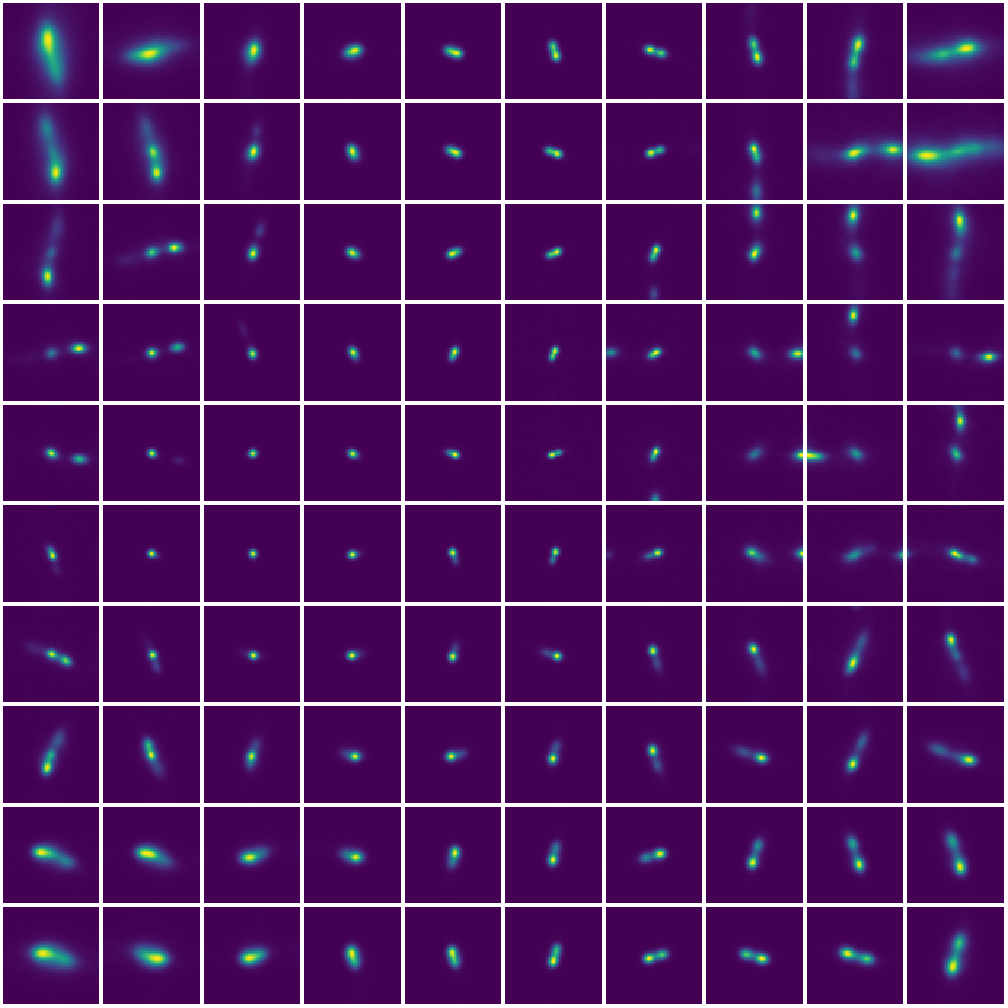}
\caption{Final $10\times10$ cyclic SOM. Each one of the 100 representative images represents a cluster of similar morphologies present in the training dataset. Topology across the representative images is well conserved: Similar shaped representative images are close to each other in the SOM. We note that the SOM is cyclic, the representative images at the right border are connected to those at the left and the representative images at the bottom are connected to the representative images at the top.} \label{fig:final_som}
\end{center}\end{figure*}

\subsection{Final $10\times10$ trained SOM}
\label{sec:presenting_the_som}

Figure \ref{fig:final_som} shows the final $10\times 10$ cyclic SOM, where cyclic indicates that its boundary conditions are periodic. Each one of the 100 representative images represents a set of similarly shaped sources from our training dataset. 
Training took place using \textsc{PINK} version 0.23 on a single Tesla K80 GPU and took 4.2 and 3.5 hours for the initial SOM and the final SOM respectively. Thus, on average, \textsc{PINK} processed roughly 37 radio images per second.

As expected, neurons that represent similarly shaped sources are close to each other in this SOM, illustrating that topology across the SOM is conserved.
The AQE on the training set is $1.65$ with a standard deviation of $1.38$ and the TE is $9.94\%$. 
These values compare to an AQE of $1.4$ with a standard deviation of $1.35$ and $\textup{TE}=5.63\%$ for the initial SOM training run that included the marginally resolved source population.
We expect the AQE to be higher than in the first trained SOM as a result of ejecting $5,057$ well represented (unresolved and barely resolved) sources from our training set.

\begin{figure*}\begin{center}
\includegraphics[width=0.9\textwidth]{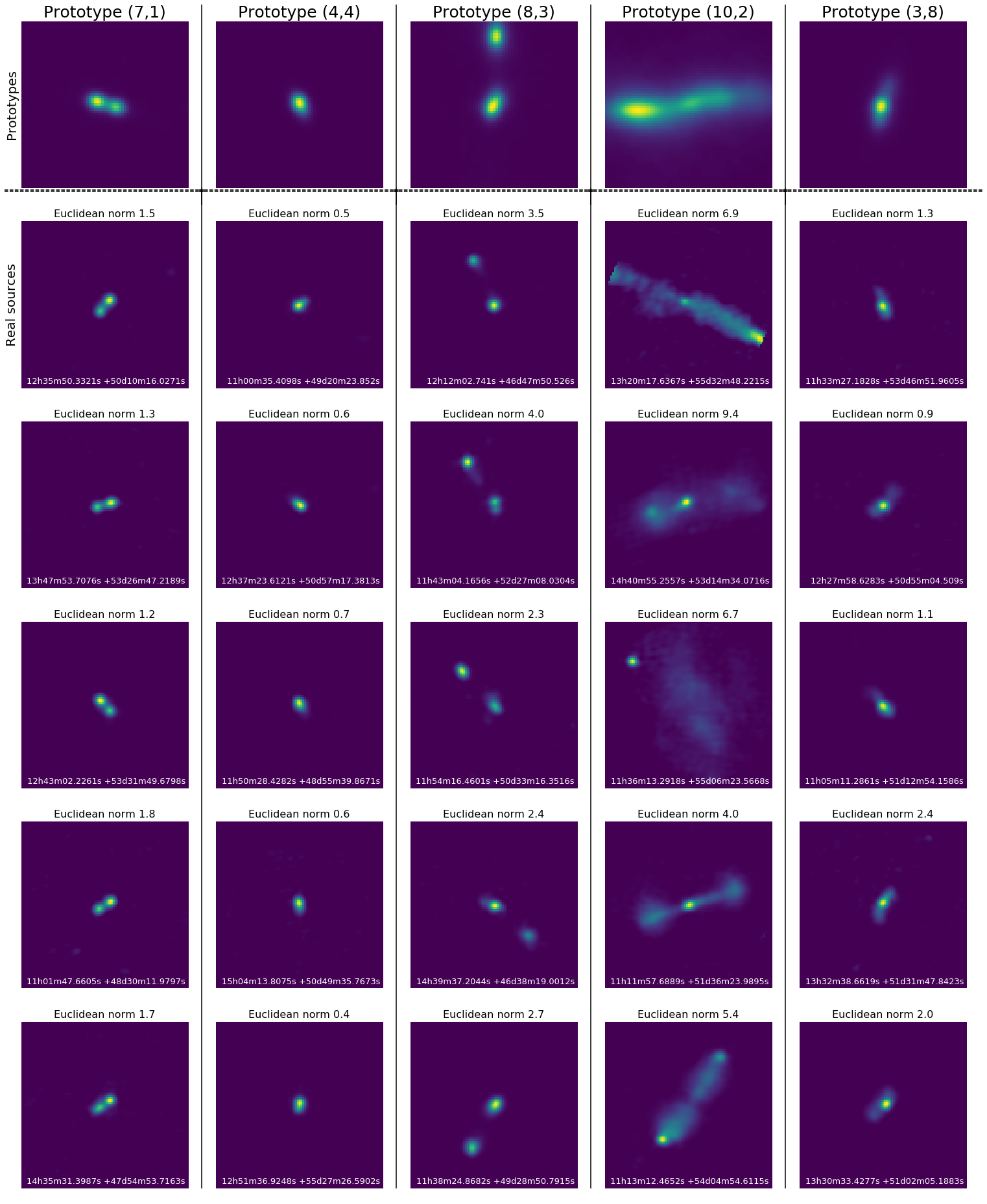}
\caption{Closer look at five representative images with distinctly different shapes. The first row shows five hand-picked representative images from the trained SOM. The location of each representative image on the SOM is indicated by (column, row), thus the first highlighted representative image in this figure is positioned in the seventh column, second row of the SOM in Fig. \ref{fig:final_som}.
In each column, we show five (randomly selected) radio sources that have been mapped to the representative images in the first row.}
\label{fig:examples_ID51}
\end{center}\end{figure*}

In Fig. \ref{fig:examples_ID51}, we see that most sources get assigned to a representative image that more or less matches their shape or contours. However, through degeneracies in the Euclidean norm similarity measure and the small number of representative clusters that we use to represent all shapes and sizes in our data, it is still possible to have a range of different shapes assigned to some representative images. The potential number of degenerate morphologies assigned to a representative image increases with the size and brightness of the representative image, as can be seen by comparing the radio sources associated with representative image (4,4) to those associated with representative image (10,2): (10,2) still contains a variety of different morphologies whereas the sources belonging to (4,4) are all very similar.

\begin{figure*}\begin{center}
\includegraphics[width=0.9\textwidth]{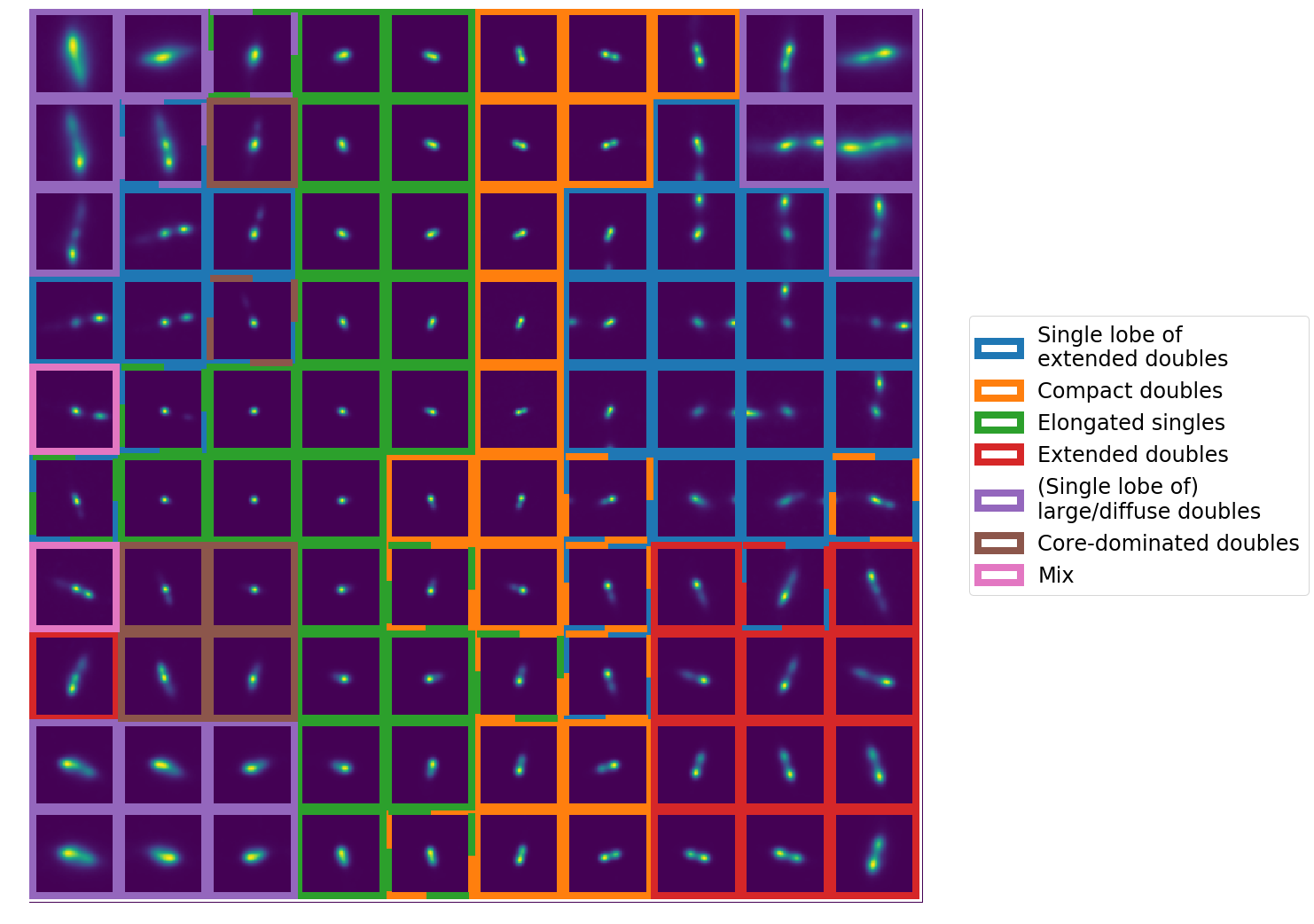}
\caption{Final $10\times10$ cyclic SOM manually labelled into seven categories. These categories describe the type of sources that are dominant or most occurring in the set of sources that best matches each of the 100 representative images. If there are multiple dominant types of sources  best matching a representative image, the representative image is labelled using multiple categories, which is visualised by the dashed multi-coloured edges.
}
\label{fig:final_som_labelled}
\end{center}\end{figure*}

In Fig. \ref{fig:final_som_labelled}, we label the SOM based on the category of known radio morphologies that each representative image most closely resembles.
Representative images that clearly belonged to more than one group were assigned to multiple groups. 
They can be seen to form fully connected groups (the map is cyclic) with the exception of two neurons labelled as `core-dominated doubles' and two neurons labelled as `Mix', indicating that the topology of the dataset is indeed conserved. 

The labelling process reveals numerous distinct populations: with `core-dominated' we indicate double-lobed AGN where the core has a higher peak flux than the lobes;
`compact doubles' indicate compact, double-lobed AGN;
`extended doubles' indicate double-lobed AGN where the hotspots are spatially separated;
`single lobe of extended doubles' indicate a (catalogue entry centred on a) single lobe of a double-lobed AGN.
`elongated singles'  indicate compact emission probably originating from unresolved or barely resolved AGN or from strongly beamed single AGN lobes.
`(single lobe of) diffuse/large doubles' indicate either fluffy, double-lobed AGN or a single lobe of a large double-lobed AGN.
Finally, `mix' is used for the two SOM cells that contain a variety of sources (these often include spurious emission from a neighbouring bright source).

A distinct morphological population that is missing in these labels is that of bent FRI type sources. We expected that the different curvatures in the lobes of these sources results in them ending up in various labelled groups within the SOM, not very well represented by any of them.
By mapping the NAT and WAT collection from \citet{mingo2019revisiting} to our SOM, we confirmed that this is true. 
With a median value of 2.07, the Euclidean norm of the NAT and WAT sources to their representative image is $4\%$ higher than that of the large non-bent FRI sources from \citet{mingo2019revisiting} and $57\%$ larger than the median Euclidean norm of all sources in our dataset, reinforcing our conclusion.

\subsection{Morphology distribution of LoTSS extended radio sources}
\begin{center}\begin{figure}
\includegraphics[width=0.5\textwidth]{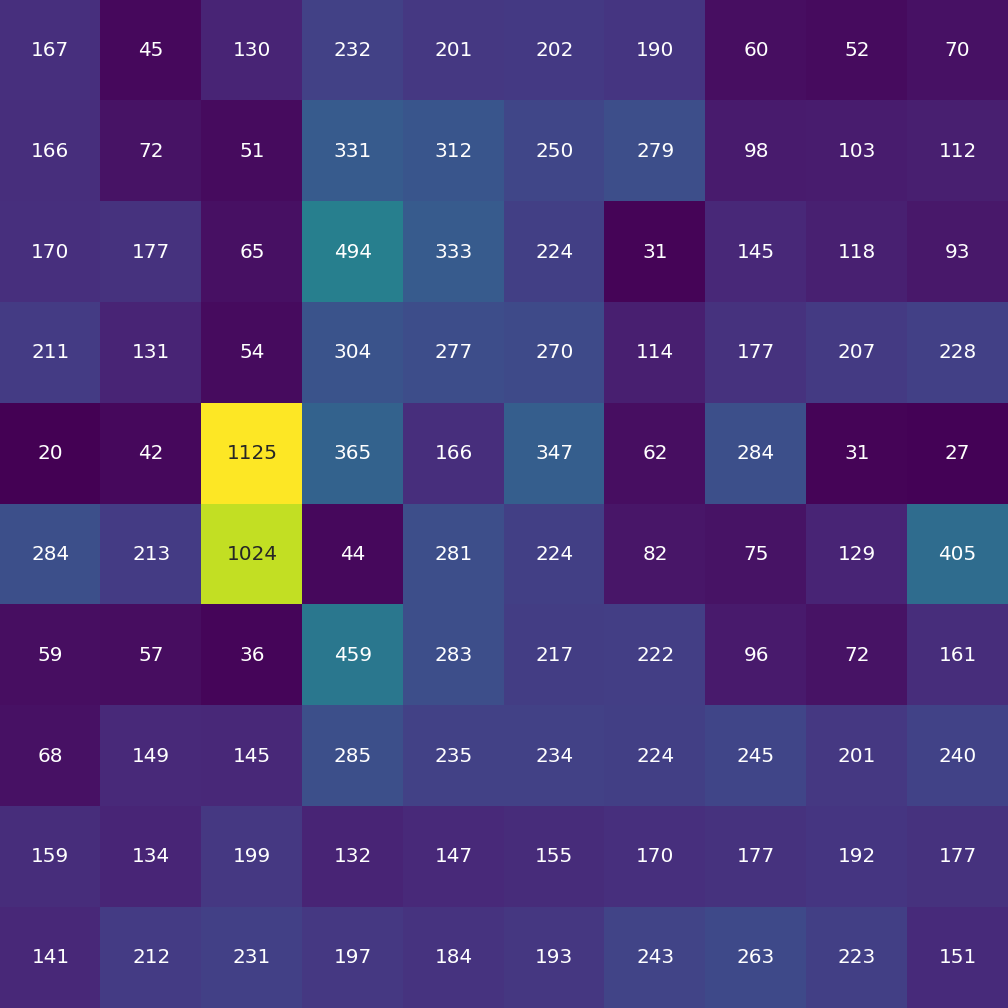}
\caption{Heatmap of the $10\times10$ cyclic SOM, indicating the number of sources from our training dataset mapped to each of the representative images shown in Fig. \ref{fig:final_som}.}
\label{fig:heatmap}
\end{figure}\end{center}

In Fig. \ref{fig:heatmap} we show the distribution of the number of best matching sources for each representative image in the SOM (a `heatmap'). 
We can see that the most common representative images are those that resemble unresolved sources or elongated single sources while the least common represent faint sources with a hard-to-distinguish shape and core-dominated sources.
 
\begin{center}\begin{figure}
\includegraphics[width=0.5\textwidth]{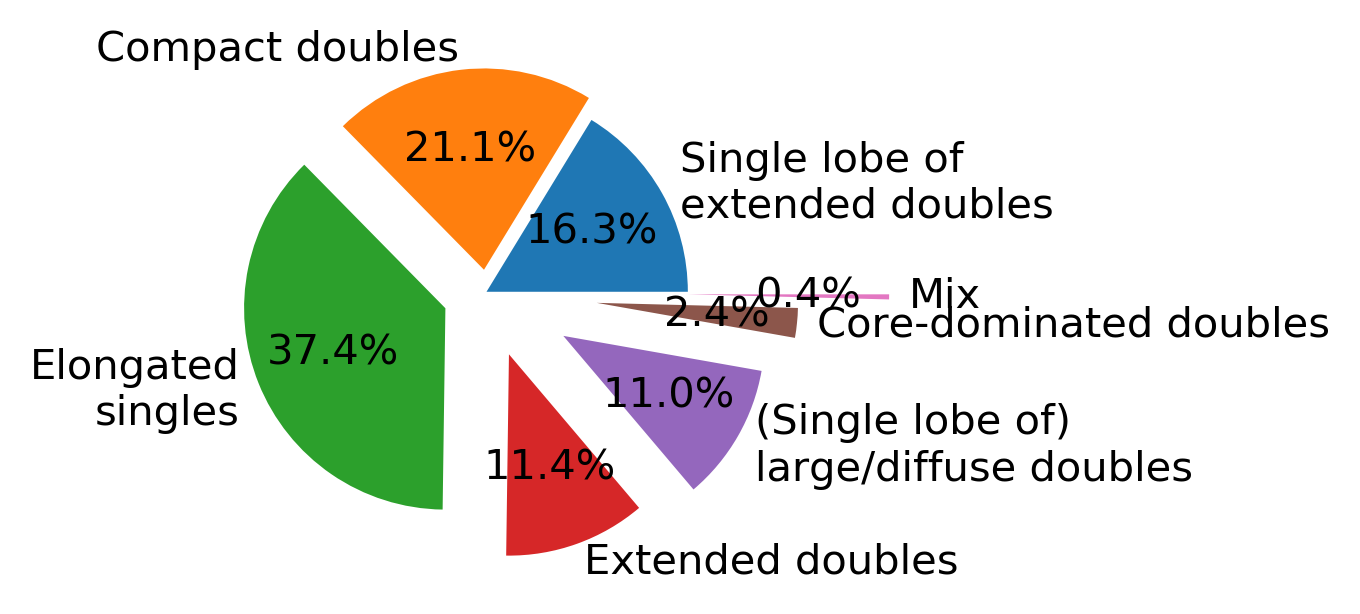}
\caption{Piechart of dominant groups in the $10\times10$ cyclic SOM, providing a rough indication of the percentage of sources best matching a neuron in each group. 
The number of sources best matching a neuron that belonged to two groups was counted half for the tally in each group. These are first-order estimates: each group will contain sources that a human will classify differently after individual inspection.}
\label{fig:sizes}
\end{figure}\end{center}

After combining the heatmap with the representative image labels assigned above, we get a picture of the overall morphology distribution of the extended sources in LoTSS DR1 (see Fig. \ref{fig:sizes}). 
We caution however that these results serve as a first order estimate and there will be individual sources best matching a given particular neuron that could be more accurately labelled by a label not previously included. 

If we count every source associated with the label `single lobe of an extended double' -- disregarding the ones associated to representative images with multiple labels and the ones associated to the label `(Single lobe of) large/diffuse doubles' -- we can get a conservative estimate of the percentage of inadequate source association by the source extraction software. 
As such, we estimate that the percentage of inadequate source association for the $19,544$ extended source entries in the catalogue is about $13\%$. This percentage is higher than the percentage of sources that were associated through visual inspection as detailed by \citet{Williams2019}: $5\%$ of these $19,544$ sources ended up combined in the value added catalogue\footnote{See the LoTSS-DR1 merged component catalogue: \url{https://lofar-surveys.org/public/LOFAR_HBA_T1_DR1_merge_ID_v1.2.comp.fits}}. 
To perform this visual inspection and correct the PyBDSF catalogue \citet{Williams2019}  use crowd-sourced manual component association through the Zooniverse platform\footnote{Web address: \url{zooniverse.org}}. This platform allowed every source (LOFAR radio contours on top of a WISE or Pan-STARRS image) to be viewed and associated by five different astronomers. In the same process, optical and or infrared host-galaxies were assigned to the radio emission if possible. Only associations agreed upon by four or five out of five people were combined in the value added catalogue. The difference between the number of source associations in \citet{Williams2019} and our estimate here shows the difficulty in distinguishing between a pair of lobe-hotspots and two unrelated, unresolved or barely resolved radio sources. 
\clearpage

\subsection{Discovering morphologically rare sources through outlier score}
\label{discover}

\begin{center}\begin{figure}
\includegraphics[width=0.49\textwidth]{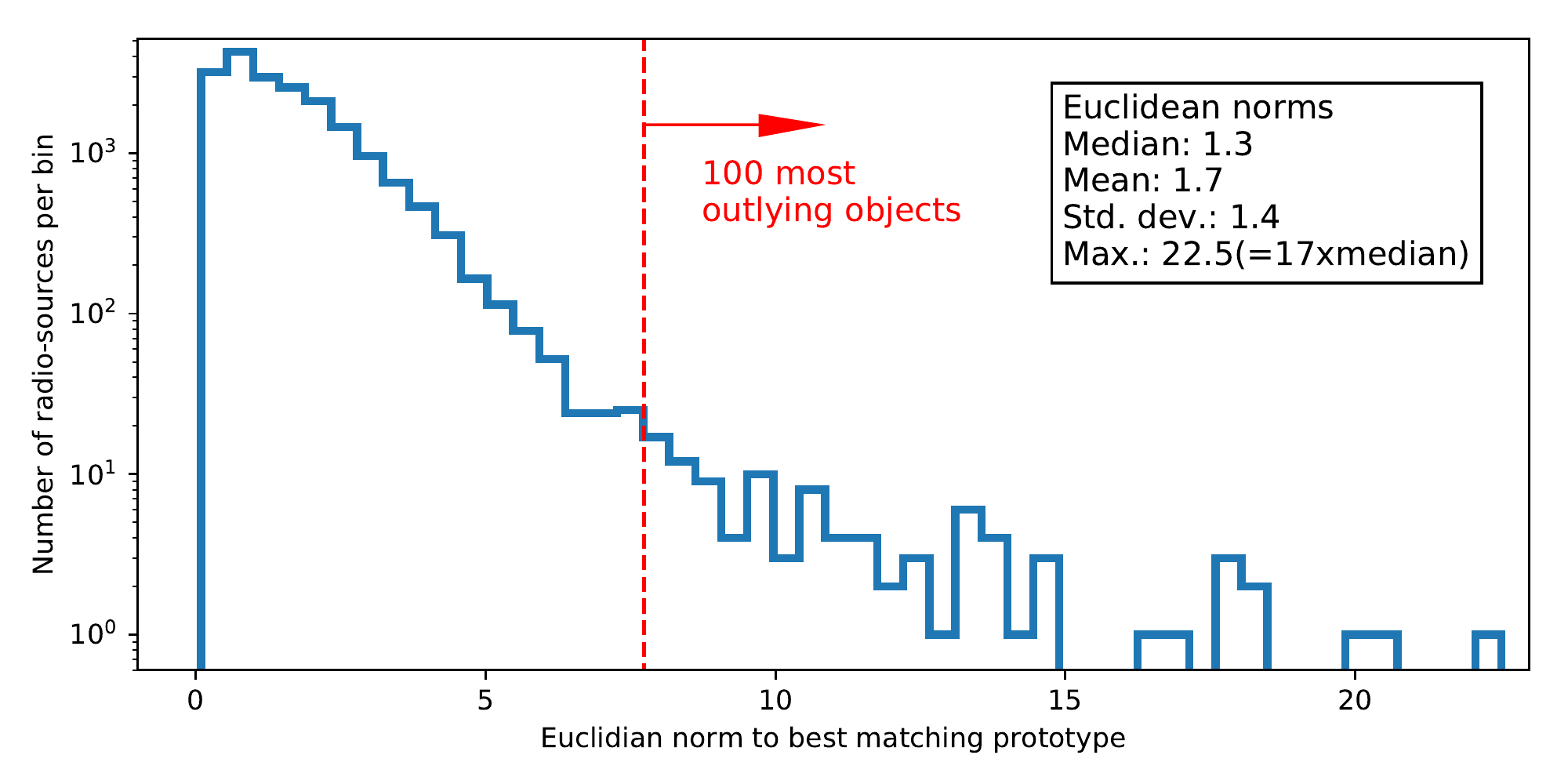}
\caption{Histogram of the Euclidean norm from each radio source to its closest representative image in the final $10\times10$ SOM, for each of the $19,544$ radio sources in our final training set. Most sources have a very small Euclidean norm to their best resembling representative image in the SOM, which means they are morphologically similar to at least one representative image in the SOM. We isolated the 100 most outlying sources and show them in Appendix~\ref{app:outliers}.}
\label{fig:outliers_hist}
\end{figure}\end{center} 

\begin{center}\begin{figure}
\includegraphics[width=0.5\textwidth]{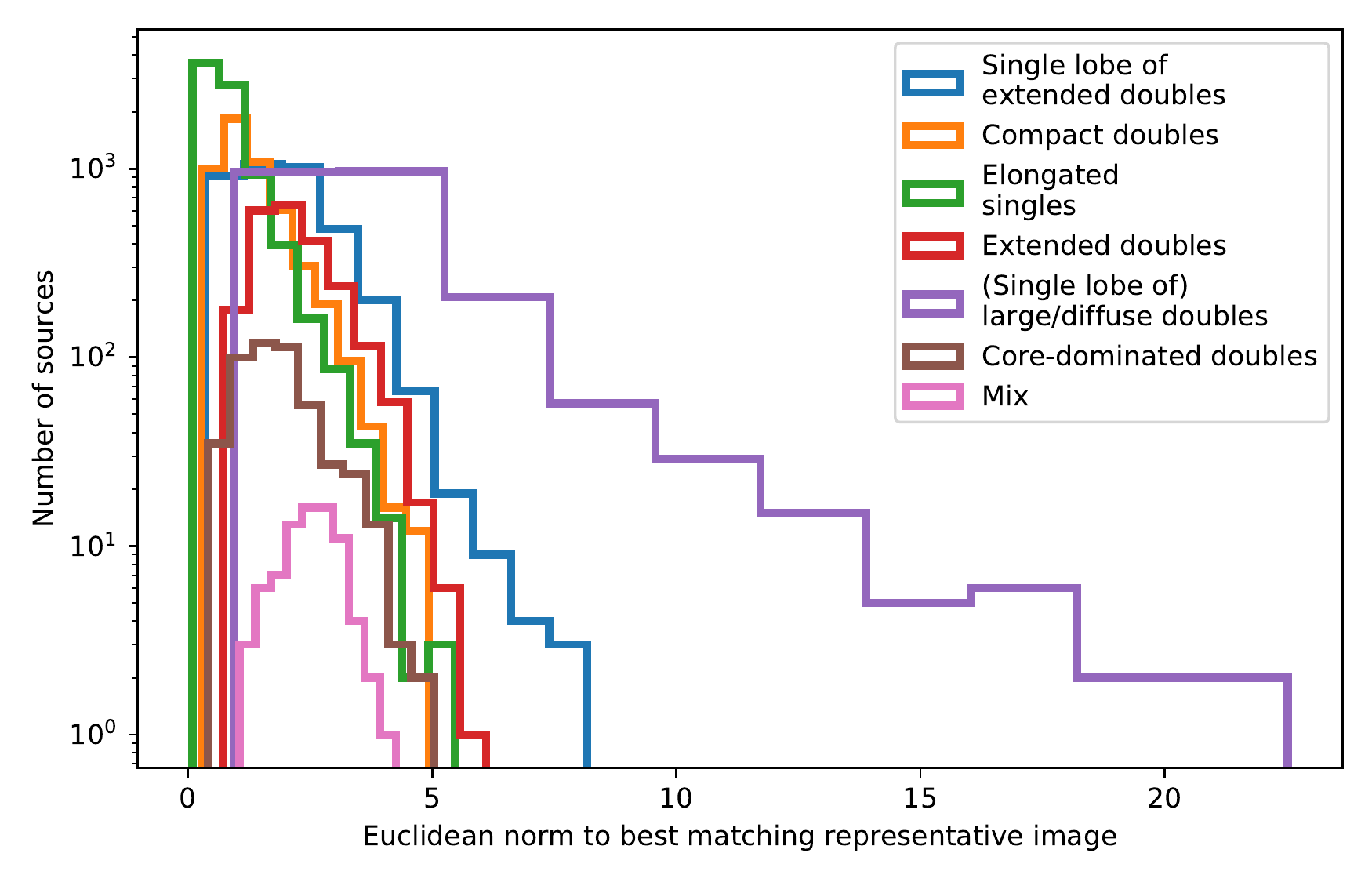}
\caption{Histogram of Euclidean norm from each radio source to its closest representative image in the final $10\times10$ cyclic SOM, aggregated by group. This shows how well each source resembles the best matching representative image in its group. The lower the Euclidean norm, the better the match. In general, a steeper declining trend indicates a better match across sources with the same label.}
\label{fig:group_hist}
\end{figure}\end{center} 

The trained SOM gives us an overview of the archetypes or dominant morphologies present in the dataset.
One can imagine that if the difference (the Euclidean norm) between a cutout and its best matching representative image is much larger than average, see Fig. \ref{fig:outliers_hist}, this means that the cutout does not match well to any representative image in the map. 
This implies that radio-objects with this angular size and morphology are not abundant in the training set: they are morphological outliers. The Euclidean norm between a cutout and its best matching representative image can therefore be seen as an `outlier score'.

Pre-processing the sources, specifically sigma-clipping or rescaling our cutouts, affects the magnitude of the Euclidean norm values.
As we scale all our pixel values to range between 0 and 1, we are biased towards having large apparent objects as outliers. 
For sources that are slightly different than their representative image, objects with a large angular size will have a larger Euclidean norm than smaller sources.
Indeed Fig. \ref{fig:group_hist} shows that the group `(single lobe of) large/diffuse sources' shows higher Euclidean norms than all other groups. 
Before we inspect the objects with high outlier score in our dataset, we discuss three options to reduce the bias in finding objects with large angular size and explain why we do not use them.

The first option is scaling the pixel values such that the sum of the pixel values add up to one in each cutout. 
This will steer the bias towards sources that have bright emission concentrated in a small area. In practice, cutouts with the highest outlier scores will then be those compact bright doubles that enter the catalogue slightly off-centre and compact objects with unrelated neighbouring compact objects within our fixed-sized window. 

A second option to reduce the bias is to look for the objects with the highest Euclidean norm per labelled group of neurons. 
As the groups represent sources of varying angular size, we can expect to find outlying morphologies for sources with smaller angular size. 
In practice, the objects with the highest Euclidean norm in the groups -- except in the `(single  lobe of) large diffuse doubles' group -- are mostly artefacts and regular FRI or FRII sources close to the noise-level. Visual inspection shows that adopting a peak flux threshold of 0.5 mJy/beam and a total flux threshold of 5 mJy, excludes most of the images containing these artefacts. We then inspect the resulting 14 objects with the highest Euclidean norm per group in each of the groups except the `(single  lobe of) large diffuse doubles' group; these 84 sources comprise 33 NAT or WAT objects, 42 non-bent FRI or FRII, and 9 sources too small or faint to clearly categorise. Many of the non-bent FRI or FRII and a number of the other objects (31 out of the 84 in total) received a high outlier score, not because of their rare morphology, but because of unrelated close-by neighbouring radio objects. %So we proceed by inspecting the objects with the overall highest Euclidean norm.

A third option to alleviate our bias towards finding outlier objects with large angular size, is by including another measure in the outlier score. We experimented with an outlier score that is the summation of the Euclidean norm and the Euclidean norm times the fraction $G$ times a tunable parameter $\gamma$. Here, $G$ is the maximum sum of pixel values attained by any neuron in the SOM divided by the sum of the pixel values of the considered neuron. Therefore, $G$ will be one for the neuron with the highest sum and proportionally bigger for neurons with smaller summed values -- thus proportionally bigger than one for neurons that represent sources with smaller angular sizes. By increasing $\gamma$, we increase the outlier score for sources with smaller angular sizes with respect to sources with larger angular sizes.
This approach yields similar results to the previous experiment: we do indeed find more small angular sized objects, but they are often positioned close to an unrelated neighbouring radio source.
As unrelated close neighbour sources do not represent physically rare objects, we do not consider this to be an optimal approach -- although ultimately the best definition to define outliers should depend on the user's specific science goals.

\begin{center}\begin{figure}%\includegraphics[width=0.49\textwidth]{"figures/outliers_ID51/FR class distribution in outliers - 10x10 cyclic SOM".png}
\includegraphics[width=0.49\textwidth]{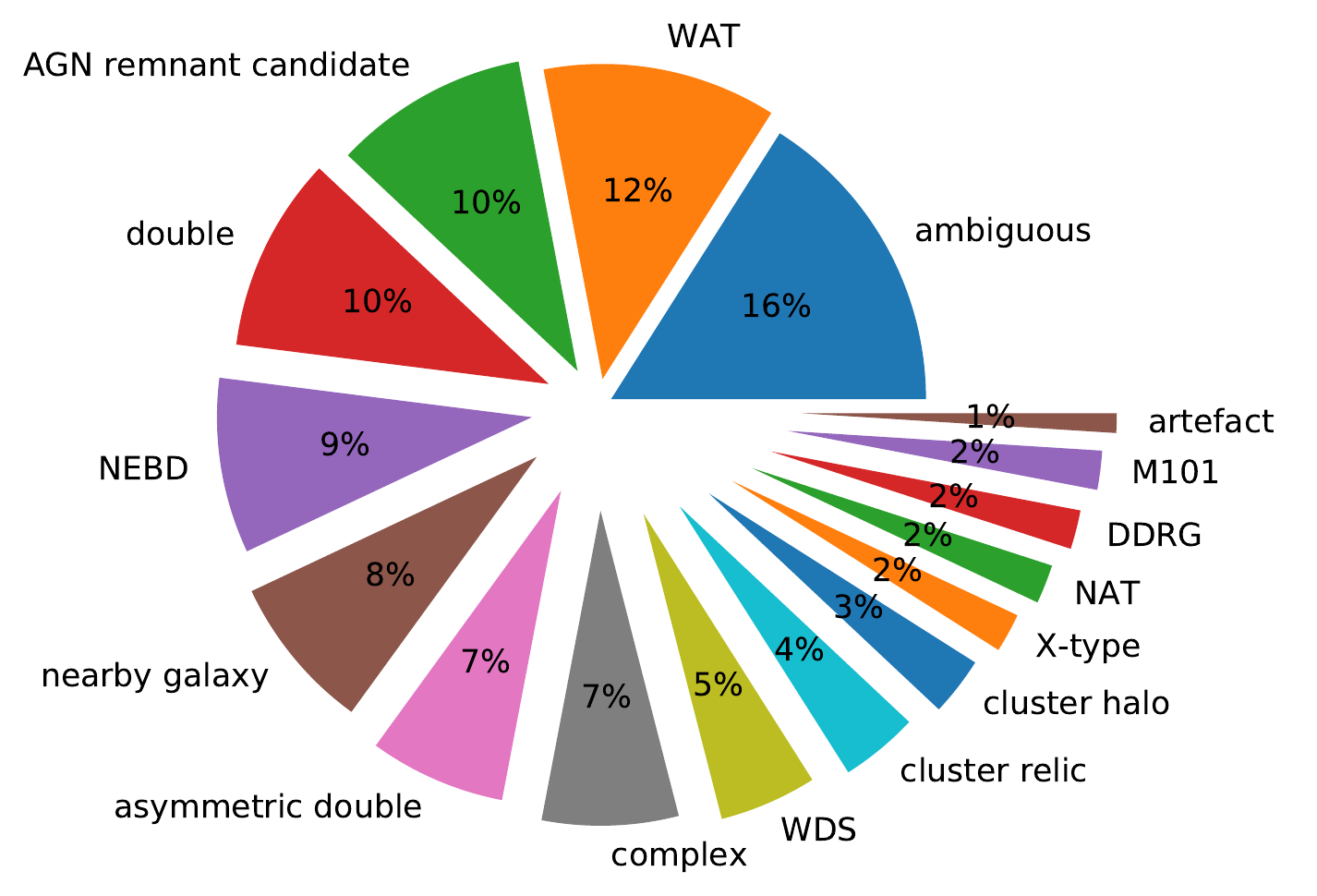}
\includegraphics[width=0.49\textwidth]{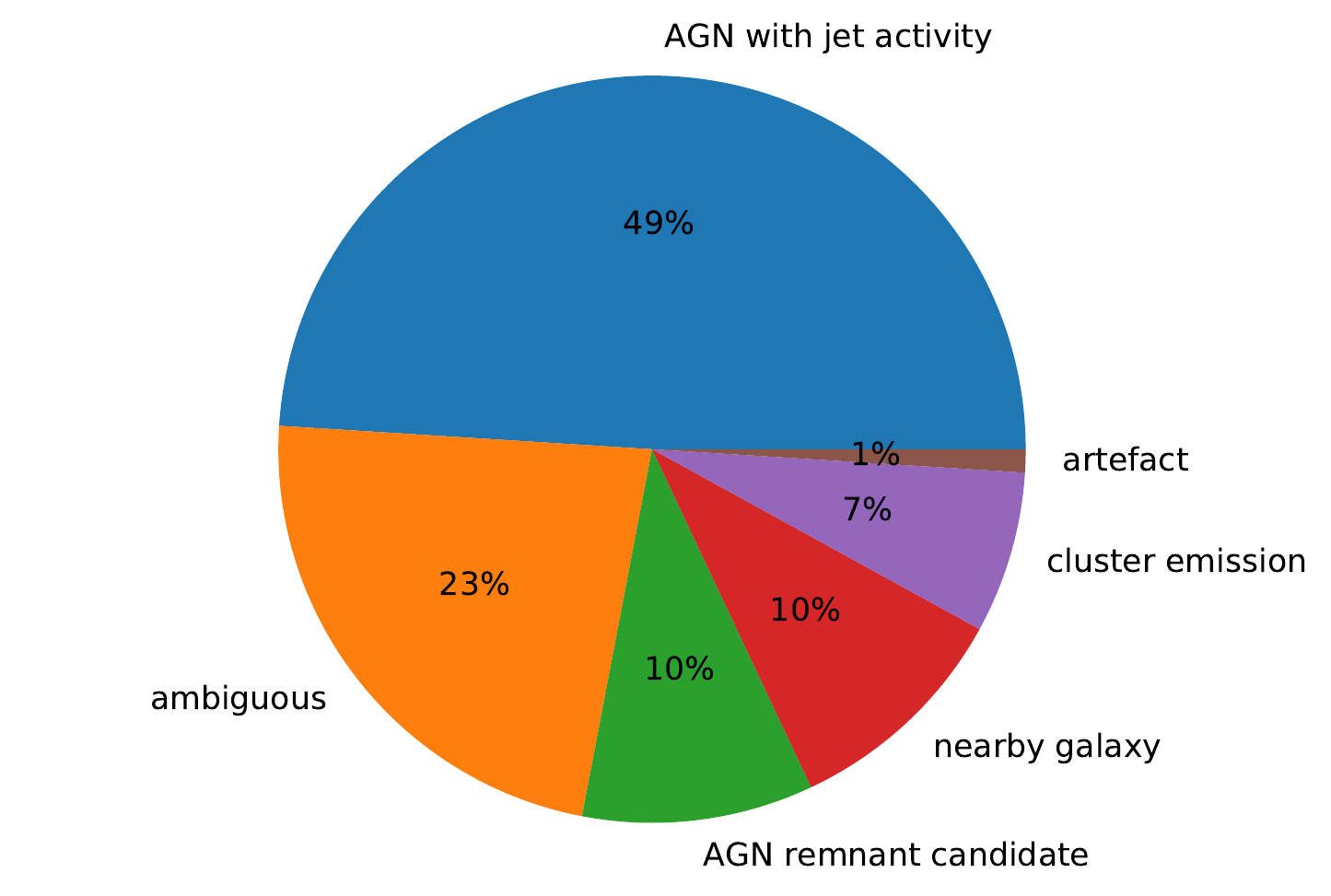}
\caption{Piechart showing the wide variety of morphologies and radio source types present in the
100 most outlying sources in the final $10\times10$ cyclic SOM. \textit{Top panel:} Descriptive labels have been established by a manual visual inspection of each source in radio (LoTSS), mid-infrared (WISE) and optical (SDSS9 colour). We labelled sources that fitted in more than one category as `ambiguous'. `X-type' sources are `X-shaped' extended doubles \citep{Rees1978}, `WATs' and `NATs' are Wide Angle Tailed and Narrow Angle Tailed sources, `WDS' are Wide Double Sources, `NEBD' are Narrow Edge Brightened Doubles \citep{Miley1980}, `DDRG' are Double-Double Radio Galaxies \citep{ddrg}. An `AGN remnant candidate' is an extended double with such relaxed morphology that the AGN jets might not be active any more. As M101 is an exceptionally large apparent object, the source detection software did not group the separate detections of this nearby spiral, as a result it features multiple times in the catalogue. \textit{Bottom panel:} We grouped subtypes together to get an overview of the dominant types of sources in the outliers: `M101' was grouped with `nearby galaxy'; `complex' with `ambiguous'; `cluster relic' and `cluster halo' were grouped into `cluster emission' and `DDRG', `NEBD',  `NAT', `WAT', `WDS', `double', `asymmetric double' and `X-type' were grouped into `AGN with jet activity'.
A figure containing each of these one hundred most outlying sources can be found in Appendix~\ref{app:outliers}.}
\label{fig:outliers_types}
\end{figure}\end{center}

In this paper, we use only the Euclidean norm as our outlier score and inspect the objects with the overall highest outlier score. 
To illustrate the potential of this method, we select the 100 sources with the greatest outlier score (all sources to the right of the red vertical line in Fig. \ref{fig:outliers_hist}) and present all of them in Appendix~\ref{app:outliers}.  
To suppress the number of duplicate objects (broken up during the automatic source-extraction) we require each table-entry to be at a distance greater than 400 arcsec from all previous table-entries.
We visually inspected each of these 100 sources to investigate the nature of these outliers and provide a physical classification in Table \ref{tab:out100}.
In Fig. \ref{fig:outliers_types} we highlight the wide range of different source-types and morphologies present.

We observe that $49\%$ of the outliers are AGN with jet activity, the lobes thereof show a range of curvature -- from wide angle tailed ($12\%$) to narrow angle tailed ($2\%$) -- indicating motion through the circum galactic medium or rotation in the case of X-type sources ($2\%$). Examples of outliers of each type can be found in Figure \ref{fig:100outliers1} by looking at the Radio description in Table \ref{tab:out100}.
The sample is also diverse with respect to edge-brightening and edge-darkening, we can see that the sample contains 29 FRI and 25 FRII type sources. For three sources the FR classification is unclear and for the remaining sources it is not applicable.
The outliers cover multiple stages of jet-activity: from active doubles, to `dying' AGN remnants, to restarted Double-double radio galaxies.

Although diverse, not all AGN that we present here can be regarded as radio objects with outlying morphologies. Specifically, the Narrow edge brigthened sources, Wide double sources and regular doubles account for $24\%$ of the sources. These objects occur less frequently in the survey compared to similar objects of smaller angular size, but their morphologies can hardly be regarded as outlying.

Ten percent of the outliers are nearby starforming galaxies, where the radio-emission largely overlaps with the optical emission in Pan-STARRS $grizy$ bands. The synchrotron emission originating from supernovas is a star-formation tracer that is unobstructed by dust. As expected, both sharp cluster relics like `bow shocks' and the more diffuse and amorphous cluster halos can also be found among the 100 objects with highest outlier scores. 

Twenty-three percent of the outliers are ambiguous or complex in nature. Most ambiguous outliers ($16\%$) are diffuse and somewhat amorphous, raising the question whether they are active AGN, AGN remnants or cluster related emission. For the complex objects ($7\%$), there seems to be additional interaction with neighbouring objects. 

In Fig. \ref{fig:outliers_highlights1}, we highlight a number of individual outliers. The red boxes within these images show the fixed size of the cutouts as they enter the SOM for training: the solid red boxes show the initial  $142.5\times 142.5$ arcsec\textsuperscript{2} cutout and the inner (dashed) boxes show the $100\times 100$ arcsec\textsuperscript{2} cutout after cropping as described in Section \ref{prepro}. The Figure shows that objects that fall entirely within the fixed-size box, and those which extend beyond it, can both be given high outlier scores, demonstrating that the fixed cutout size does not prevent us from finding morphological outliers of different apparent sizes, even when relying on imperfect source extraction.

\newcommand{\halfpage}{0.3}
\begin{figure*}\begin{center}
\includegraphics[width=\halfpage\textwidth]{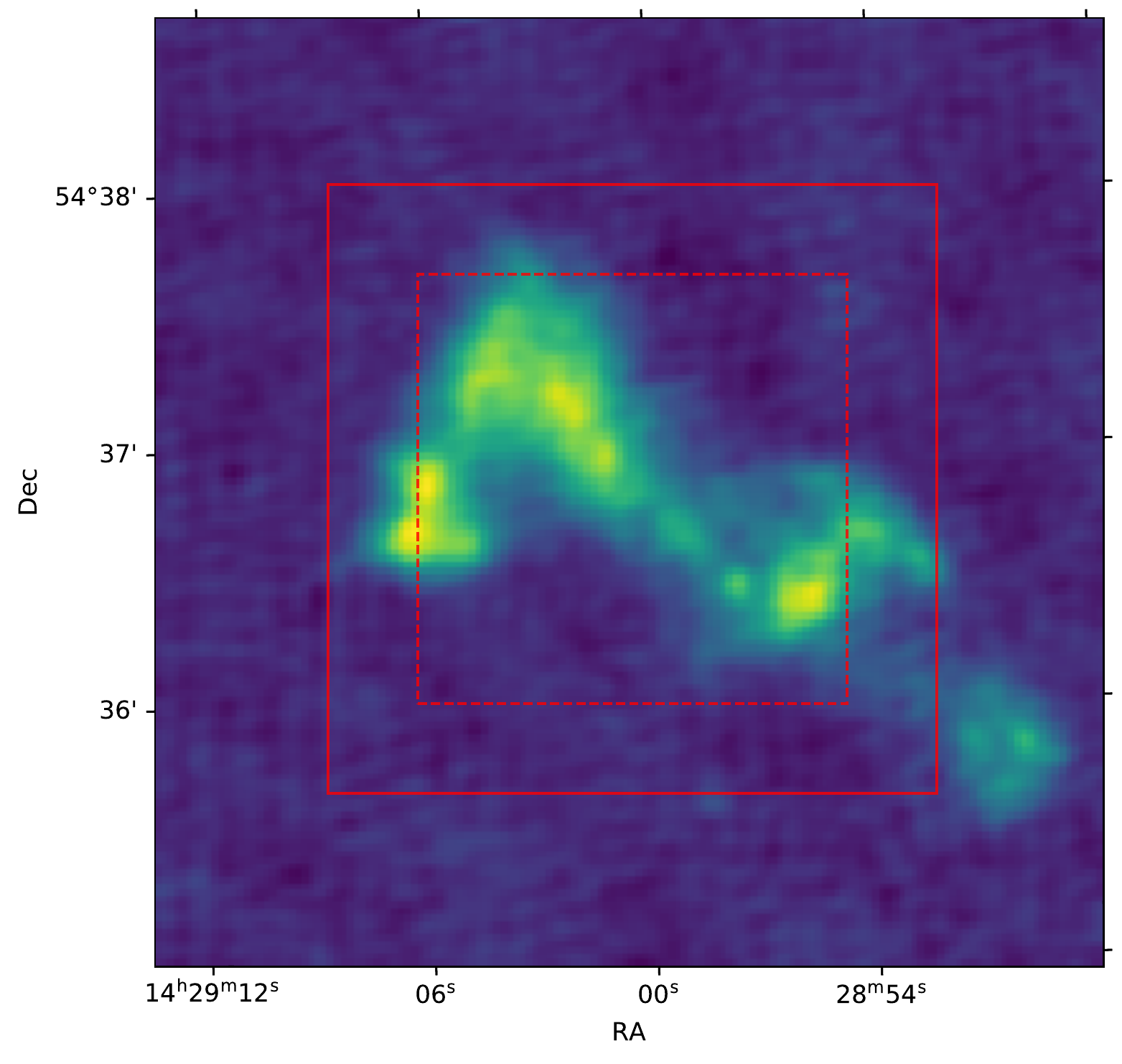}
\includegraphics[width=\halfpage\textwidth]{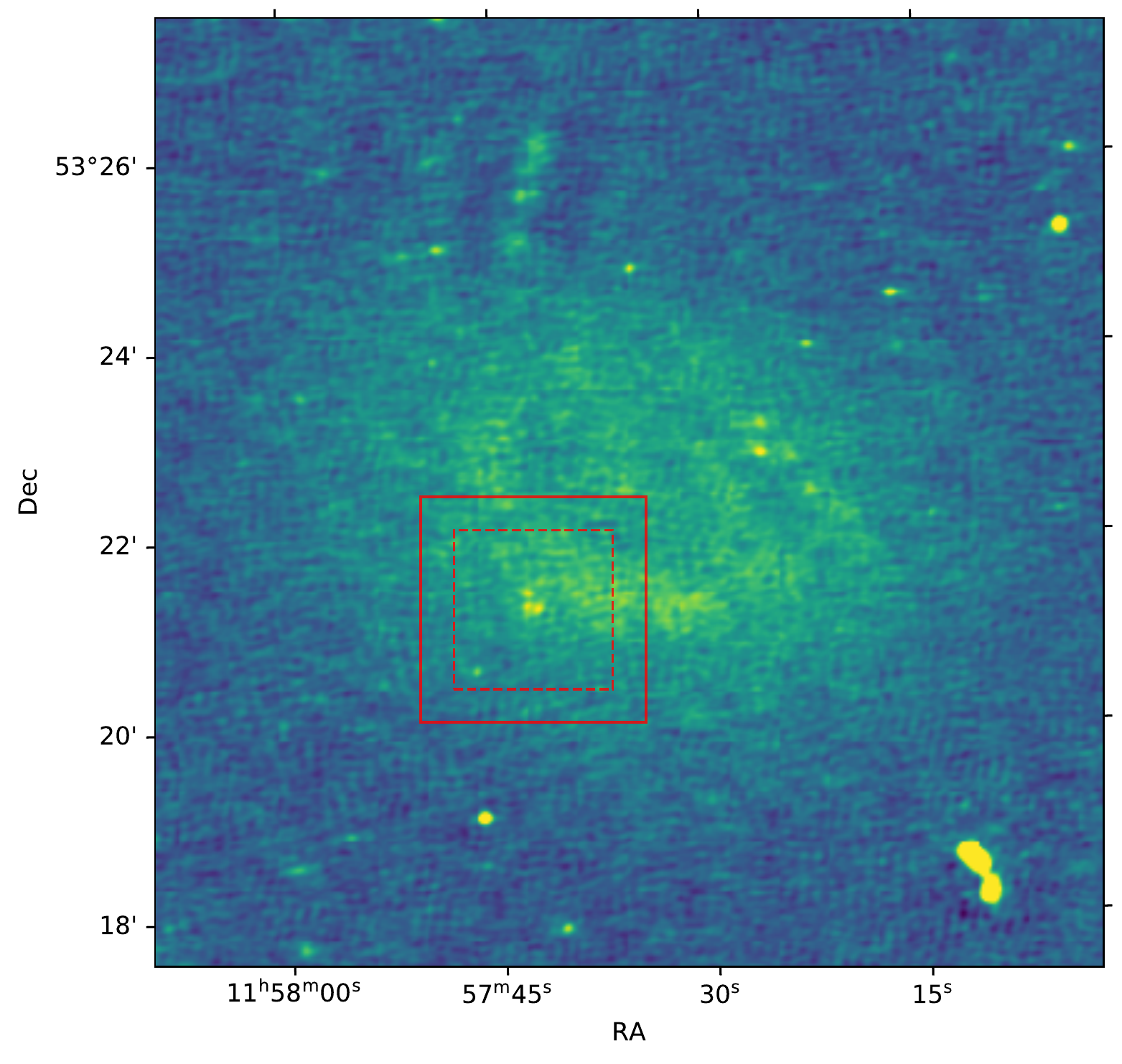}
\includegraphics[width=\halfpage\textwidth]{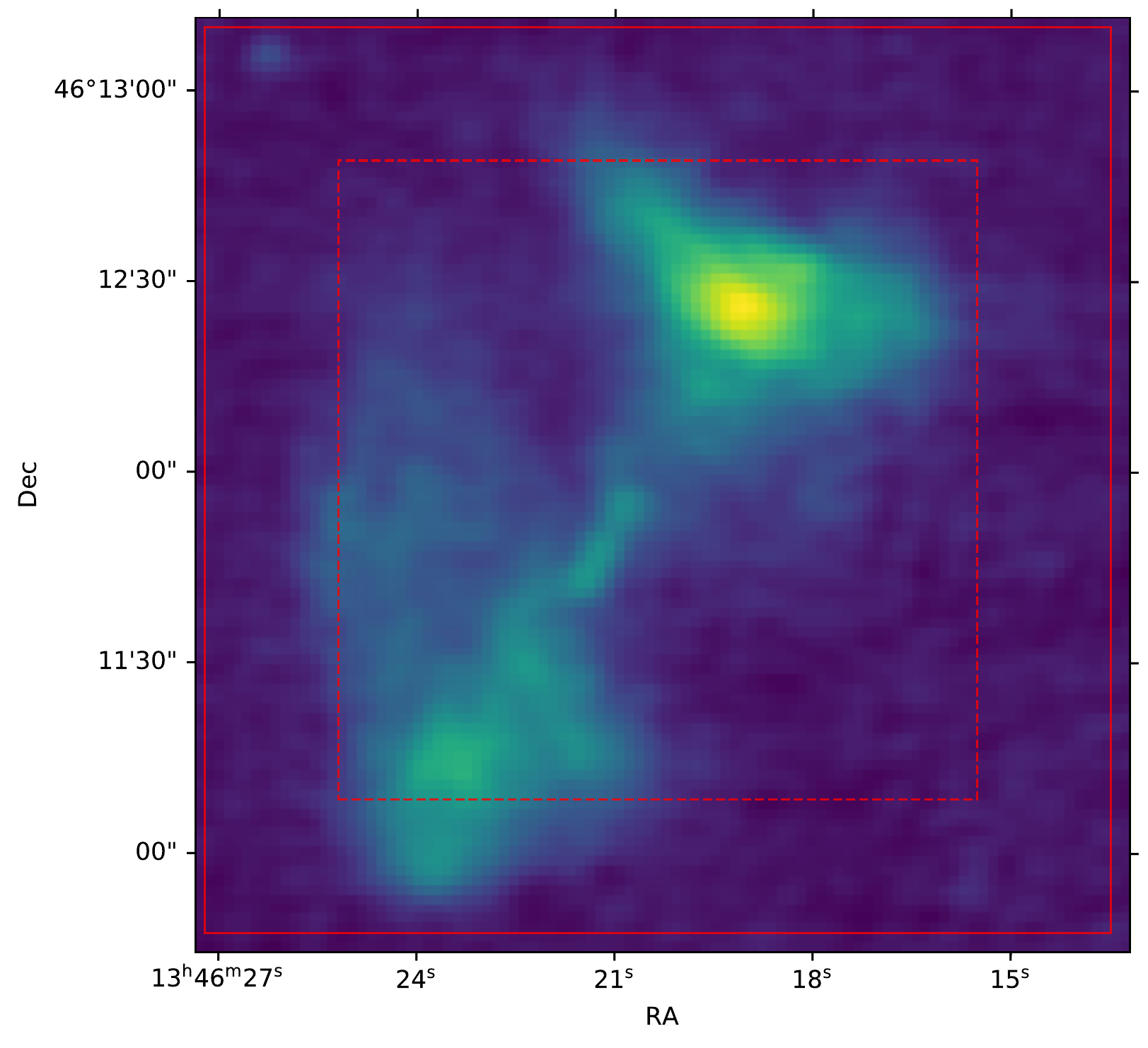}
\includegraphics[width=\halfpage\textwidth]{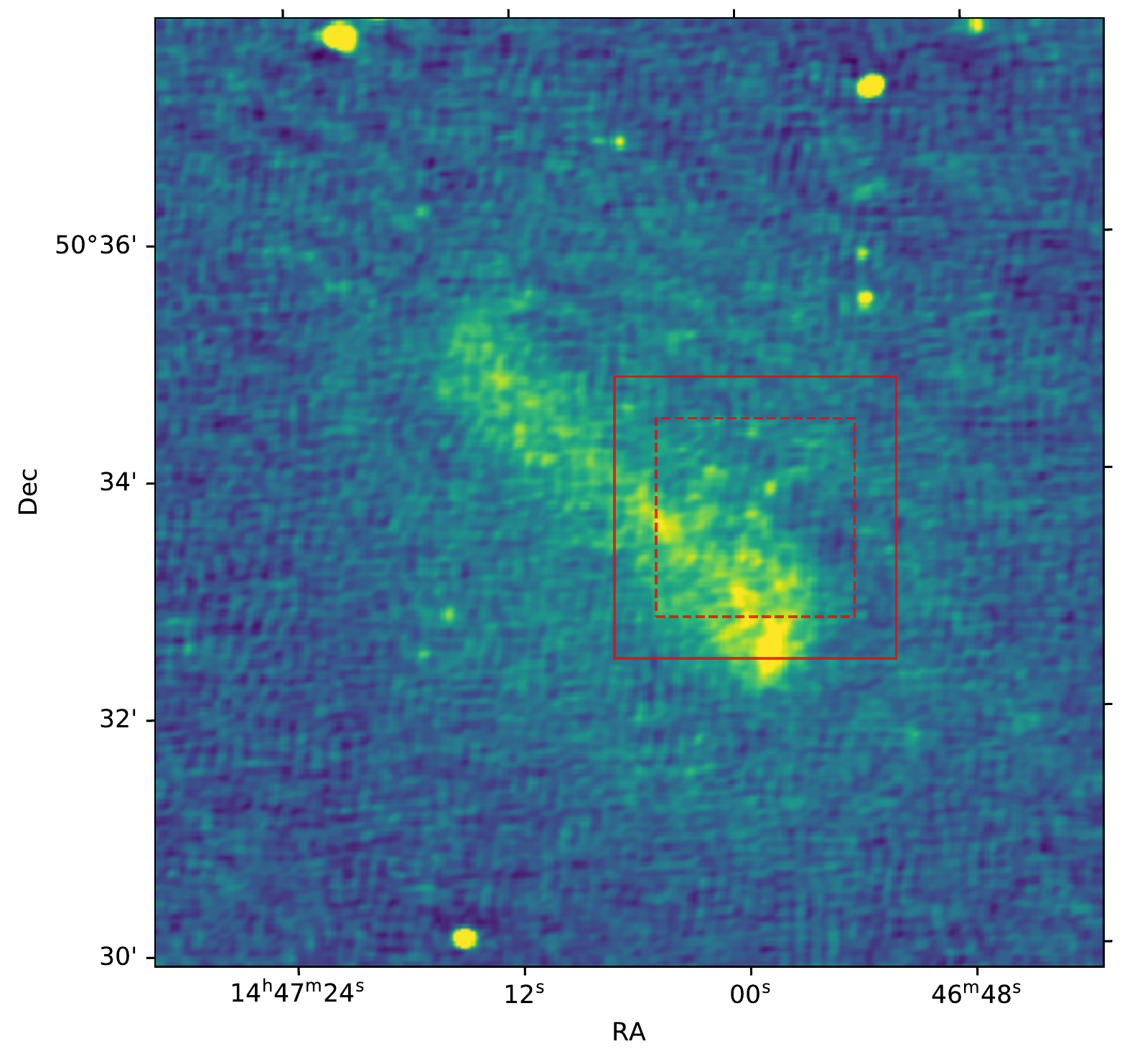}
\includegraphics[width=\halfpage\textwidth]{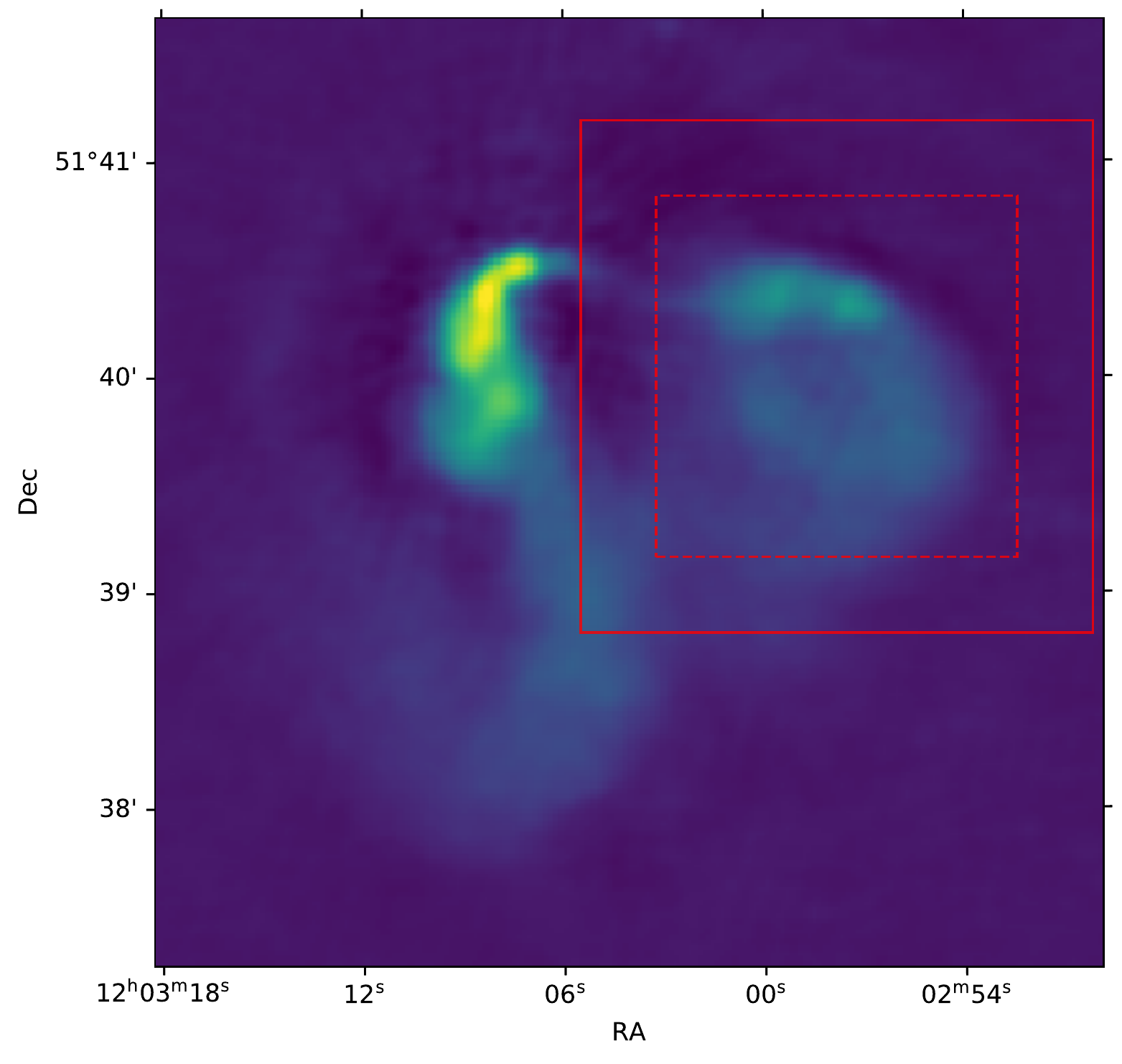}
\includegraphics[width=\halfpage\textwidth]{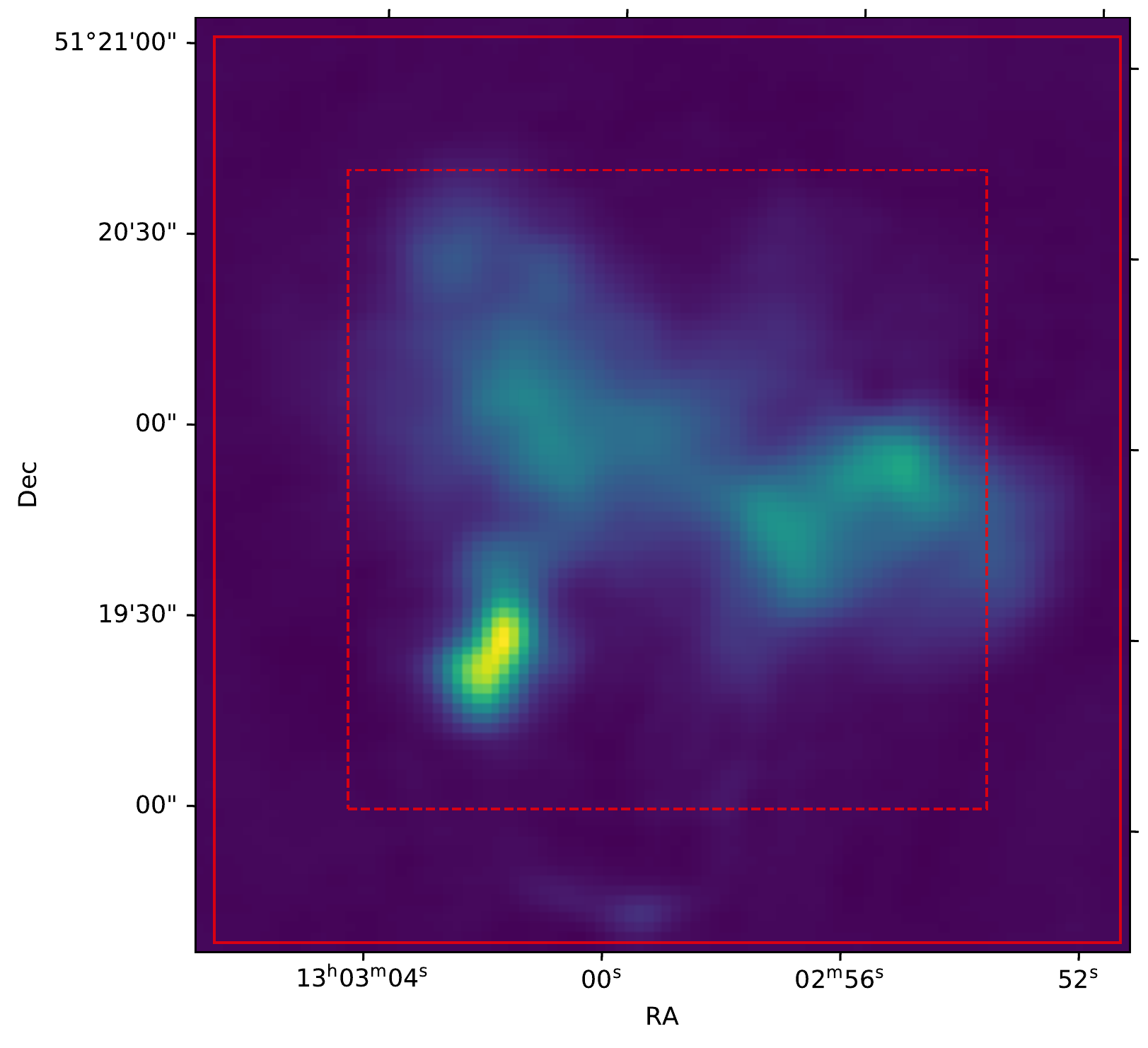}
\includegraphics[width=\halfpage\textwidth]{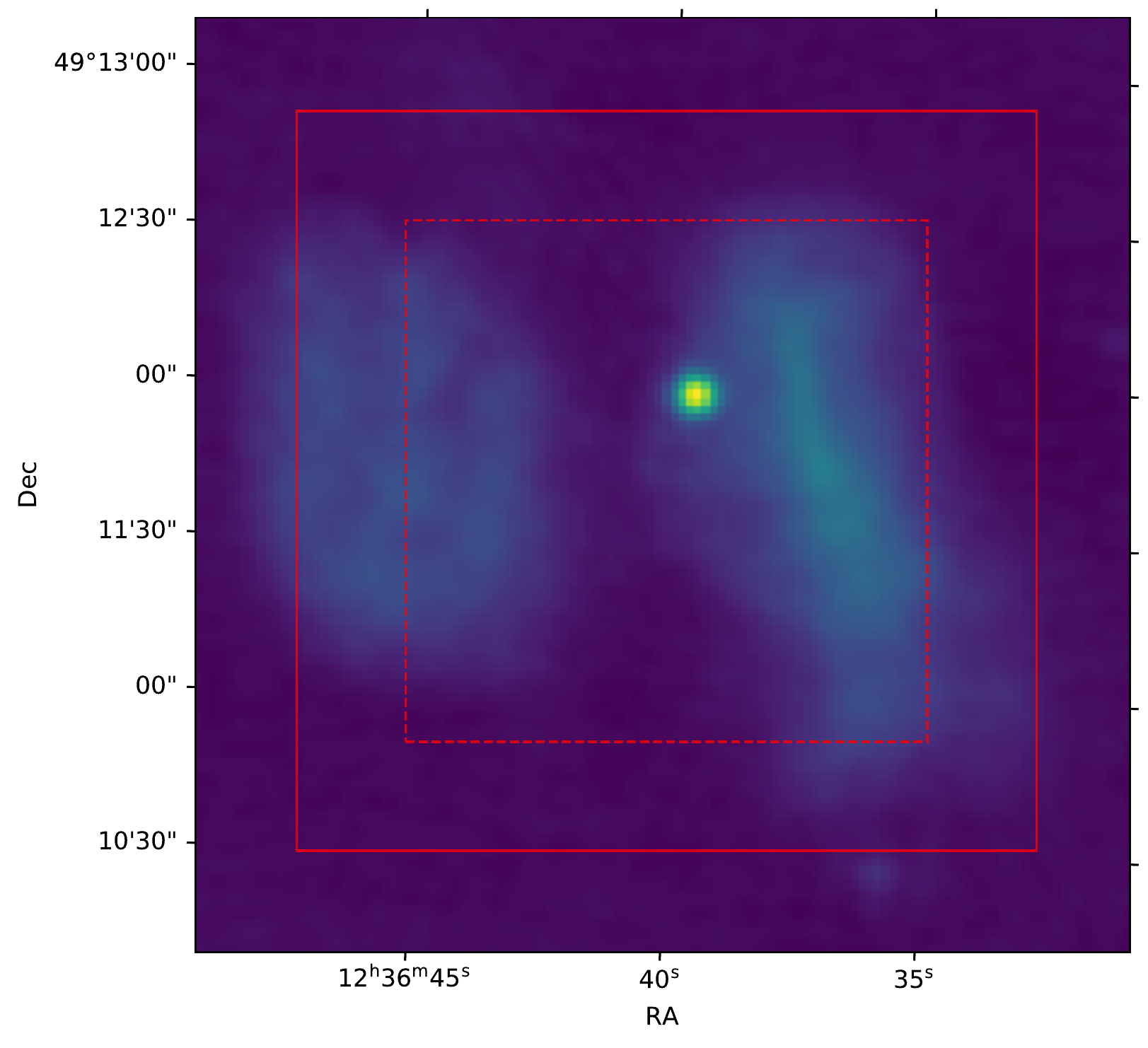}
\includegraphics[width=\halfpage\textwidth]{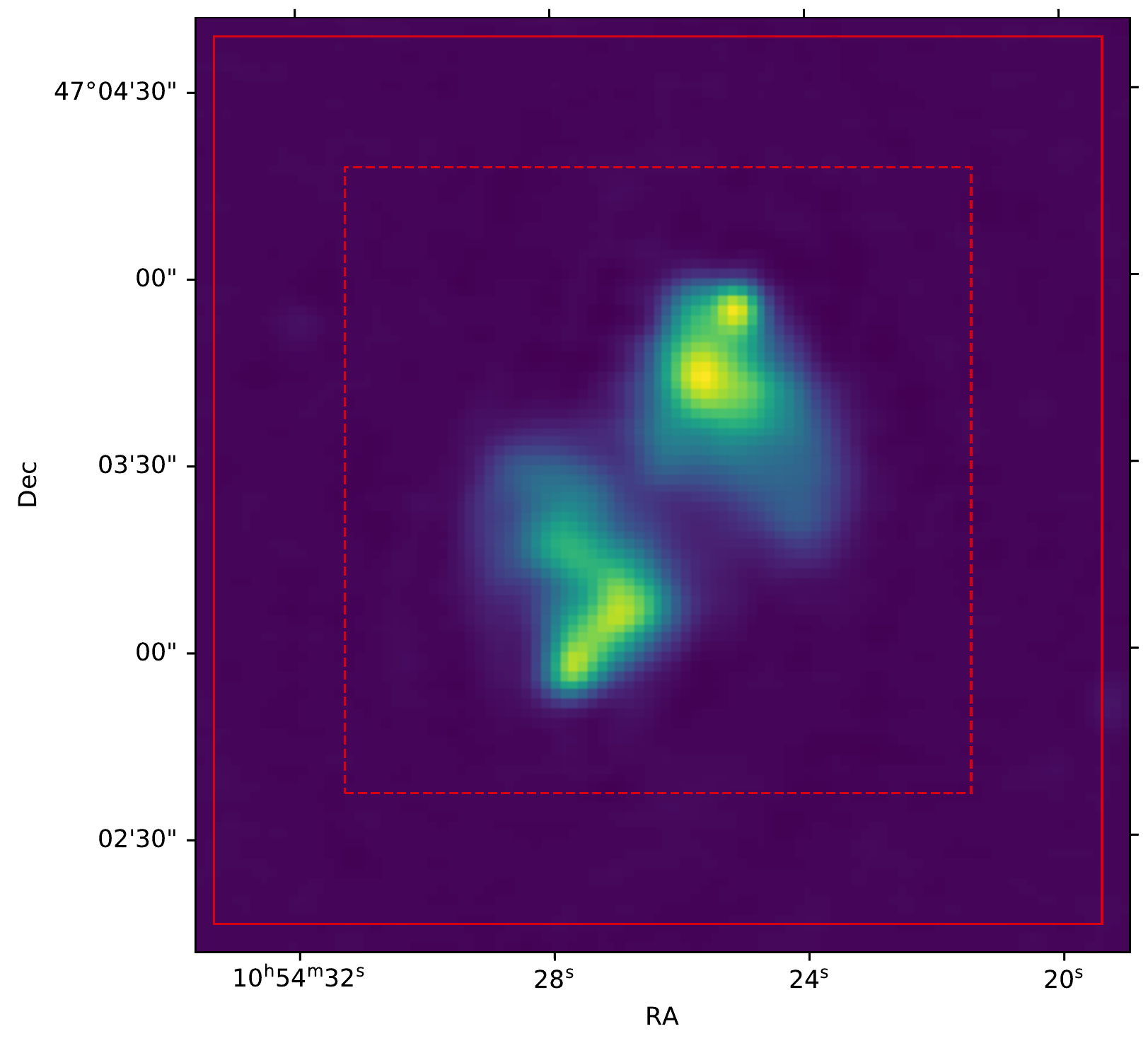}

\caption{Handpicked diverse examples taken from the list of 100 morphological outliers created by using the SOM. The red boxes indicate the size of the cutouts fed to the SOM, the inner (dashed) red boxes indicate the size of the cutouts after rotation and cropping. In this figure, we manually zoomed out to show the full extent of the objects. The outliers list contains a wide range of objects. Going row by row from left to right: a rotating extended double, a nearby spiral (M109), an asymmetric double, an AGN remnant candidate, a WAT,  a head-tail radio source, an AGN remnant candidate with spatially separated lobes and an x-shaped radio source. The complexity of some of the objects means that these descriptions are preliminary. Follow-up observations are necessary to understand the nature of some of these objects.}
\label{fig:outliers_highlights1}
\end{center}\end{figure*}

\begin{figure*}
\ContinuedFloat
\begin{center}
\includegraphics[width=\halfpage\textwidth]{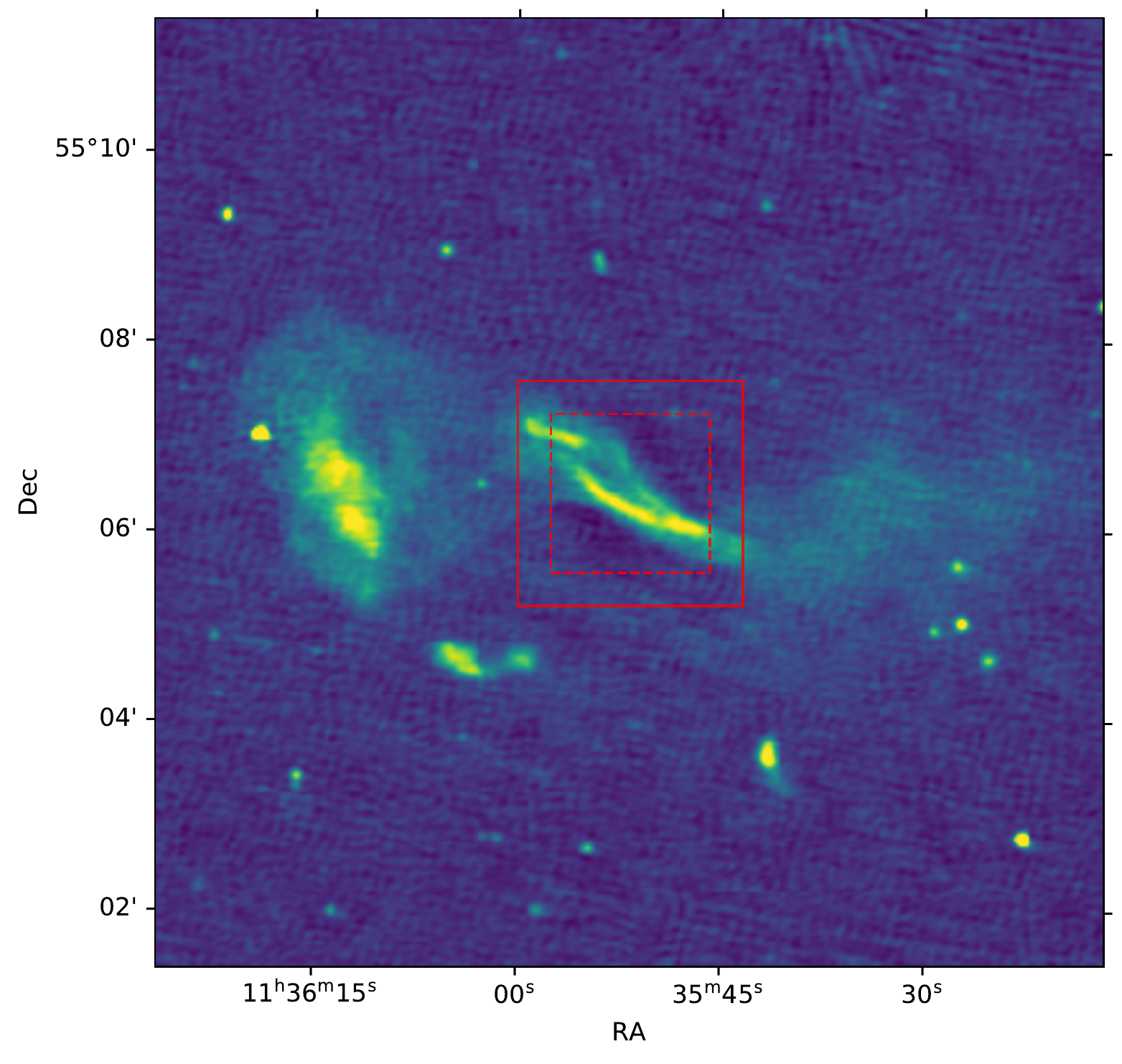}
\includegraphics[width=\halfpage\textwidth]{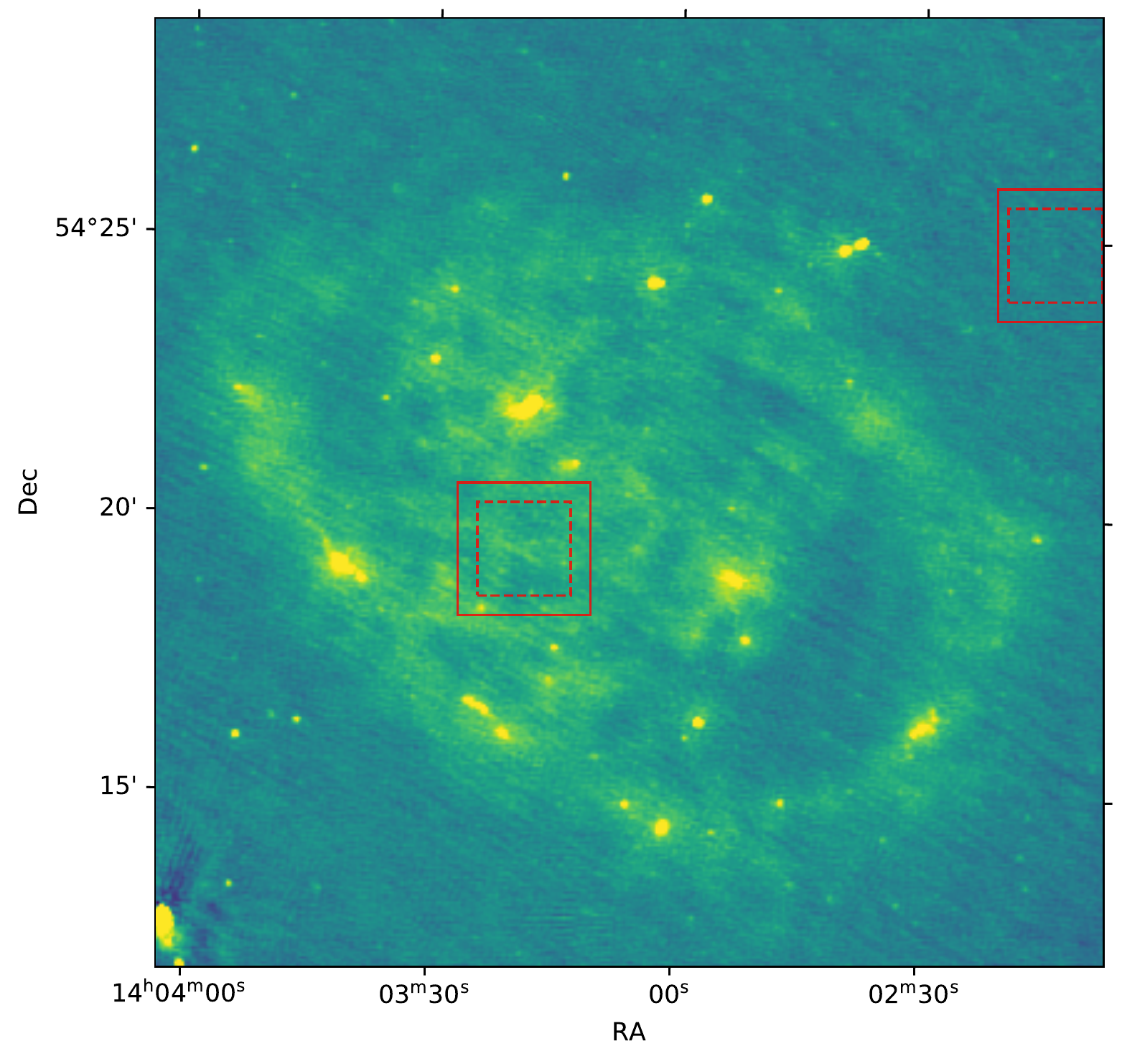}
\includegraphics[width=\halfpage\textwidth]{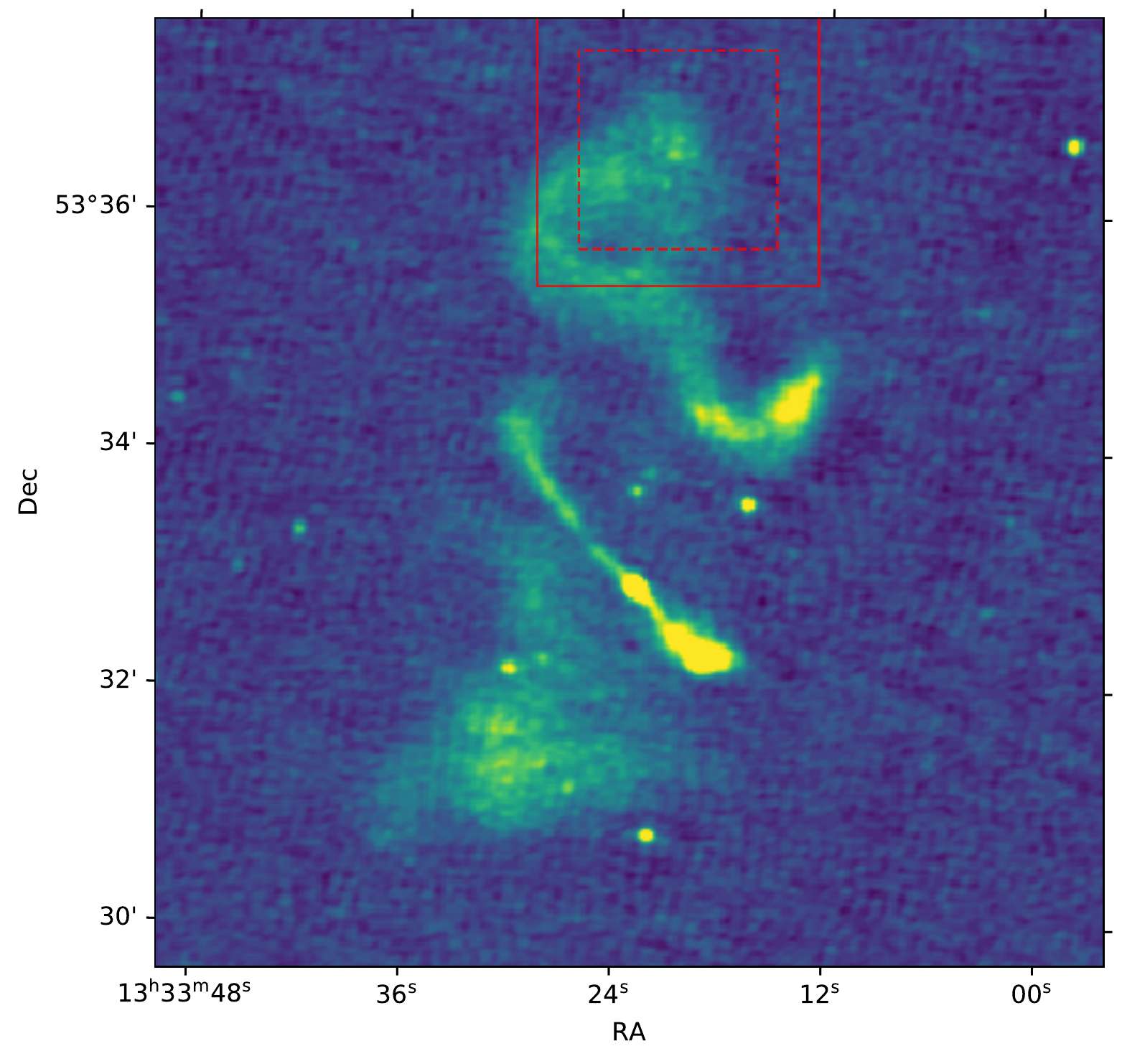}
\includegraphics[width=\halfpage\textwidth]{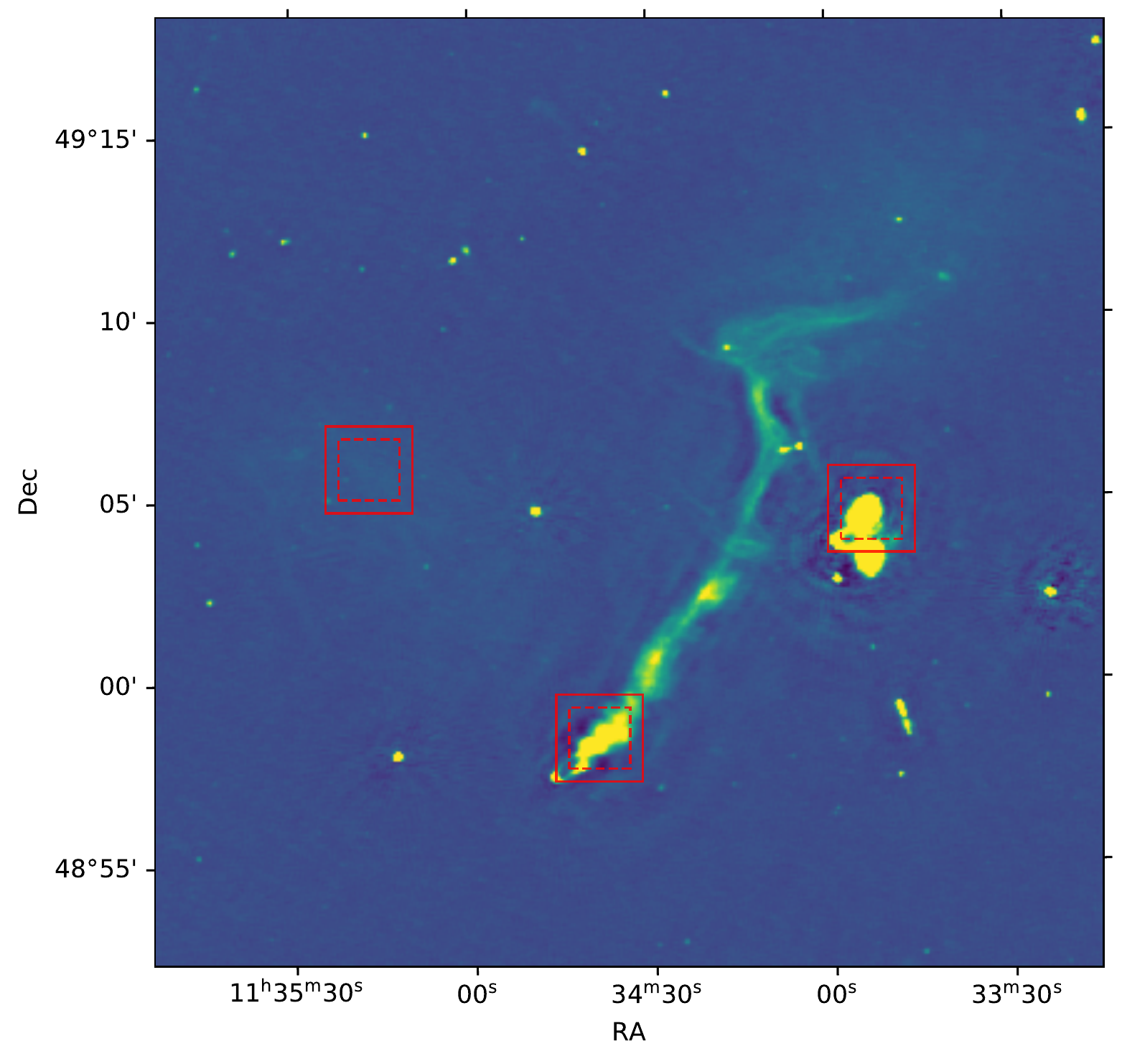}

\caption{\textit{Continued:} The outliers list contains a wide range of objects. Going row by row from left to right: an AGN remnant candidate (Shulevski et al in prep.), M101, diffuse emission combined with AGN emission, diffuse emission next to a giant head-tail AGN next to a WAT \citep{wilber2019evolutionary}. As the source extraction and association software PyBDSF did not associate radio emission of objects with a large (greater than a few hundred arcsec) apparent size, these appear as multiple separate entries in the source catalogue. The SOM can thus give different parts of the same astronomical object a high outlier score. We adopt a minimum separation distance between outlying objects of $400$ arcsec.}
\label{fig:outliers_highlights2}
\end{center}\end{figure*}  

It is possible to map a dataset to an SOM that is trained on a similar dataset.
We can thus quickly find outliers in a newly observed parts of the sky within the same survey.
We showcase this concept by looking for outliers in a new dataset covering $5,296$ ($5,720-424$; LoTSS-DR2 minus the LoTSS-DR1 area) square degrees.
To avoid distortion by the synthesised beam, we used all objects within a radius of $2.5$ degree around the centre of each field.
Analogous to the approach for LoTSS-DR1, we suppress the number of duplicate objects by imposing a minimum separation distance of $200$ arcsec between the listed outliers.

\begin{table*}[t]
\tiny
  \centering
\caption{Excerpt of the table containing the 1000 sources in the DR2 area (excluding the LoTSS-DR1 area) with the largest Euclidean norm (outlier score) to their best matching representative image in the SOM shown in Fig. \ref{fig:final_som}. The field name refers to the LoTSS field name. The complete table is available at \url{rafaelmostert.com/lotss-dr2-outliers.html}.}
\label{tab:DR2outliers}
\begin{tabular}{rllllll}
\toprule
  \# & Outlier score &       RA [deg] &     DEC [deg] & Field name &    Rep. image & 2MASX or SDSS galaxy cluster \\
\midrule
  1 &         32.02 &  193.894 &  27.172 &    P192+27 &  (1,1) &                              \\
  2 &         27.69 &   15.003 &  30.041 &    P015+31 &  (1,1) &                              \\
  3 &         27.66 &  193.873 &  27.224 &    P192+27 &  (1,1) &                              \\
  4 &         27.57 &  239.386 &  54.739 &    P241+55 &  (1,1) &                              \\
  5 &         27.46 &  339.267 &  34.415 &    P337+36 &  (1,1) &       2MASX 22370410+3424573 \\
  6 &         27.46 &   16.921 &  32.226 &    P018+31 &  (1,1) &                              \\
  7 &         27.40 &  214.642 &  37.802 &    P213+37 &  (1,1) &         WHL J141837.7+374625 \\
  8 &         26.34 &  157.948 &  57.038 &    P160+57 &  (1,1) &         WHL J103201.9+570318 \\
  9 &         26.20 &  142.027 &  30.017 &    P141+29 &  (1,1) &                              \\
 10 &         26.14 &  151.258 &  35.034 &    P151+35 &  (1,1) &                              \\
\bottomrule
\end{tabular}
\end{table*}

Table \ref{tab:DR2outliers} shows an excerpt of the table of the 1000 most outlying objects in this dataset.
Again, the outliers include objects that span a large range in apparent size and morphology.
The apparent sizes range from less than a hundred arcsec to a degree.
The morphologies include soft-edged (diffuse) emission as well as sharp-edged (compact) emission.
Cross-matching against the extended source catalogue \citep[2MASX;][]{jarrett20002mass} from the Two Micron All Sky Survey \citep[2MASS;][]{skrutskie2006two}, we see that 93 (9.3\%) of the first 1000 outliers align with a 2MASX object (with radius >15 arcsec), indicating that the outliers contain numerous nearby starforming galaxies.

In Fig. \ref{fig:outliers_types} we see that for the 100 outliers of LoTSS-DR1, based on their radio morphology, a significant percentage can be classed as cluster emission (either as relic or halo emission). Furthermore, the movement of the AGN in  the intra-cluster medium is known to effect the morphology of AGN radio emission and can for example lead  to long bent tails. As such, many of the WAT identified by the outlier selection will likely also be associated with galaxy cluster or group environments. Indeed for LoTSS-DR2 we find that 303 (30.3\%) of the first 1000 outliers align with known SDSS galaxy clusters \citep{wen2012catalog}.

The LoTSS-DR2 outlier source types include many AGN with bent and or asymmetric jets, restarted AGN (DDRGs) and even phoenixes (revived fossil plasma in a galaxy cluster).
A radio phoenix (source \#181) features in \citet{mandal2020revived}; a tailed AGN (source \#320) features in \citet{hardcastle2019ngc}; and a symmetric set of arcs around Abel 2626 (source \#413) features in \citet{ignesti2017new, ignesti2018mystery, kadam2019merging}, proving that our method of finding morphological outliers is able to automatically identify classes of rare sources with genuine scientific importance.

\section{Discussion}

\subsection{Linear invariance and the challenge of radio source association}
\label{linear-invariance}
Without correctly associated radio sources and corresponding correct apparent size estimates, we are not able to linearly rescale each radio source in size to make the SOM linearly size invariant.
This means that objects that appear to be large on the sky will also appear as larger shapes in the SOM.

Given a large enough SOM, some neurons will start to represent parts of larger objects such as a single lobe of an AGN with large angular size. In practice, these neurons may also resemble the disk-shaped emission of a nearby starforming galaxy, limiting the SOM's ability to separate intrinsically different sources. 
Sources much smaller than our fixed image size of 100 arcsec with closeby (within 50 arcsec) neighbouring sources are also poorly analysed by our method. 
Indeed, inspection of the outlying objects per group shows that many images with a high Euclidean norm are often occurring morphologies like regular compact FRII with closeby unrelated radio sources. The group of neurons labelled as `(single lobe of) large diffuse doubles' are the exception to this observation.

At this moment, no automatic source extraction code comes close to a human ability to associate spatially separated radio emission. However, source association performed by humans is a tedious, time intensive process.

An ongoing LOFAR Galaxy Zoo project tackles the association problem for LoTSS-DR1 \citep{Williams2019} and LoTSS-DR2 and can thereafter serve as labelled training sample for supervised learning attempts to the radio component association problem.
\citet{Dieleman2015} and \citet{Dai2018} were successfully able to recreate the labels assigned to galaxies by volunteers in the Optical and Radio Galaxy Zoo projects respectively. 
\citet{Wu2019} use FIRST and WISE images combined with Radio Galaxy Zoo crowd-sourced associations \citep{Banfield2015} to combine radio blob detection, association and classification using a promising deep learning approach. 

In this paper we showed that it is possible to select morphologically rare objects in large datasets with no previous human created labels or associations.
Our selection works across a range of apparent sizes and for broken up sources.
We did not find an upper limit to the angular size of objects for our ability to mark them as having a rare morphology. 
However, we can conclude that our SOM approach to finding outliers is ill-suited to find sources with outlying morphologies with angular sizes much smaller than our fixed view window. 
Specifically using the Euclidean norm as outlier score and a fixed view window of 100 arcsec, we find no sources with an angular size smaller than 45 arcsec in our list of 100 outliers (Appendix~\ref{app:outliers}).

Furthermore, SOM extracted features (i.e. the location of the best matching representative image of a source on the SOM lattice or its outlier score) can be used as complementary input for datasets analysed using classical machine learning techniques, for example for source cross-identification (cf. Alegre et al., in prep).

\subsection{SOM for data-exploration. The LOFAR-PINK Visualisation Tool: an interactive webtool}
\label{som-vis}
SOMs can be used as the basis for data-exploration and visualisation tasks.
As such we build the \textsc{LOFAR-PINK Visualisation Tool}\footnote{\url{https://github.com/RafaelMostert/lofar-pink-visualization-tool}} an interactive website for SOM exploration: \url{https://rafaelmostert.com/lofar/ID51/som.php}.
The tool is interactive, accessible by any webbrowser, shareable and easy to set up on a newly trained SOM. It enables users to do their own data-exploration without in-depth prior knowledge of the code or algorithms involved (see Appendix~\ref{app:visualisation}). 
It has succesfully been used for outreach projects and is an approachable way to showcase the quality of LoTSS.
 
A trained SOM used for dimensionality reduction in combination with the webtool means that one can represent the diversity of a dataset in a very compact way. As such, this combination can be used for both validation of new data and for systematic searches within large data sets based on more than one-dimensional catalogue values.

\subsection{Future work}
\label{sec:future}
In this paper, we scraped the surface of what is possible with SOMs and 2D astronomical image-data. We used an SOM trained on data from a single survey in combination with a catalogue without corrected radio component associations and without (optical or spectral) information about the host-galaxies.  

Future work could focus on the addition of complementary surveys to the LOFAR high-band survey to provide a dimensionality reduction and visualisation of multi-wavelength data. The ability to fold in multiple layered input (multiple cutouts with the same dimensions as one input) is already present in the \textsc{PINK} software. This would break degeneracies between representative images and input images and allow a more diverse set of unsupervised clusters to appear in the SOM. 

A two-layer SOM works as follows: a single image with two layers -- for example a radio image taken at 150 MHz and at 1.4 GHz -- will be compared to an SOM in which each neuron also has two layers. The first layer of the neuron will represent the 150 MHz emission and the second layer the 1.4 GHz emission and the distance metric will simply be the sum of the Euclidean norm between the image's first layer and the neuron's first layer and the Euclidean norm between the image's second layer and the neuron's second layer.

\citet{galvin2019radio} demonstrate this principle with an SOM trained on both radio \citep[FIRST;][]{Becker1995} and infrared \citep[WISE;][]{2010AJ....140.1868W} cutouts.
In the same way, our LoTSS data could be complemented by cutouts from other large infrared sky-surveys like WISE, or optical surveys such as the Panoramic Survey Telescope and Rapid Response System \citep[Pan-STARSS;][]{Kaiser2000,chambers2016pan} survey or the DESI Legacy Imaging Surveys \citep[Legacy\footnote{\url{http://legacysurvey.org}};][]{legacy2019}. 
At 3GHz the Very Large Array Sky Survey \citep[VLASS\footnote{\url{https://science.nrao.edu/science/surveys/vlass}};][]{vlass2020} can complement the LoTSS data and eventually the LOFAR Low-band Sky Survey (LoLSS\footnote{\url{https://lofar-surveys.org}}), 40-70 MHz, can complement our data at frequencies that are just above the atmospheric cut-off. 
With this additional spectral information that helps to reduce the number of morphological degeneracies, an SOM can be used as a selection tool to query a set of objects based on morphology.

Beyond visualisation and exploration, the SOM could be used to improve catalogue (labels) by aiding the association of extended radio source components. 
The SOM presented in this paper can be used to fix the simplest associations. 
A rule-based heuristic can associate all sources labelled `single lobe of extended doubles' with the closest source with the same label with either a fixed limit or a flexible maximum limit on their separation distance. \citet{galvin2020cataloging} show that in a similar approach, with a larger ($40\times40$) SOM, even more sources can be automatically associated.

\section{Summary}
Current (LoTSS, VLASS) and future surveys of the radio sky such as Evolutionary Map of the Universe \citep[EMU;][]{Norris2011}, APERTIF survey\footnote{\url{https://old.astron.nl/radio-observatory/apertif-surveys}}, \citet[MIGHTEE;][]{mightee}, reveal objects that have an extraordinary array of different morphologies which tell us about the environment of the source or the origin of the emission. There are too many sources to visually inspect and there is a risk that some of the most interesting sources can stay hidden.

Here we use a dimensionality reduction technique altered for the use on (radio) sources. We apply the rotation and flipping invariant self-organising map, \textsc{PINK}, developed by \cite{Polsterer2016} to LoTSS to reveal the different morphologies present in this dataset.

The source-detection software PyBDSF provides us with a catalogue of source coordinates and enables us to discard most of the morphologically undistinguishable unresolved radio sources. A first trained SOM allows us to eliminate even more unresolved radio sources that are still present in this subset.
A final SOM trained on these sources reveals six dominant morphological groups and allows us to estimate the number of sources belonging to each group in the dataset, which serves as an estimate for the number of sources of this \label{fig:outliers_highlights1}
archetype in the completed LoTSS survey which will eventually cover the full Northern sky.

As the SOM is a model for the (dominant) morphologies of the resolved sources in the data, we can use the algorithm to find morphological outliers: objects with a rare morphology that are not well represented by the map stand out.
We present the 100 most outlying sources and highlight a range of physically different objects. Both large diffuse objects with relaxed morphology (such as cluster halos, AGN remnants and nearby starforming galaxies) and sharp-edged complex structures (such as cluster relics and weirdly shaped extended doubles) are found.
We also show that the SOM, trained on a small region (424 square degrees) of the sky can be used to find sources with outlying morphologies in a much larger ($\sim$5300 square degrees) region of the sky without retraining.

Finally, we show how a survey can be accessed, visualised and shared by using an interactive web tool build for this purpose. Using the \textsc{LOFAR-PINK Visualisation Tool} one can intuitively explore the dataset through the SOM, the corresponding radio sources and its outliers.

Future work will focus on the combination of complementary surveys to LoTSS to provide a dimensionality reduction and visualisation of multi-wavelength data. 
The visualisation software could be expanded and improved to enable searches by morphology. 

\begin{acknowledgements}
We kindly thank the referee for the insightful comments.
This research has made use of the python Astropy package \citep{astropy}; the SIMBAD database,
operated at CDS, Strasbourg, France; the "Aladin sky atlas" developed at CDS, Strasbourg Observatory, France \citep{Bonnarel2000} and \citep{Boch2014}; and 
the VizieR catalogue access tool, CDS, Strasbourg, France (DOI: 10.26093/cds/vizier). 
The original description of the VizieR service was published by \citet{ochsenbein2000vizier}. 
R.M. also acknowledges the help of Berndt Doser at the Heidelberg Institute for Theoretical Studies, without whom this research would not have been possible. 
K.L.P. gratefully acknowledges the generous and invaluable  support  of  the  Klaus Tschira Foundation.
LOFAR is the Low Frequency Array designed and constructed by ASTRON. It has observing, data processing, and data storage facilities in several countries, which are owned by various parties (each with their own funding sources), and which are collectively operated by the ILT foundation under a joint scientific policy. The ILT resources have benefited from the following recent major funding sources: CNRS-INSU, Observatoire de Paris and Universit\'e d'Orl\'eans, France; BMBF, MIWF-NRW, MPG, Germany; Science Foundation Ireland (SFI), Department of Business, Enterprise and Innovation (DBEI), Ireland; NWO, The Netherlands; The Science and Technology Facilities Council, UK; Ministry of Science and Higher Education, Poland; The Istituto Nazionale di Astrofisica (INAF), Italy. 
M.Br. acknowledges support from the ERC-Stg DRANOEL, no 714245. 
M.Bo. acknowledges support from INAF under PRIN SKA/CTA FORECaST and from the Ministero degli Affari Esteri della Cooperazione Internazionale - Direzione Generale per la Promozione del Sistema Paese Progetto di Grande Rilevanza ZA18GR02. 
P.N.B. is grateful for support from the UK STFC via grant ST/R000972/1.
B.M. acknowledges support from the UK Science and Technology Facilities Council (STFC) under grants ST/R00109X/1 and ST/R000794/1. 
K.J.D., H.J.A.R. and W.L.W. acknowledge support from the ERC Advanced Investigator programme NewClusters 321271. W.L.W. also acknowledges support from the CAS-NWO programme for radio astronomy with project number 629.001.024, which is financed by the Netherlands Organisation for Scientific Research (NWO). 
\end{acknowledgements}
\bibliographystyle{aa}
\bibliography{major_bib}

\begin{appendix}
\section{First run self-organised map}
\label{app:firstrun}
\begin{figure}\begin{center}
\includegraphics[width=0.49\textwidth]{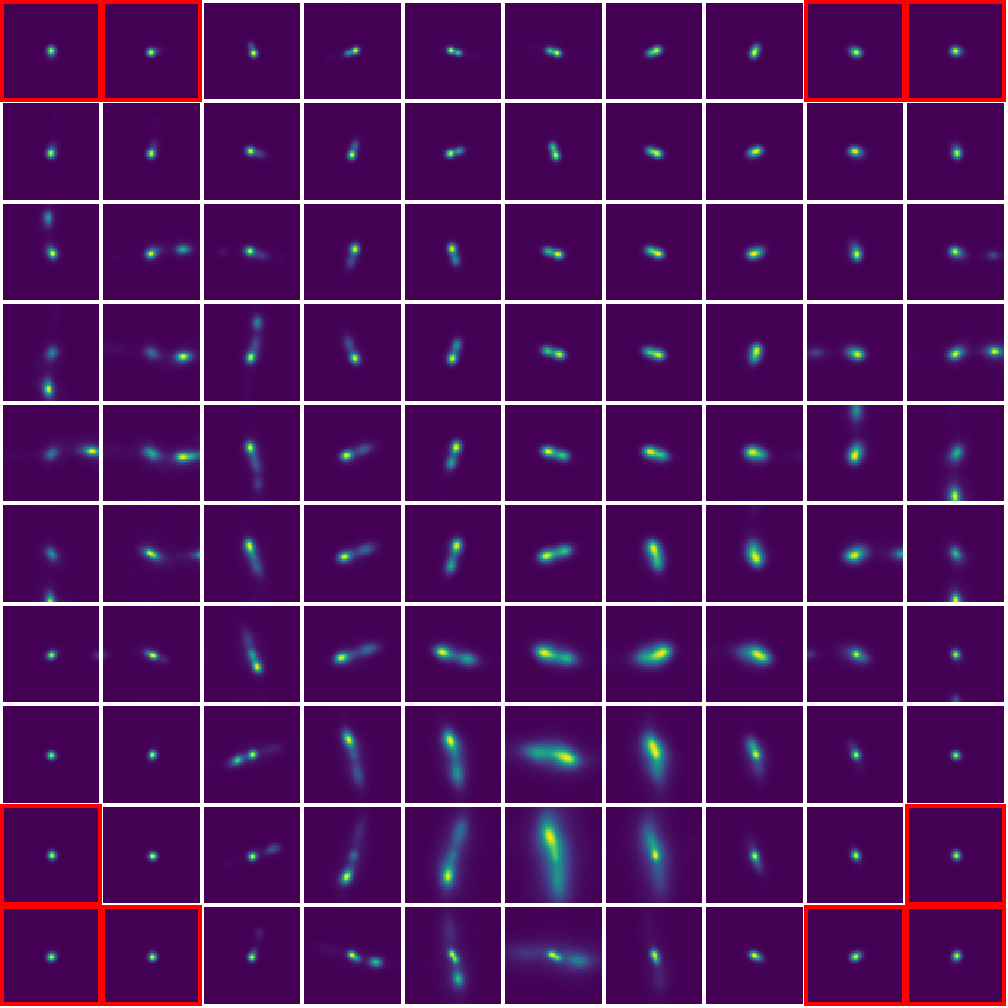}
\includegraphics[width=0.49\textwidth]{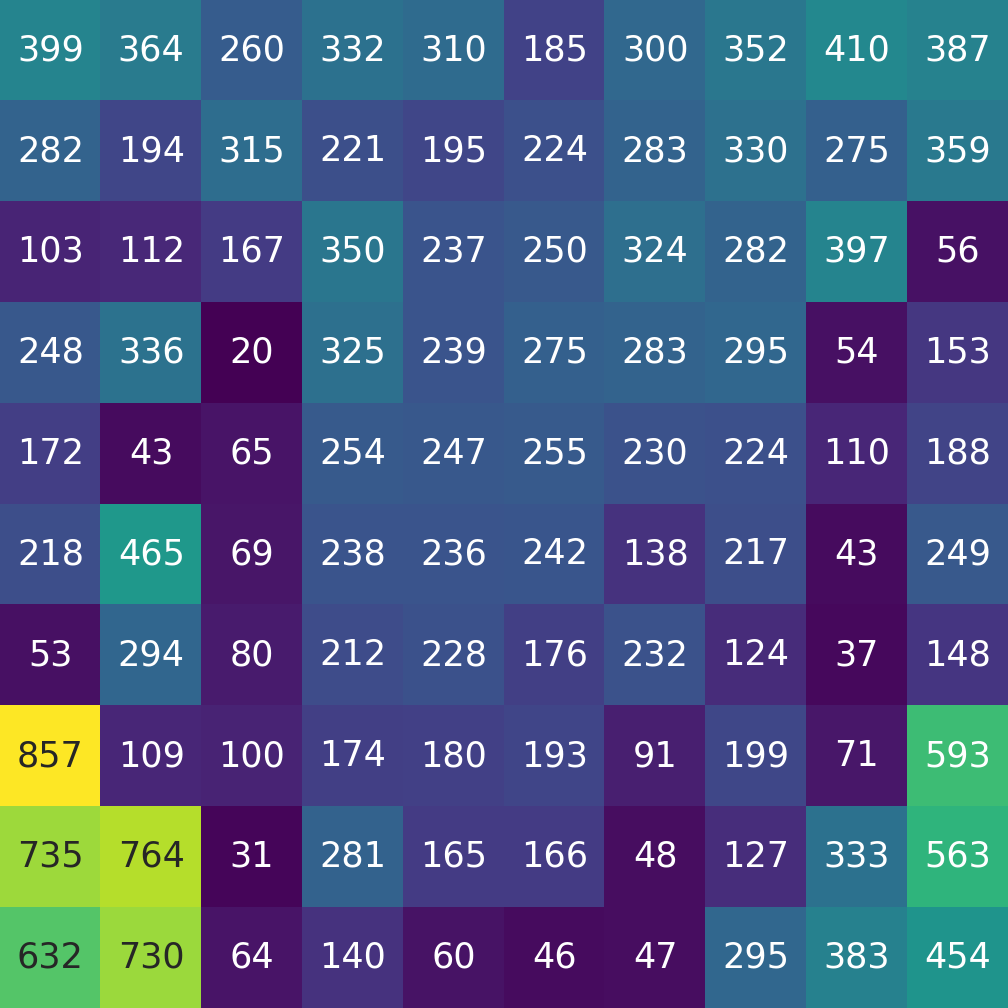}
\caption{\textit{Top panel:} First run $10\times10$ cyclic SOM. The first run reveals that our preprocessed dataset still contains a large fraction of unresolved sources. We remove these from the training dataset by removing all radio sources that map to one of the red highlighted representative images. 
\textit{Bottom panel:} Heatmap of the first run SOM, indicating the number of sources mapped to each of the representative images. }
\label{fig:first_run_som}
\end{center}\end{figure}

Figure \ref{fig:first_run_som} shows the output of our first run SOM, prior to the removal of largely unresolved sources.
The red highlighted representative images in the SOM have been selected for their similarity to their neighbouring representative images using a U-matrix. See \citet{Ultsch1990} for details on the U-matrix.

\section{Sample of 100 morphological outliers}
\label{app:outliers}

\begin{figure*}\begin{center}
\includegraphics[width=0.19\textwidth]{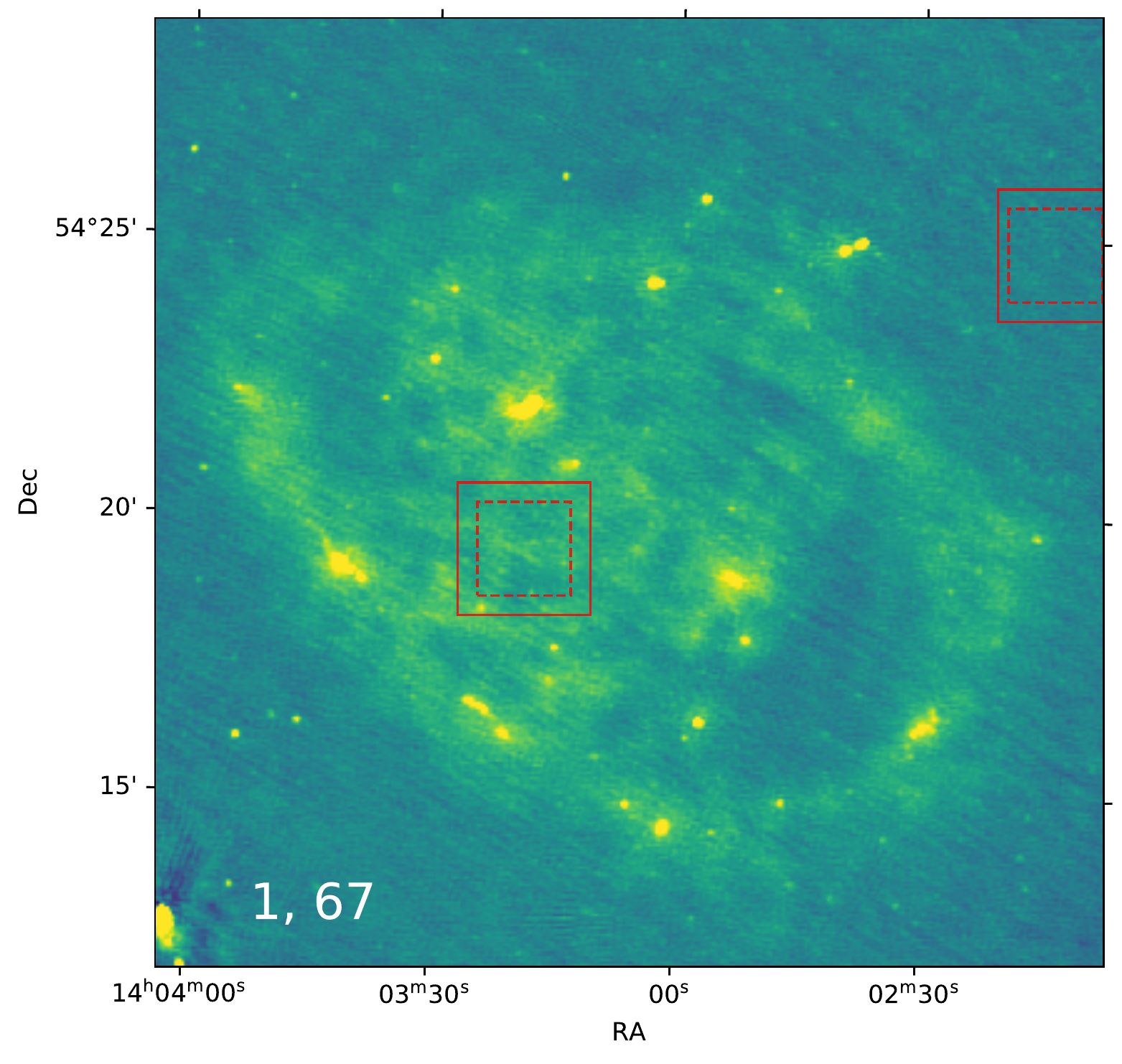}
\includegraphics[width=0.19\textwidth]{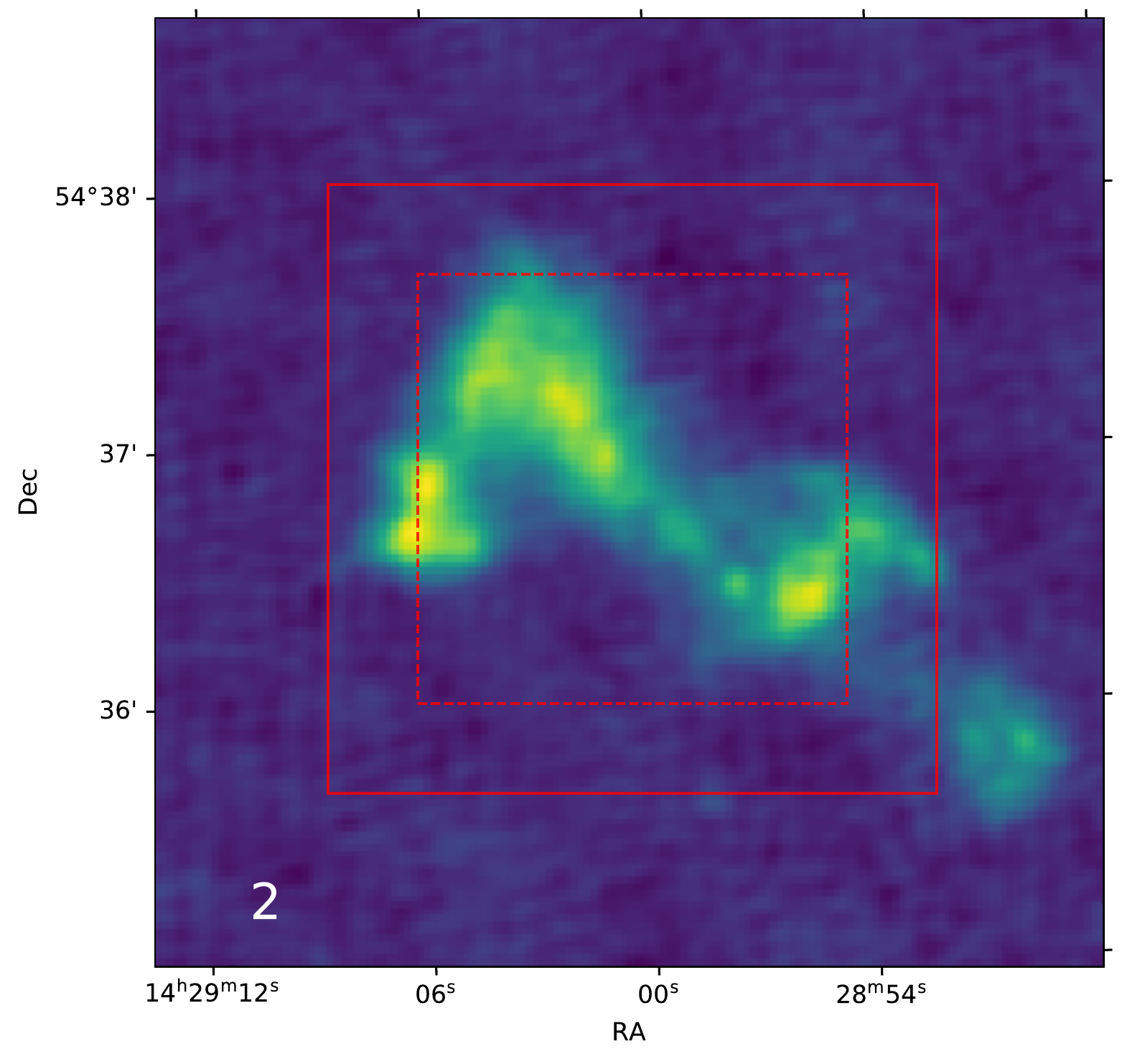}
\includegraphics[width=0.19\textwidth]{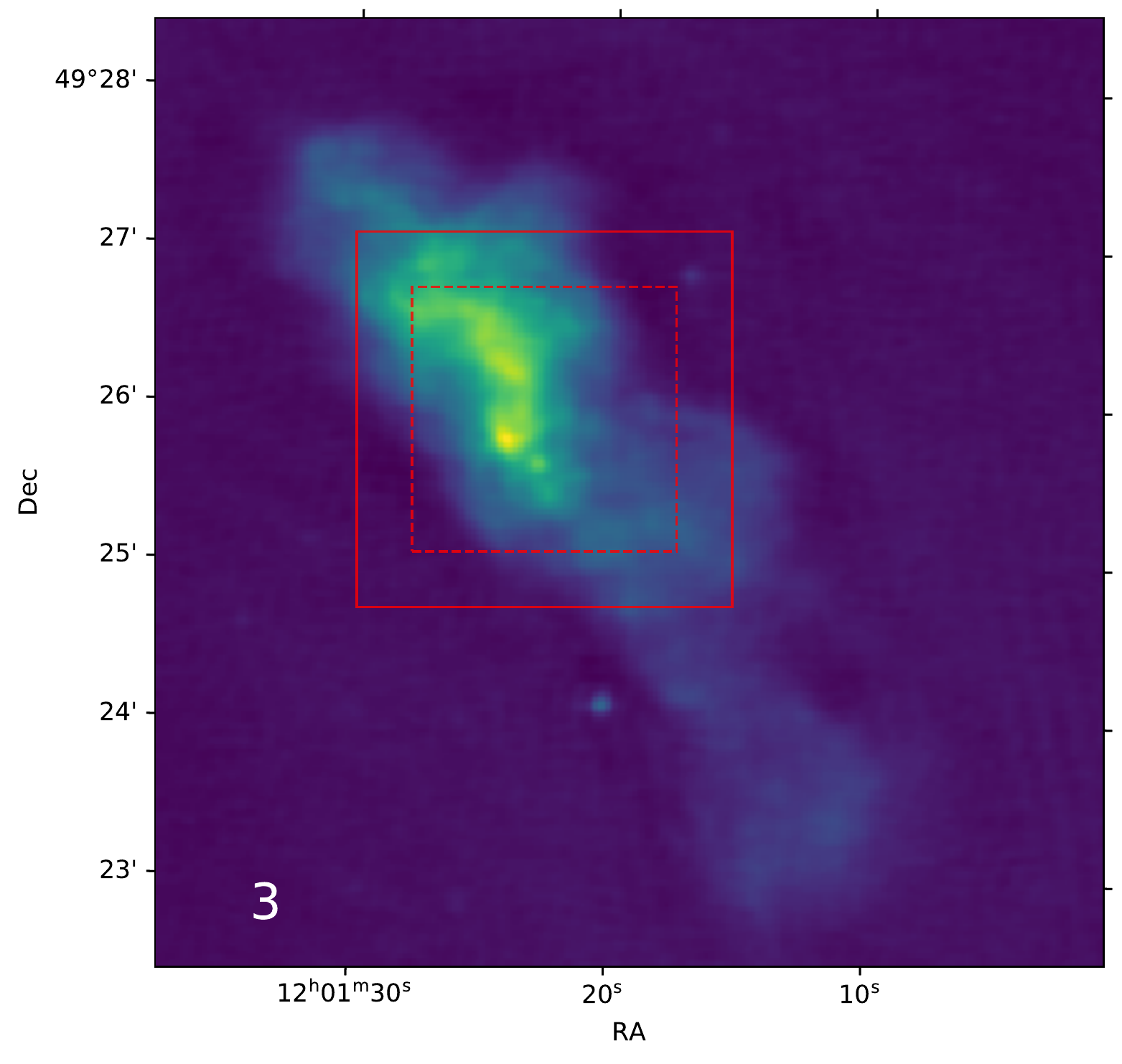}
\includegraphics[width=0.19\textwidth]{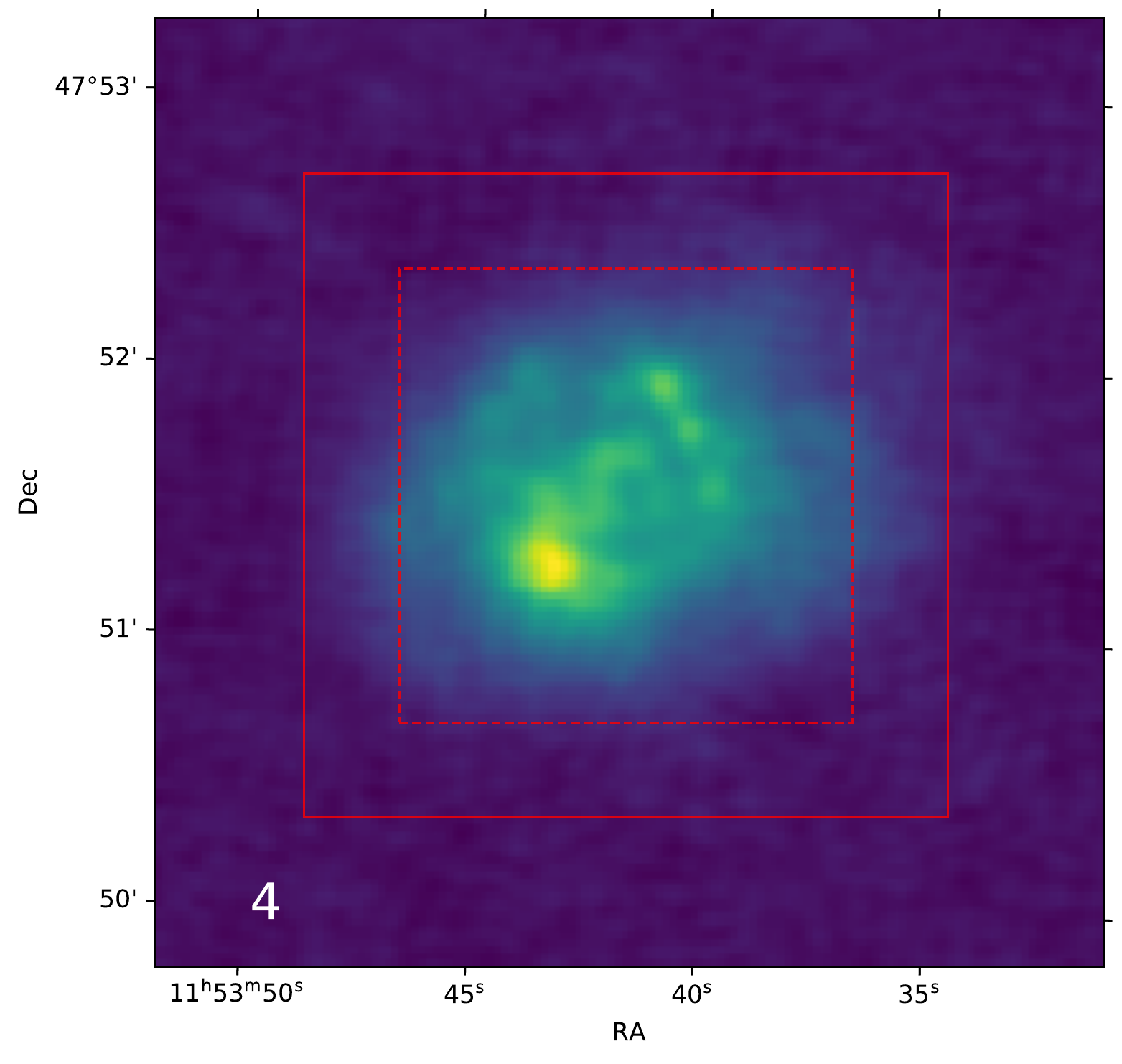}
\includegraphics[width=0.19\textwidth]{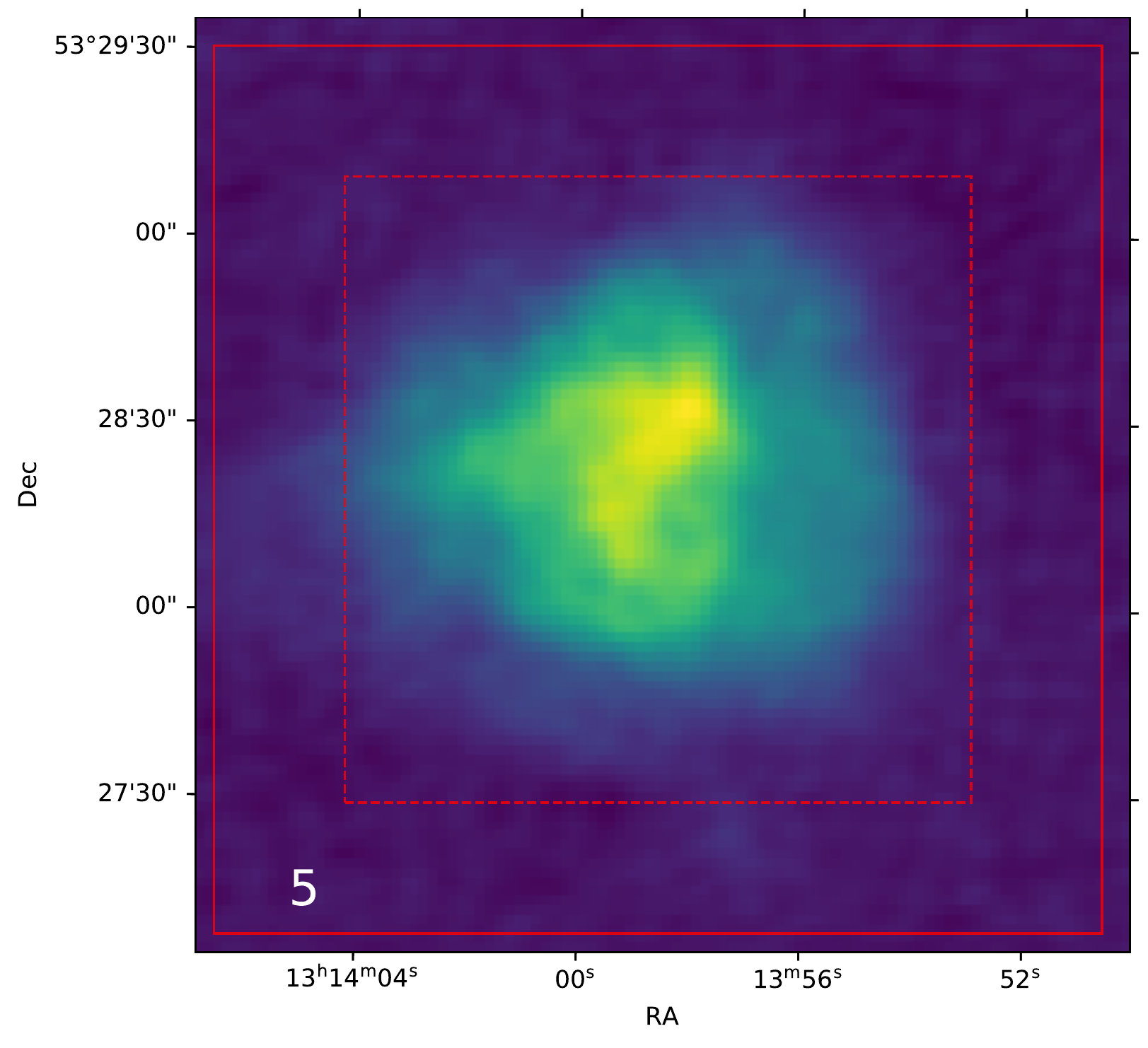}
\includegraphics[width=0.19\textwidth]{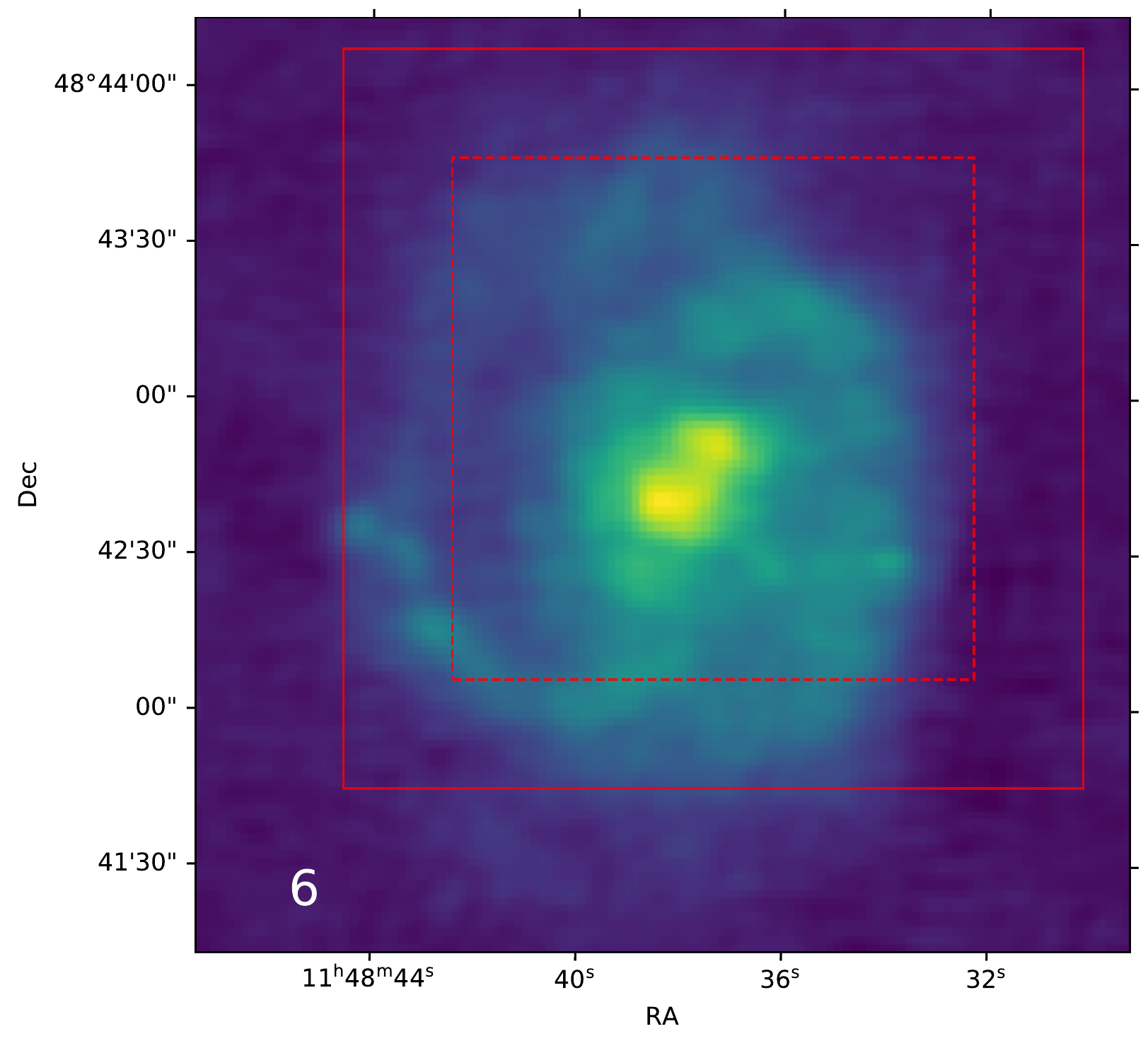}
\includegraphics[width=0.19\textwidth]{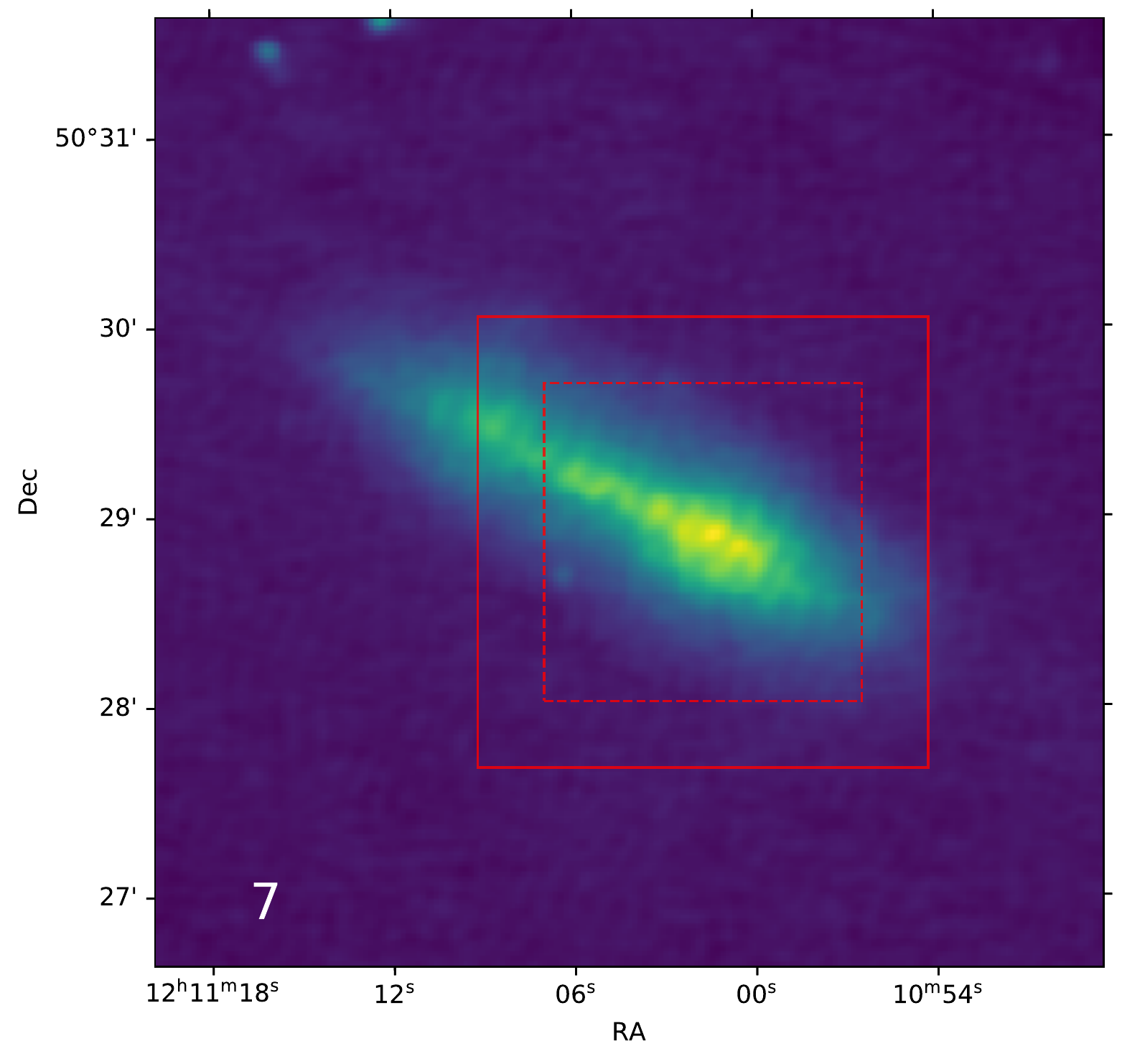}
\includegraphics[width=0.19\textwidth]{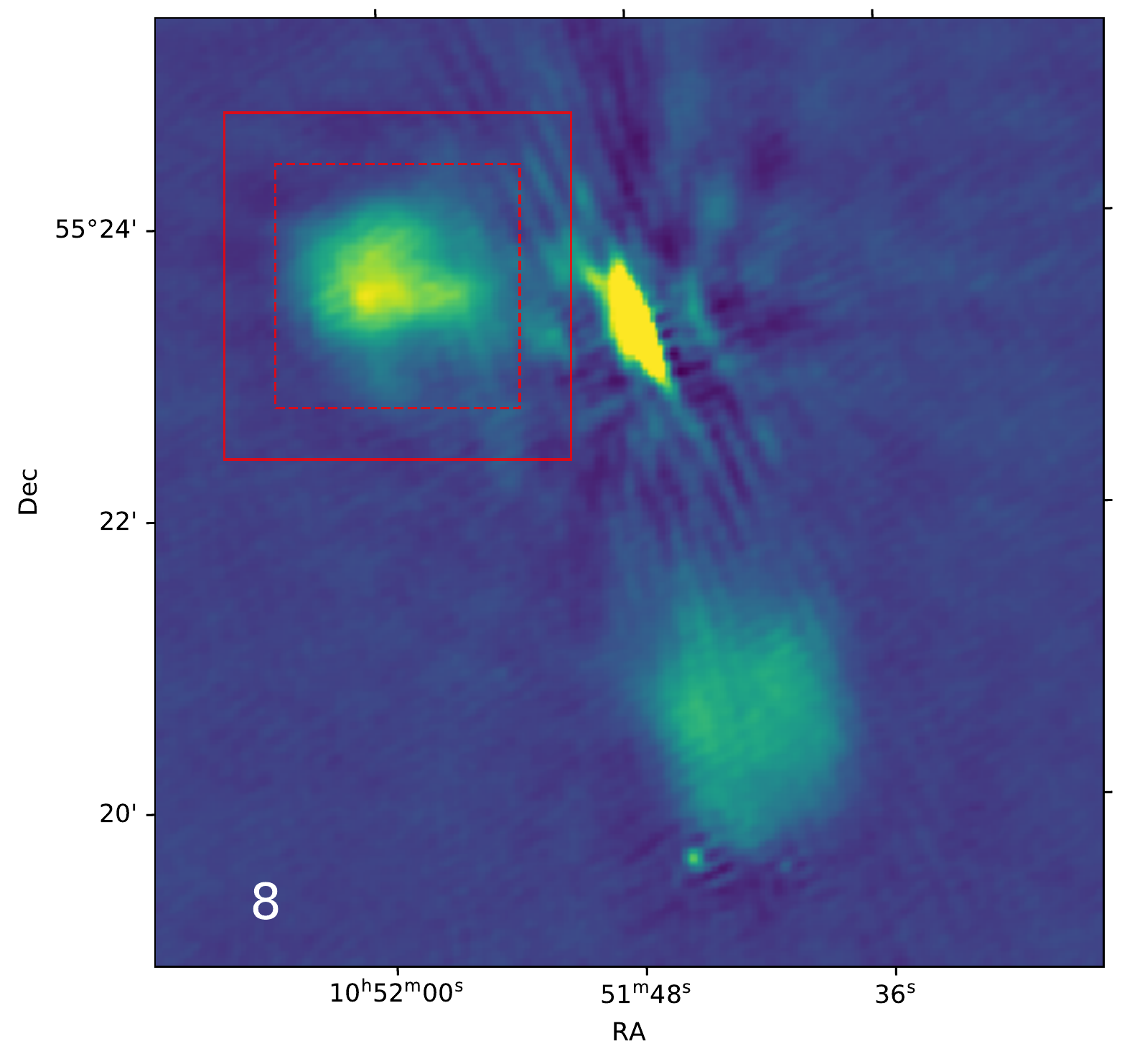}
\includegraphics[width=0.19\textwidth]{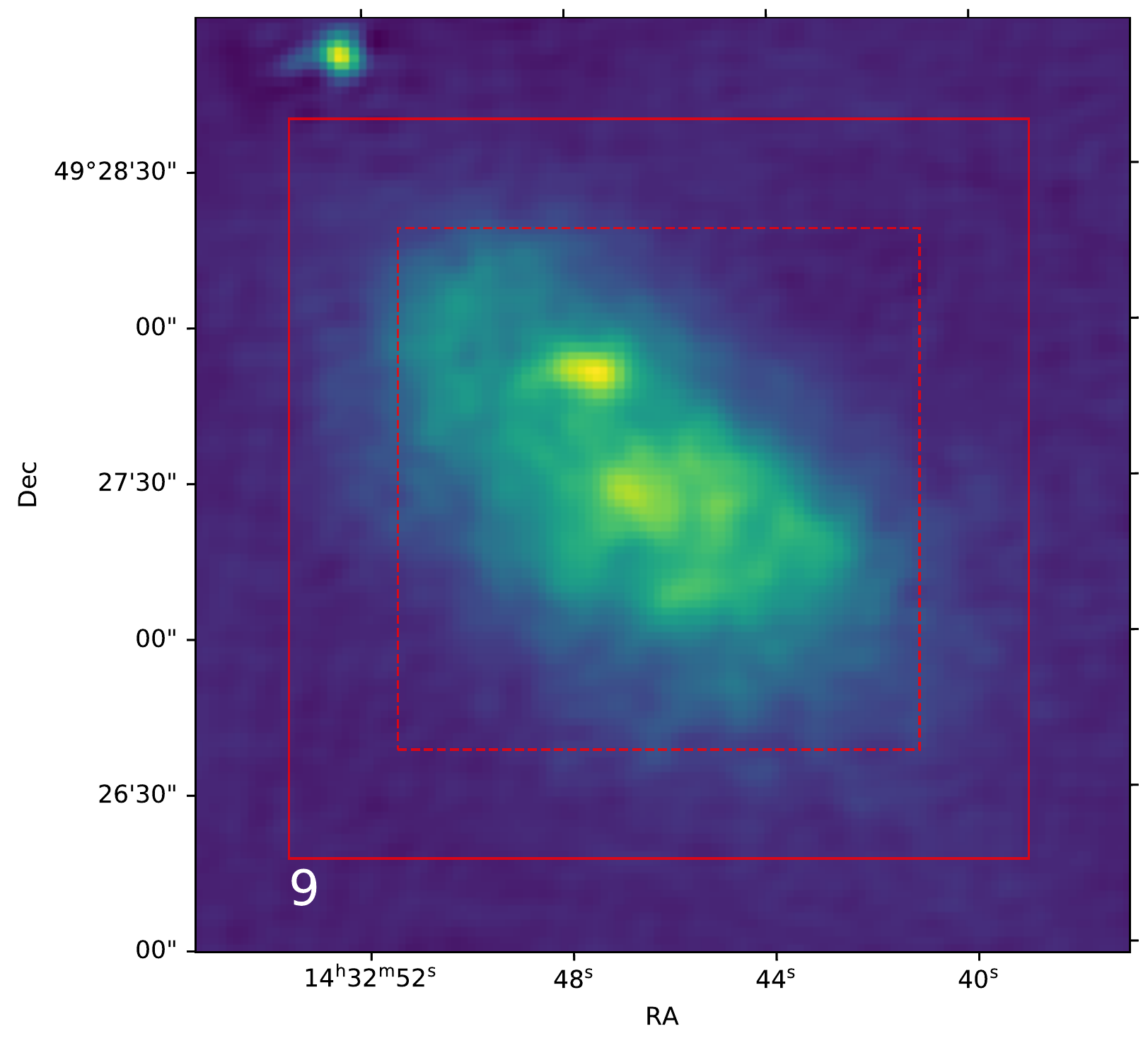}
\includegraphics[width=0.19\textwidth]{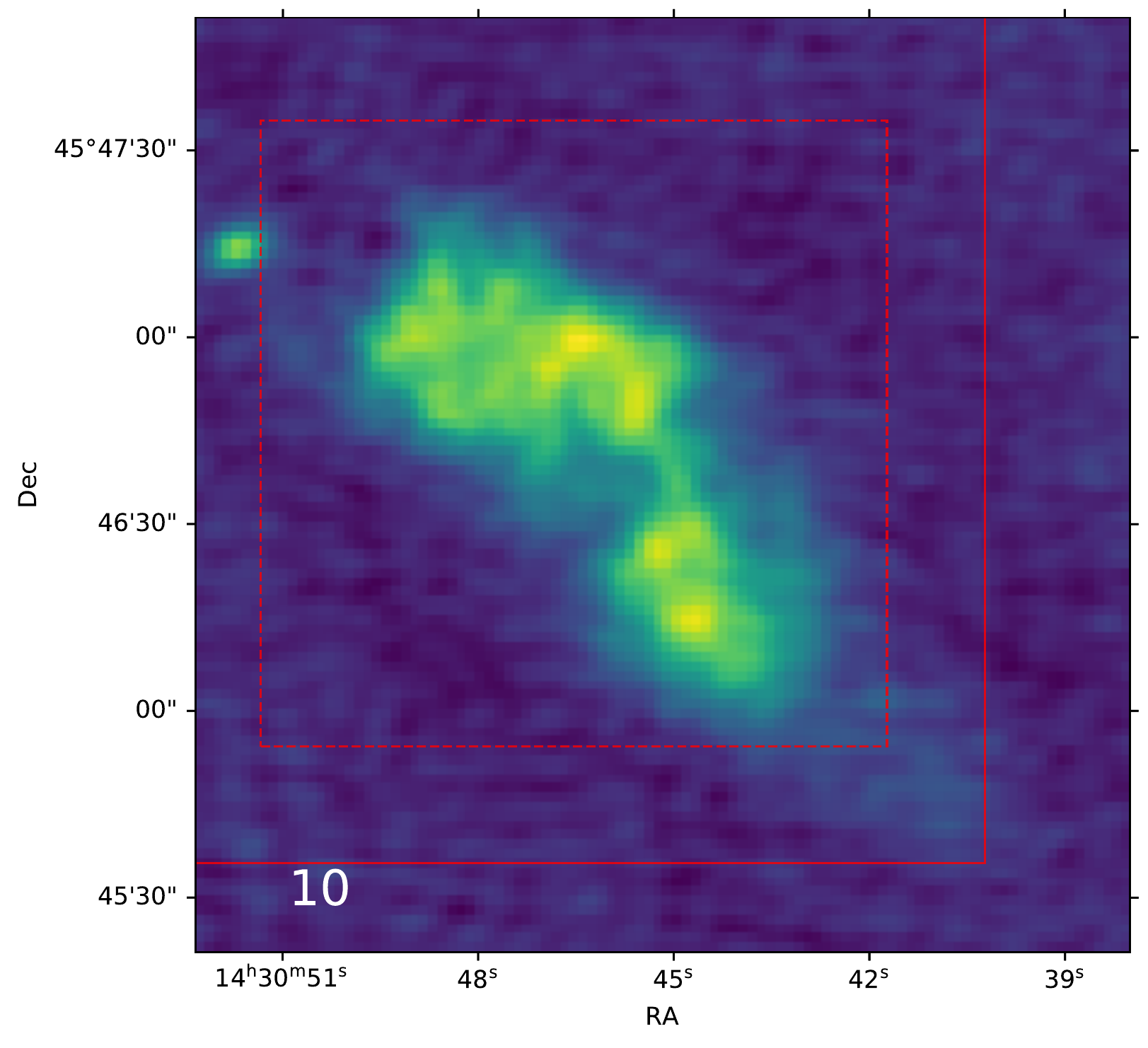}
\includegraphics[width=0.19\textwidth]{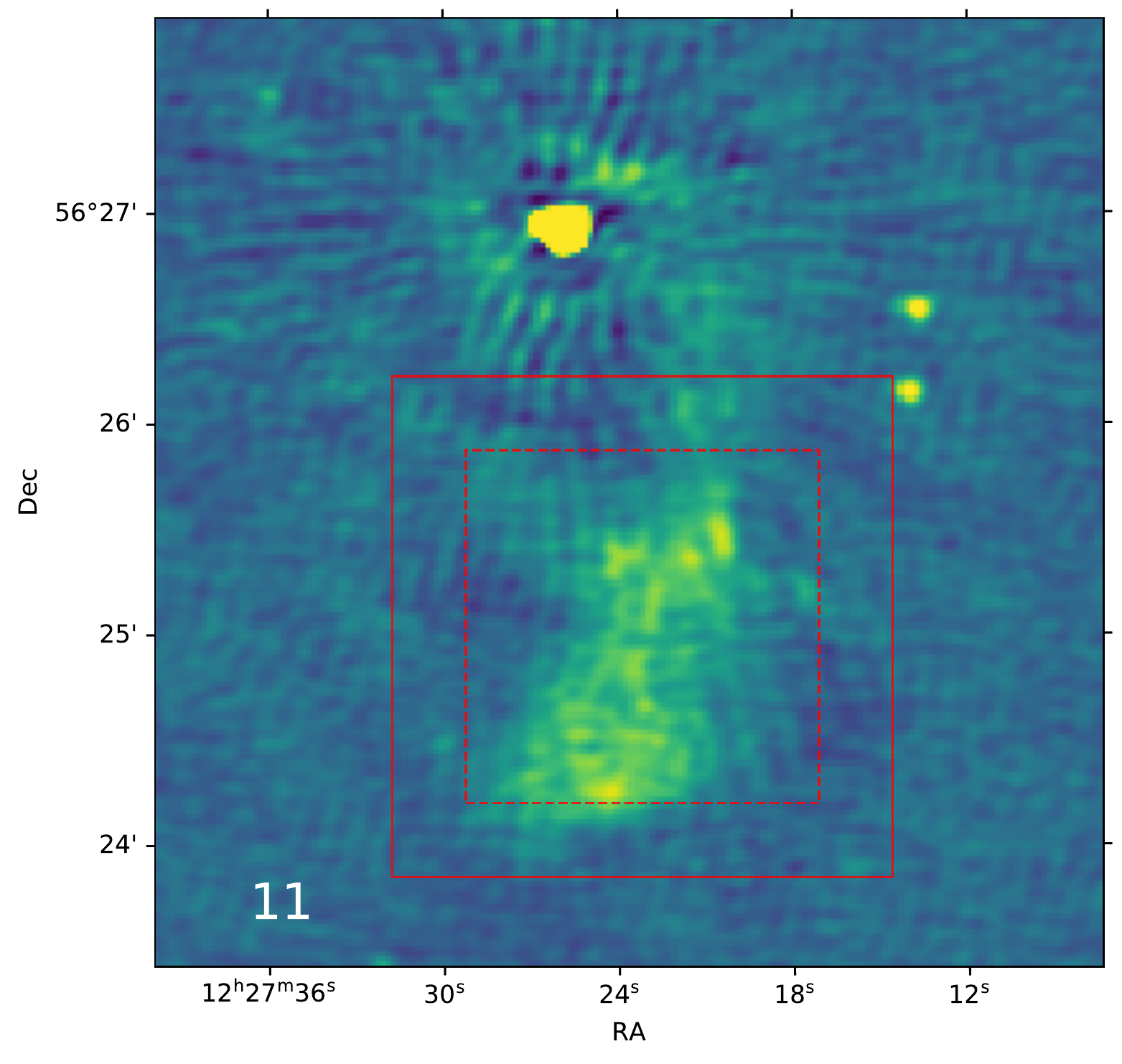}
\includegraphics[width=0.19\textwidth]{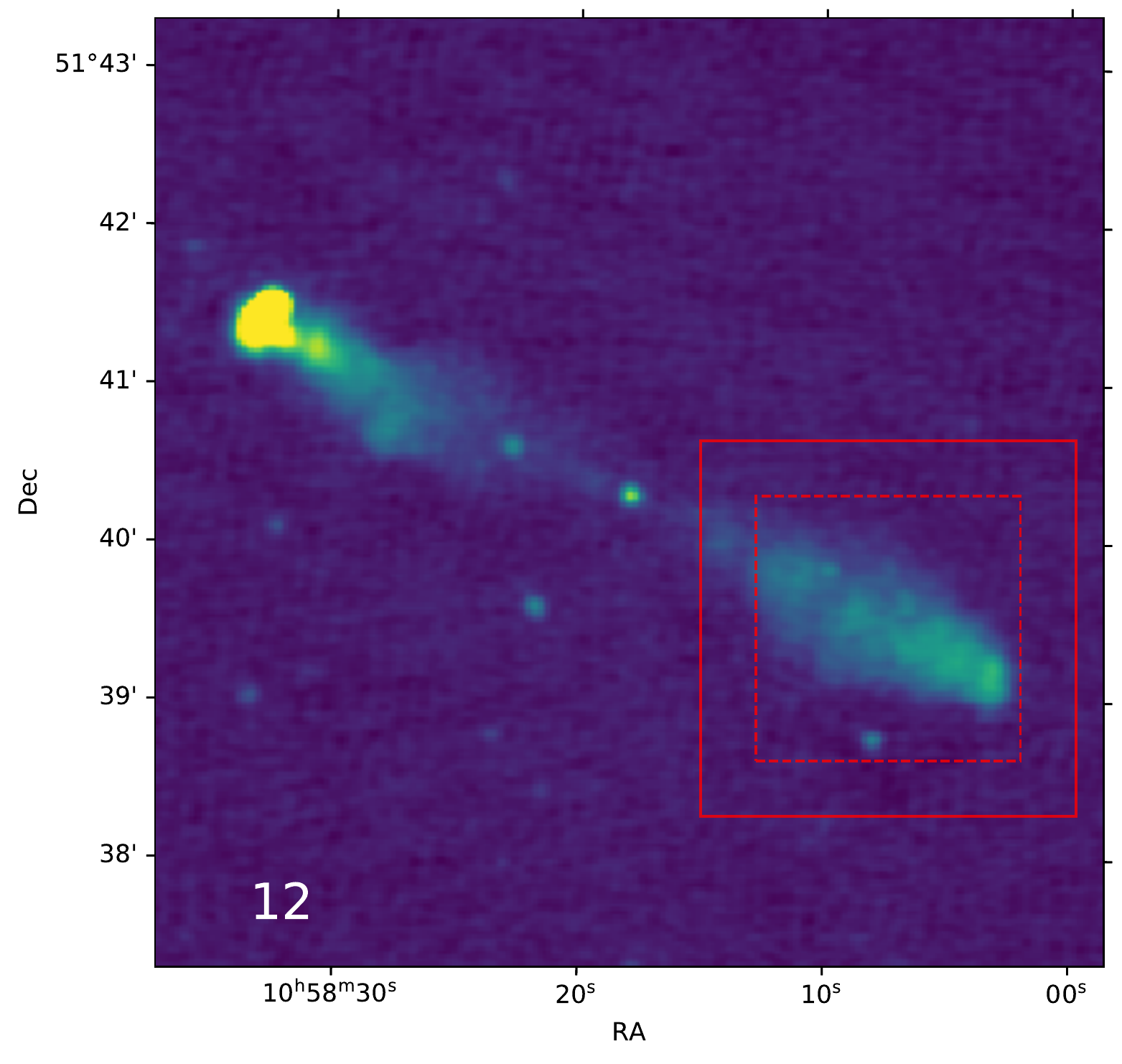}
\includegraphics[width=0.19\textwidth]{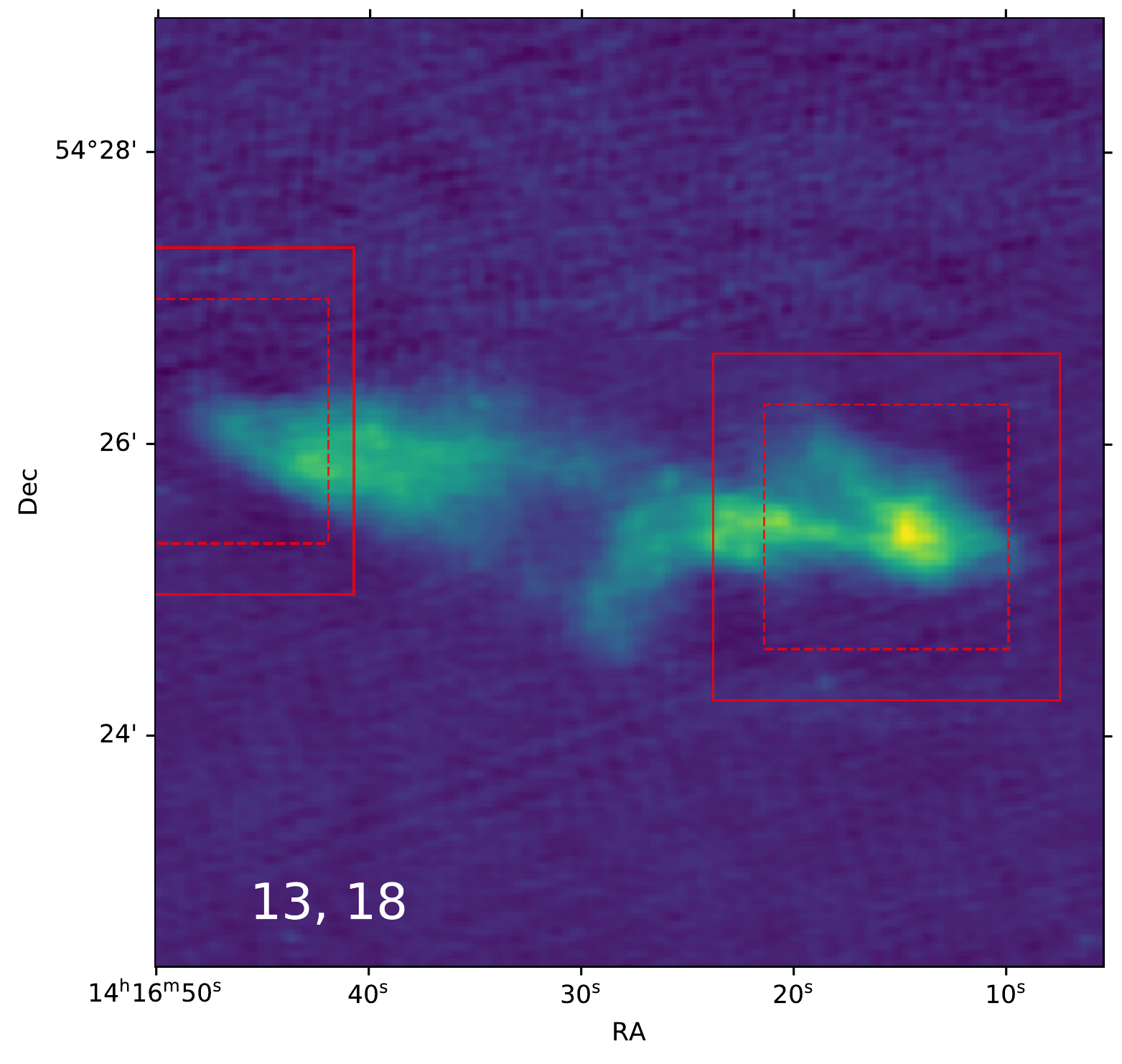}
\includegraphics[width=0.19\textwidth]{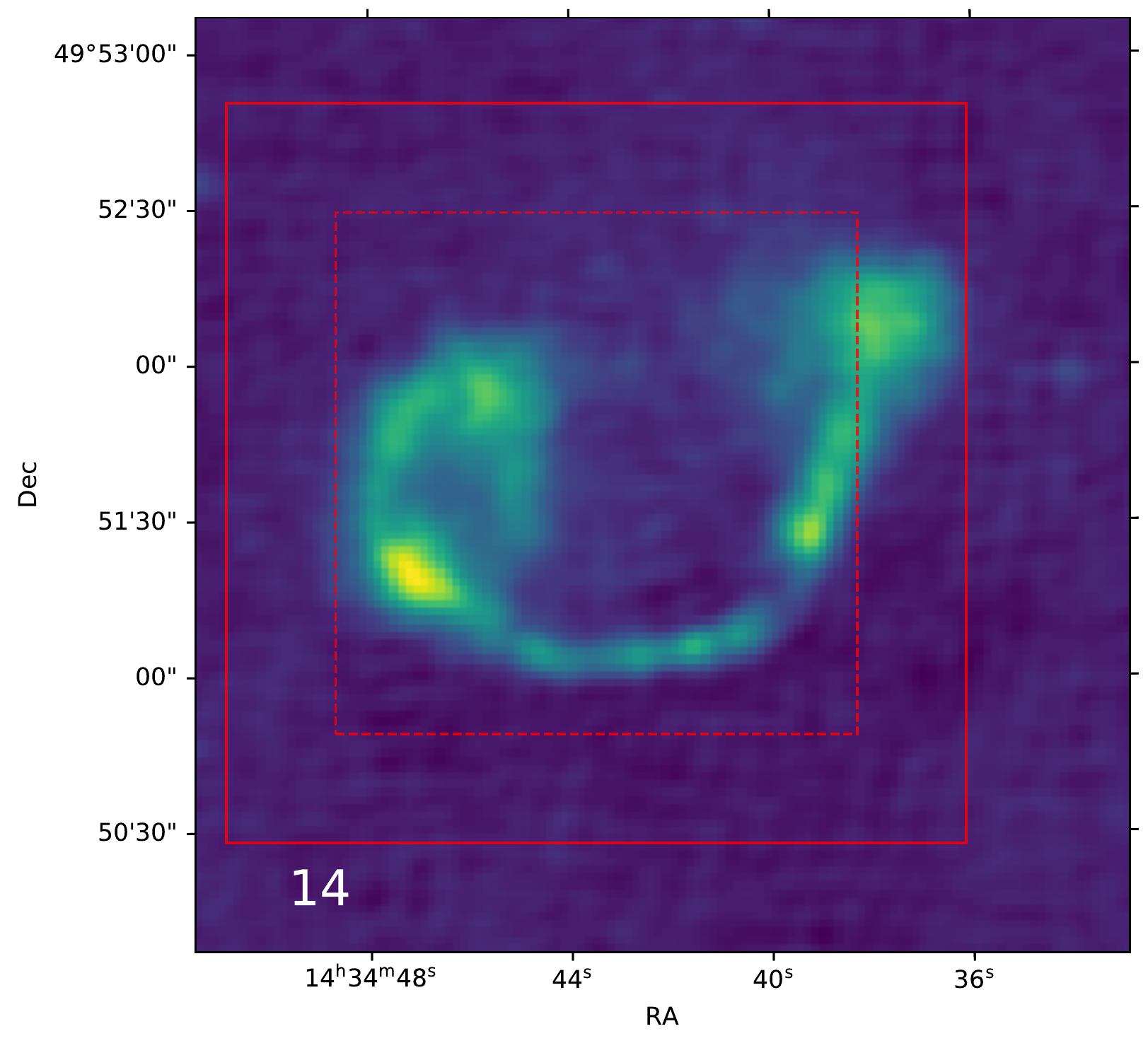}
\includegraphics[width=0.19\textwidth]{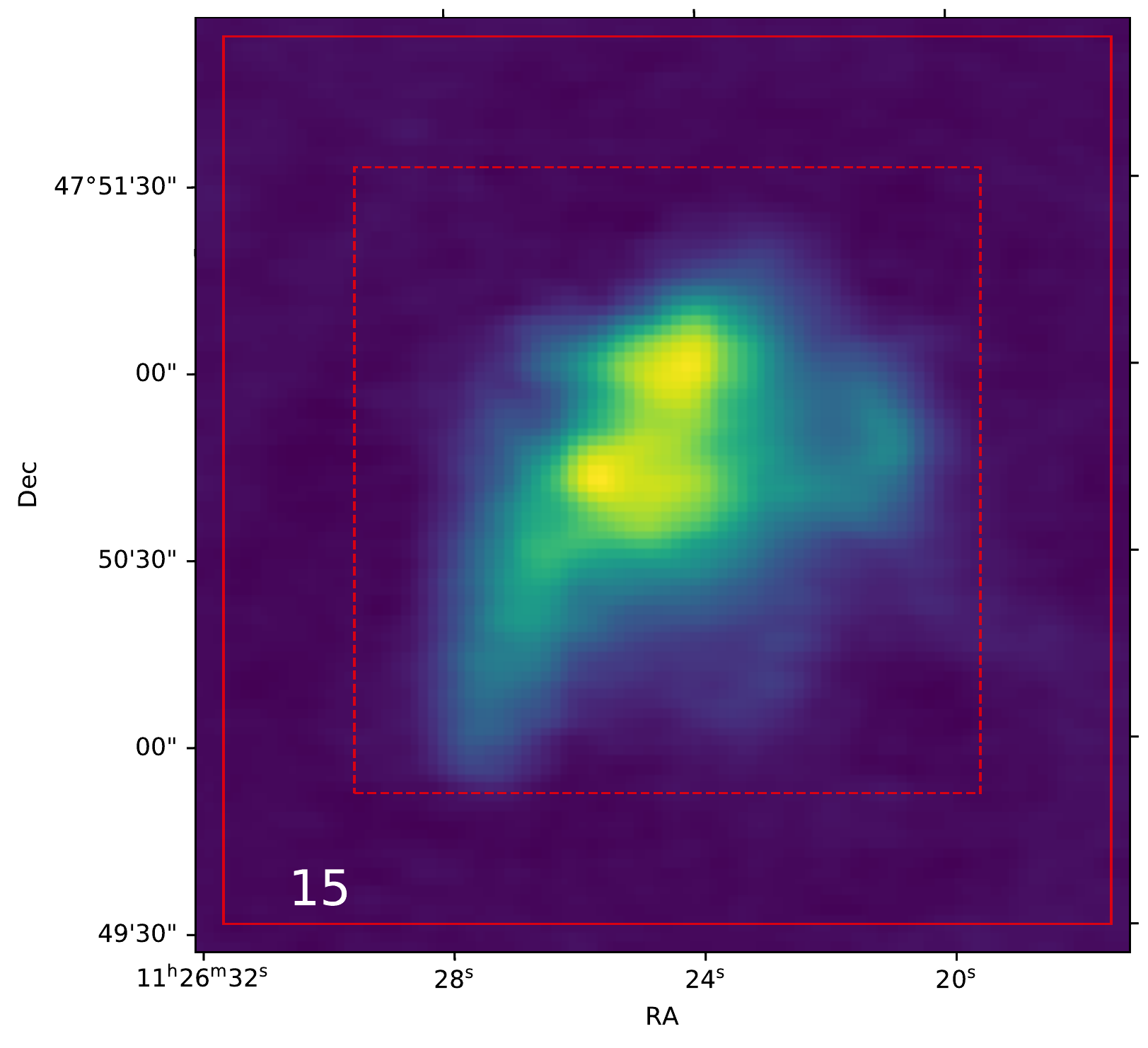}
\includegraphics[width=0.19\textwidth]{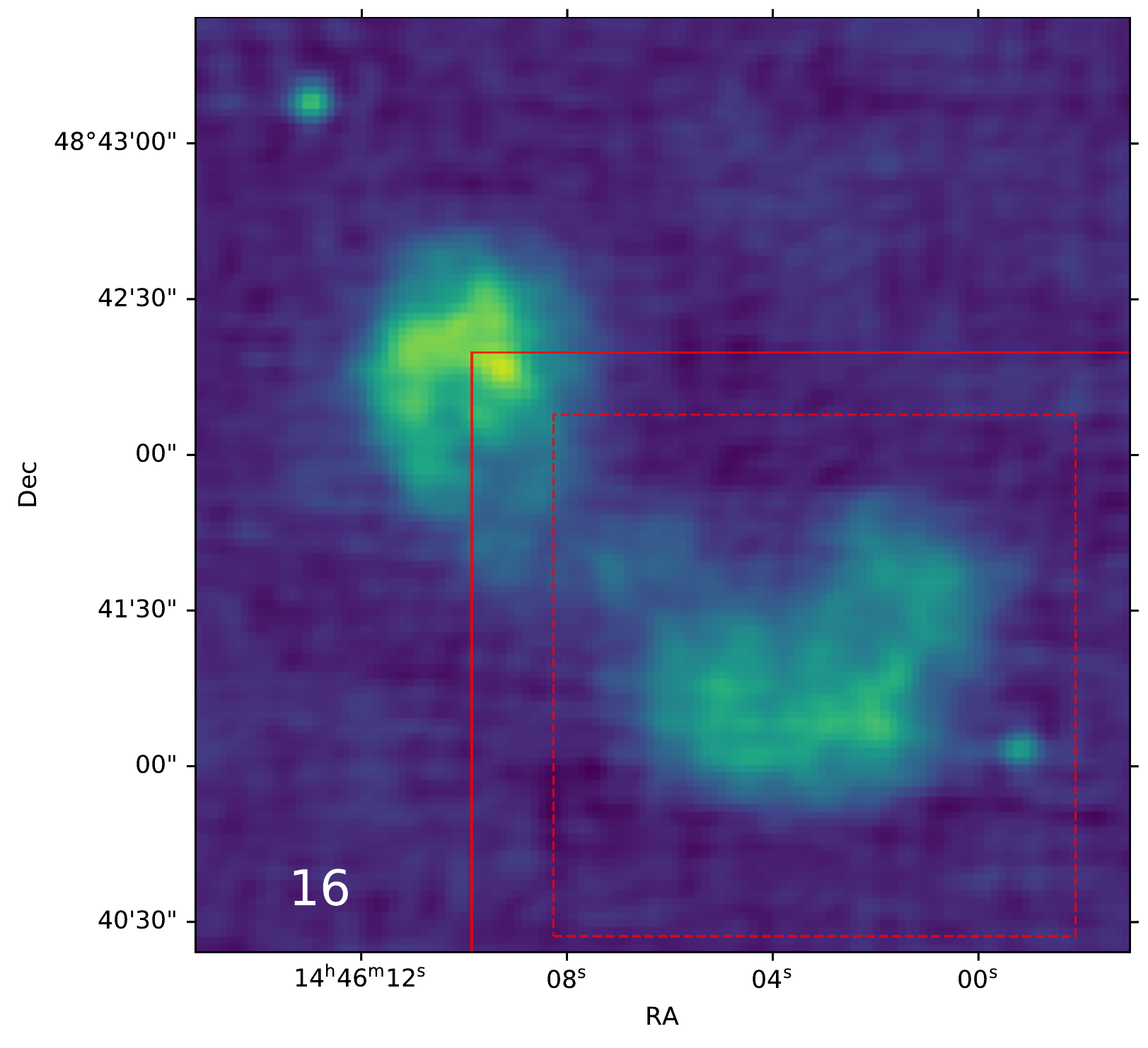}
\includegraphics[width=0.19\textwidth]{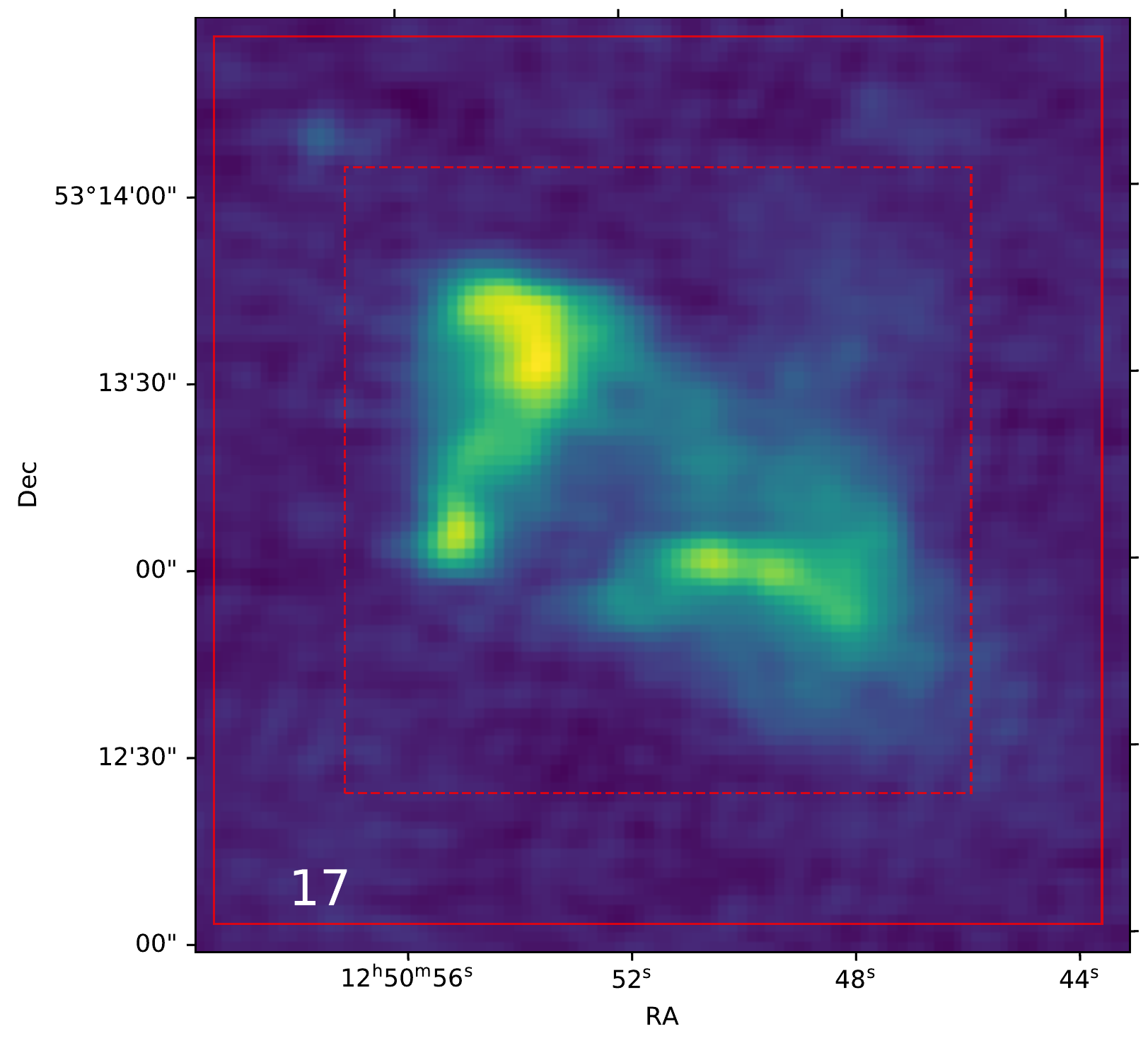}
\includegraphics[width=0.19\textwidth]{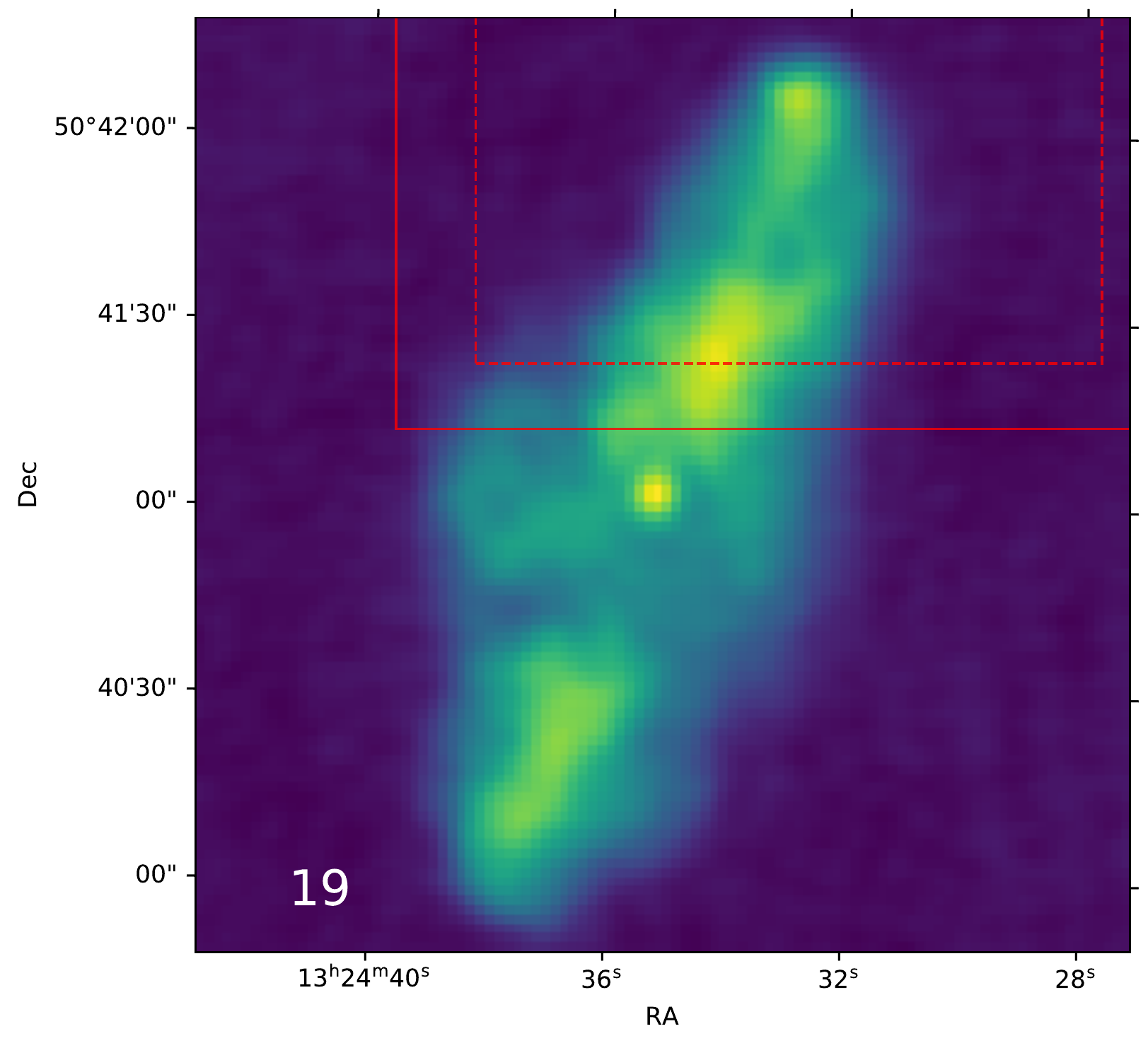}
\includegraphics[width=0.19\textwidth]{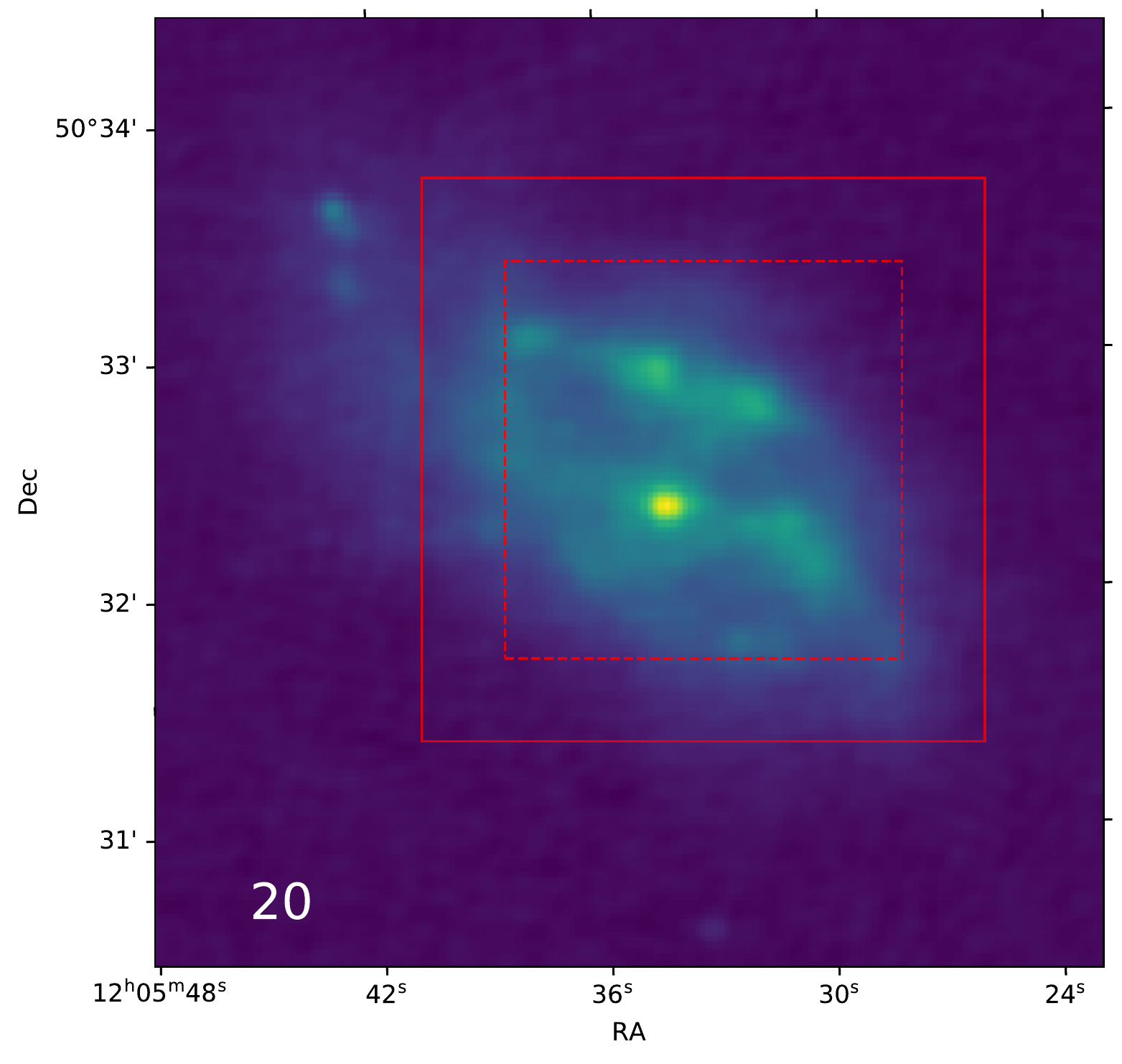}
\includegraphics[width=0.19\textwidth]{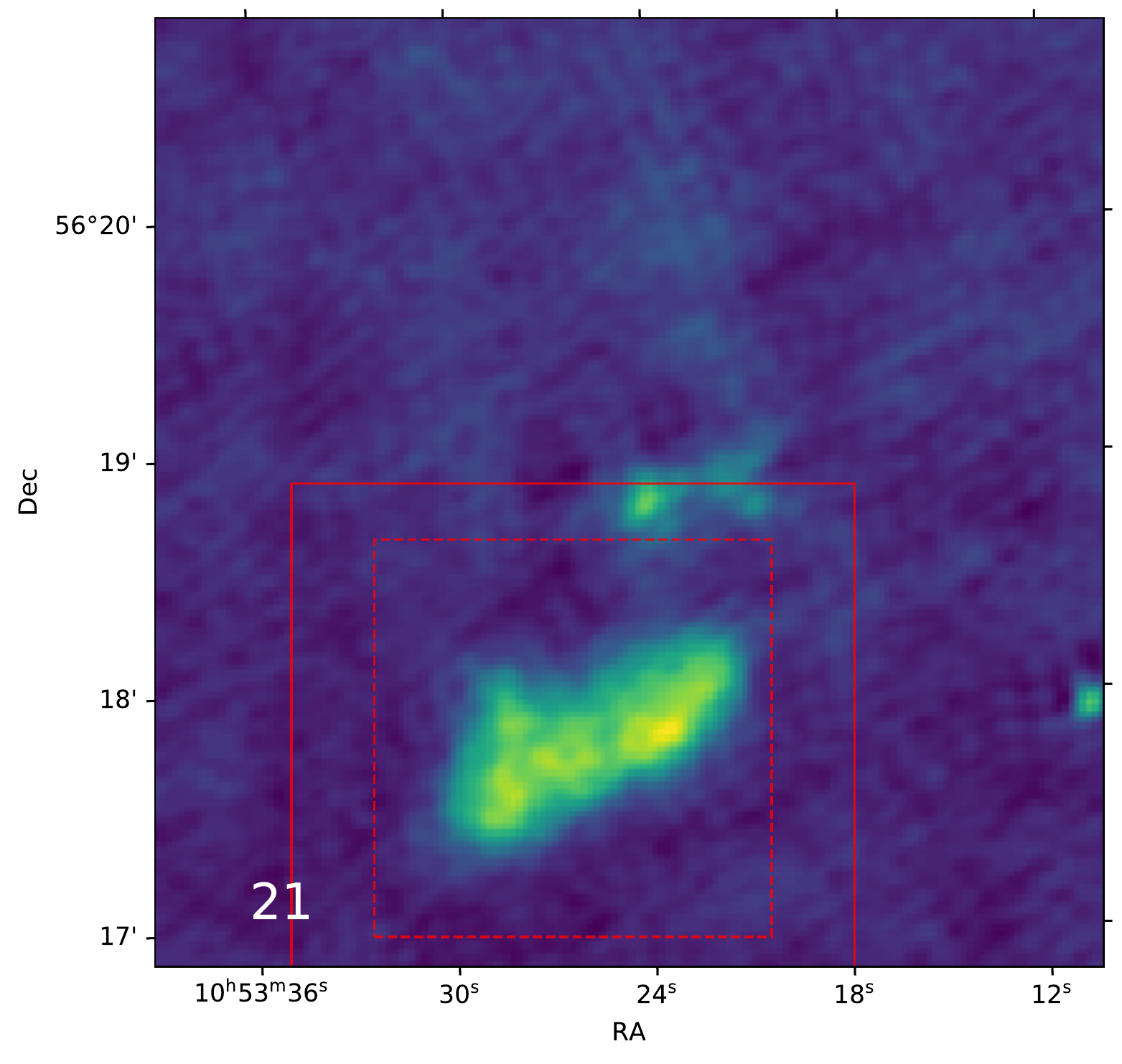}
\includegraphics[width=0.19\textwidth]{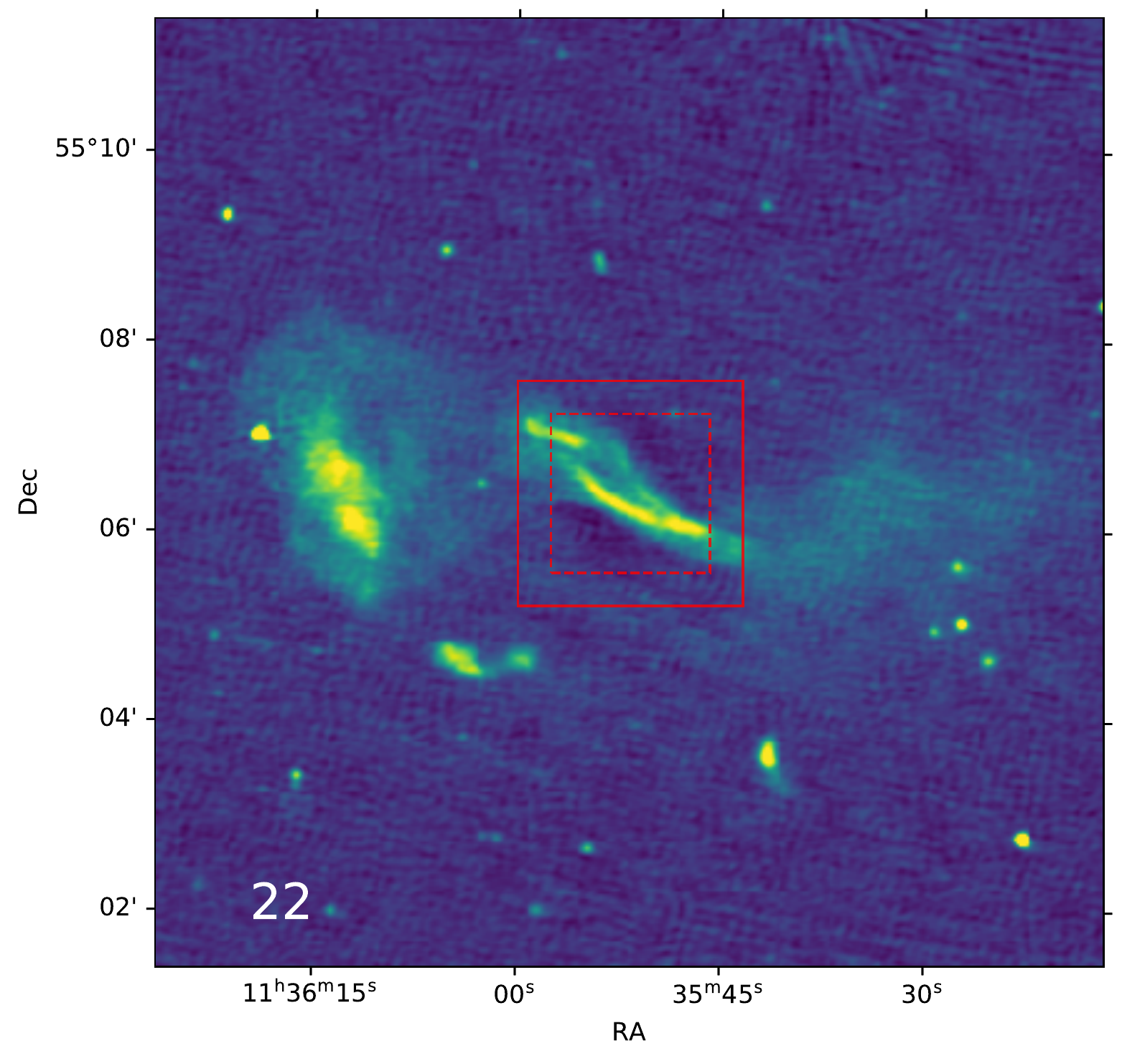}
\includegraphics[width=0.19\textwidth]{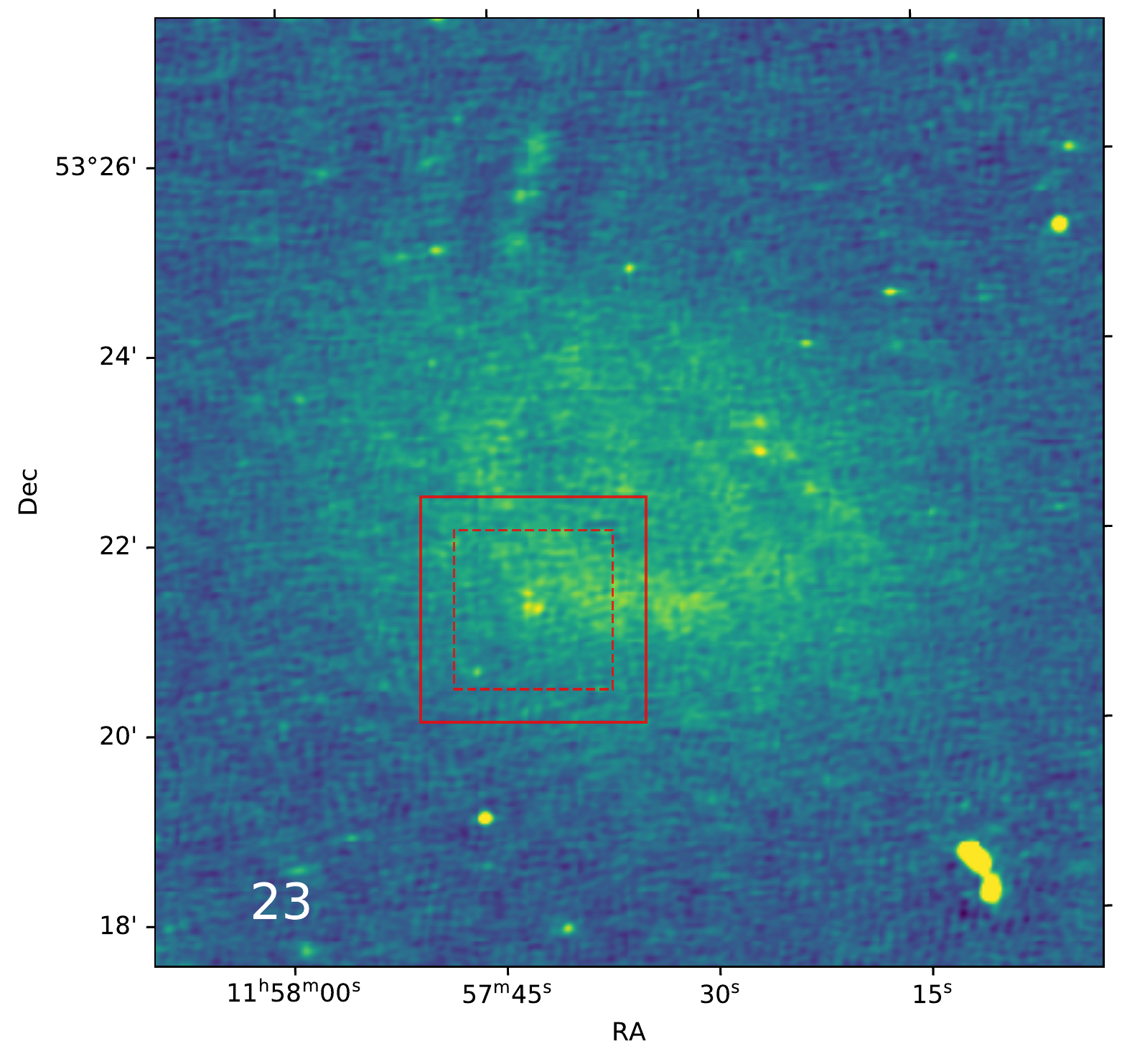}
\includegraphics[width=0.19\textwidth]{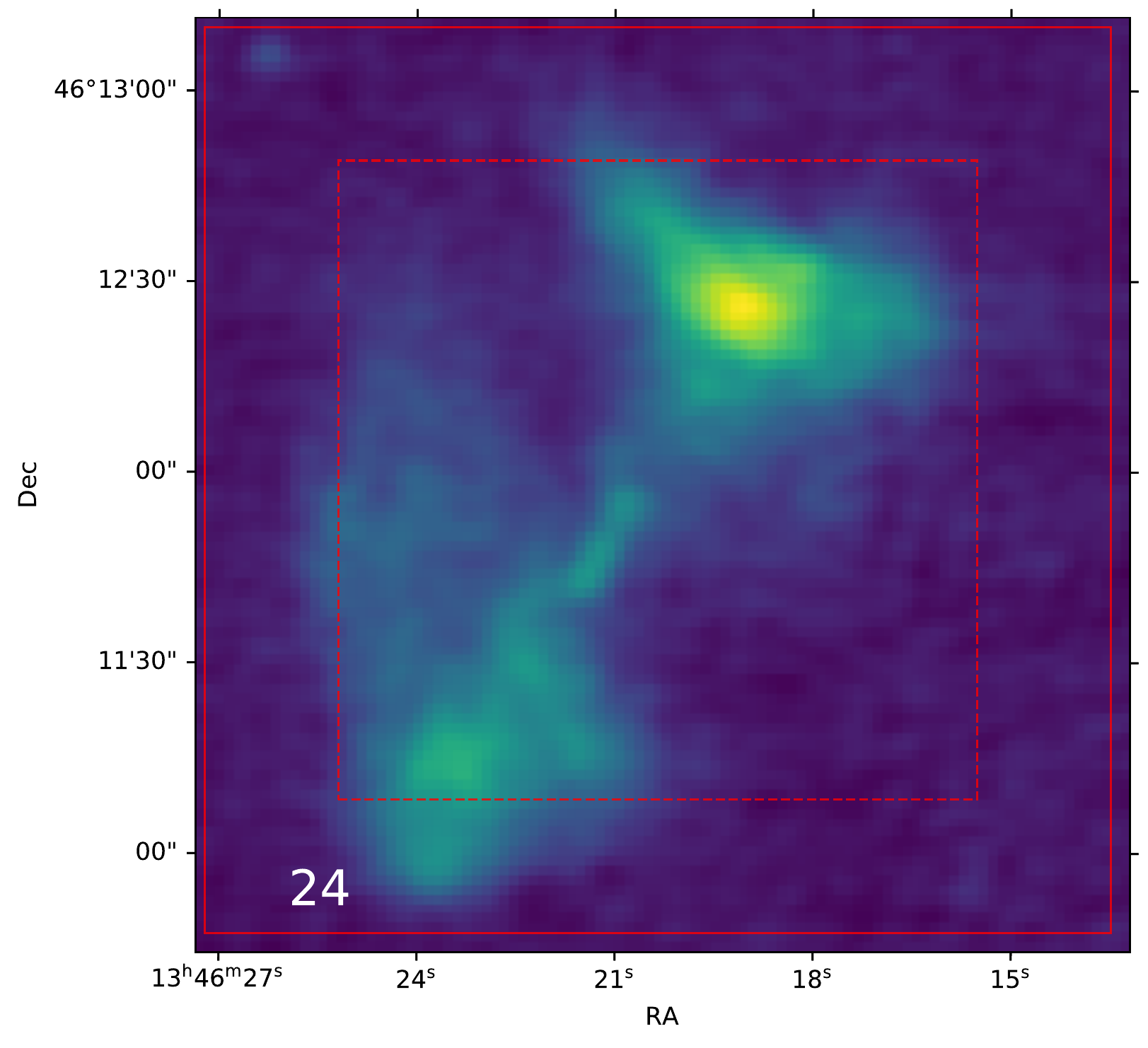}
\includegraphics[width=0.19\textwidth]{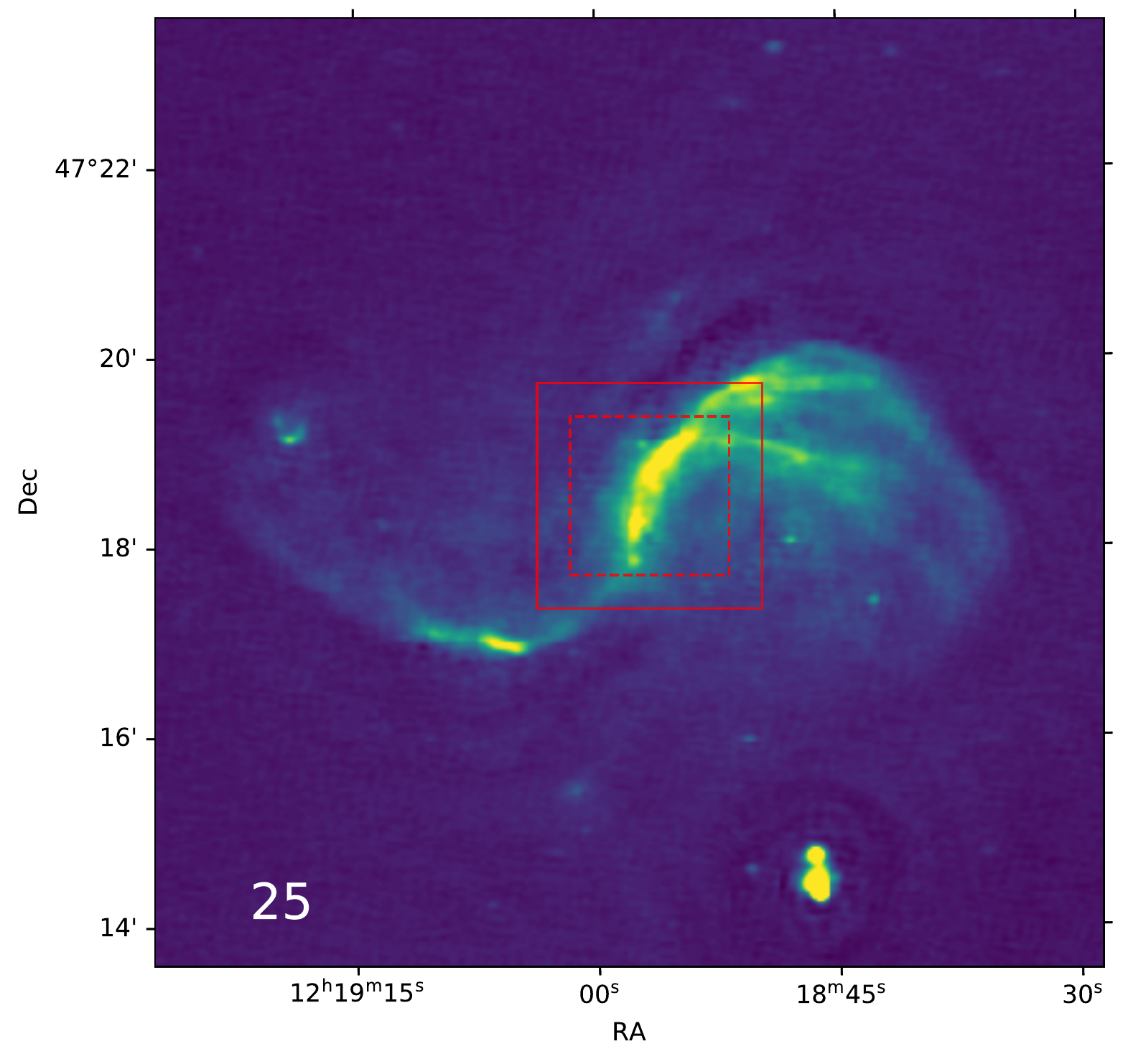}
\includegraphics[width=0.19\textwidth]{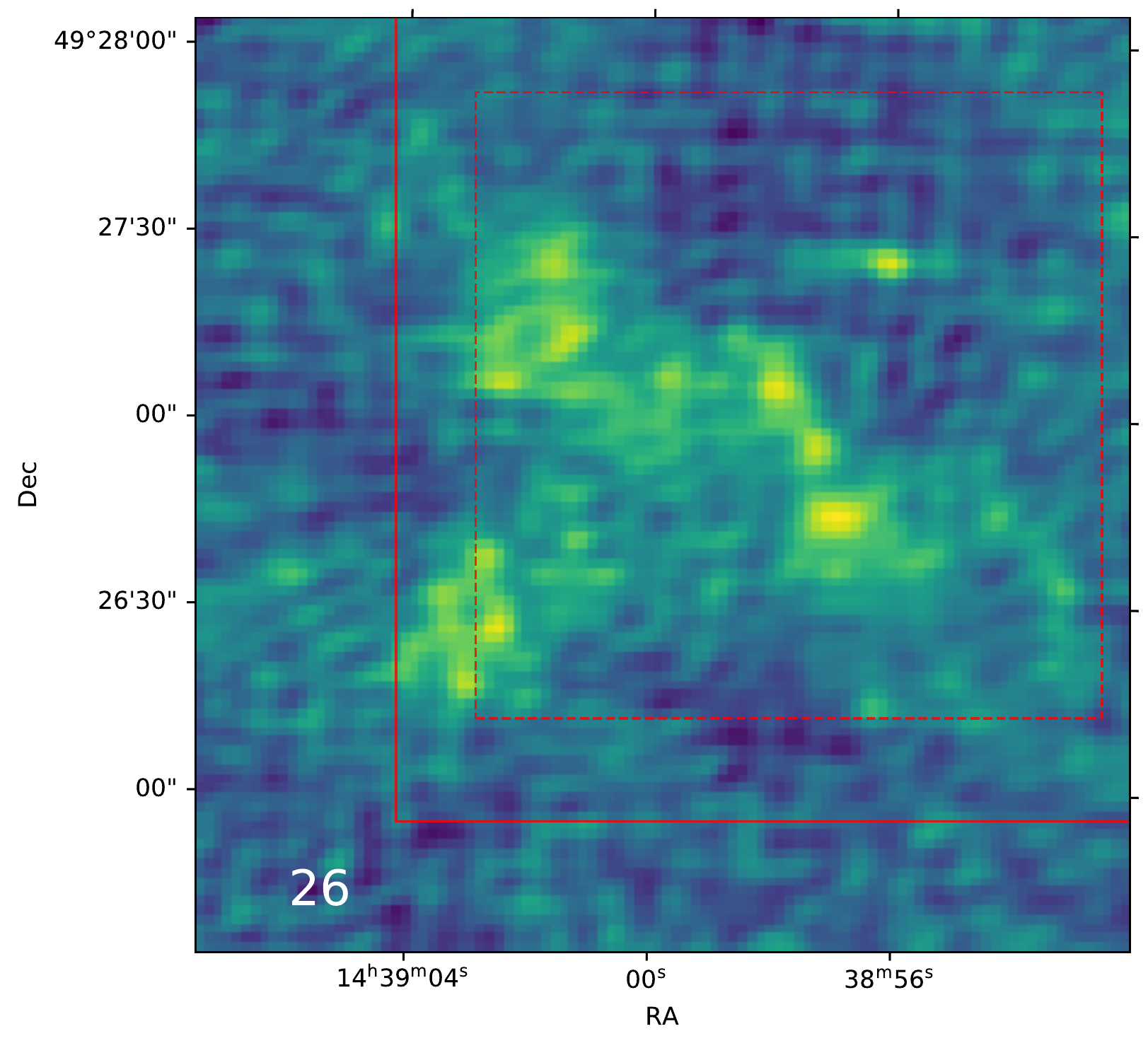}
\includegraphics[width=0.19\textwidth]{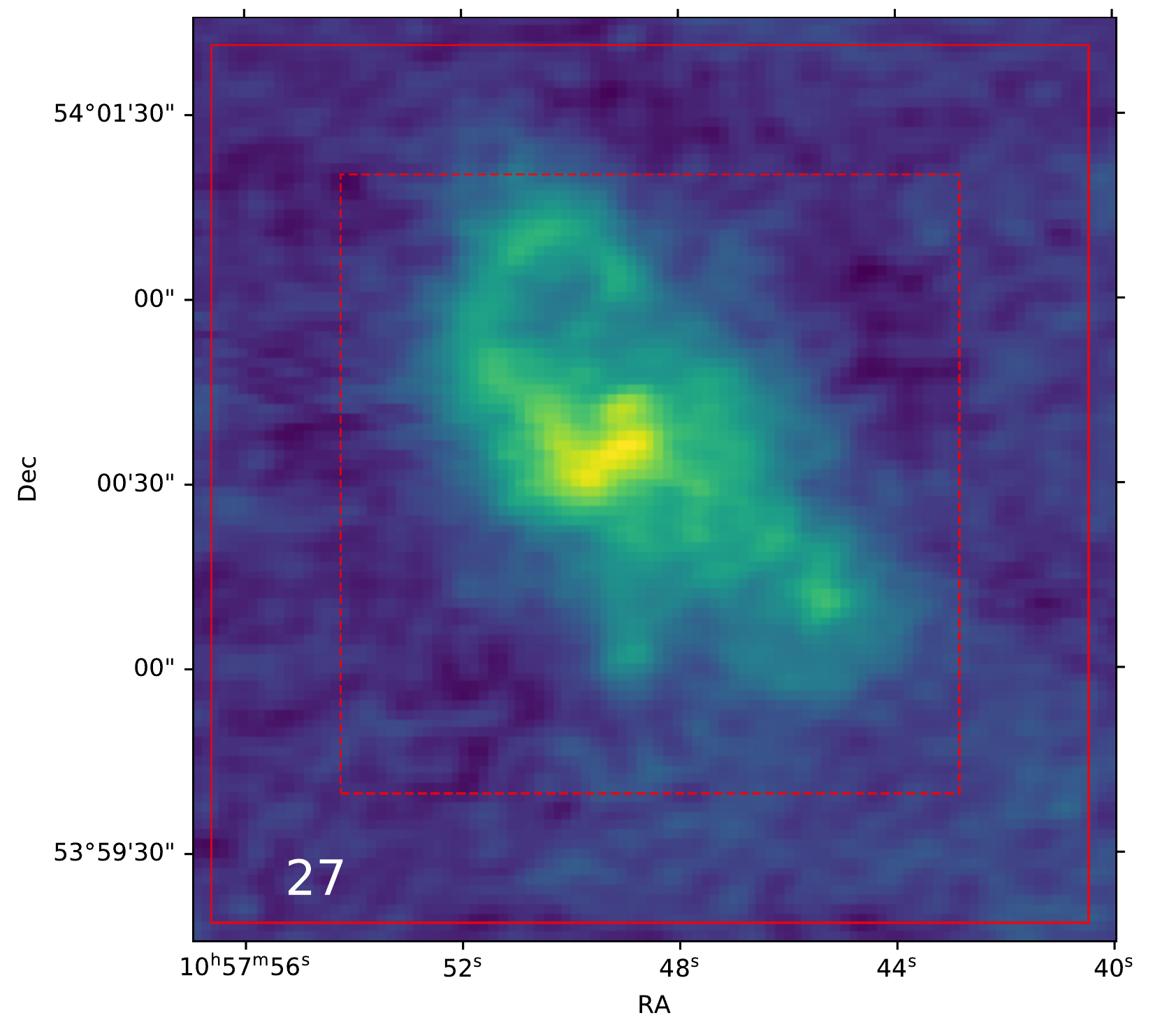}
\includegraphics[width=0.19\textwidth]{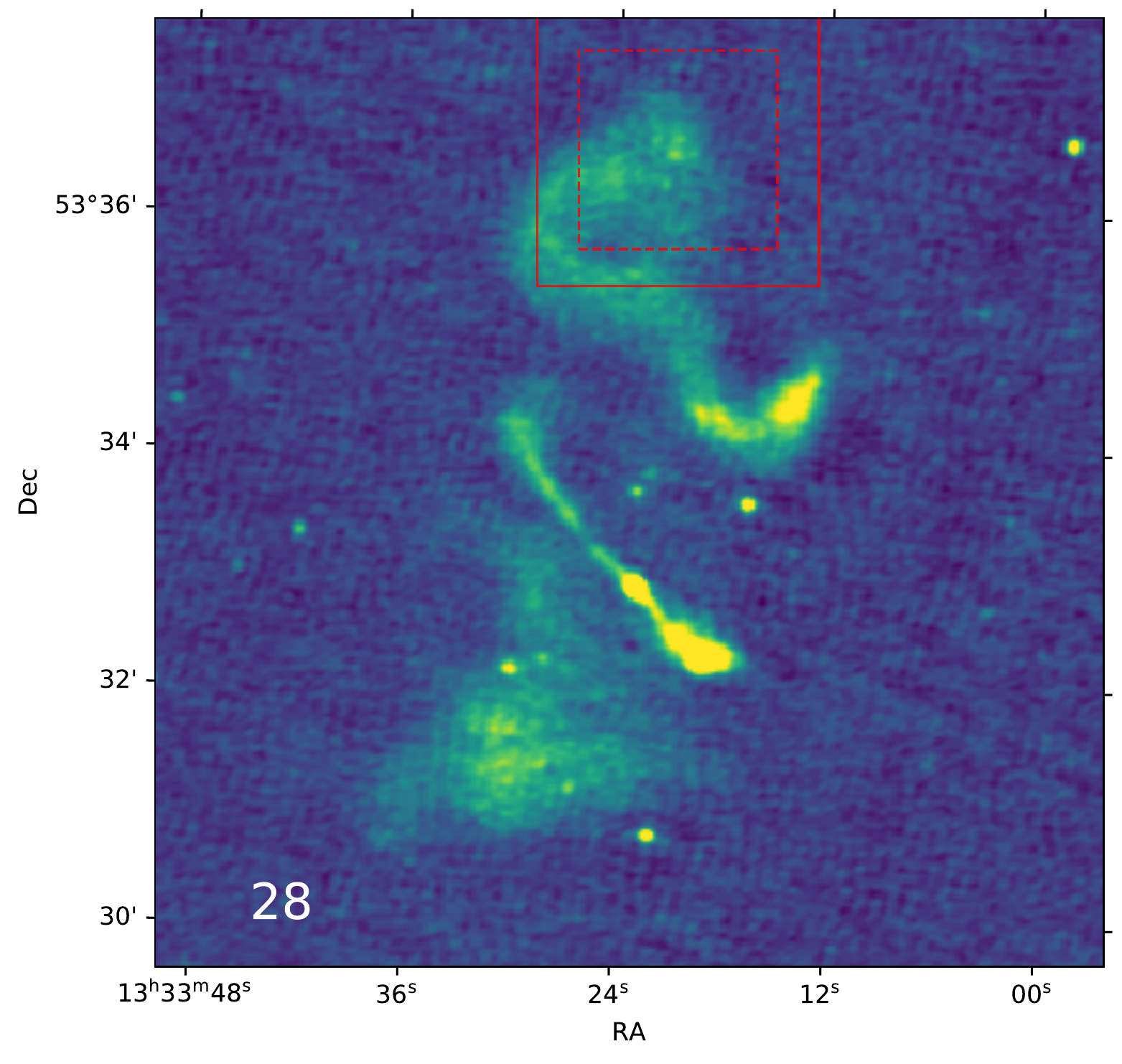}
\includegraphics[width=0.19\textwidth]{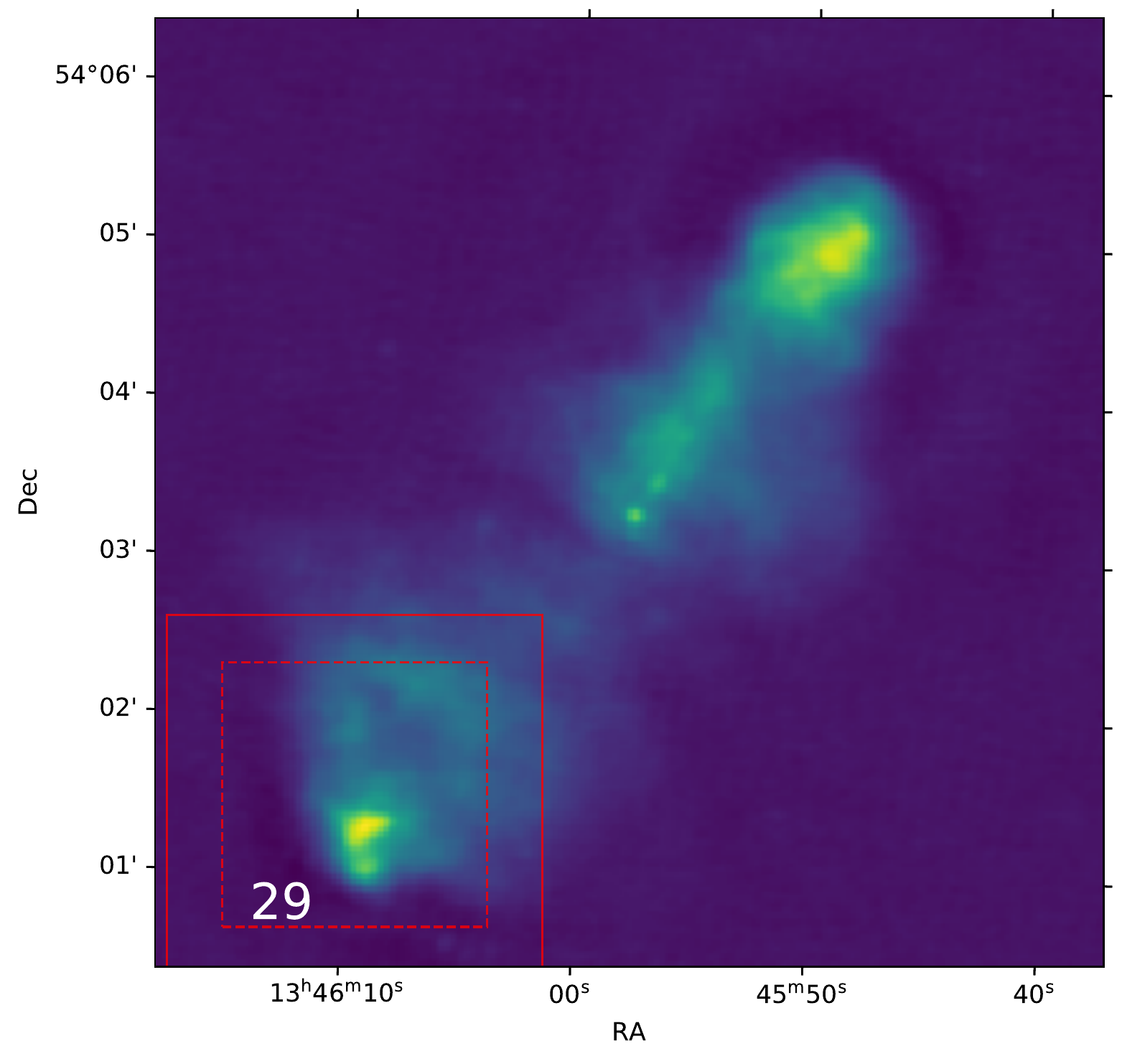}
\includegraphics[width=0.19\textwidth]{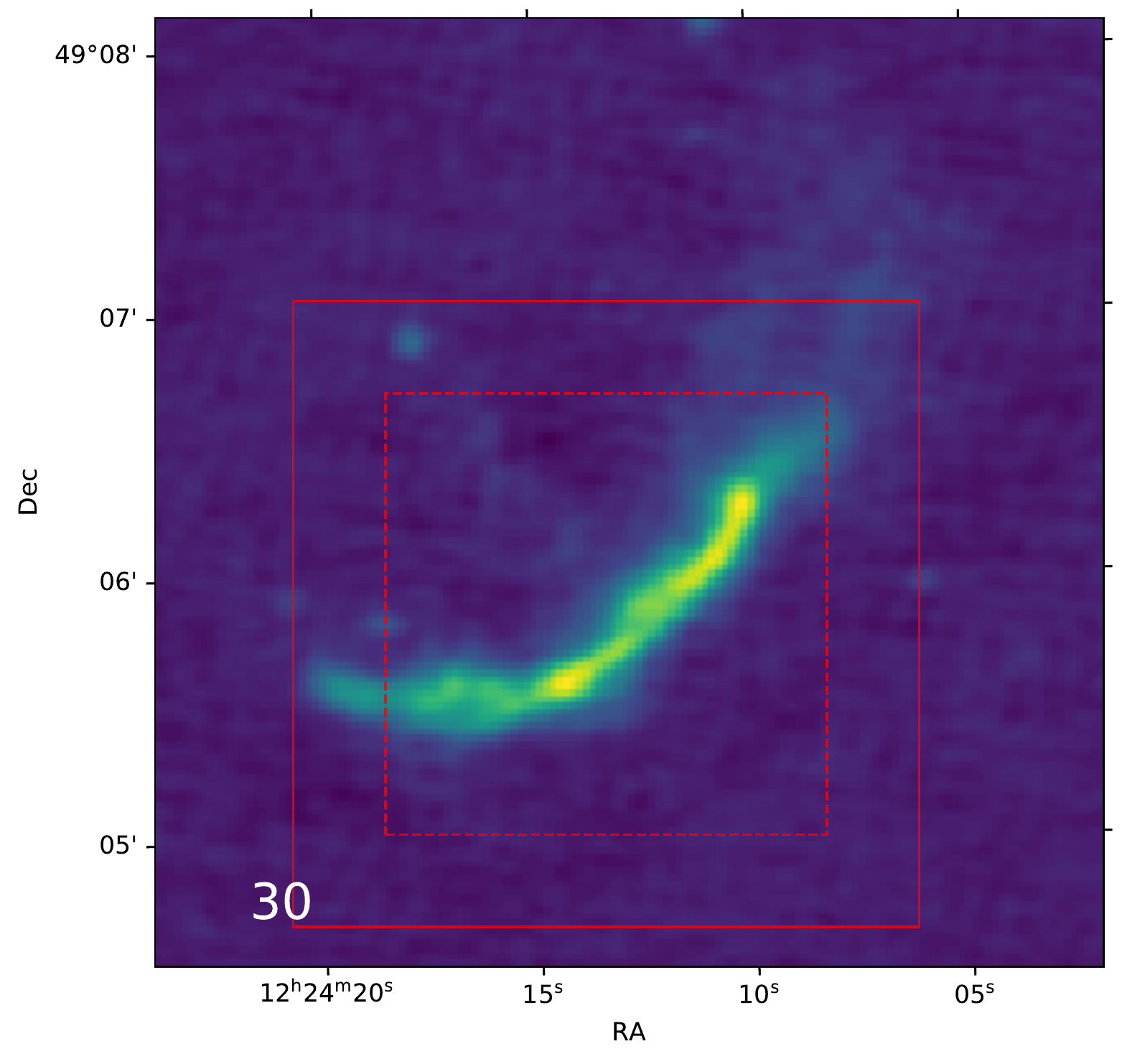}
\includegraphics[width=0.19\textwidth]{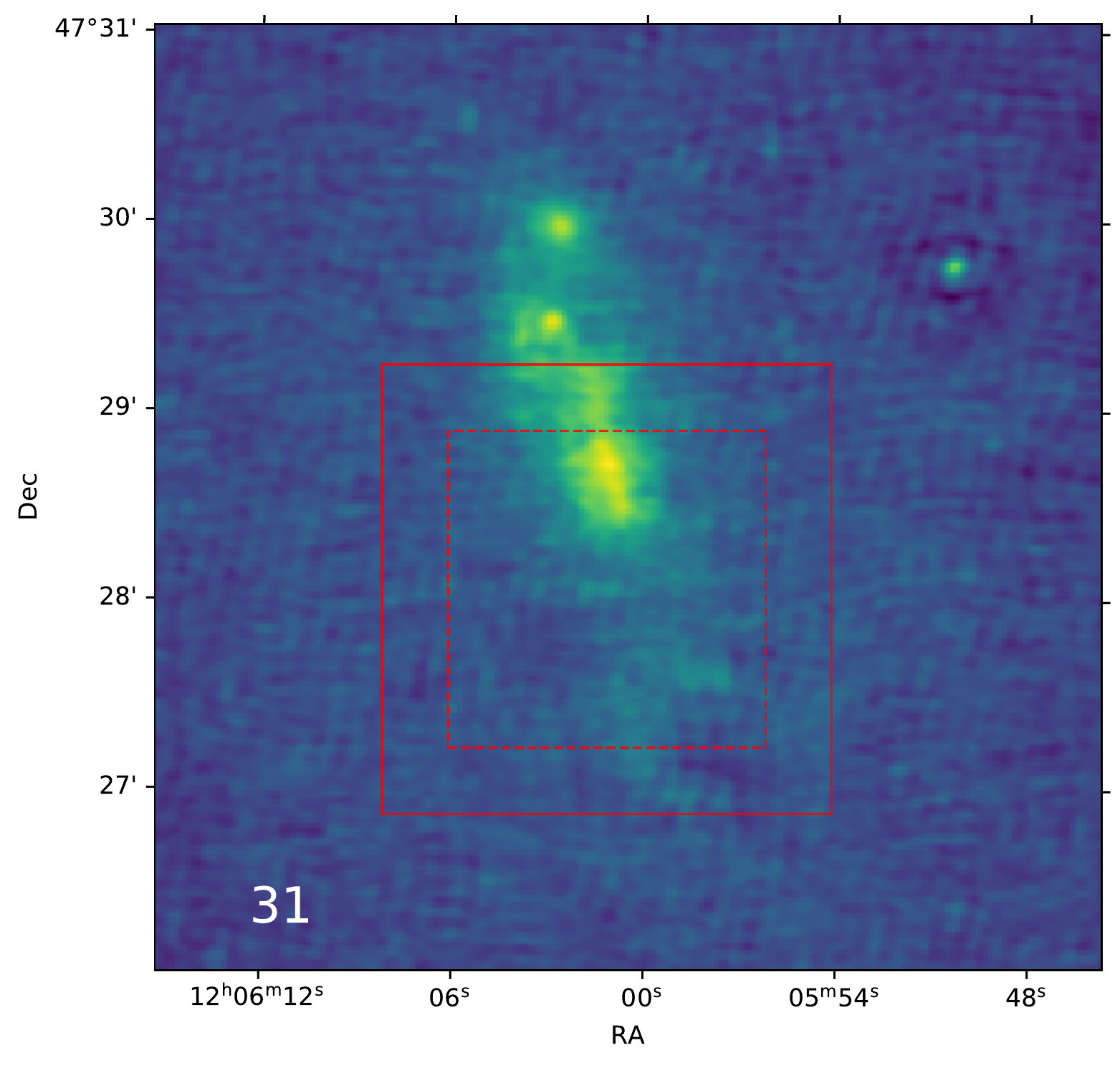}
\includegraphics[width=0.19\textwidth]{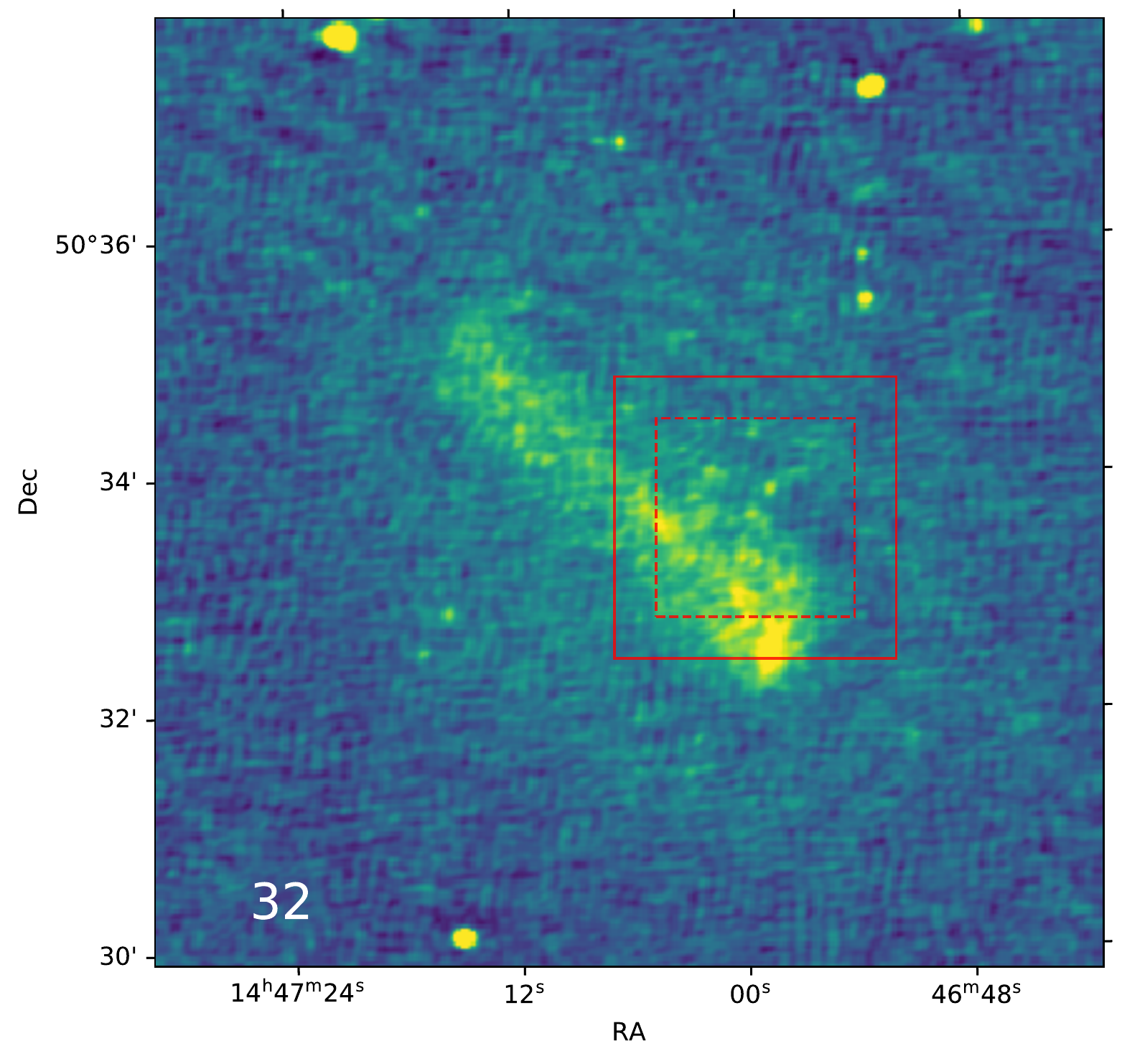}
\includegraphics[width=0.19\textwidth]{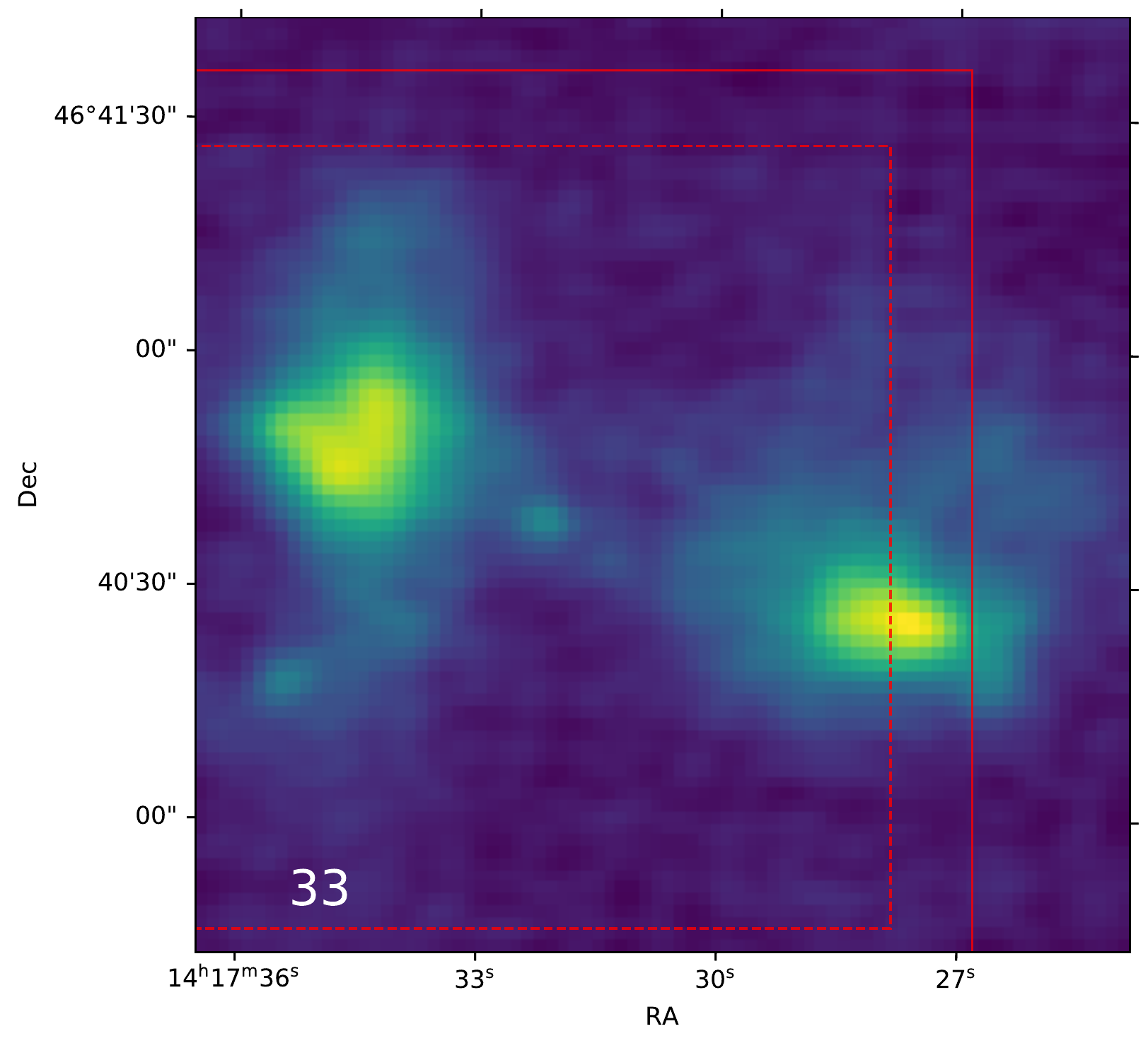}
\includegraphics[width=0.19\textwidth]{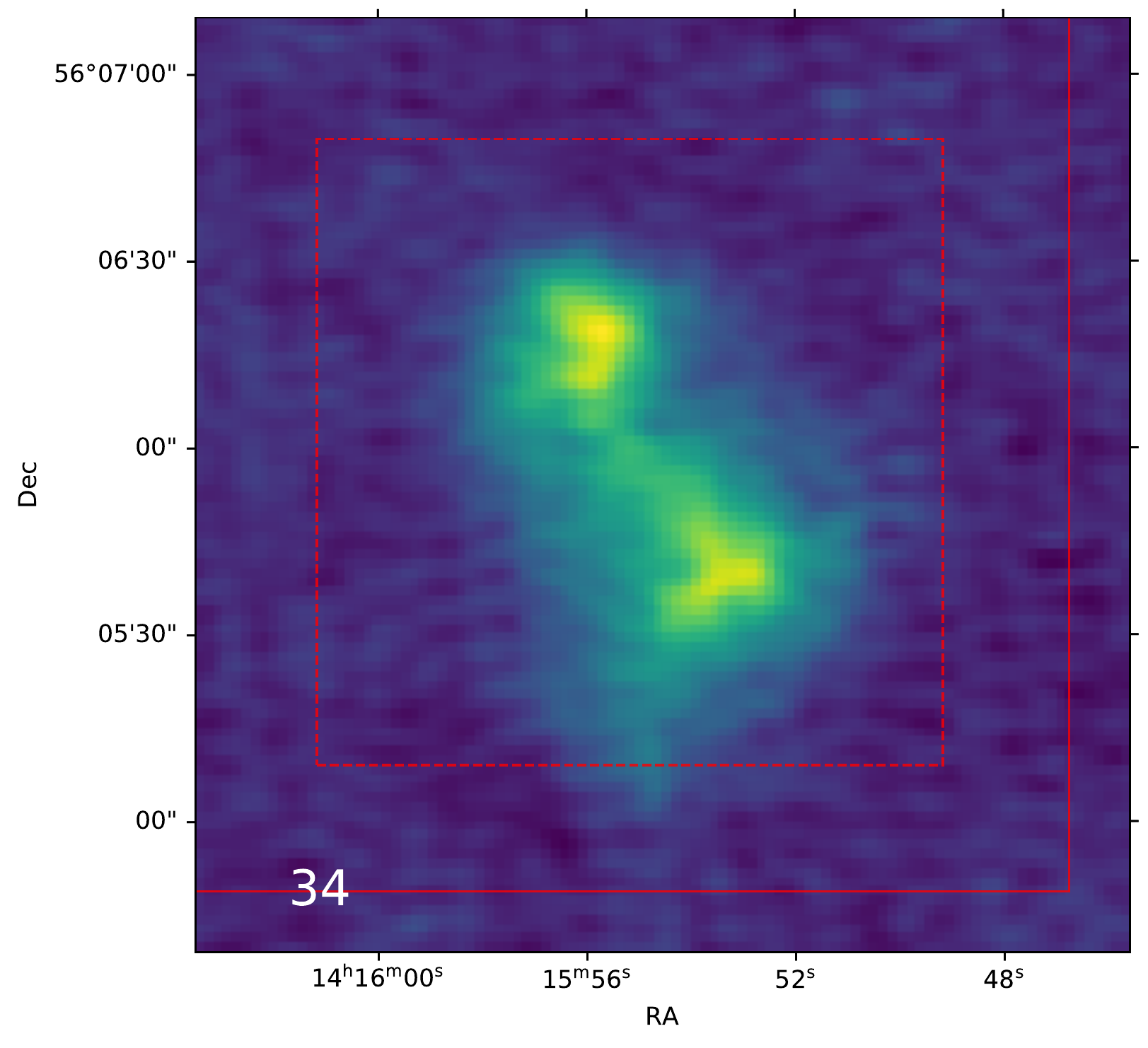}
\includegraphics[width=0.19\textwidth]{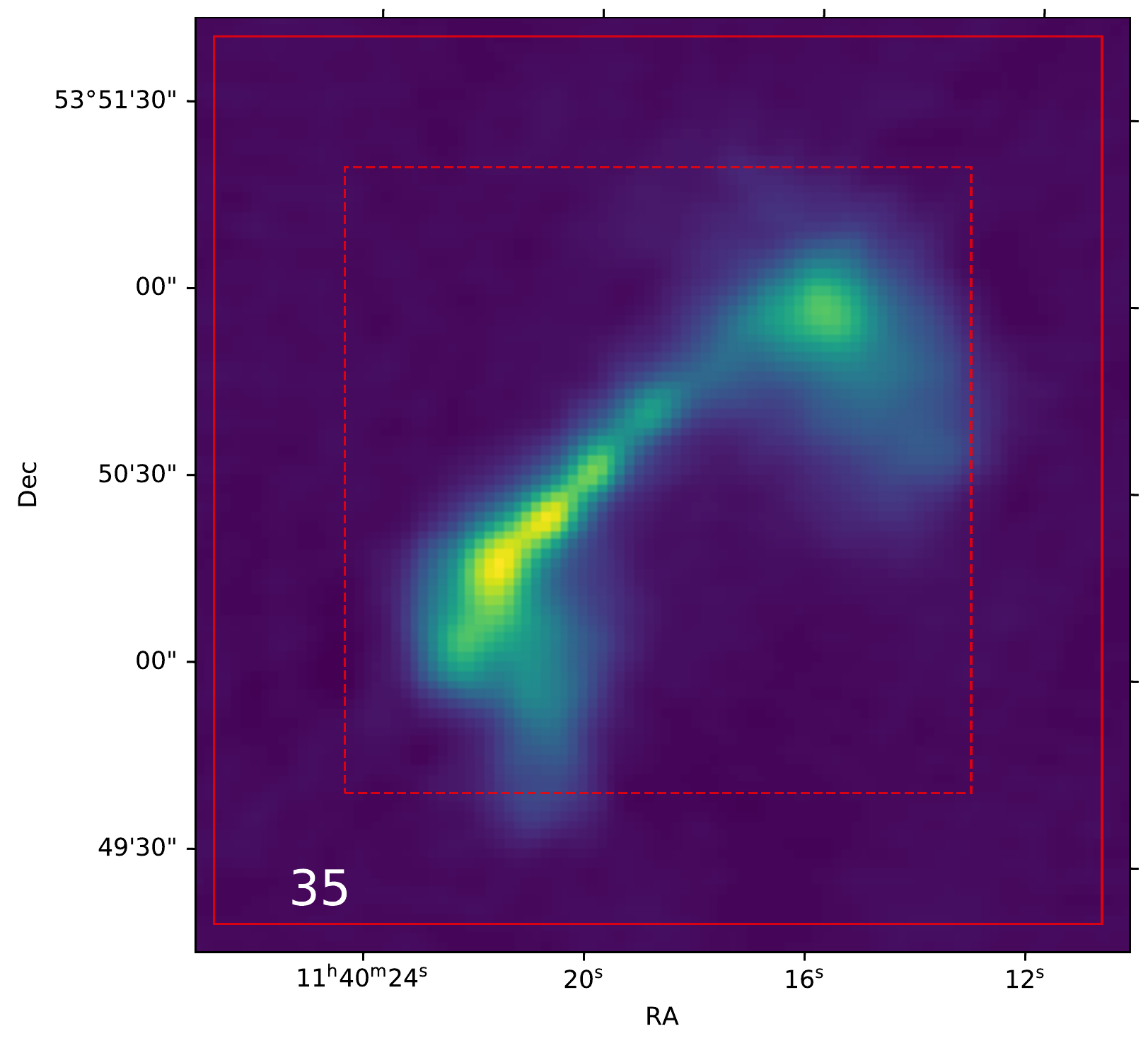}
\includegraphics[width=0.19\textwidth]{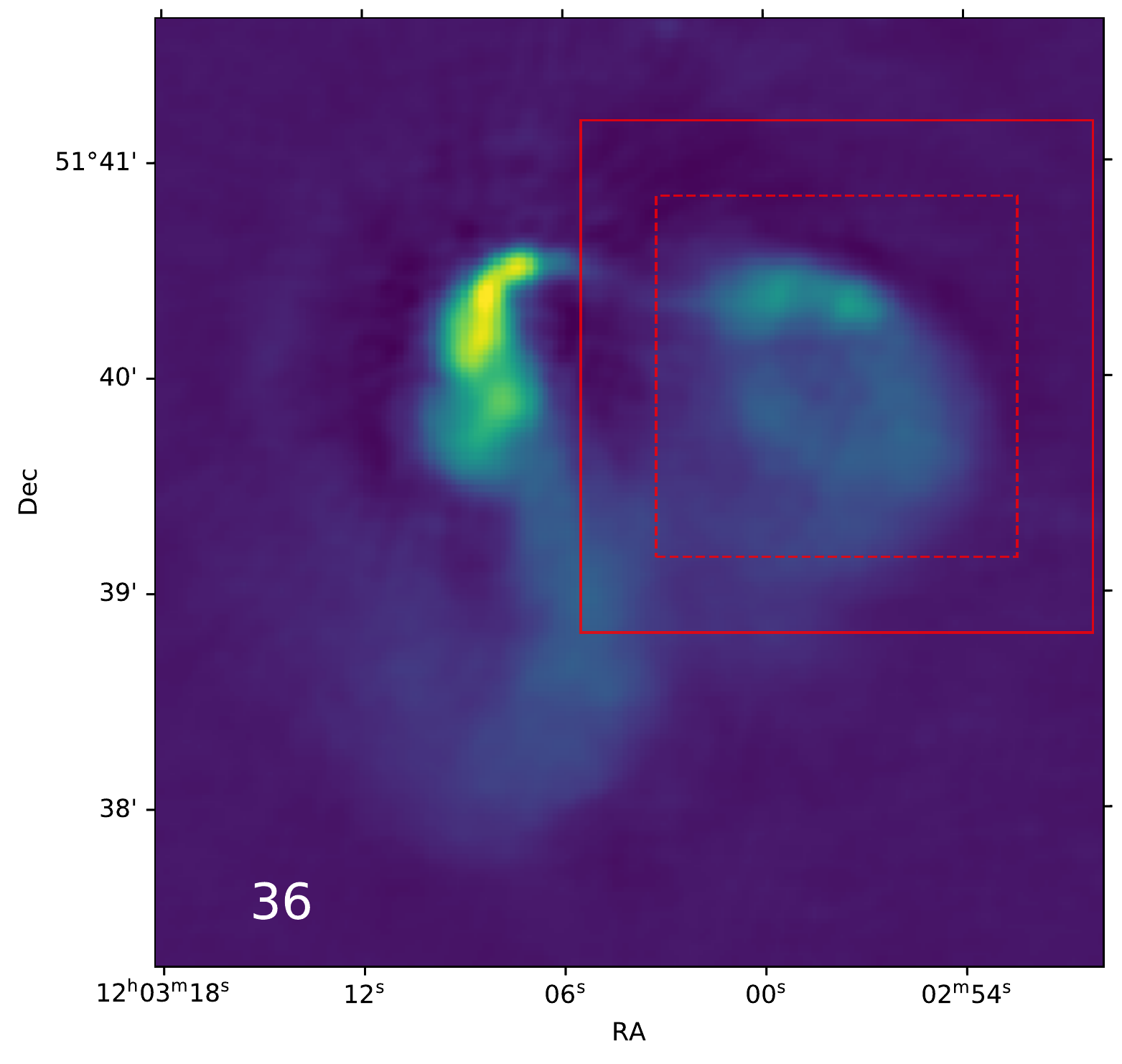}
\caption{100 morphological outliers. The image shows cutouts from the LoTSS survey for sources detected based on the large dissimilarity between the source and the trained self-organising map: we picked the 100 sources that have the largest Euclidean norm to their best matching representative image.  We note that four sources were detected multiple times.}
\label{fig:100outliers1}
\end{center}\end{figure*}

\begin{figure*}\begin{center}
\ContinuedFloat

\includegraphics[width=0.19\textwidth]{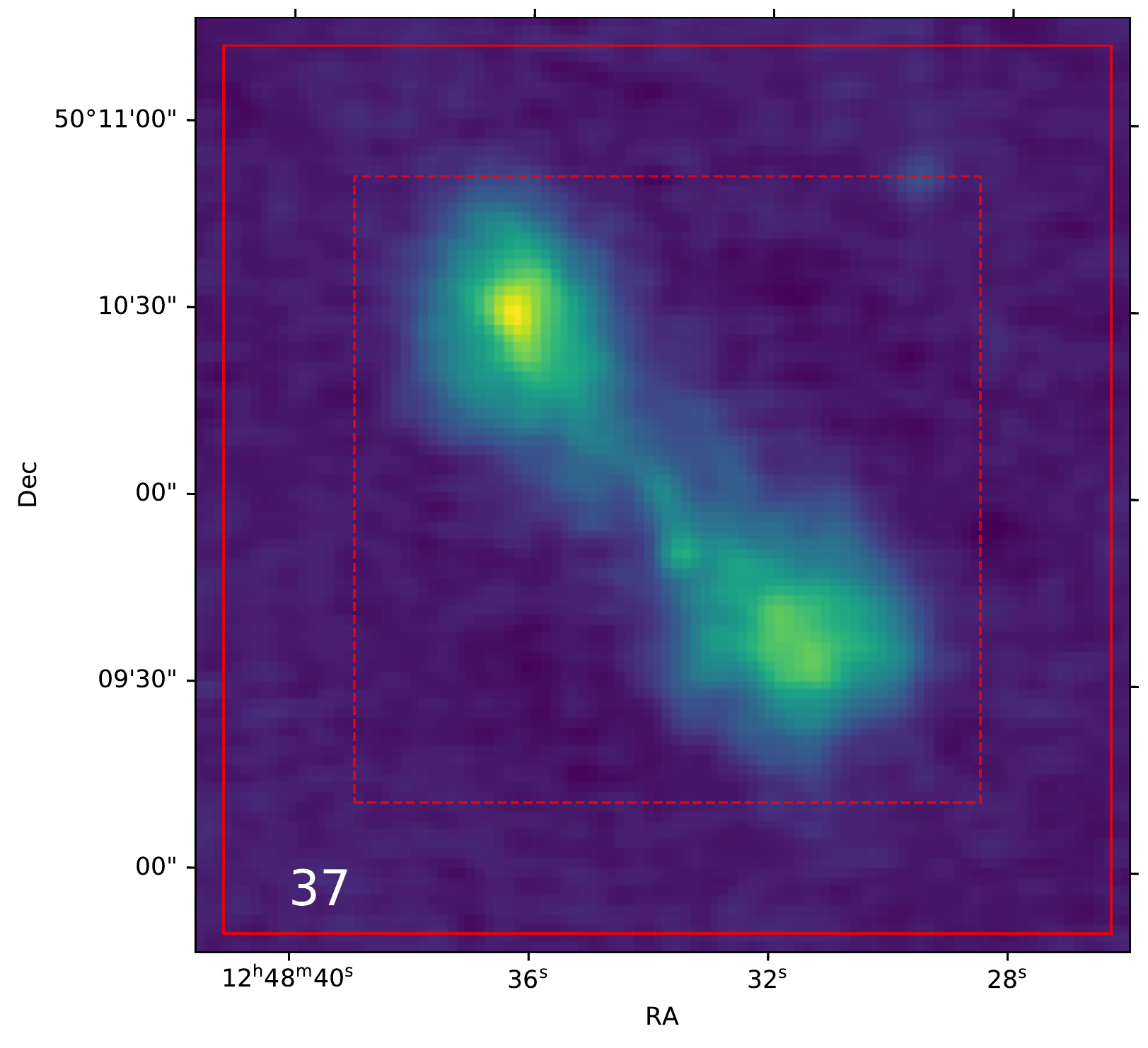}
\includegraphics[width=0.19\textwidth]{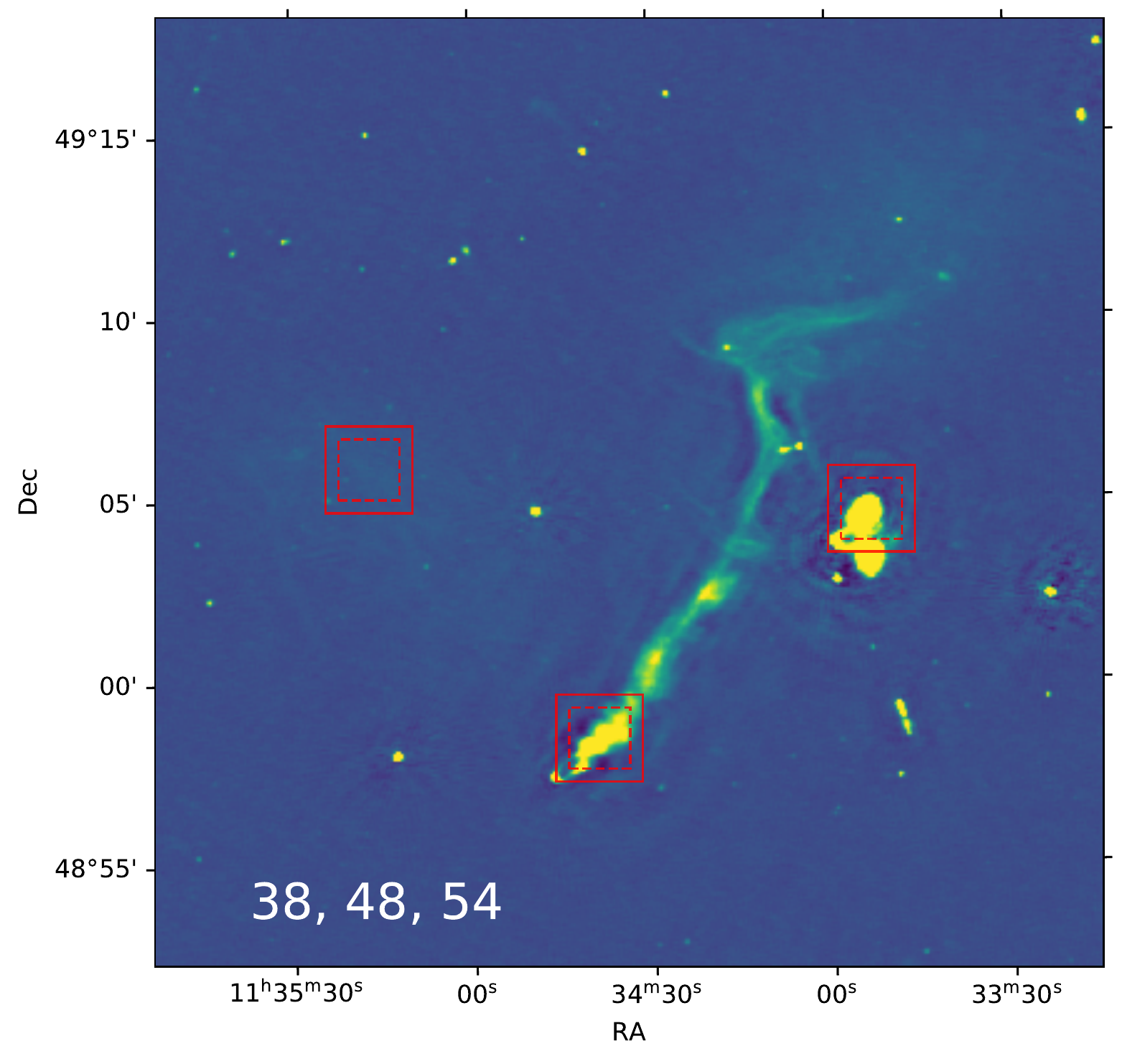}
\includegraphics[width=0.19\textwidth]{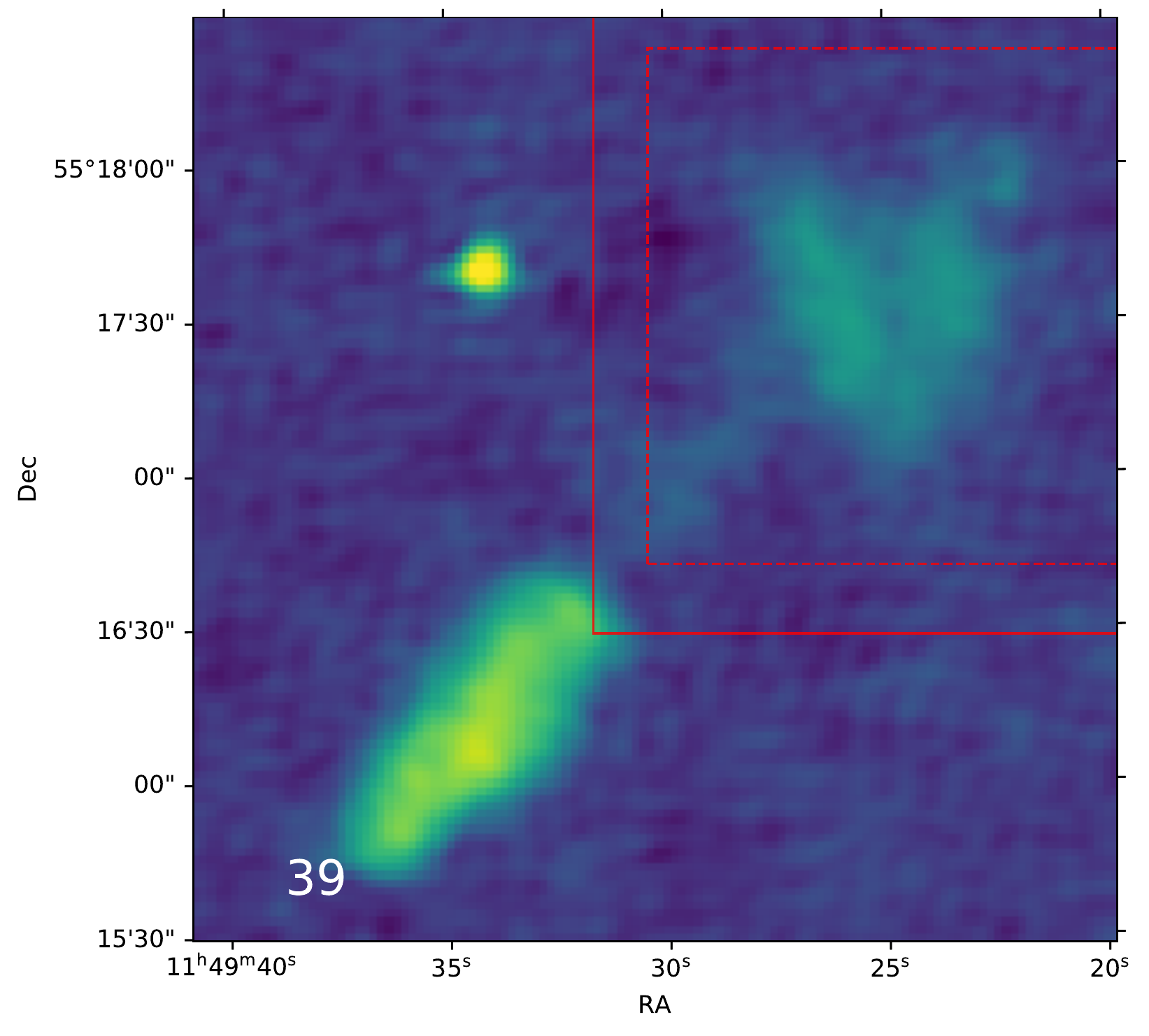}
\includegraphics[width=0.19\textwidth]{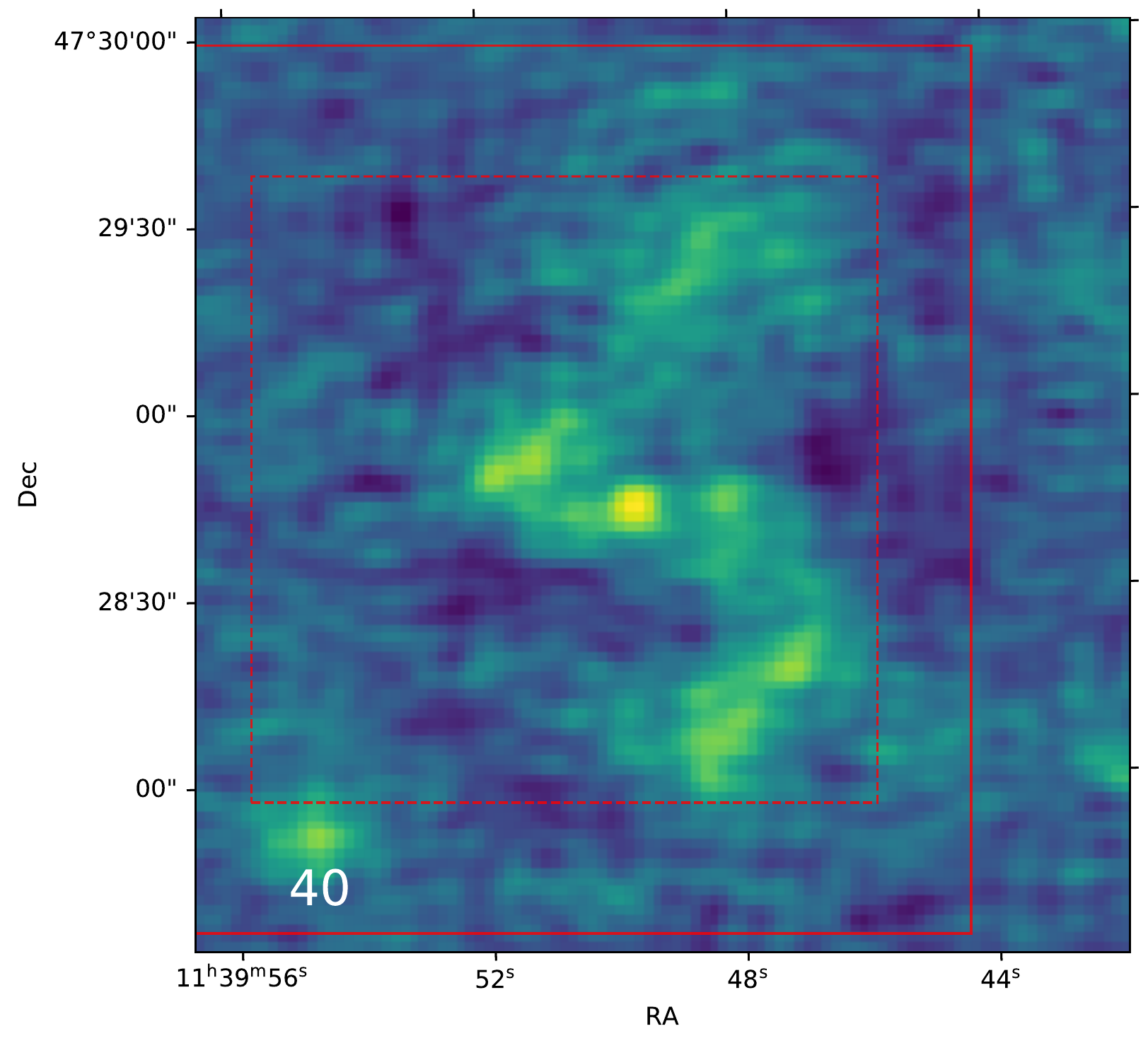}
\includegraphics[width=0.19\textwidth]{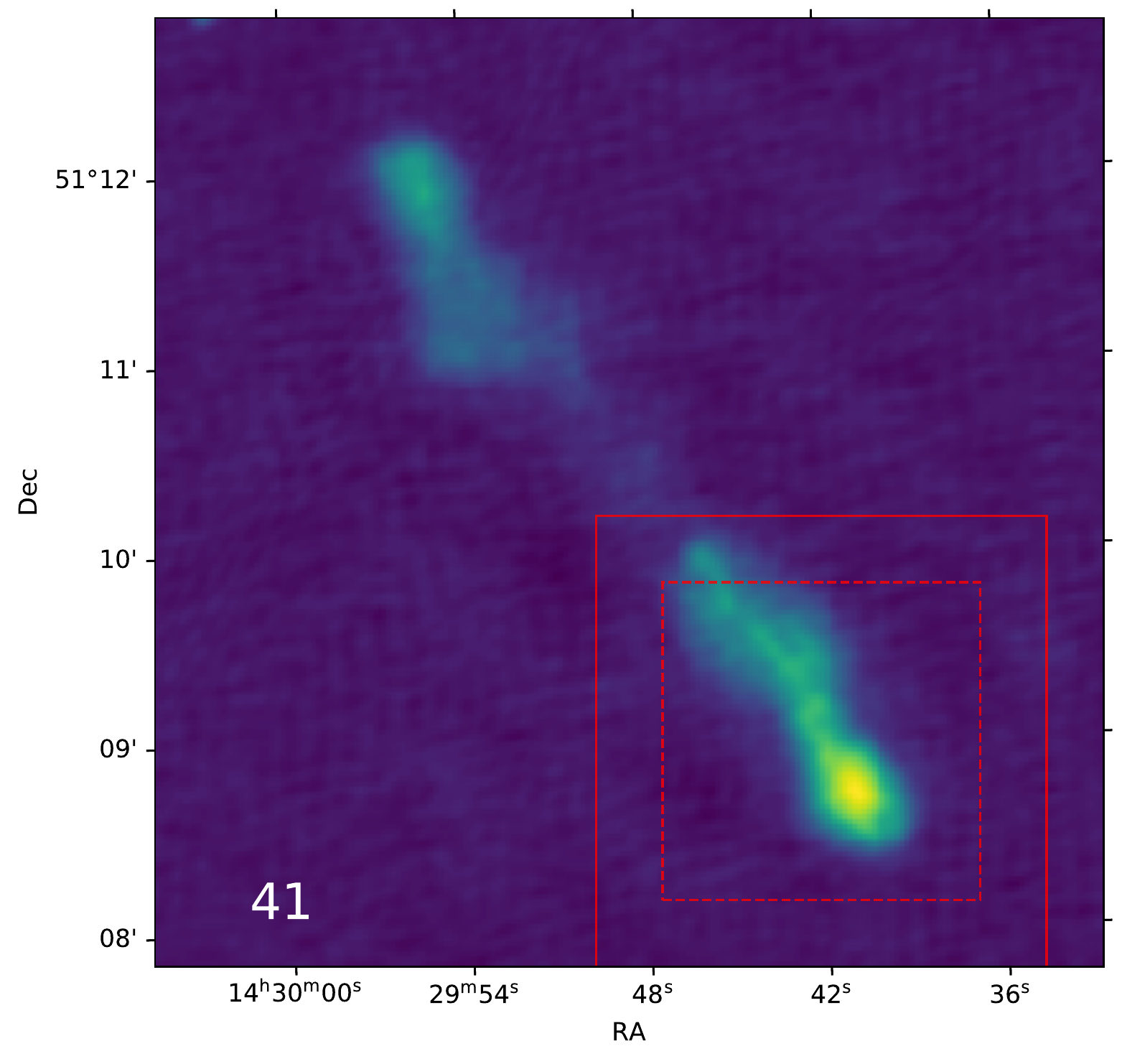}
\includegraphics[width=0.19\textwidth]{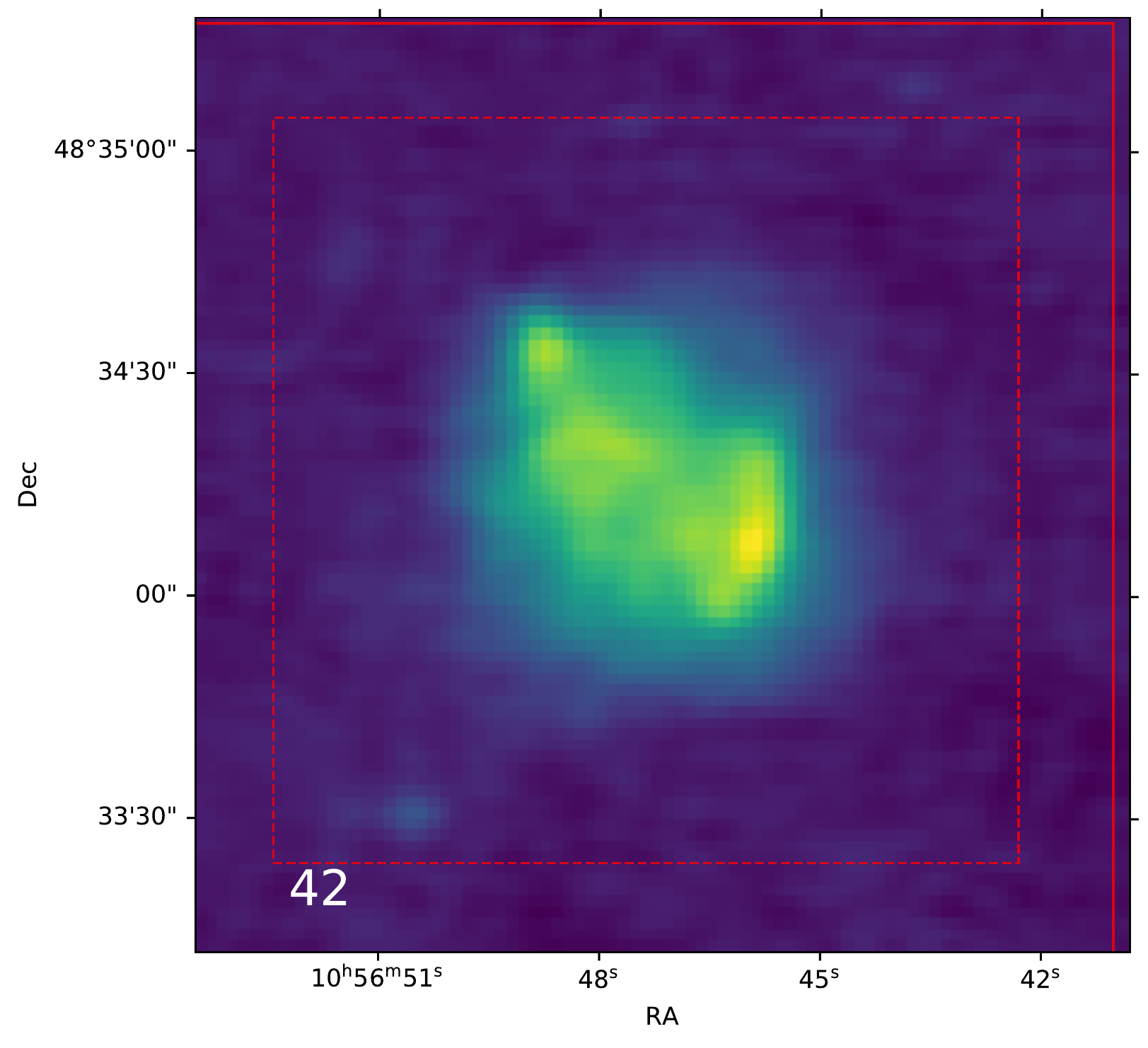}
\includegraphics[width=0.19\textwidth]{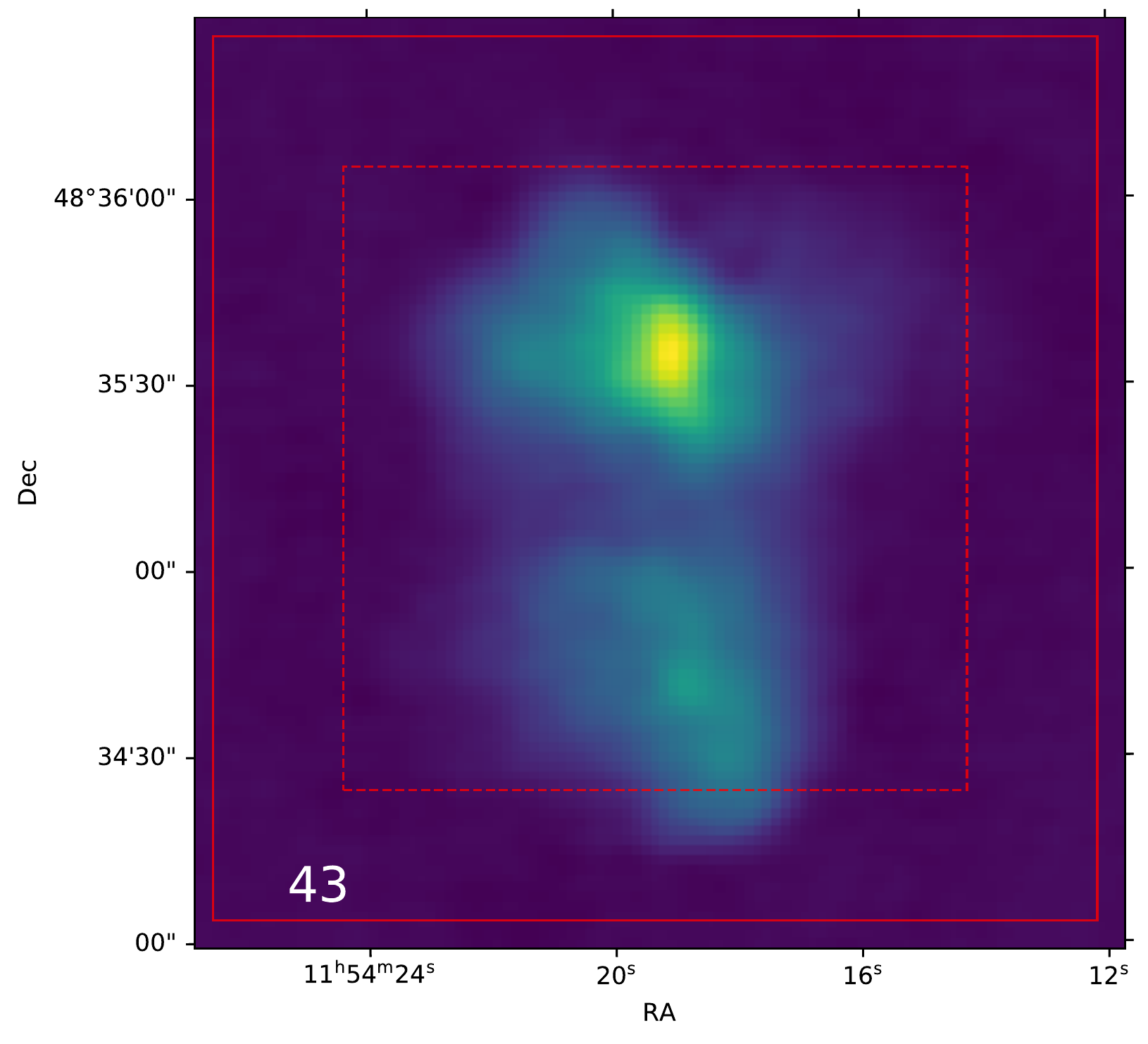}
\includegraphics[width=0.19\textwidth]{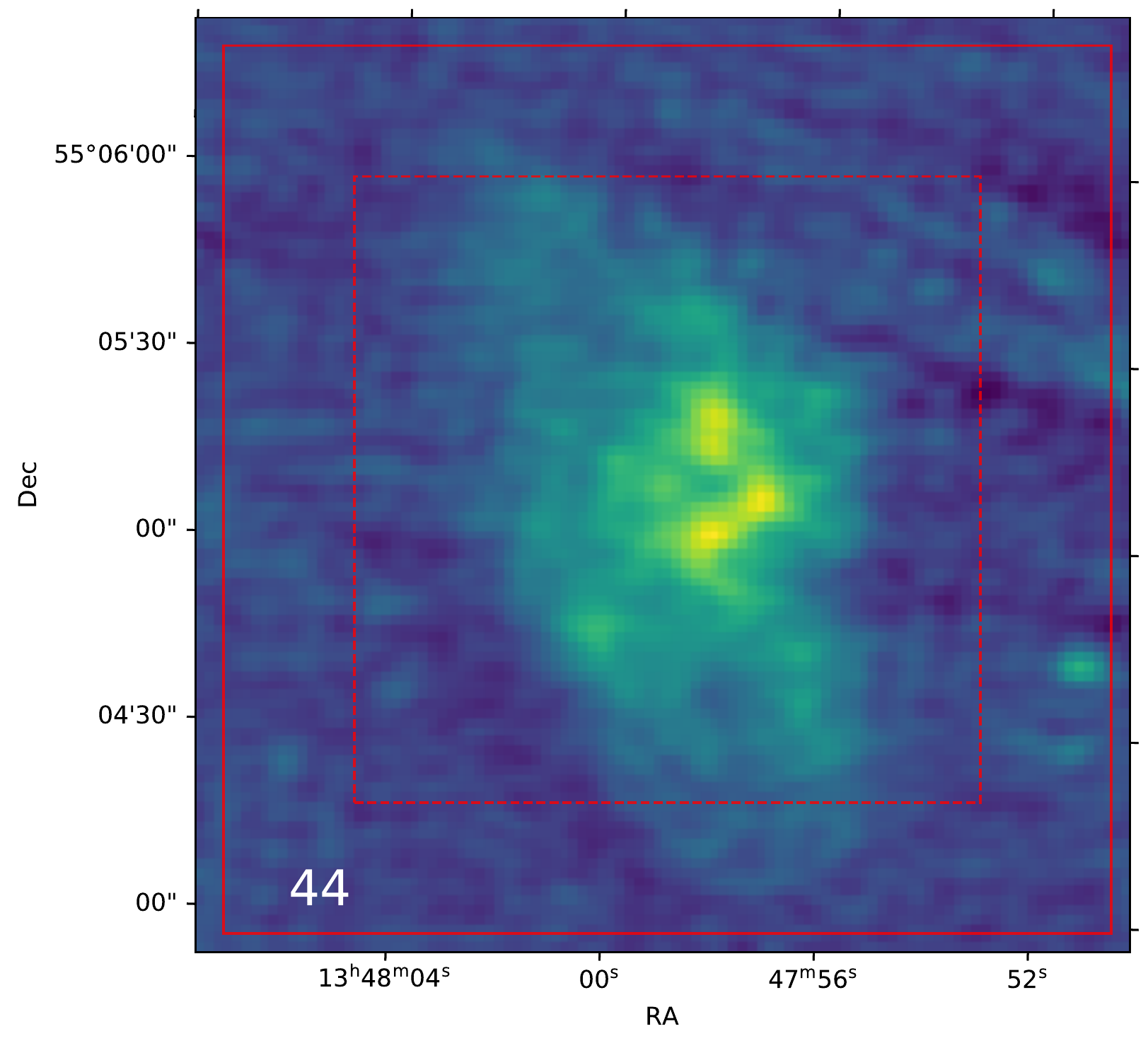}
\includegraphics[width=0.19\textwidth]{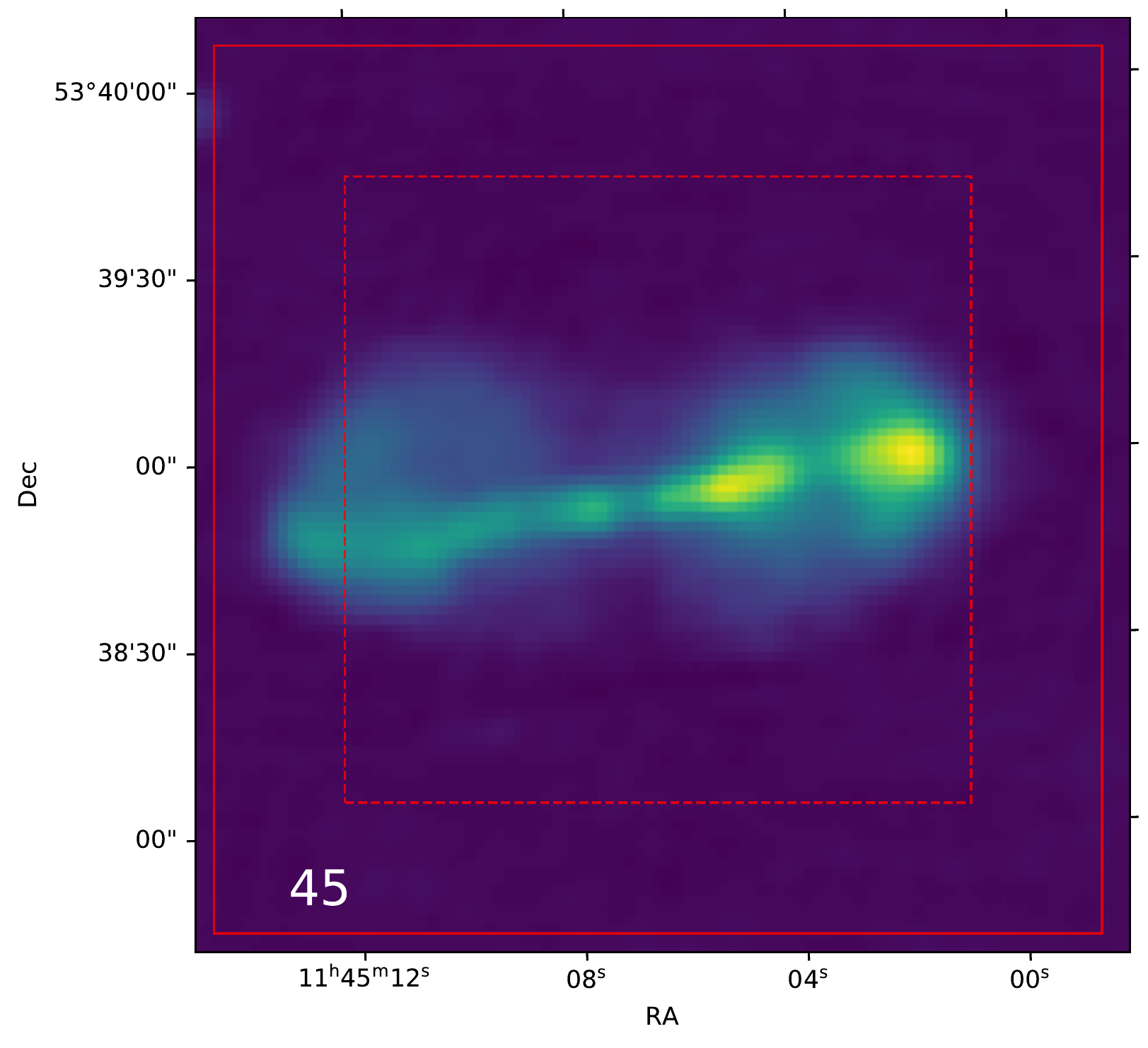}
\includegraphics[width=0.19\textwidth]{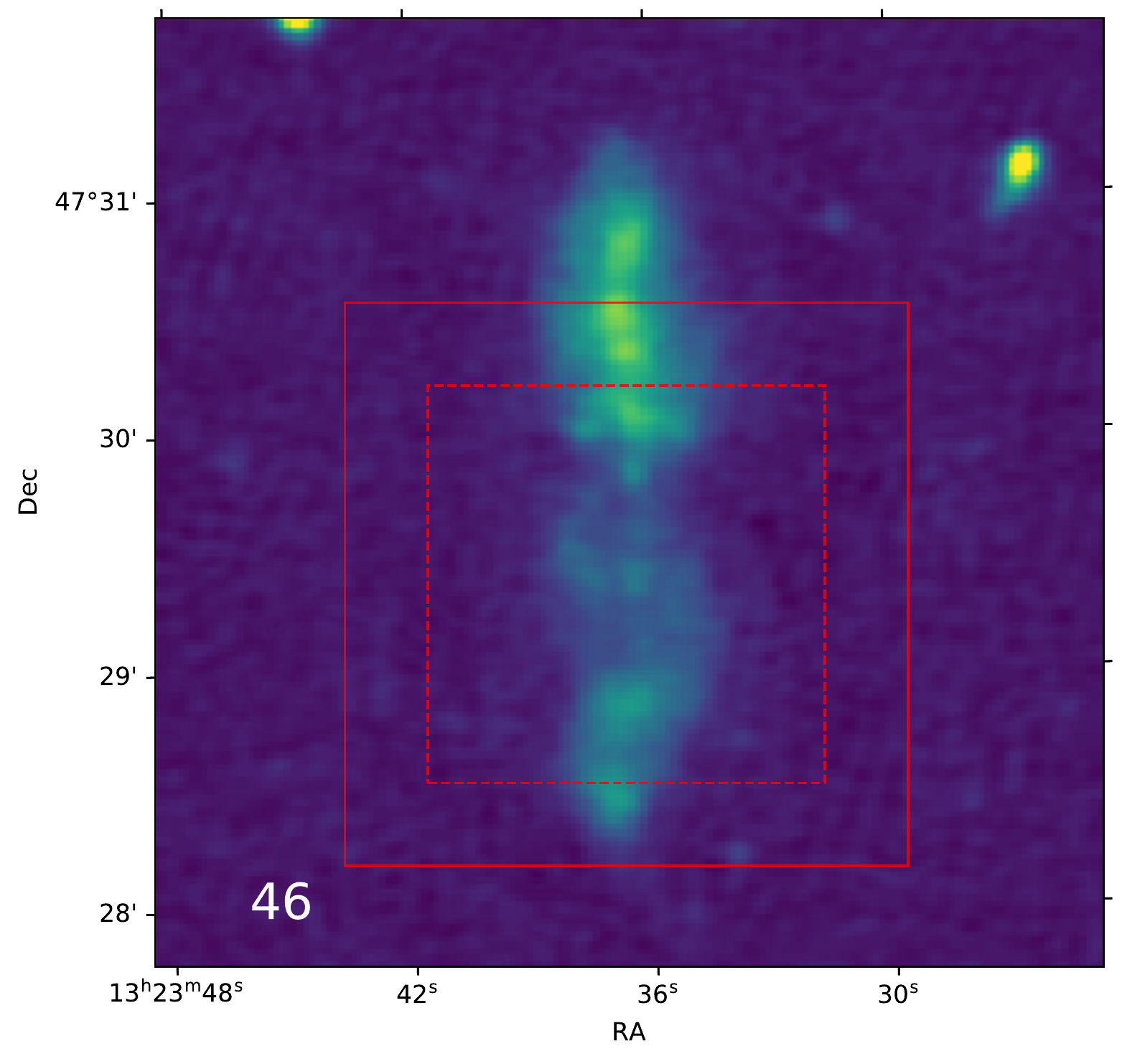}
\includegraphics[width=0.19\textwidth]{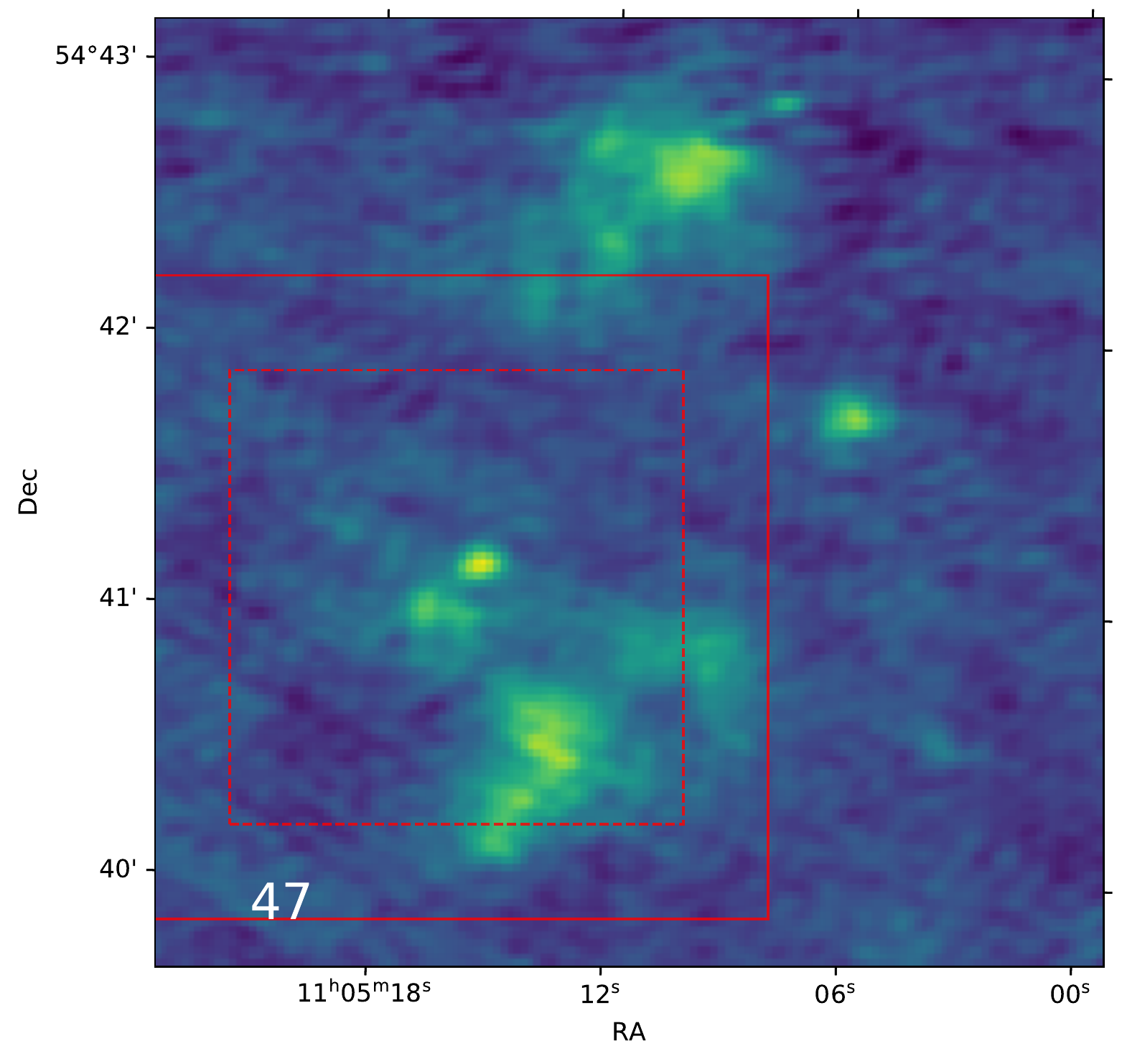}
\includegraphics[width=0.19\textwidth]{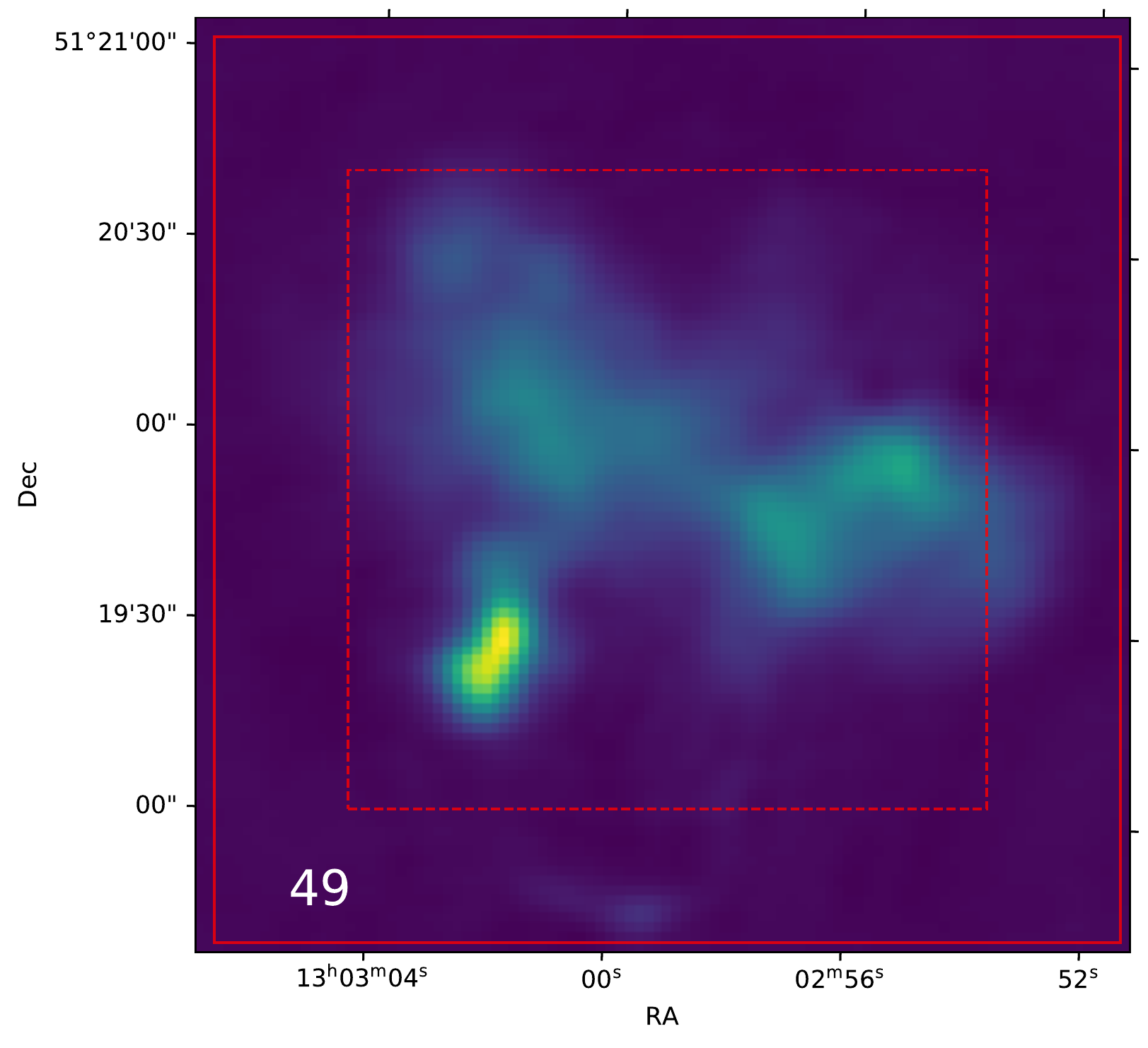}
\includegraphics[width=0.19\textwidth]{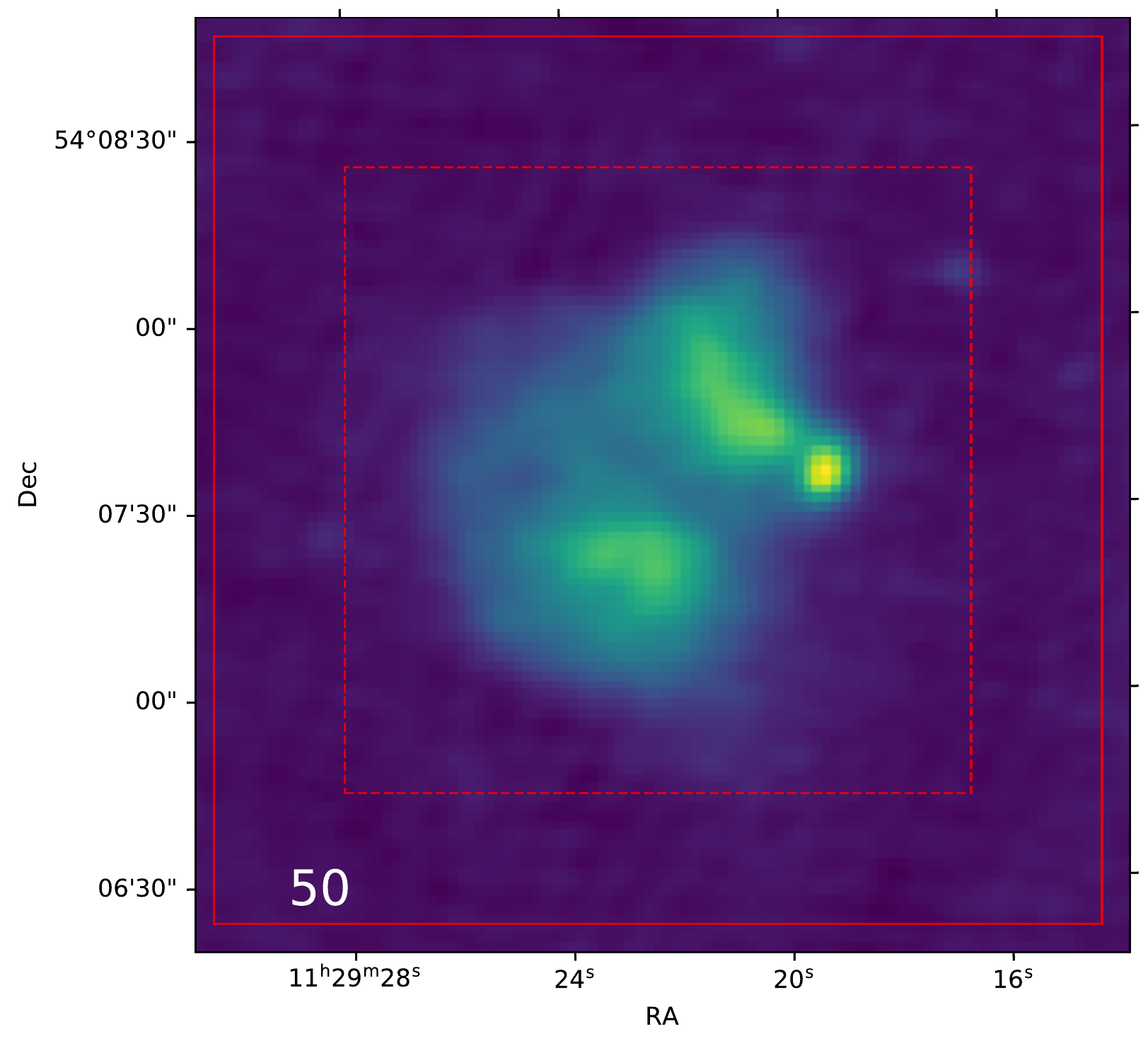}
\includegraphics[width=0.19\textwidth]{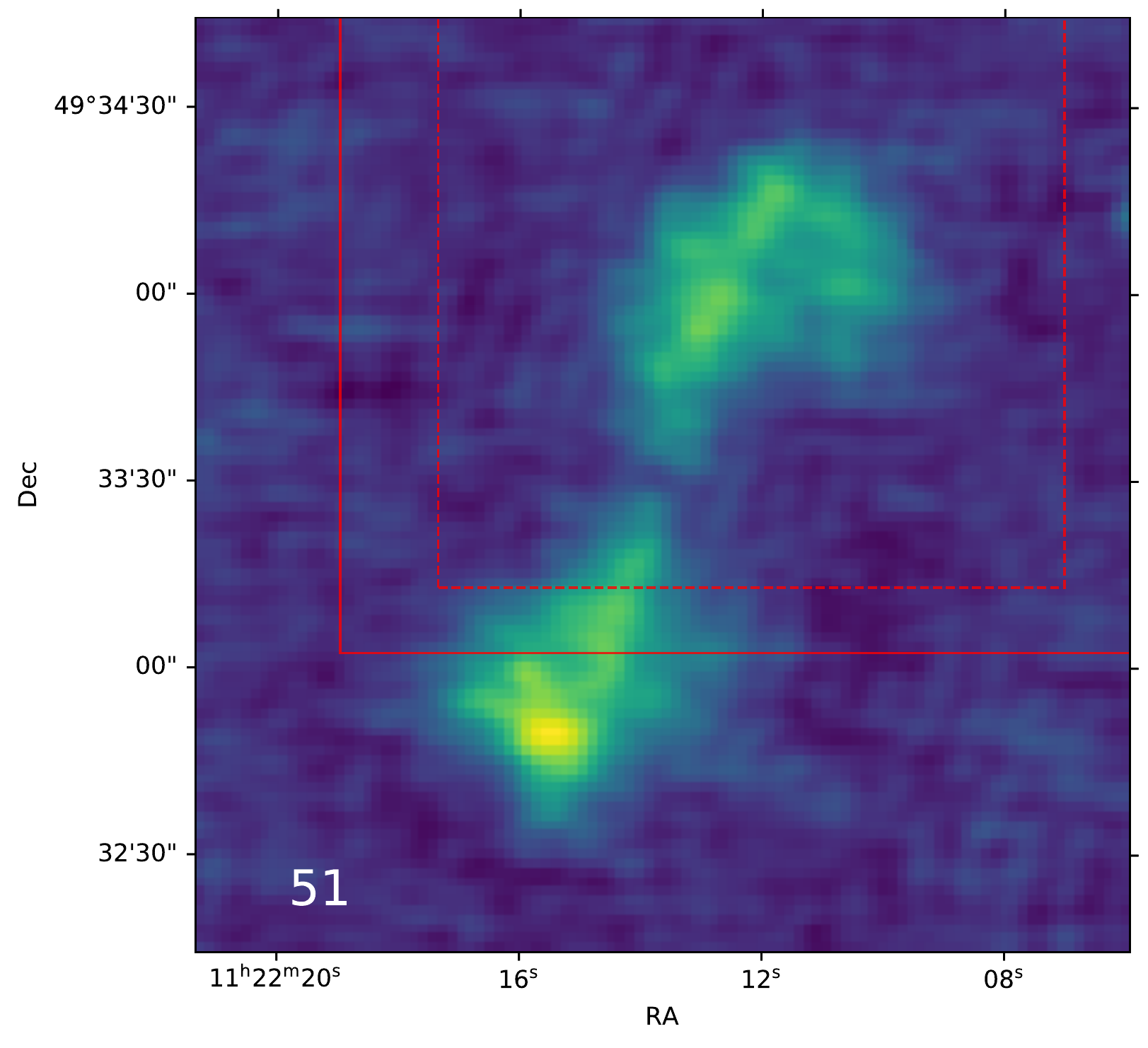}
\includegraphics[width=0.19\textwidth]{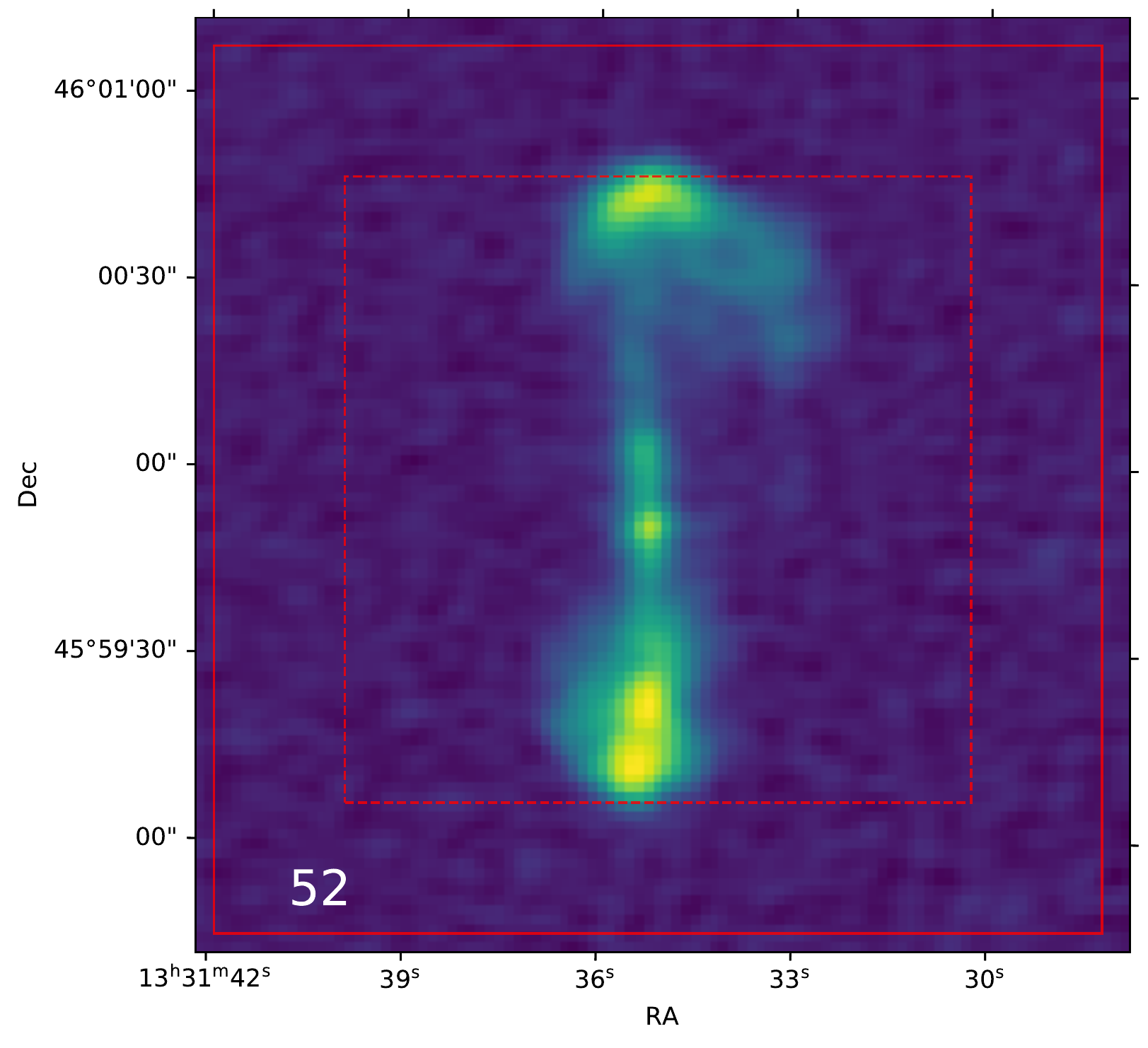}
\includegraphics[width=0.19\textwidth]{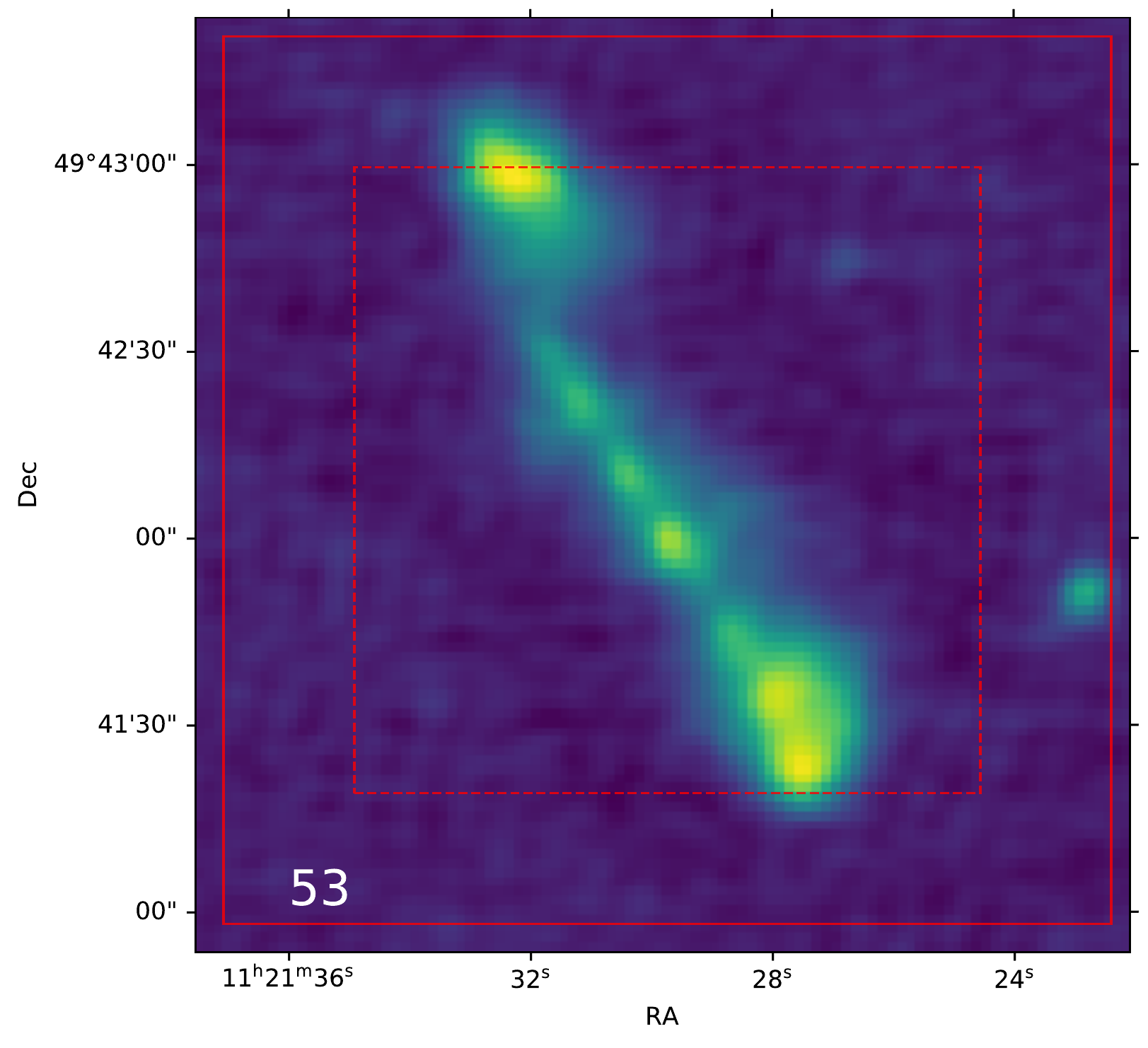}
\includegraphics[width=0.19\textwidth]{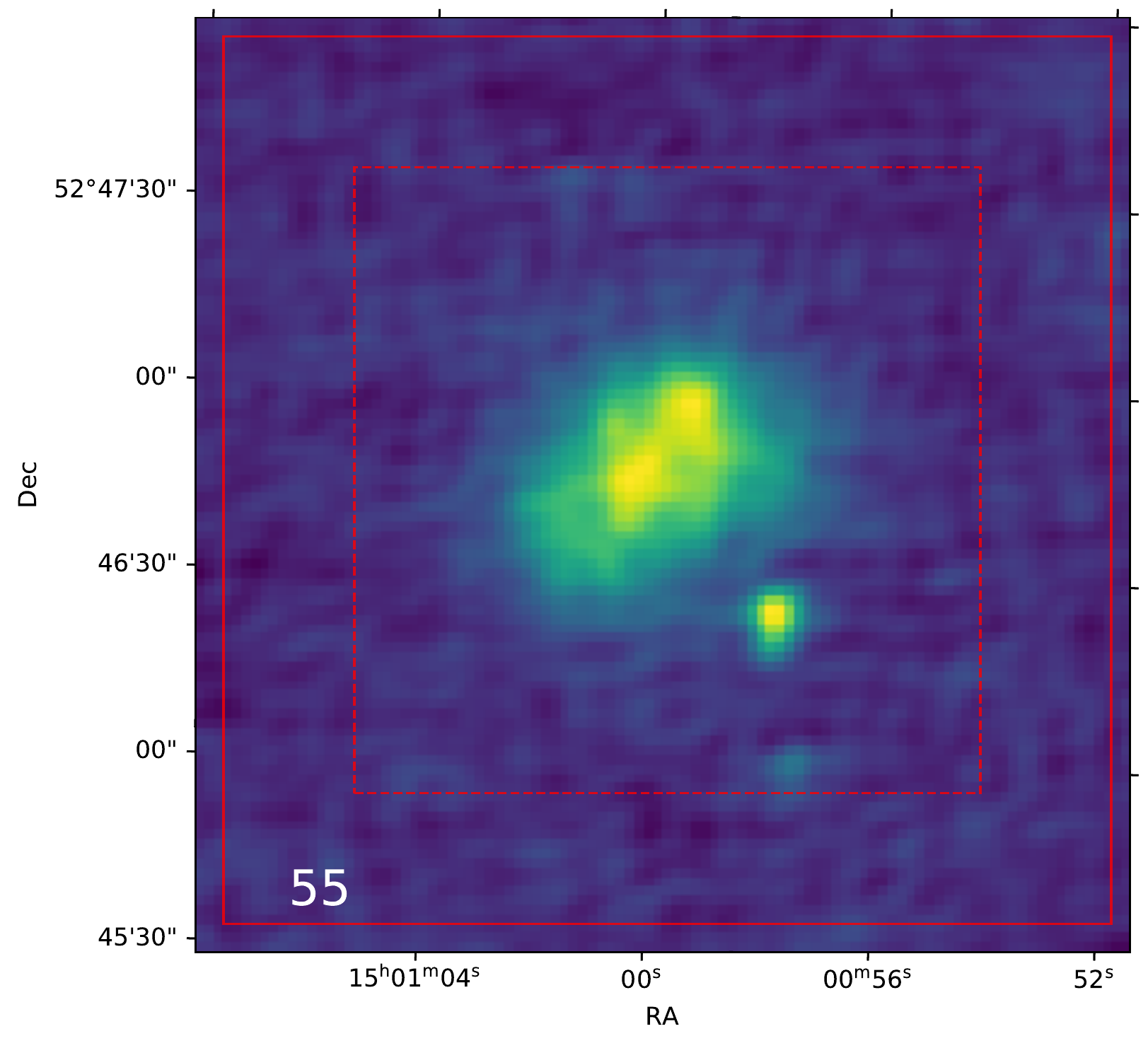}
\includegraphics[width=0.19\textwidth]{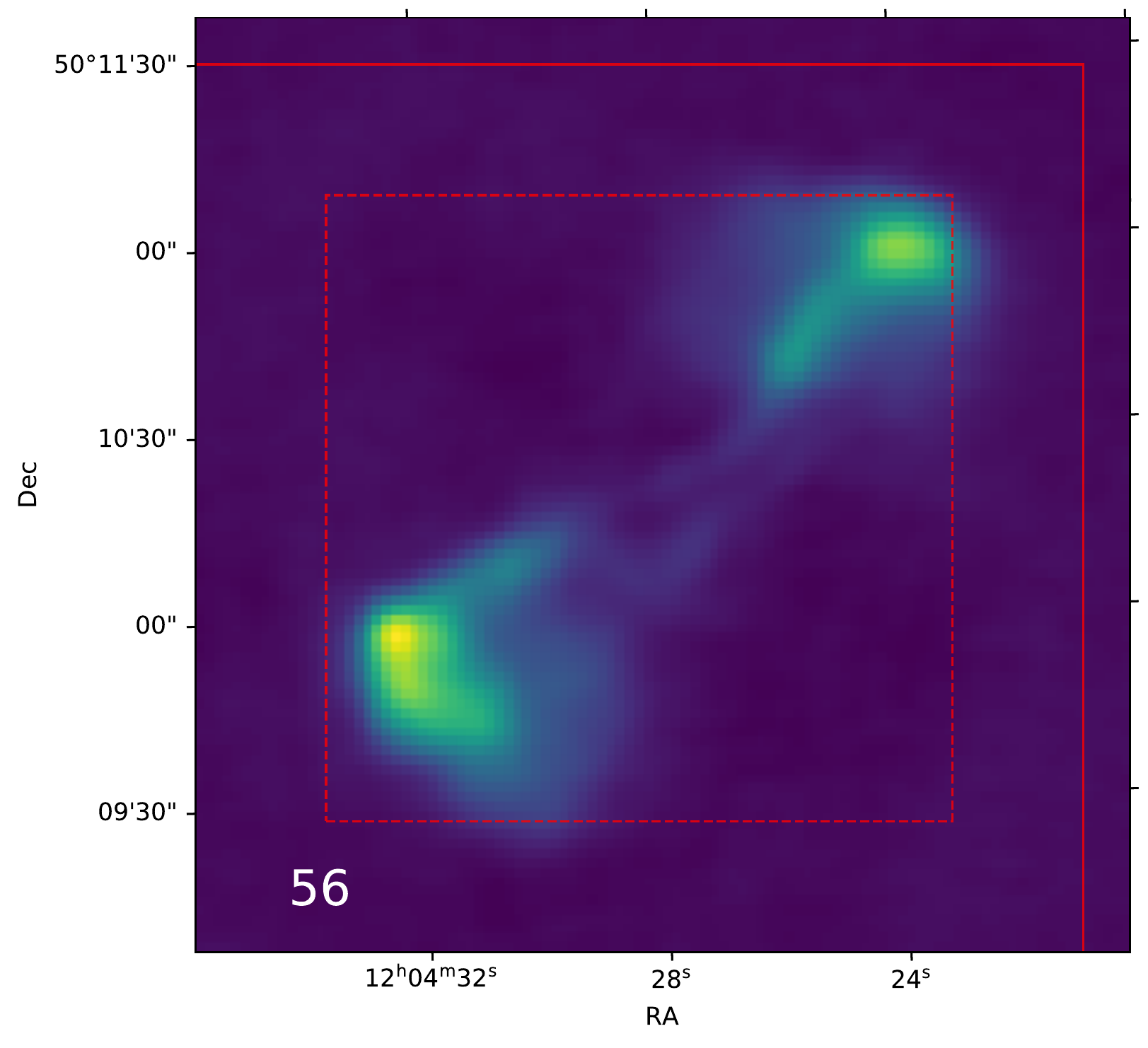}
\includegraphics[width=0.19\textwidth]{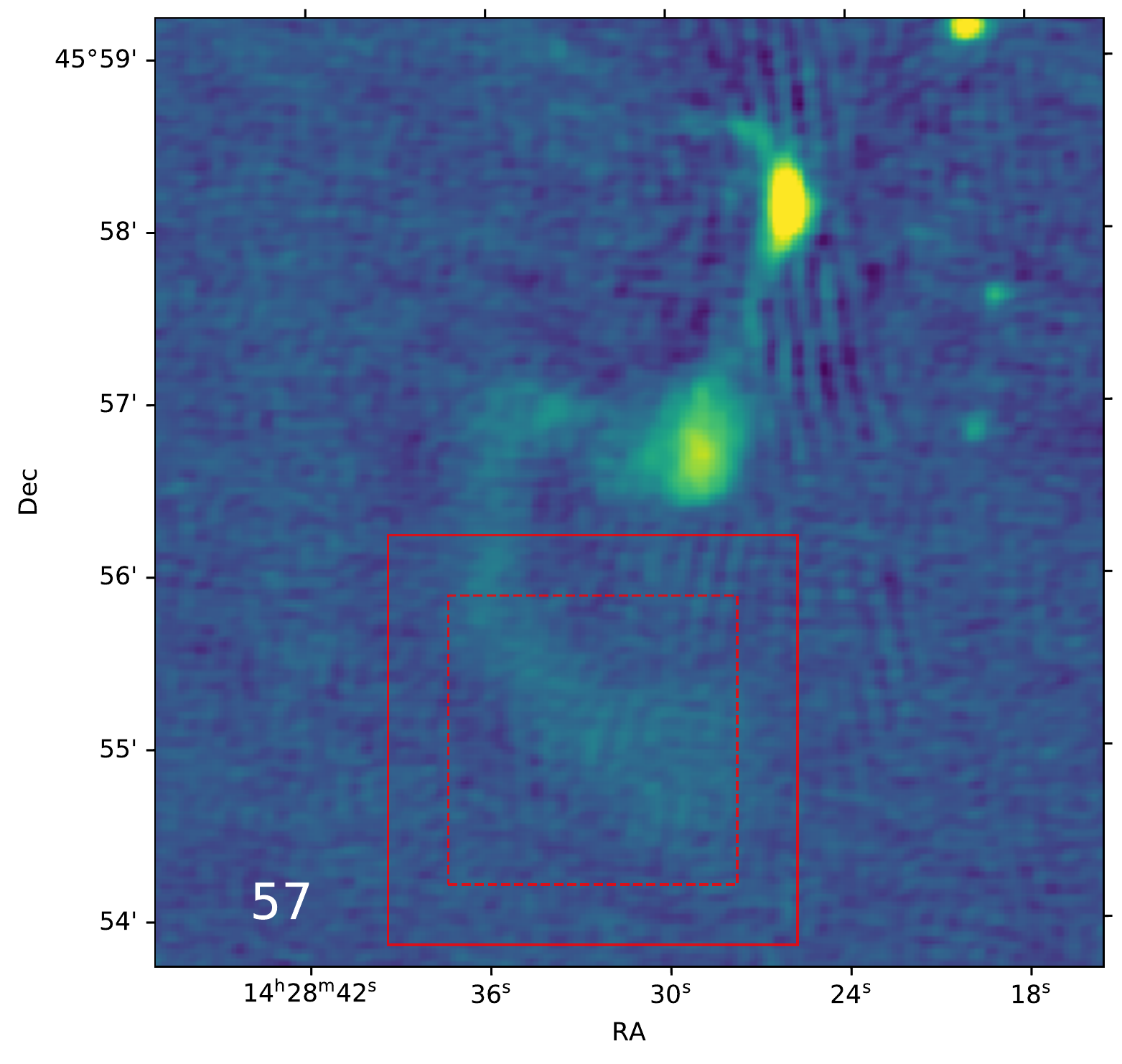}
\includegraphics[width=0.19\textwidth]{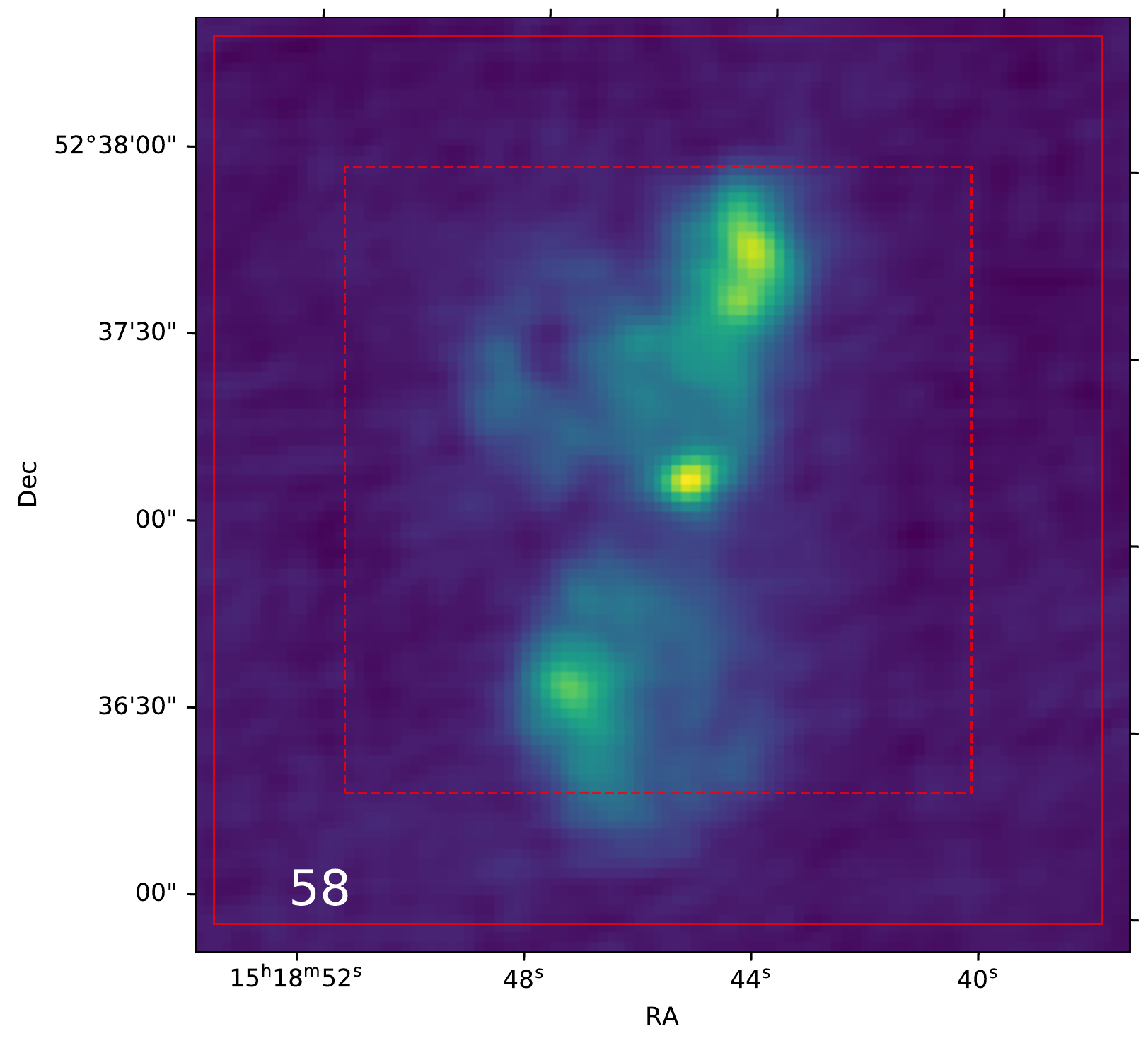}
\includegraphics[width=0.19\textwidth]{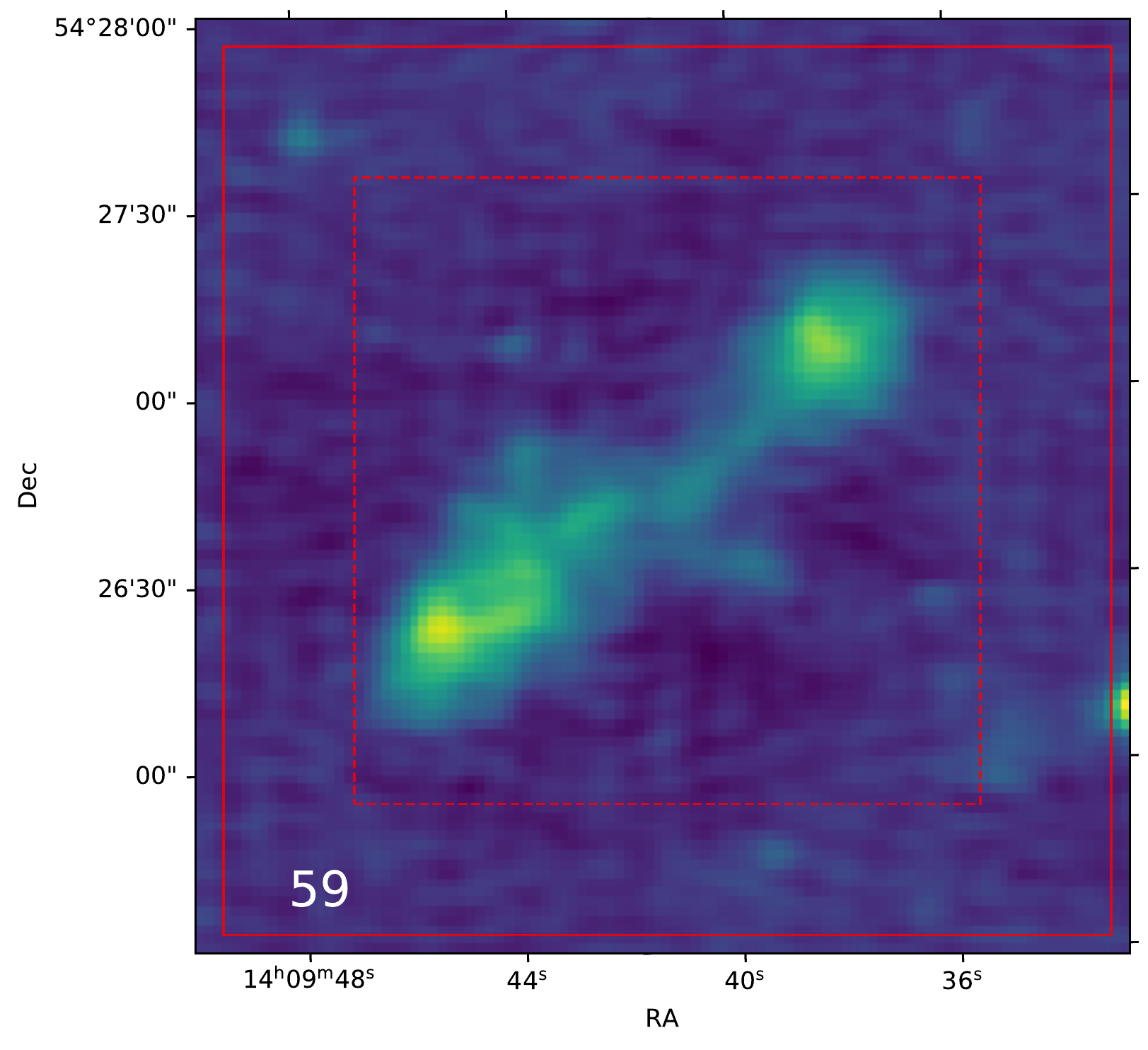}
\includegraphics[width=0.19\textwidth]{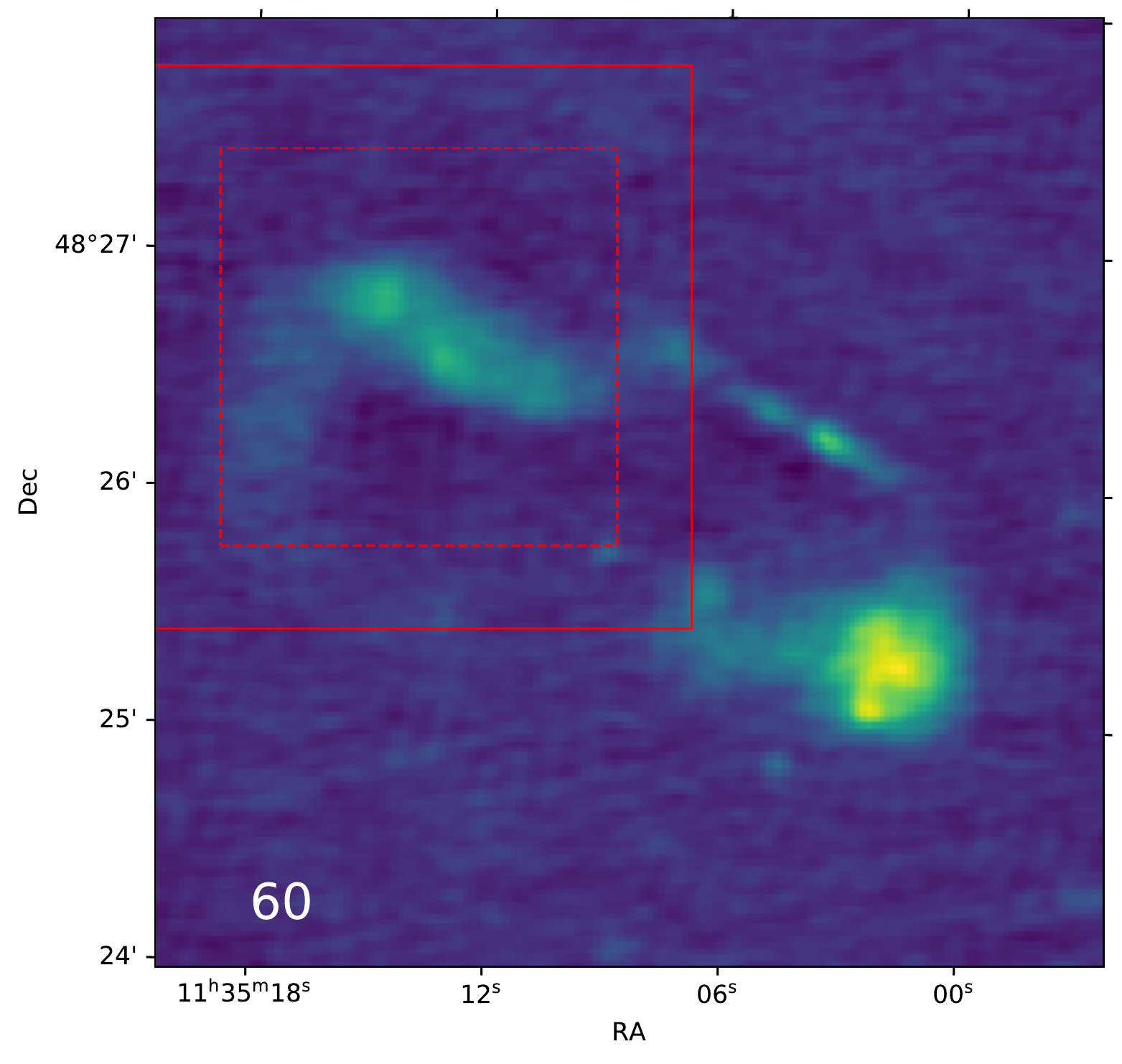}
\includegraphics[width=0.19\textwidth]{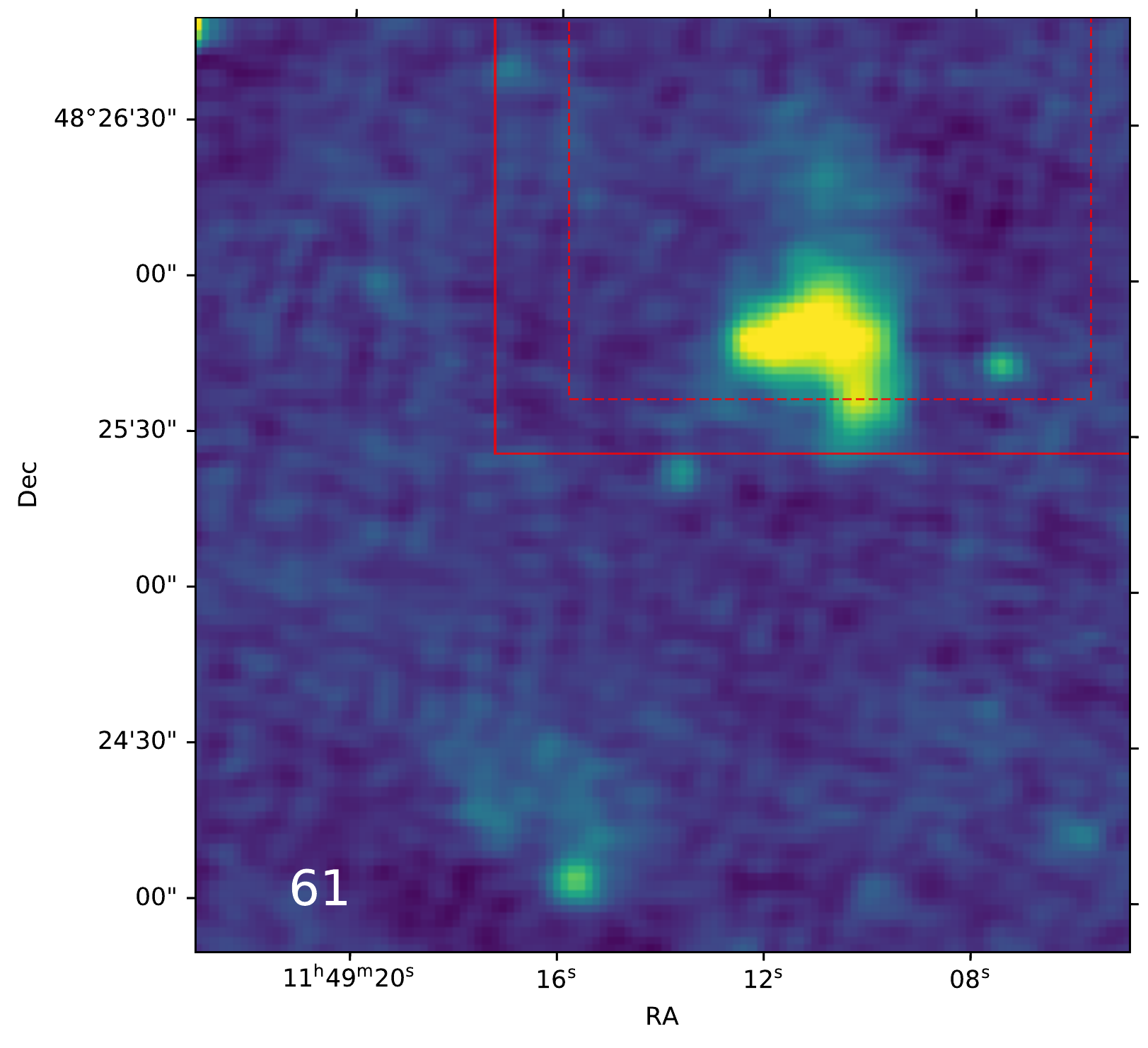}
\includegraphics[width=0.19\textwidth]{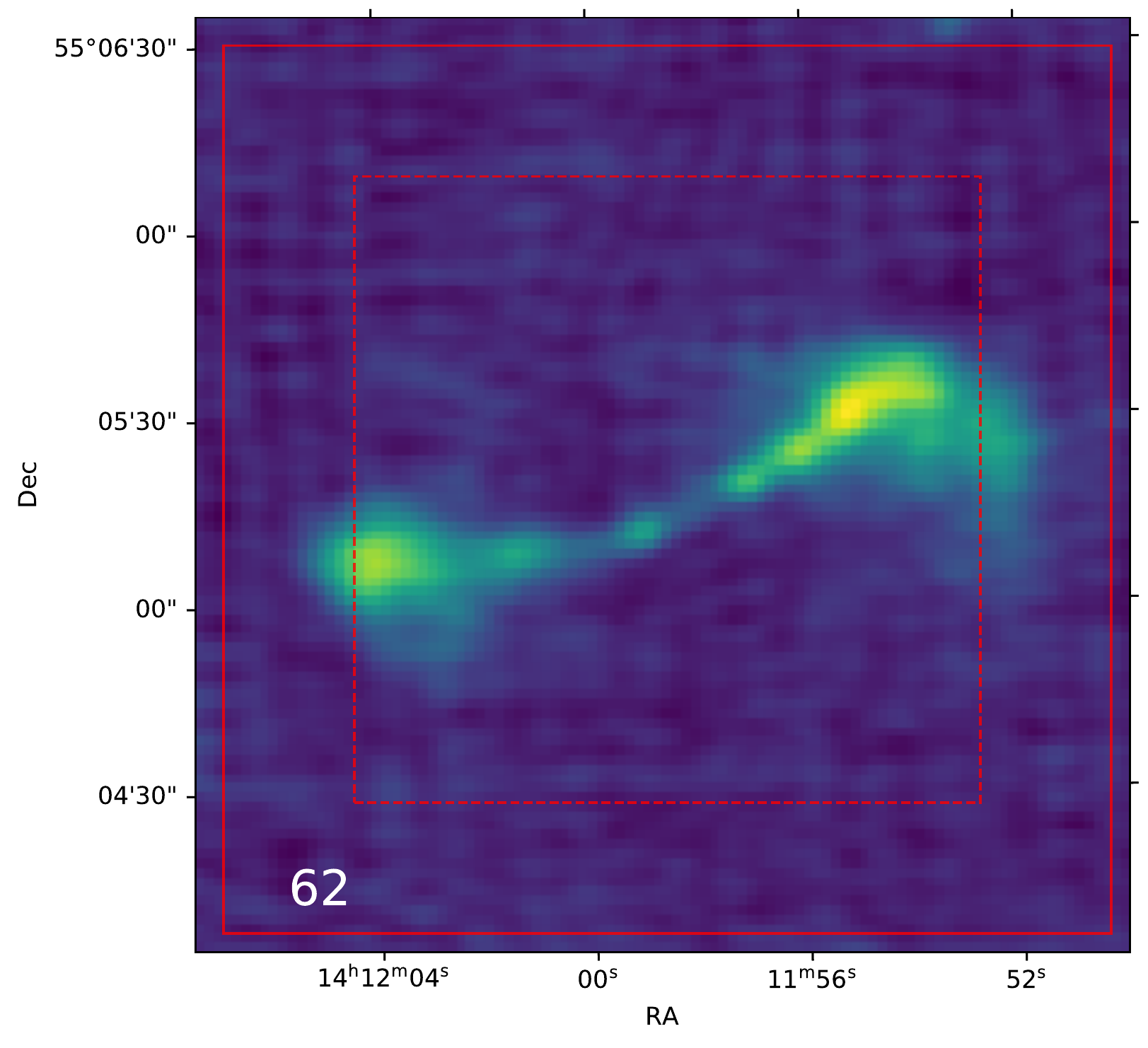}
\includegraphics[width=0.19\textwidth]{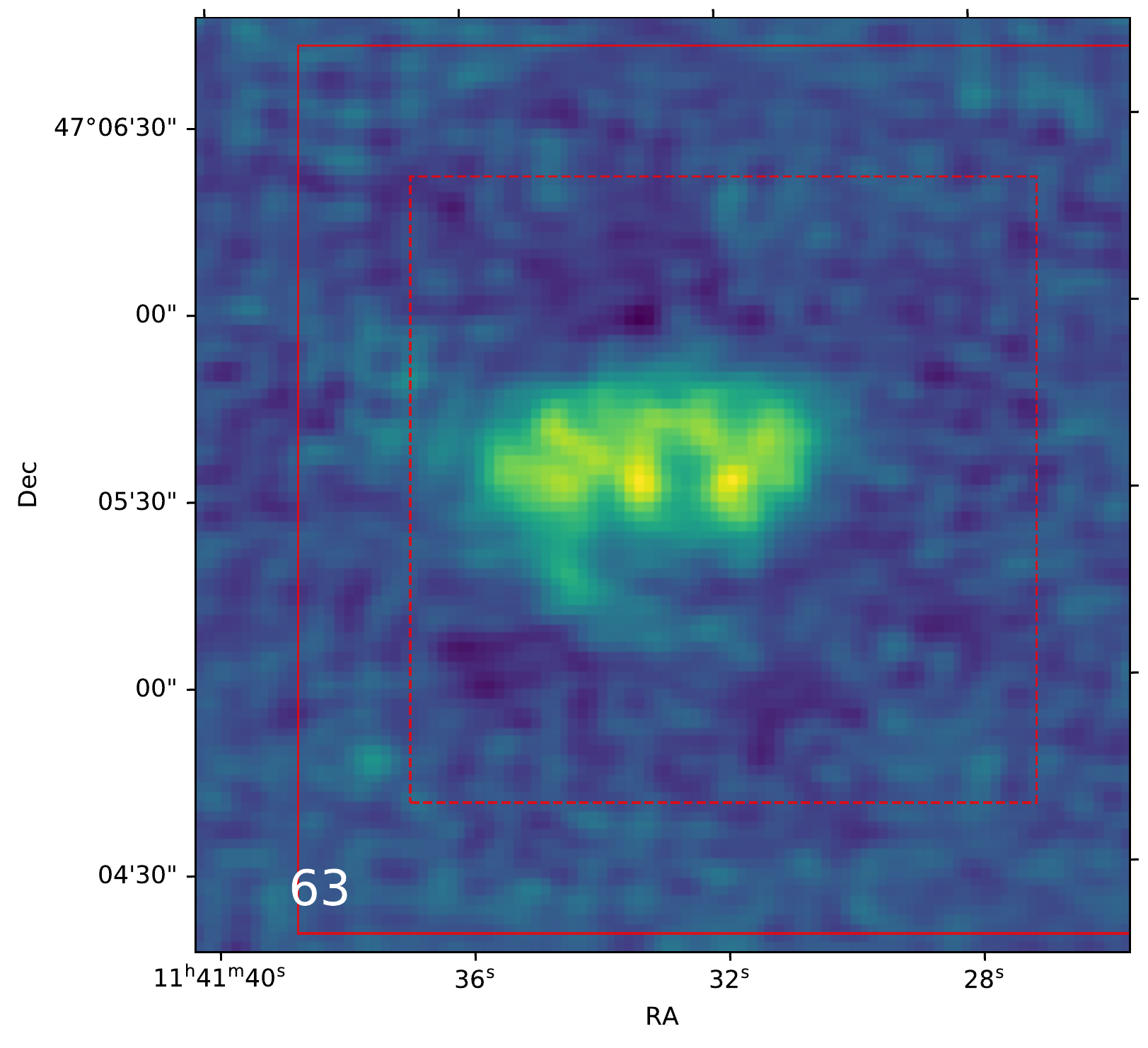}
\includegraphics[width=0.19\textwidth]{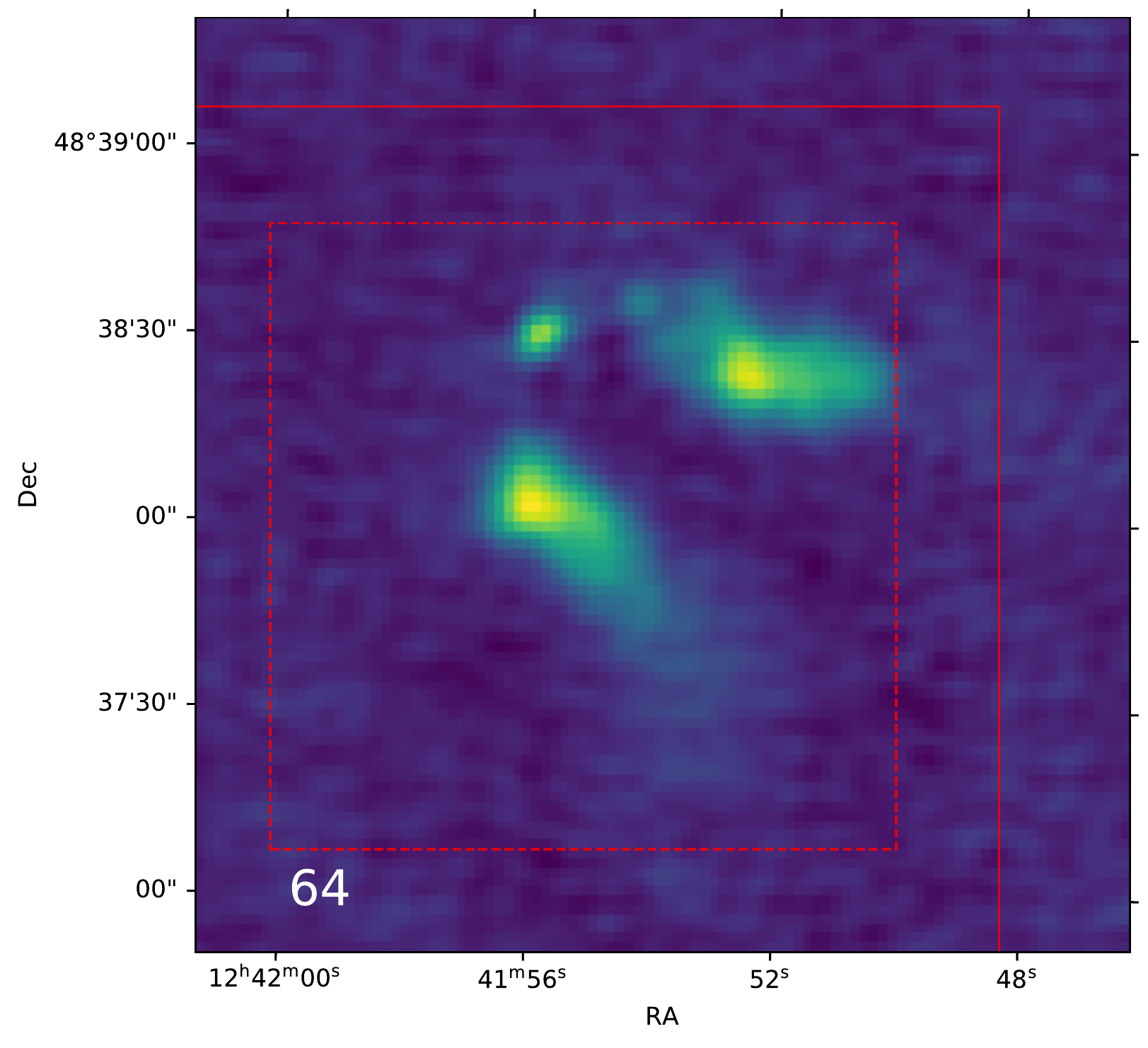}
\includegraphics[width=0.19\textwidth]{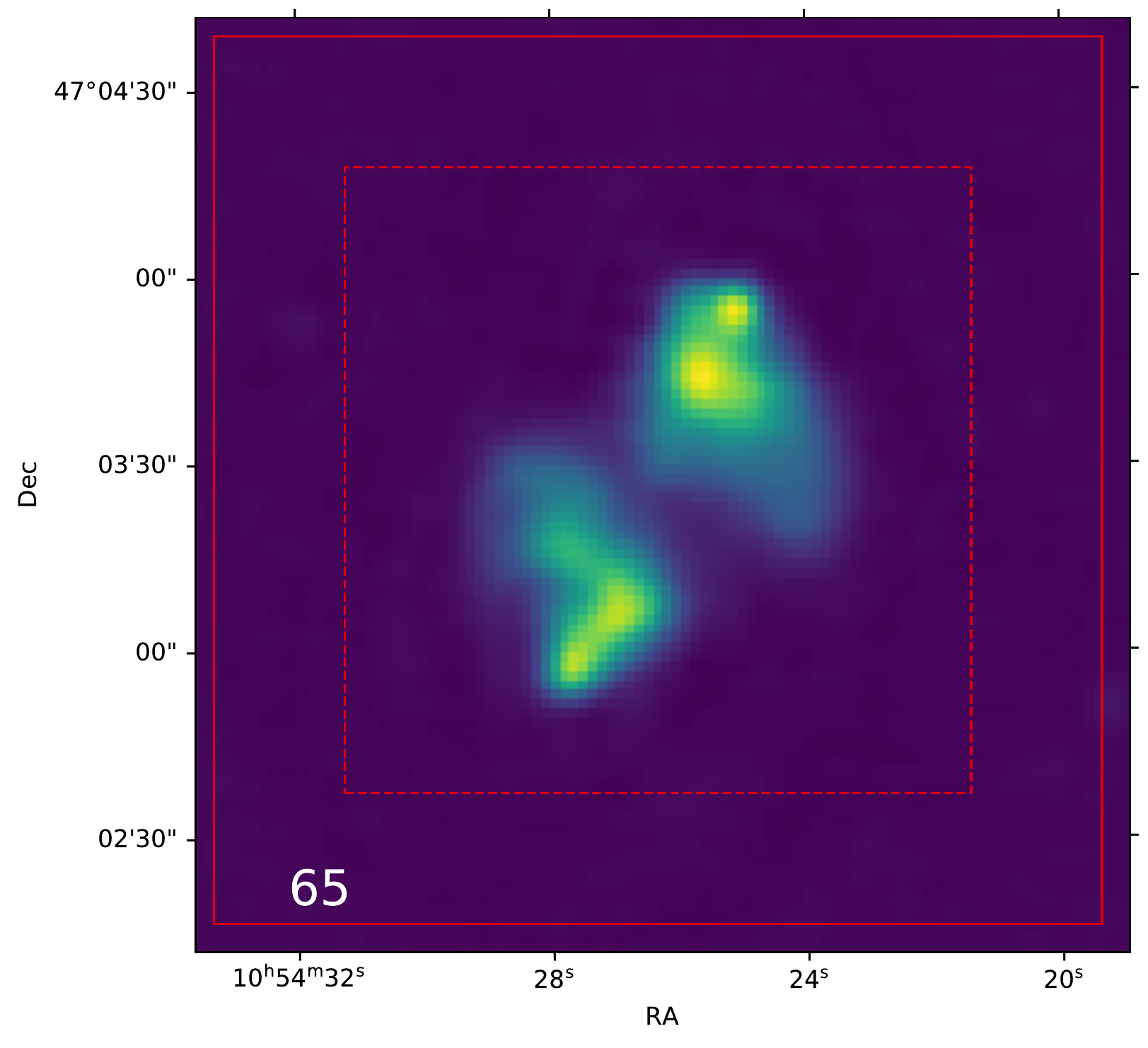}
\includegraphics[width=0.19\textwidth]{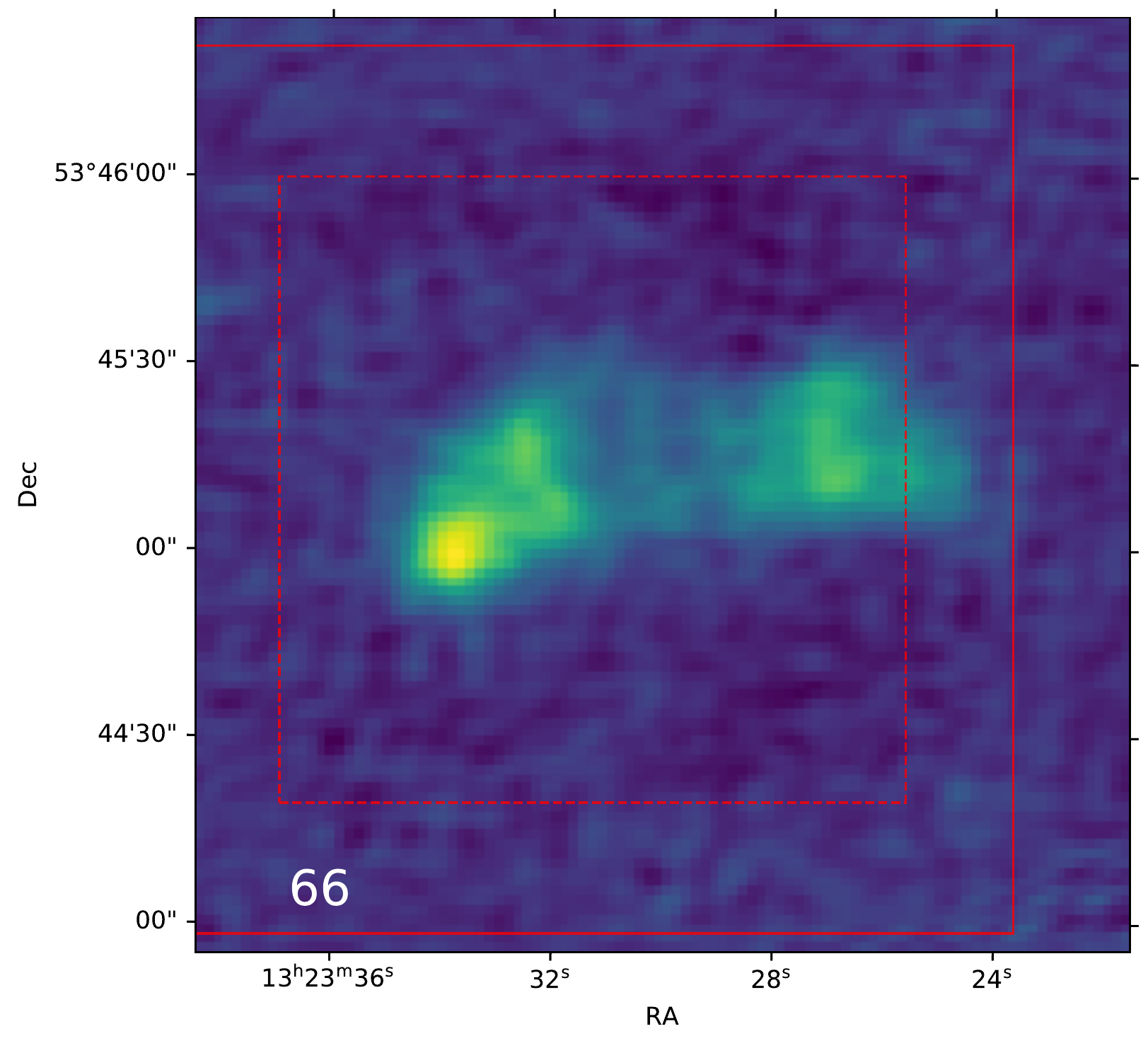}
\includegraphics[width=0.19\textwidth]{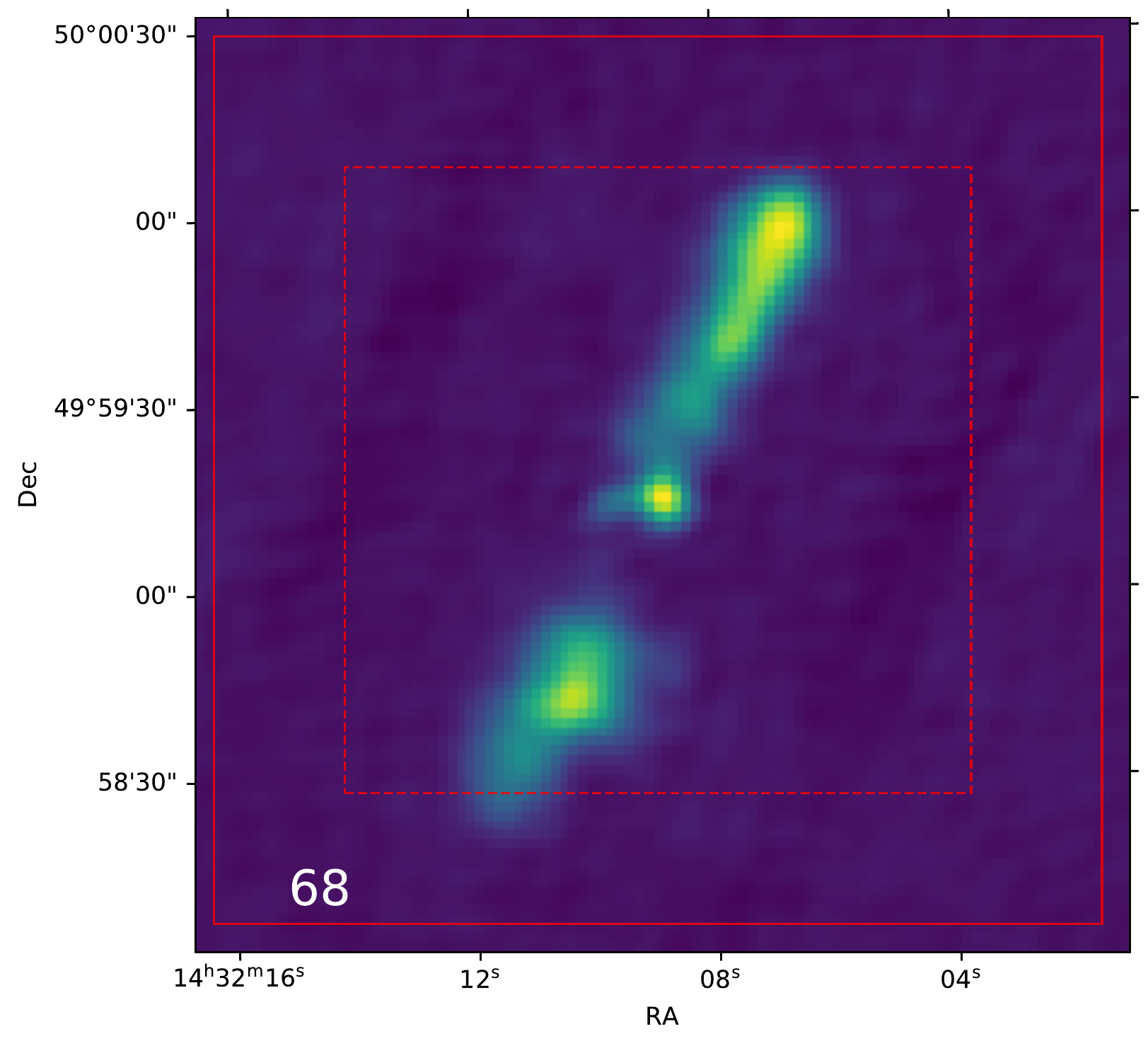}
\includegraphics[width=0.19\textwidth]{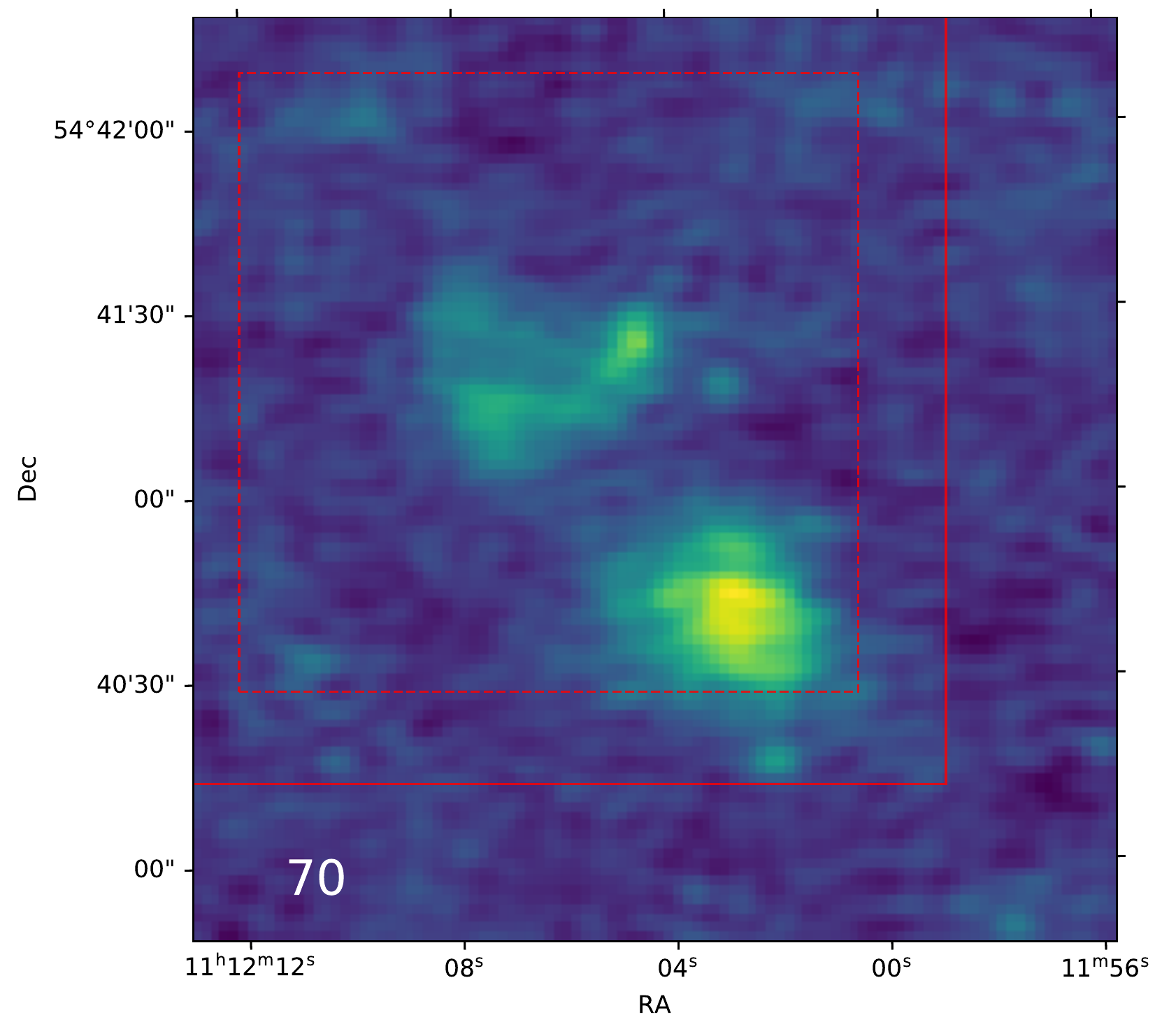}
\includegraphics[width=0.19\textwidth]{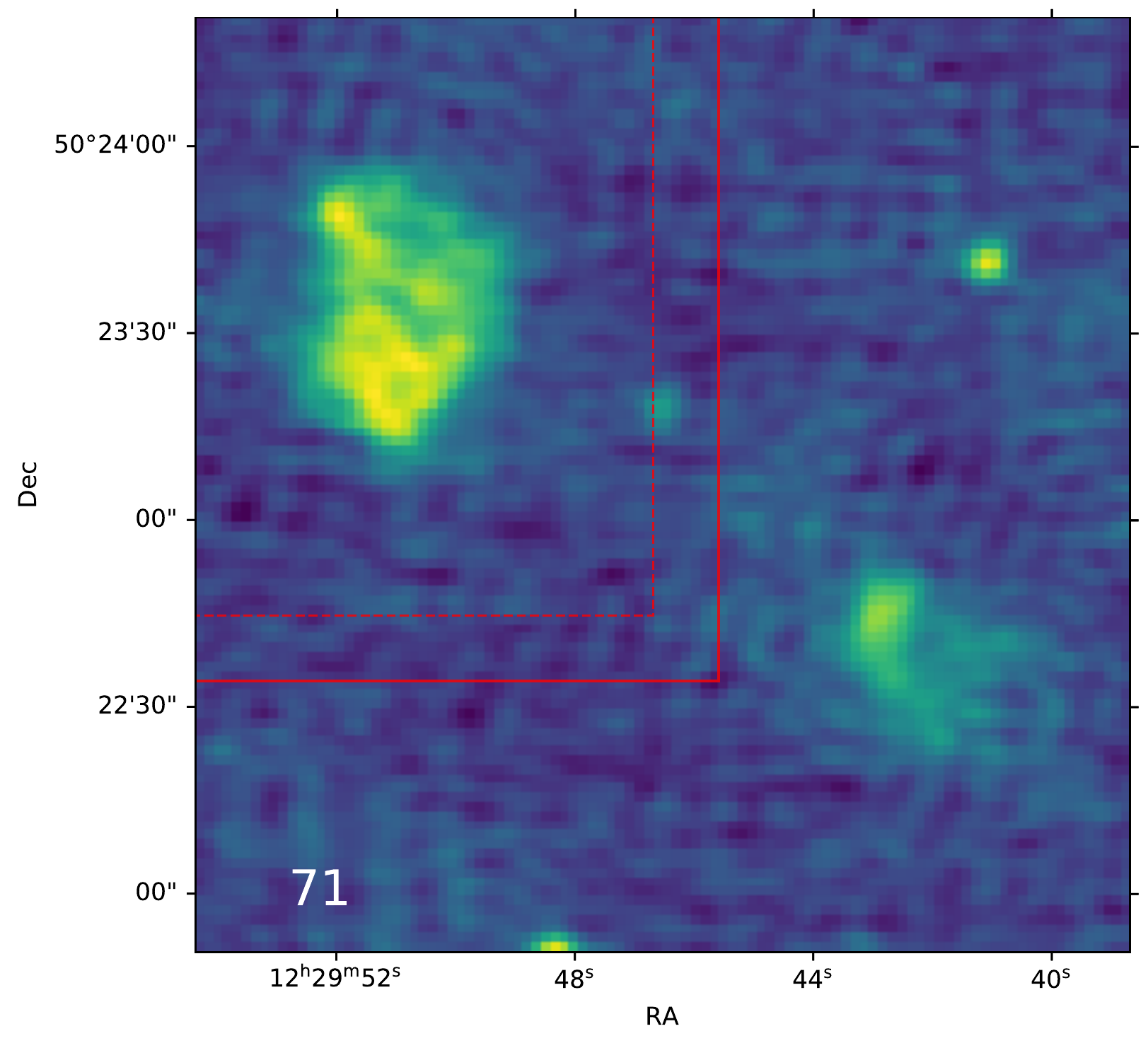}
\includegraphics[width=0.19\textwidth]{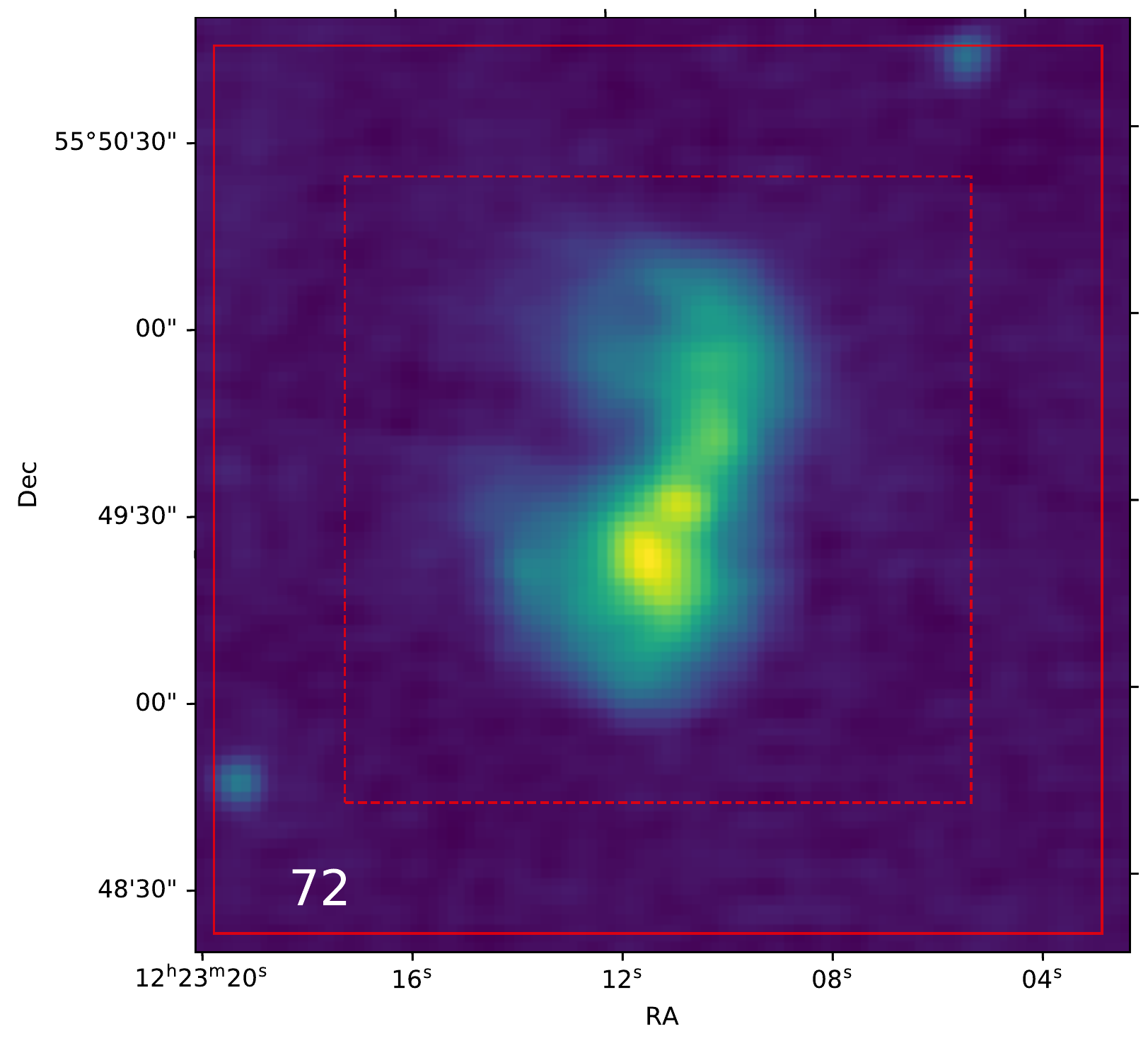}
\includegraphics[width=0.19\textwidth]{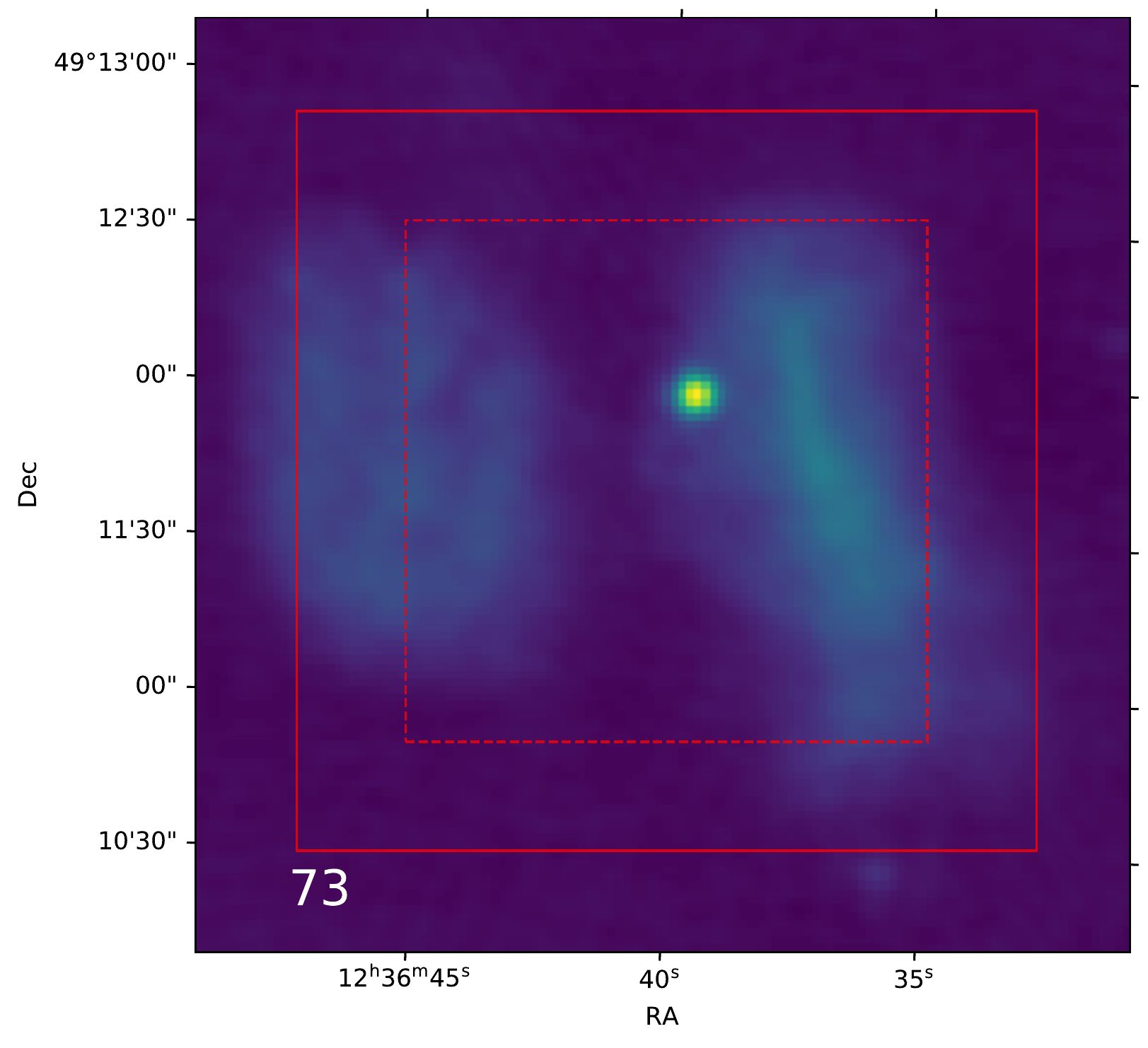}
\includegraphics[width=0.19\textwidth]{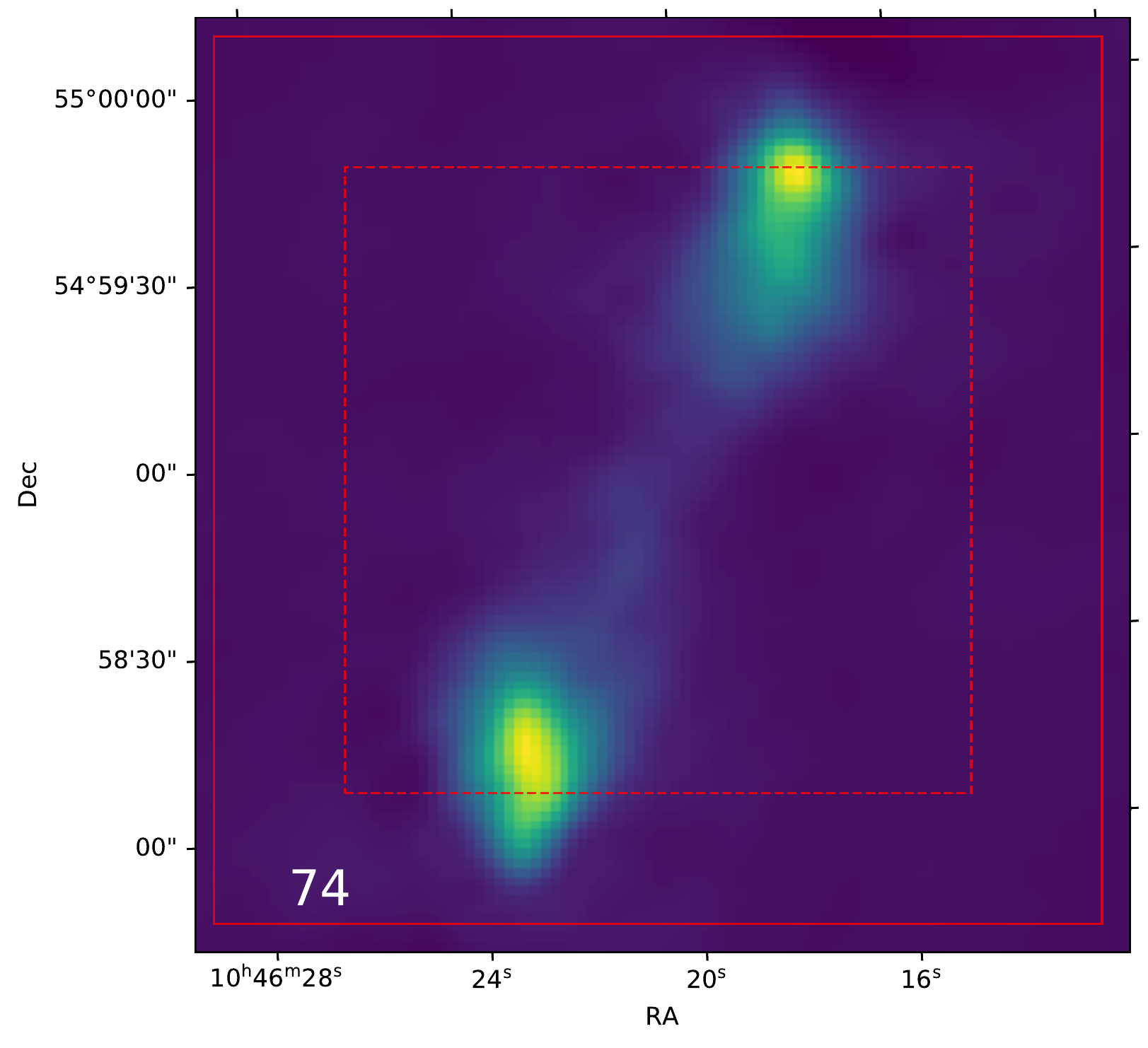}
\includegraphics[width=0.19\textwidth]{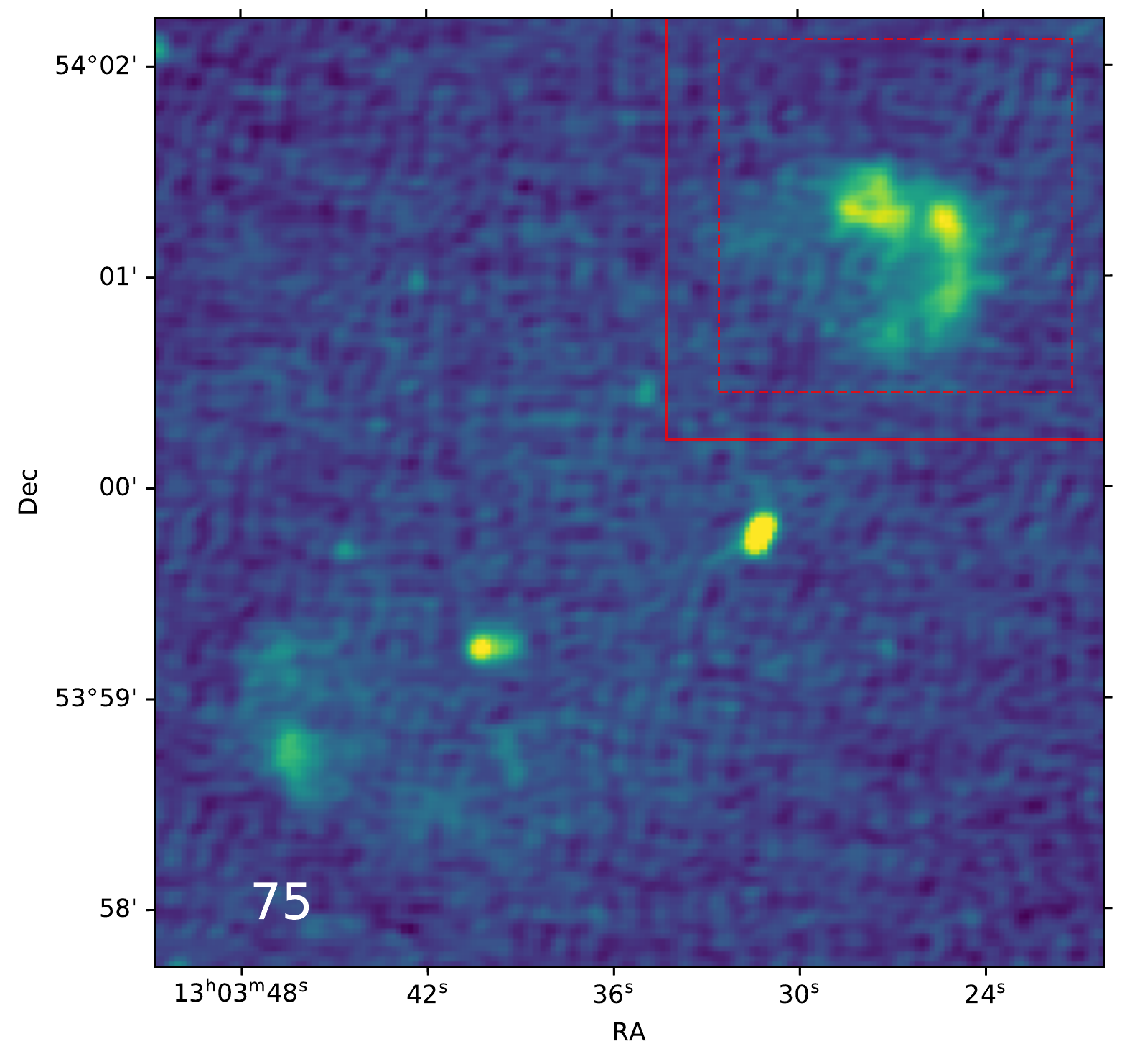}

\caption{\textit{Continued:} See previous page for the figure caption.}
\end{center}\end{figure*} 

\begin{figure*}\begin{center}
\ContinuedFloat

\includegraphics[width=0.19\textwidth]{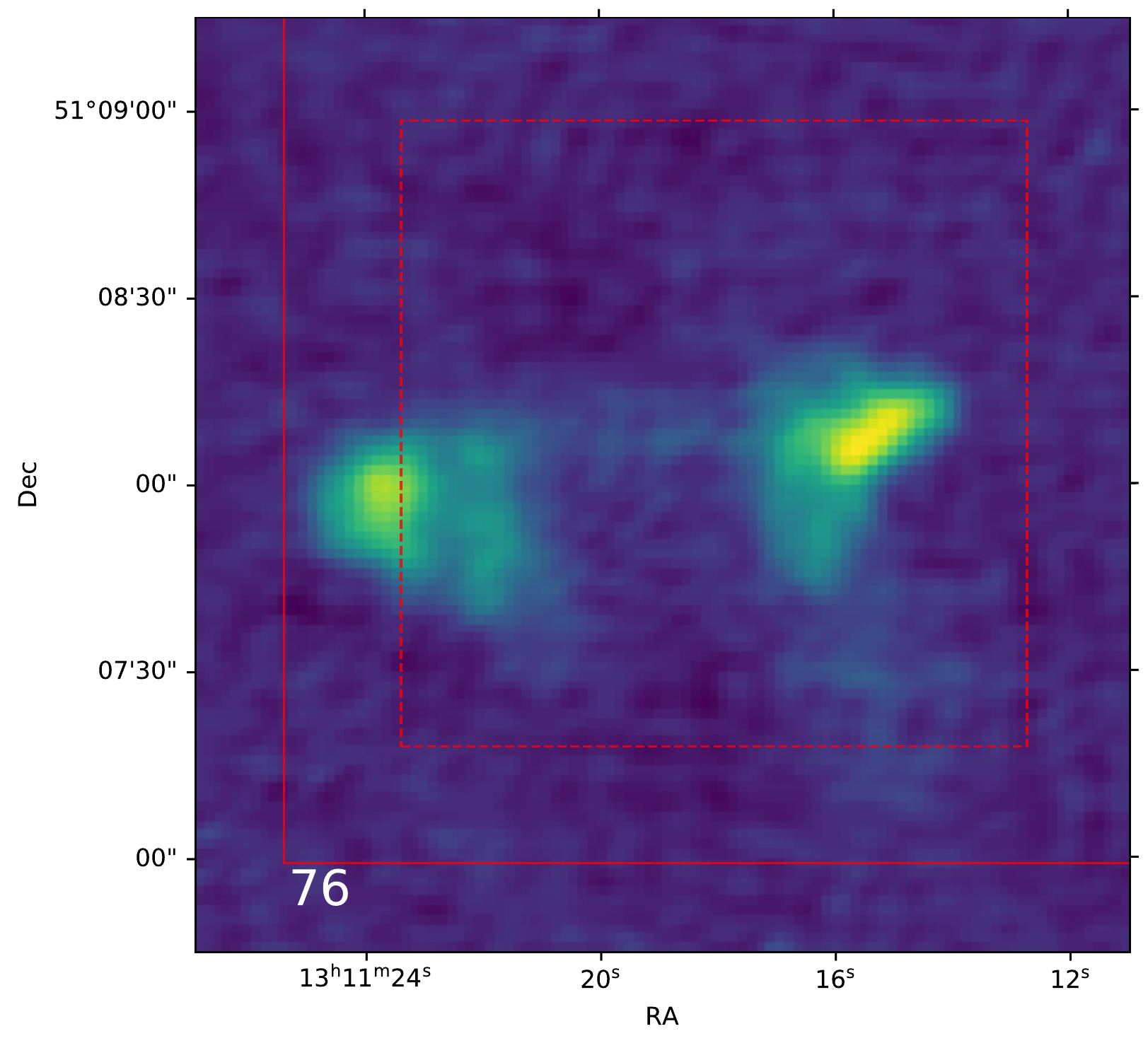}
\includegraphics[width=0.19\textwidth]{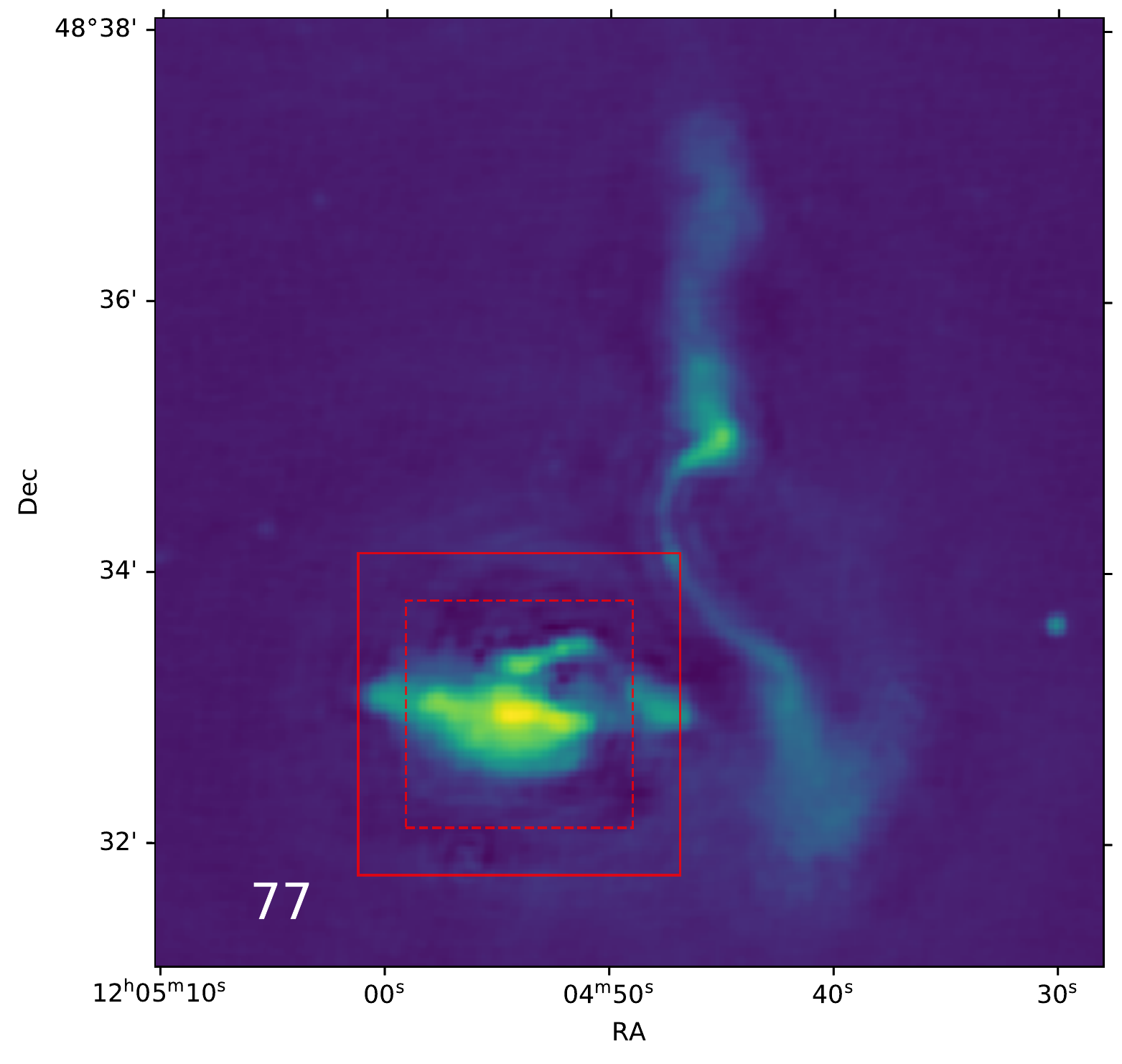}
\includegraphics[width=0.19\textwidth]{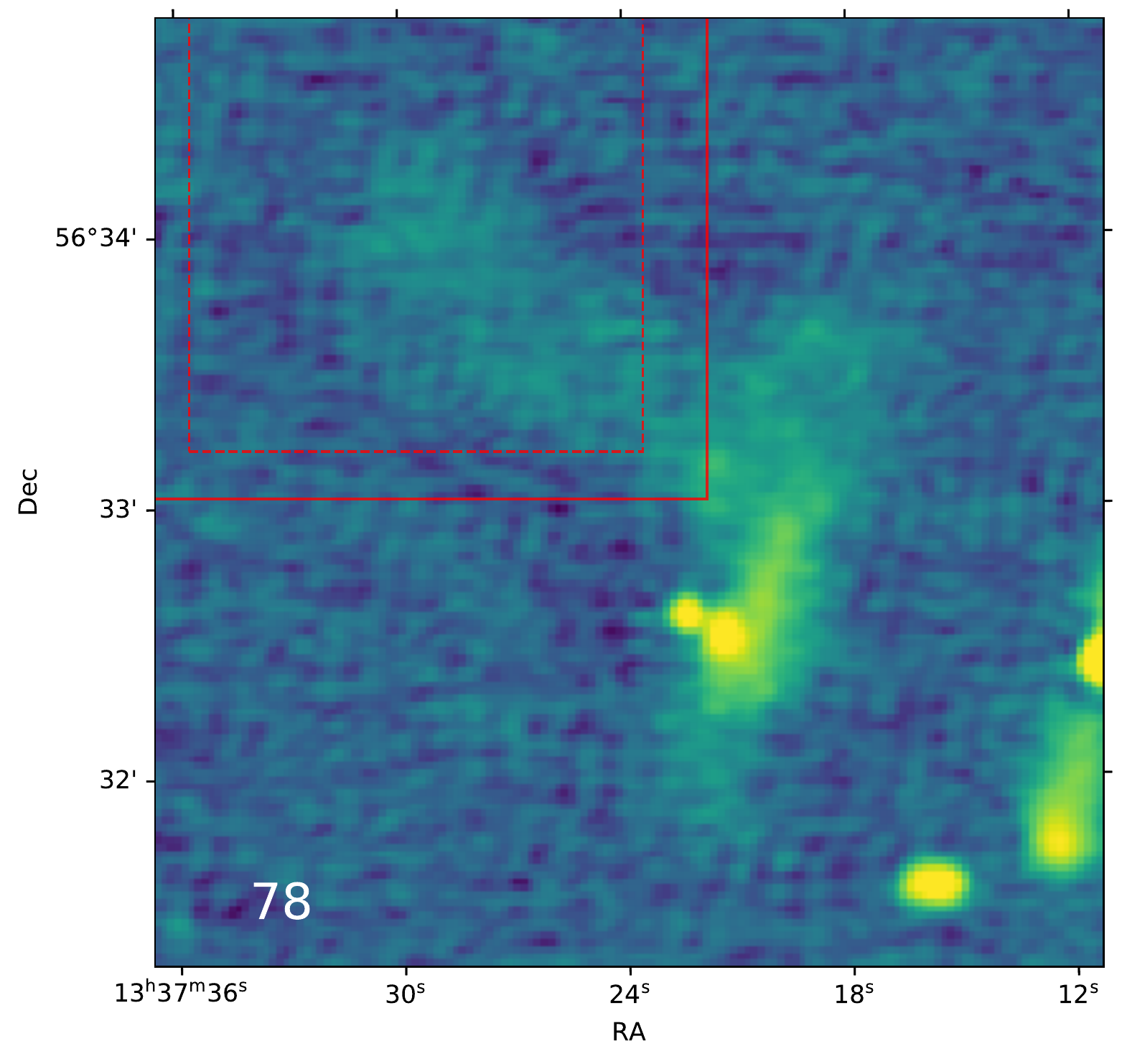}
\includegraphics[width=0.19\textwidth]{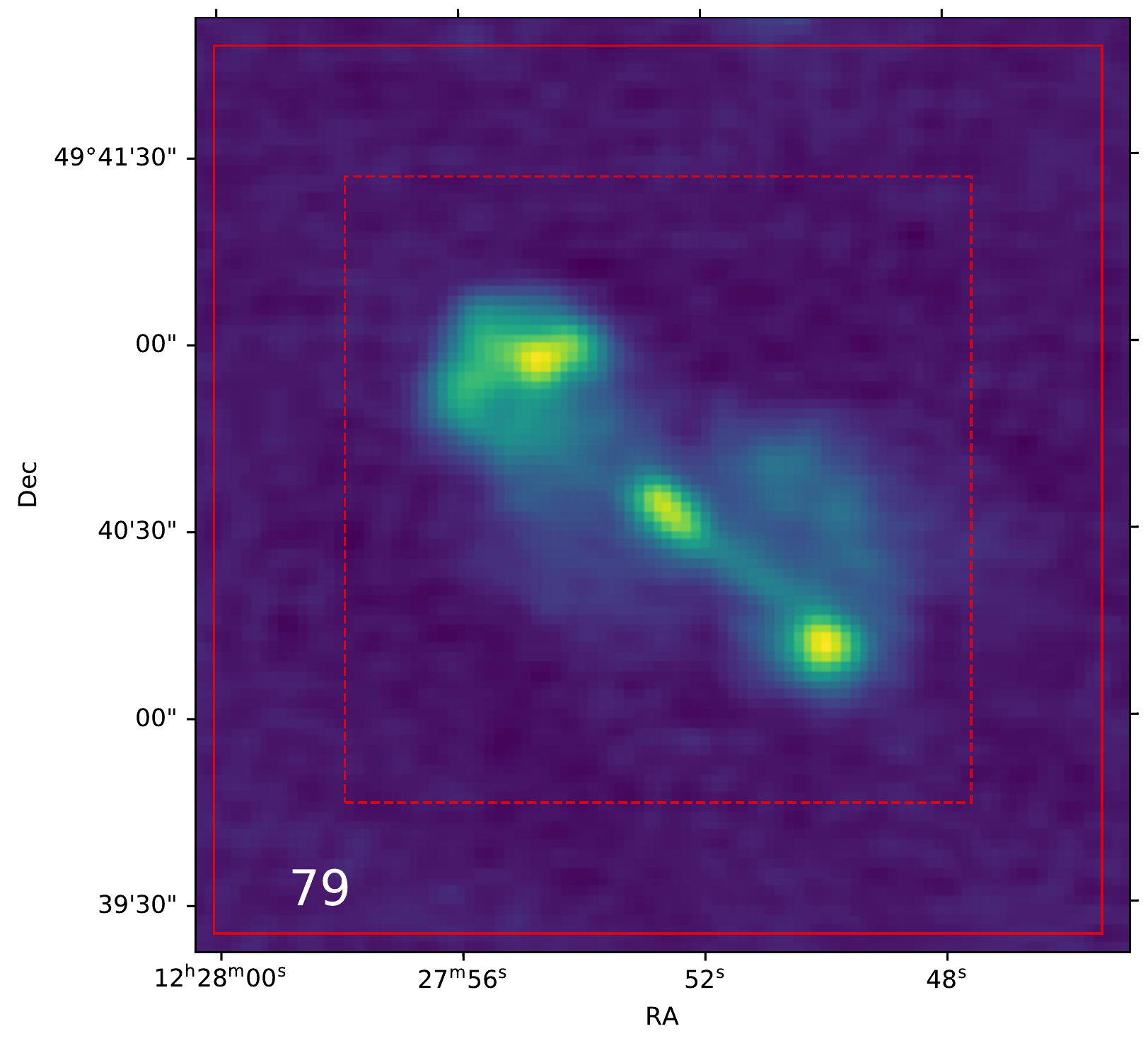}
\includegraphics[width=0.19\textwidth]{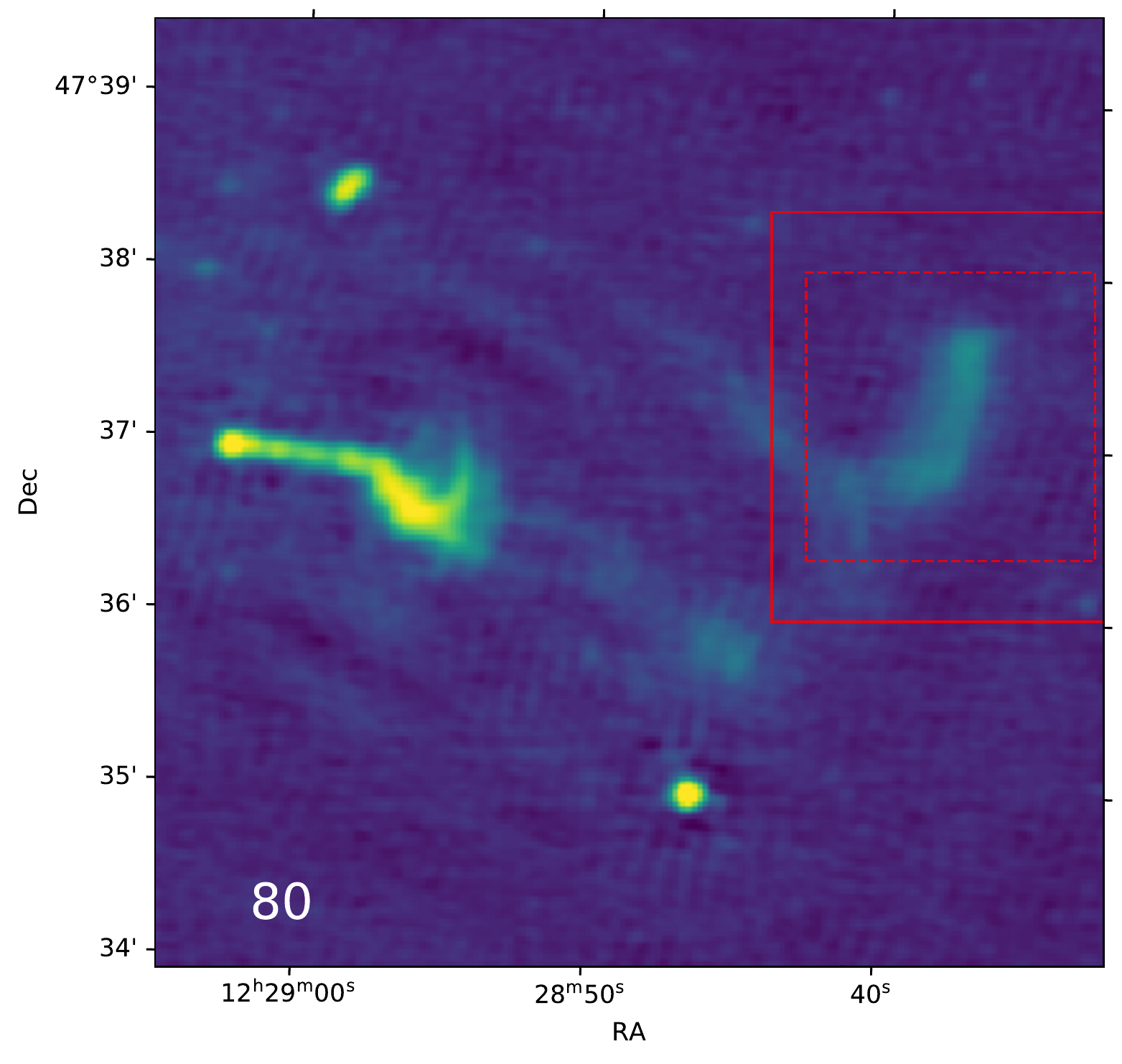}
\includegraphics[width=0.19\textwidth]{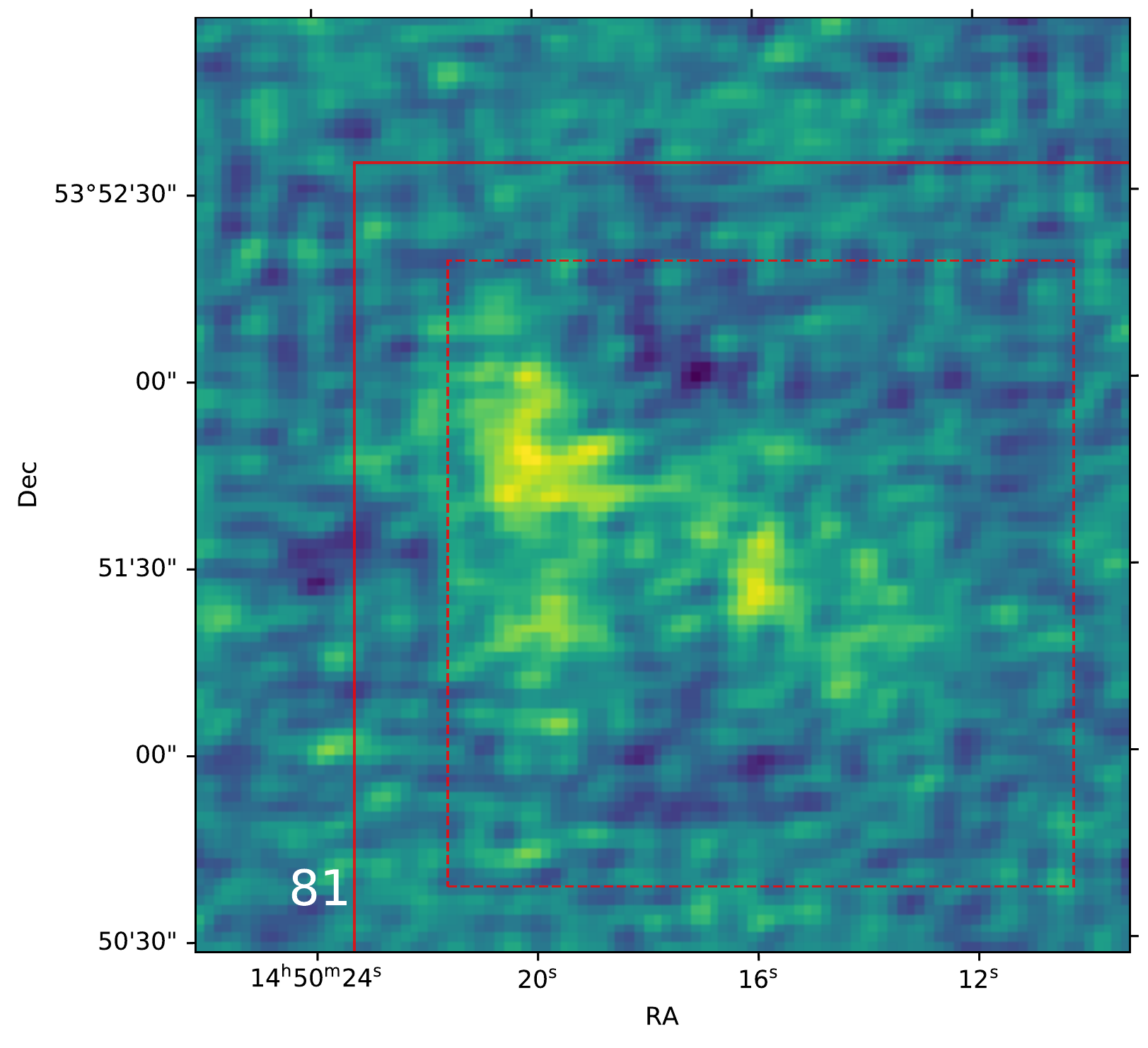}
\includegraphics[width=0.19\textwidth]{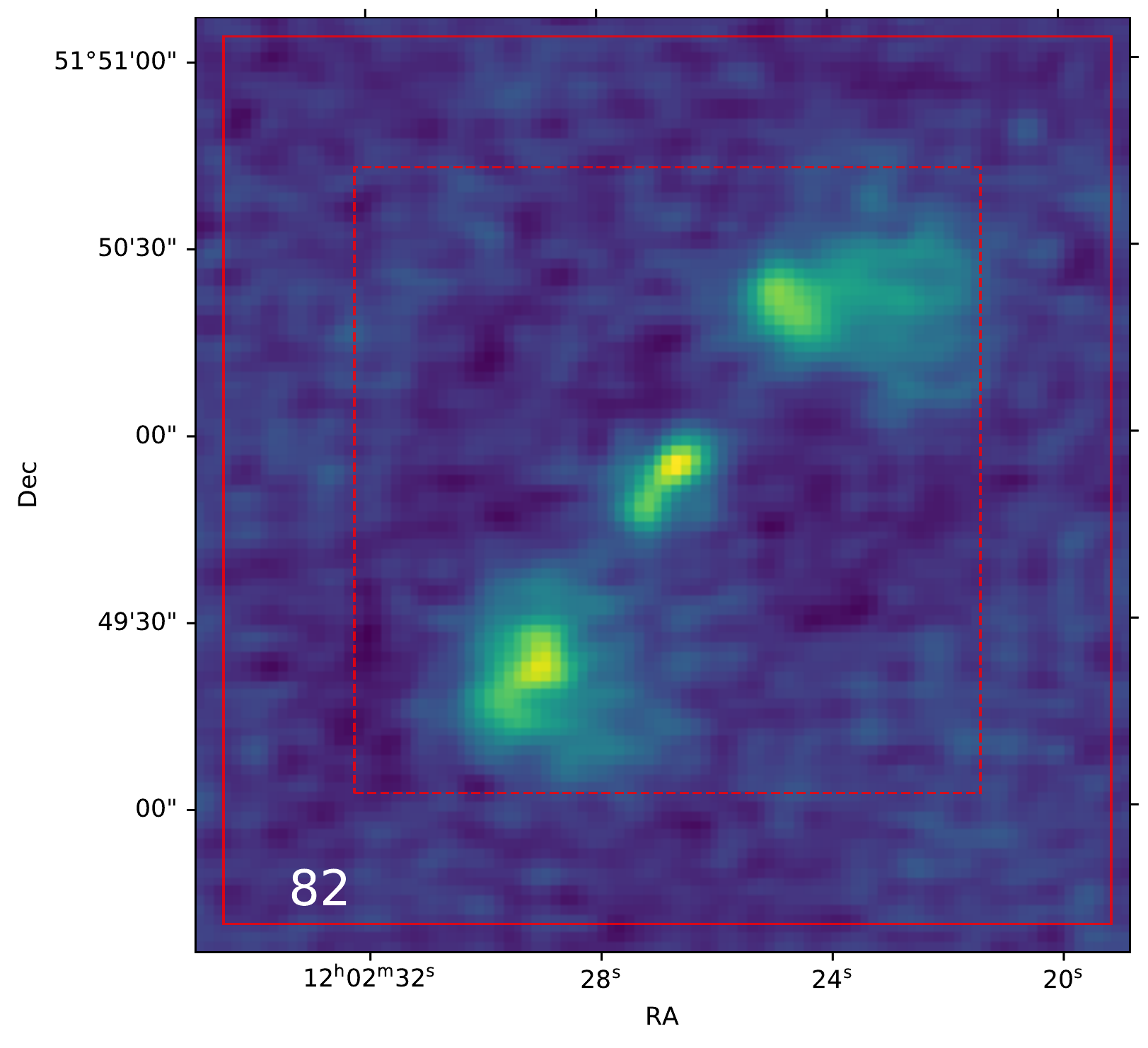}
\includegraphics[width=0.19\textwidth]{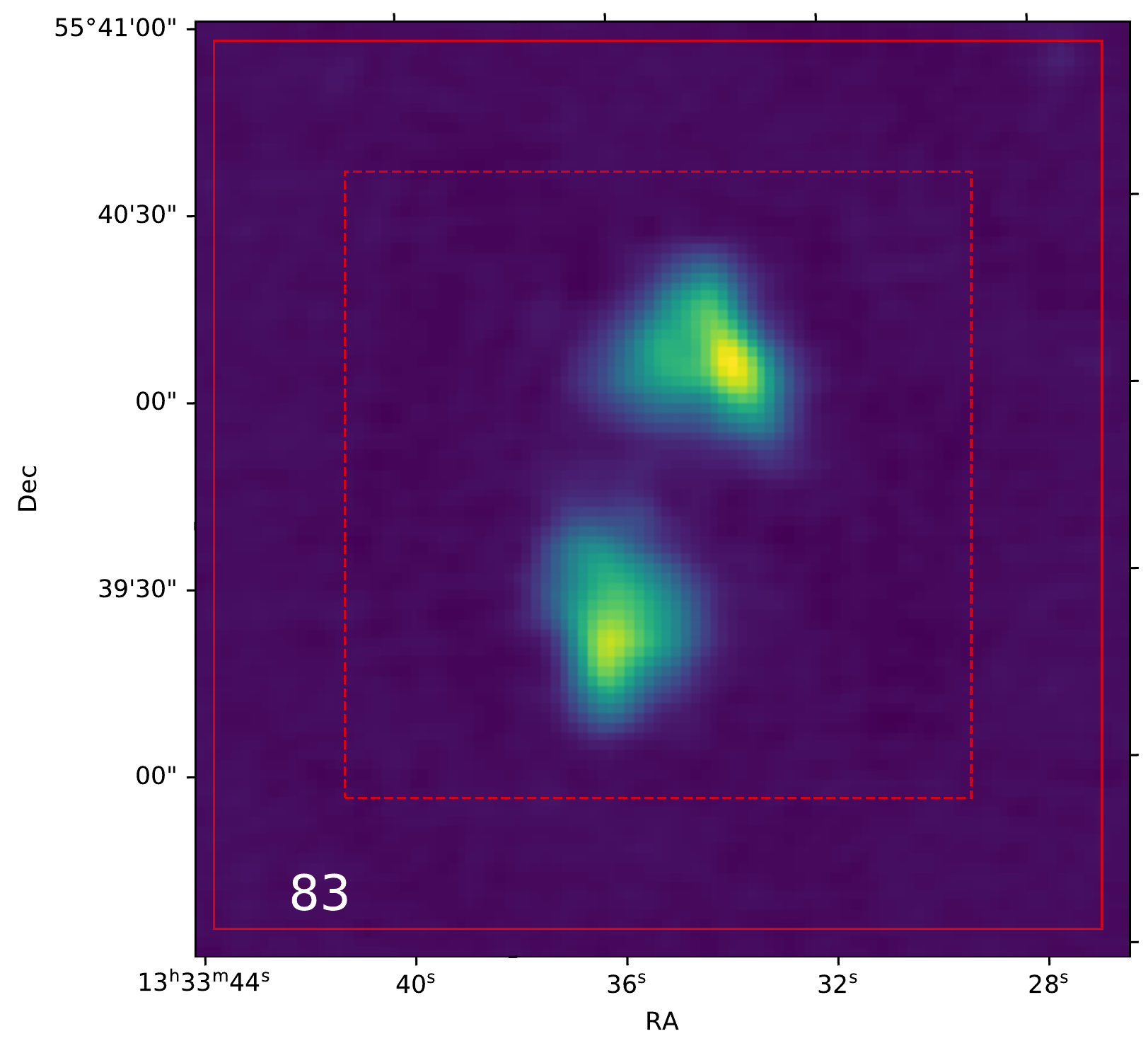}
\includegraphics[width=0.19\textwidth]{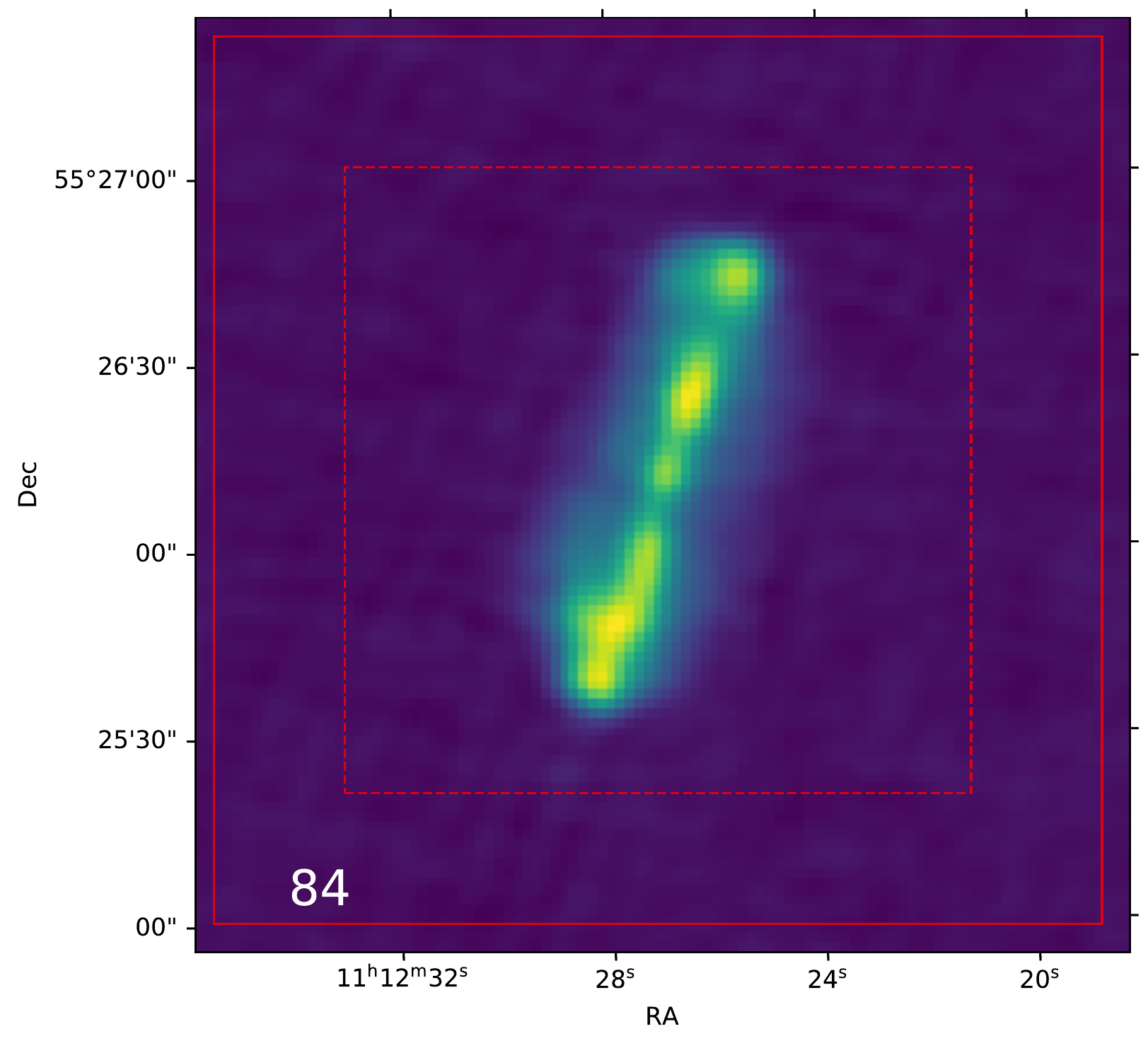}
\includegraphics[width=0.19\textwidth]{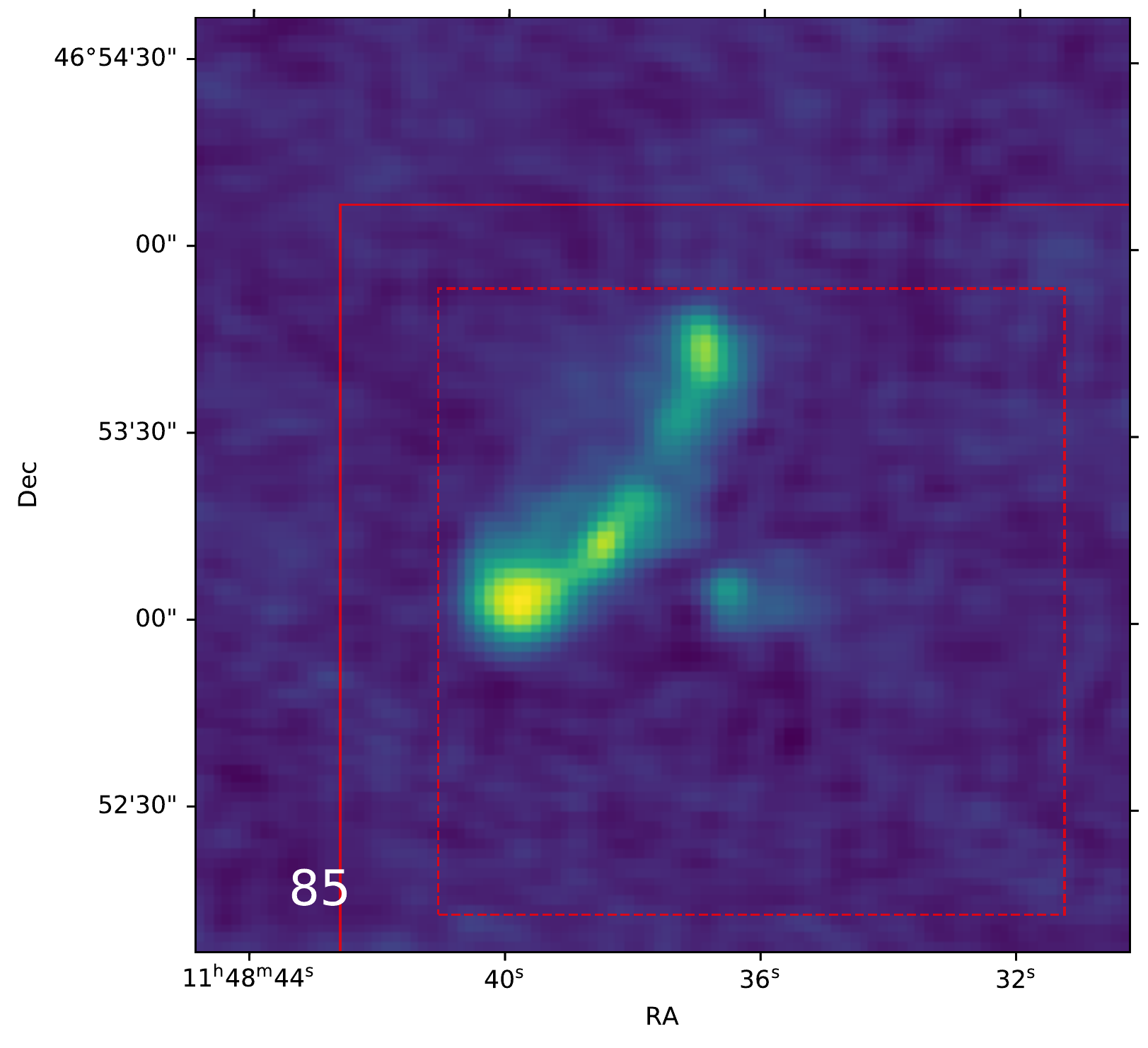}
\includegraphics[width=0.19\textwidth]{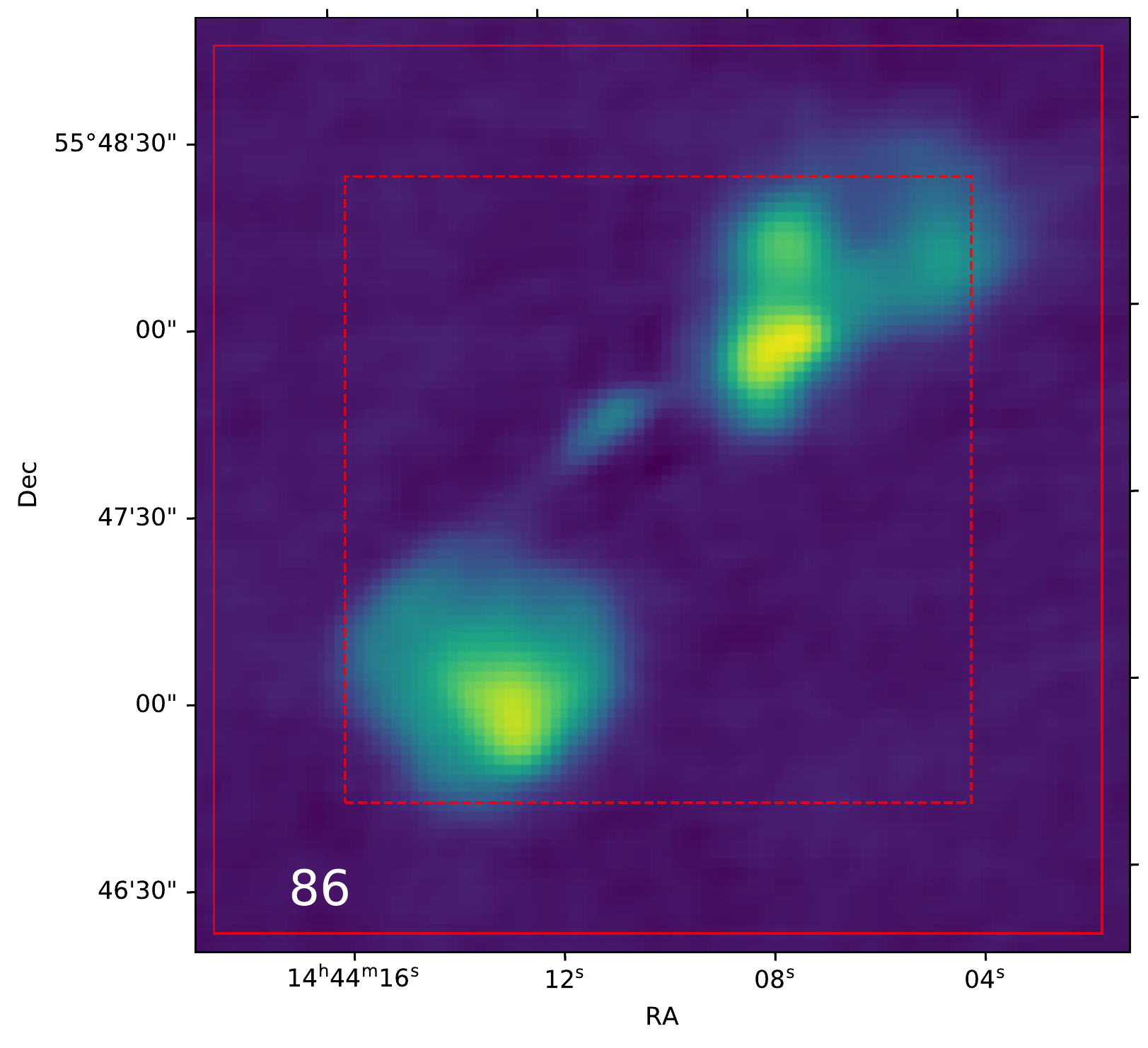}
\includegraphics[width=0.19\textwidth]{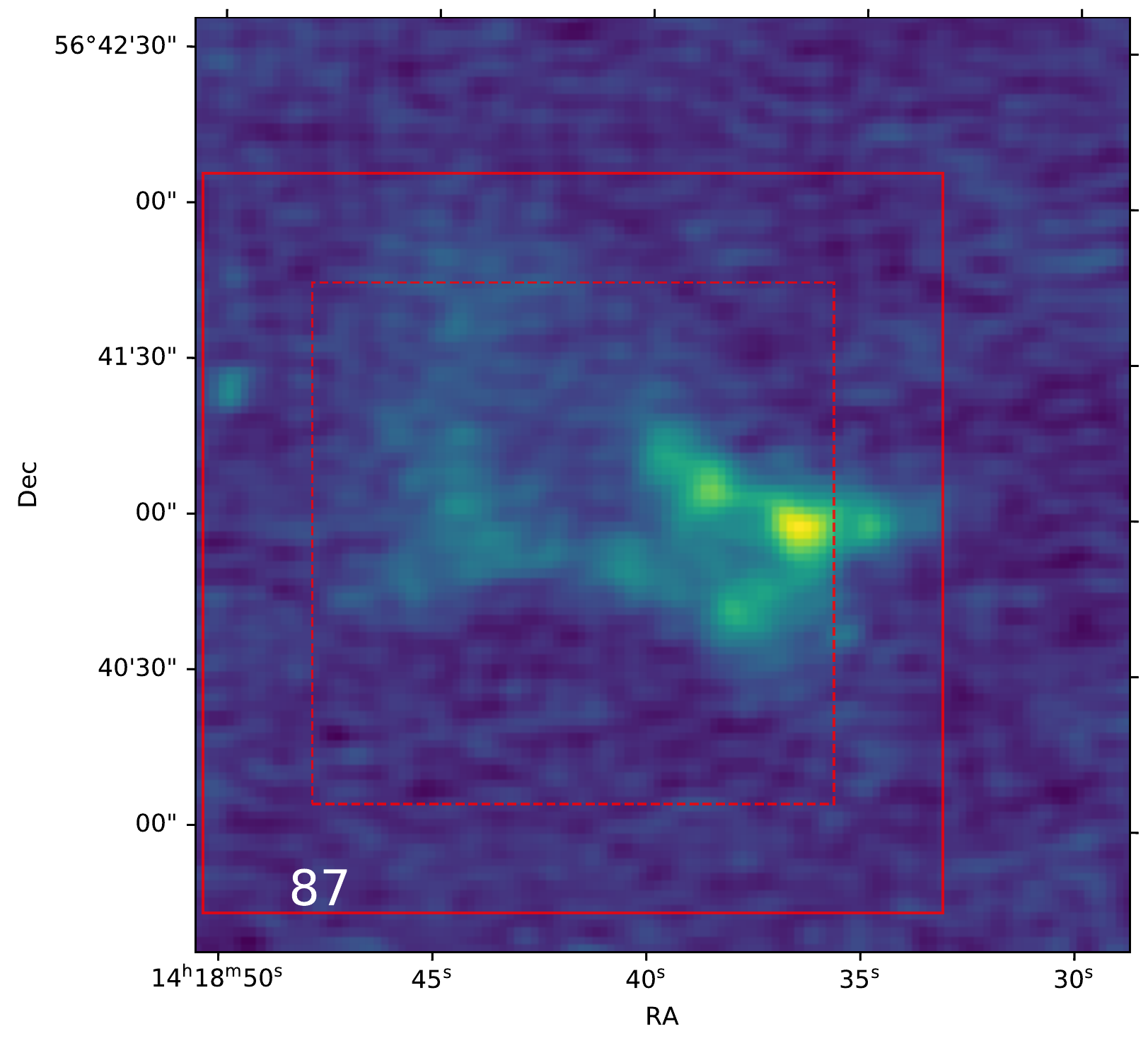}
\includegraphics[width=0.19\textwidth]{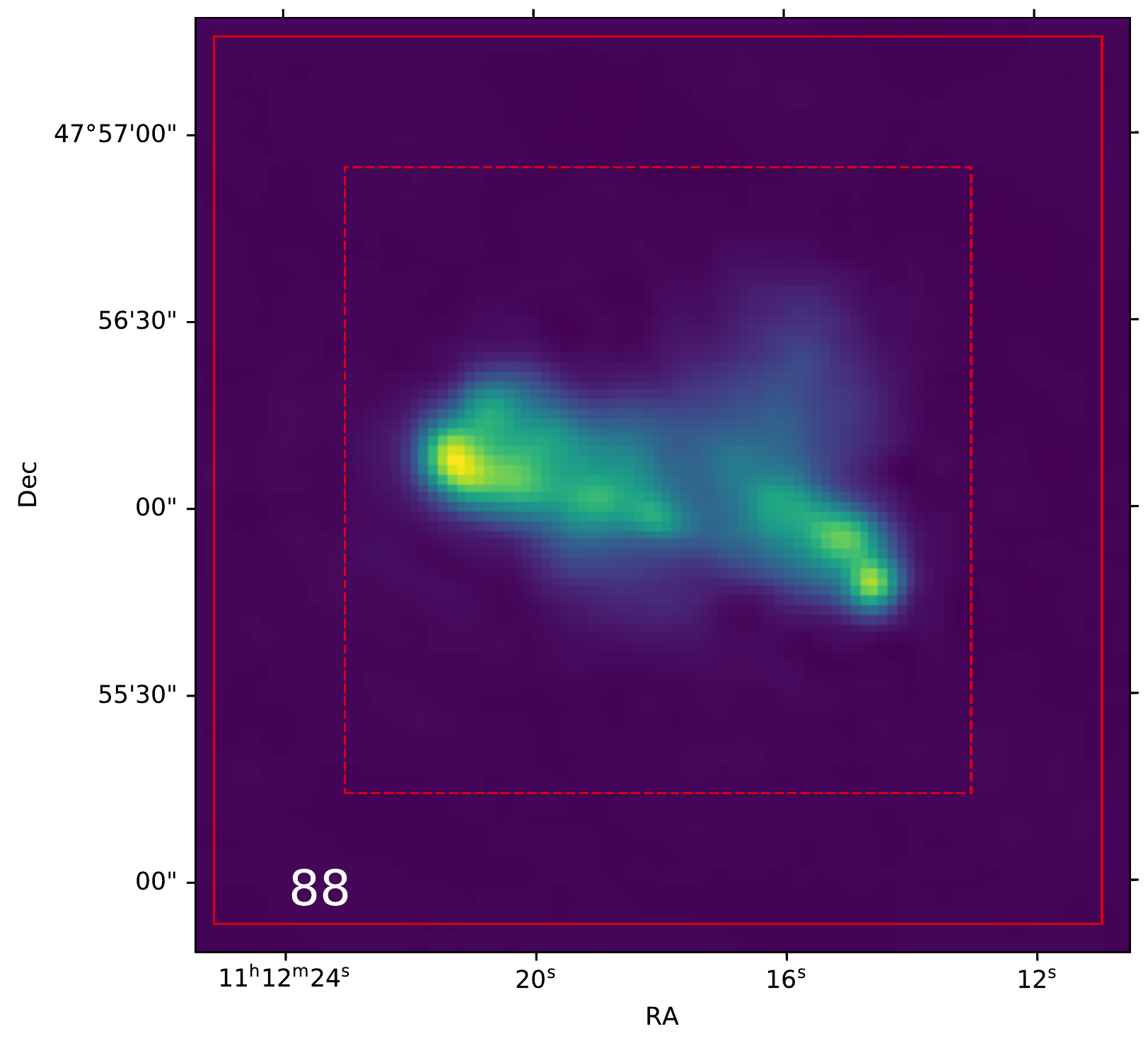}
\includegraphics[width=0.19\textwidth]{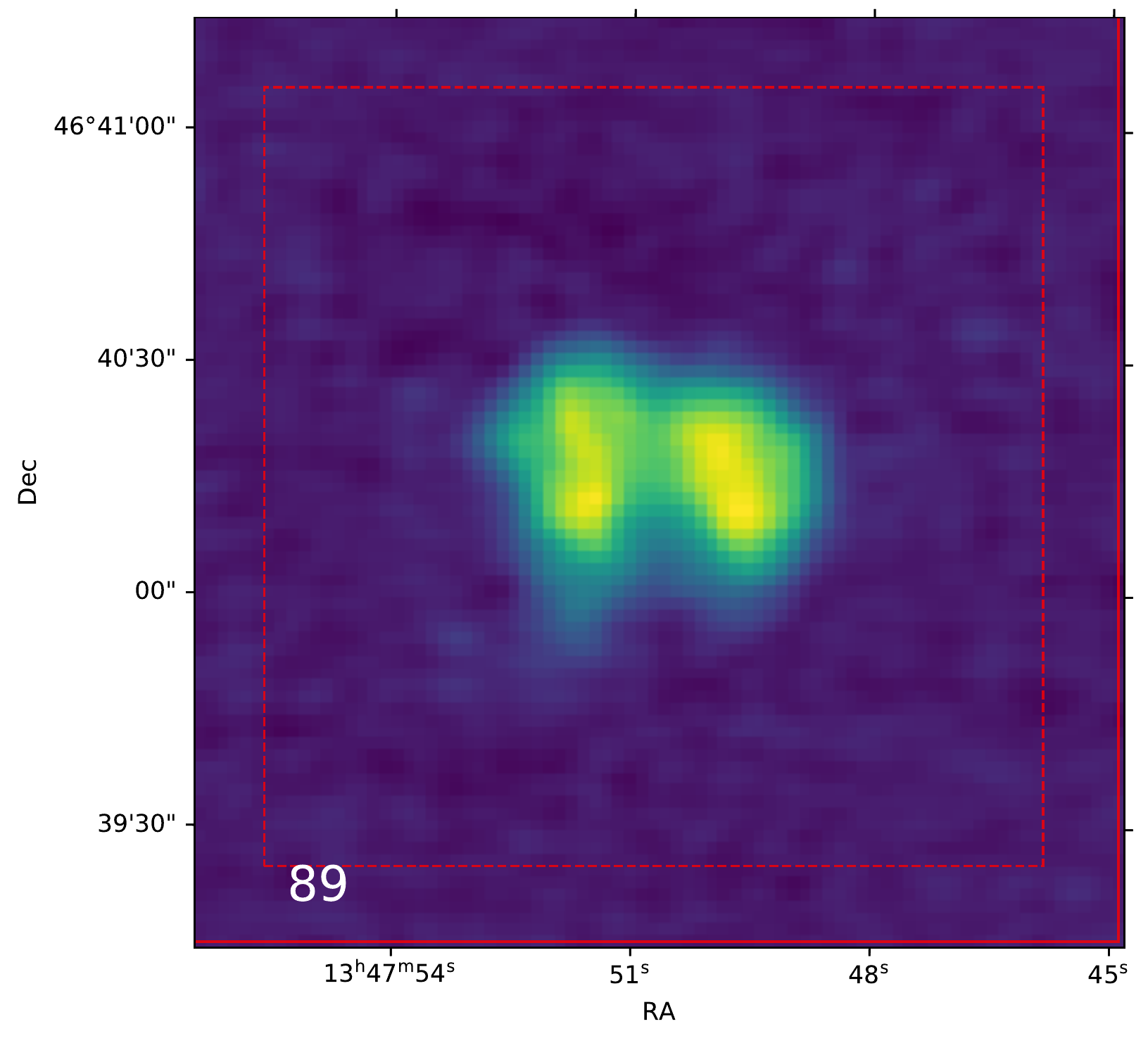}
\includegraphics[width=0.19\textwidth]{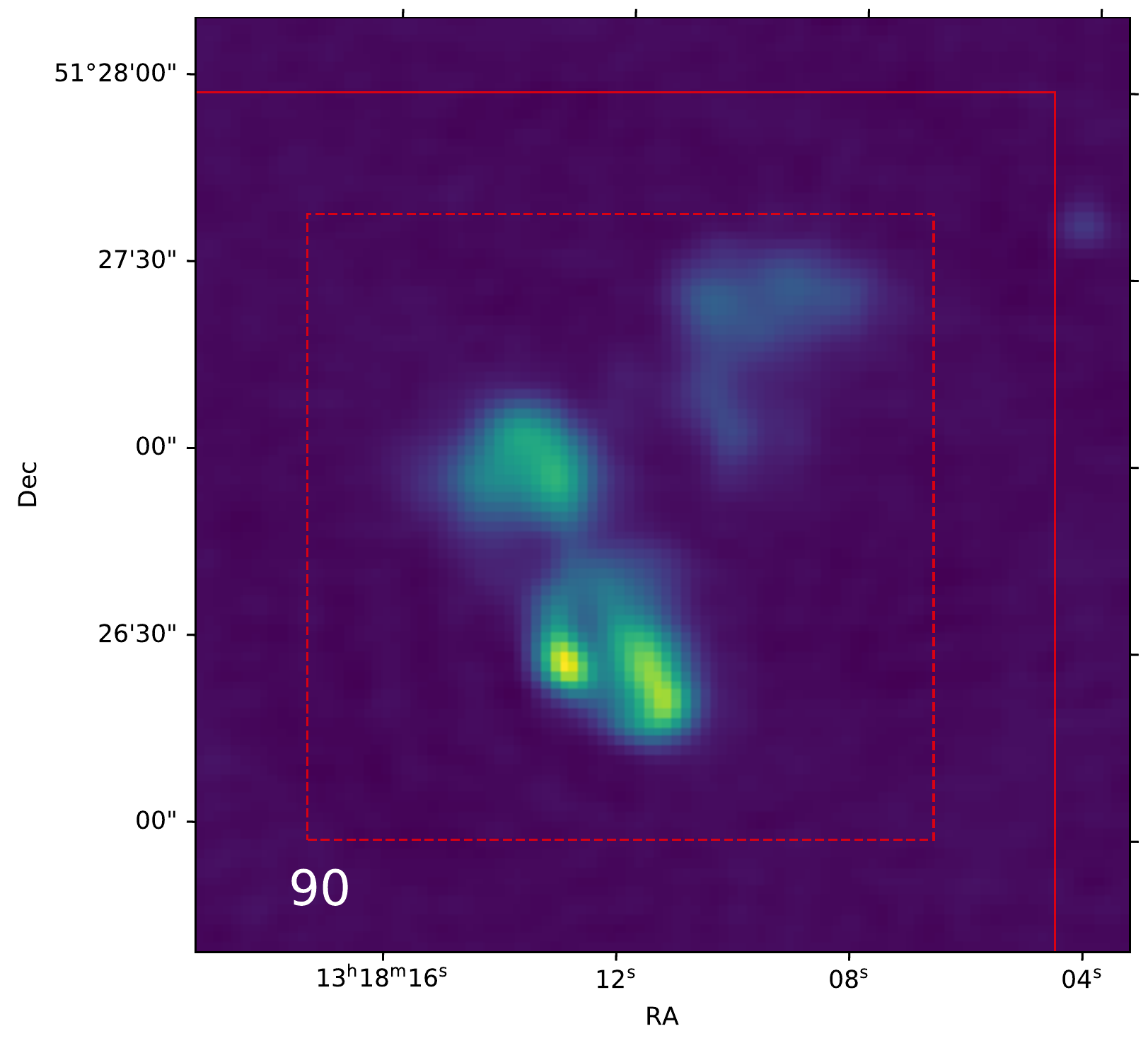}
\includegraphics[width=0.19\textwidth]{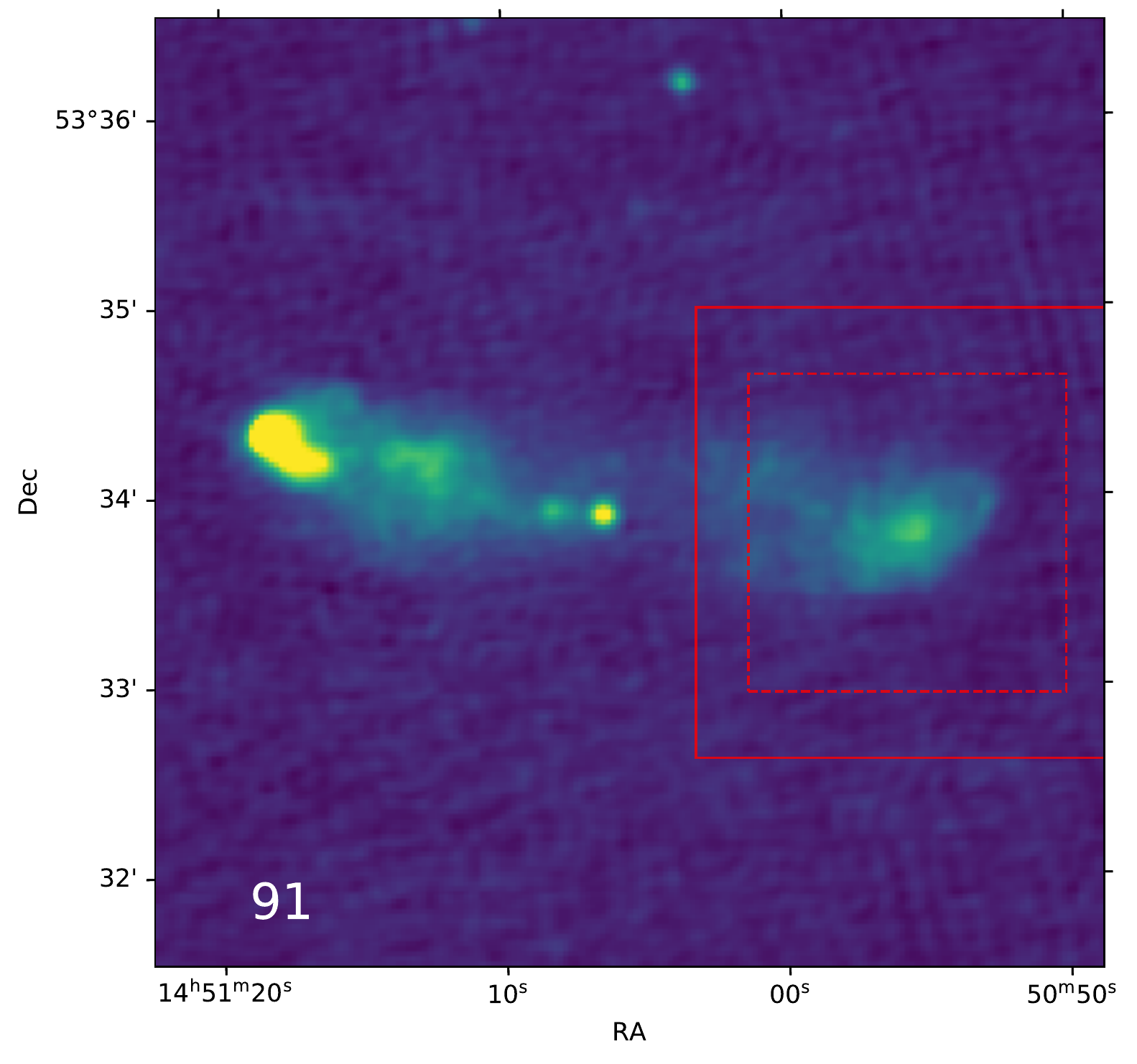}
\includegraphics[width=0.19\textwidth]{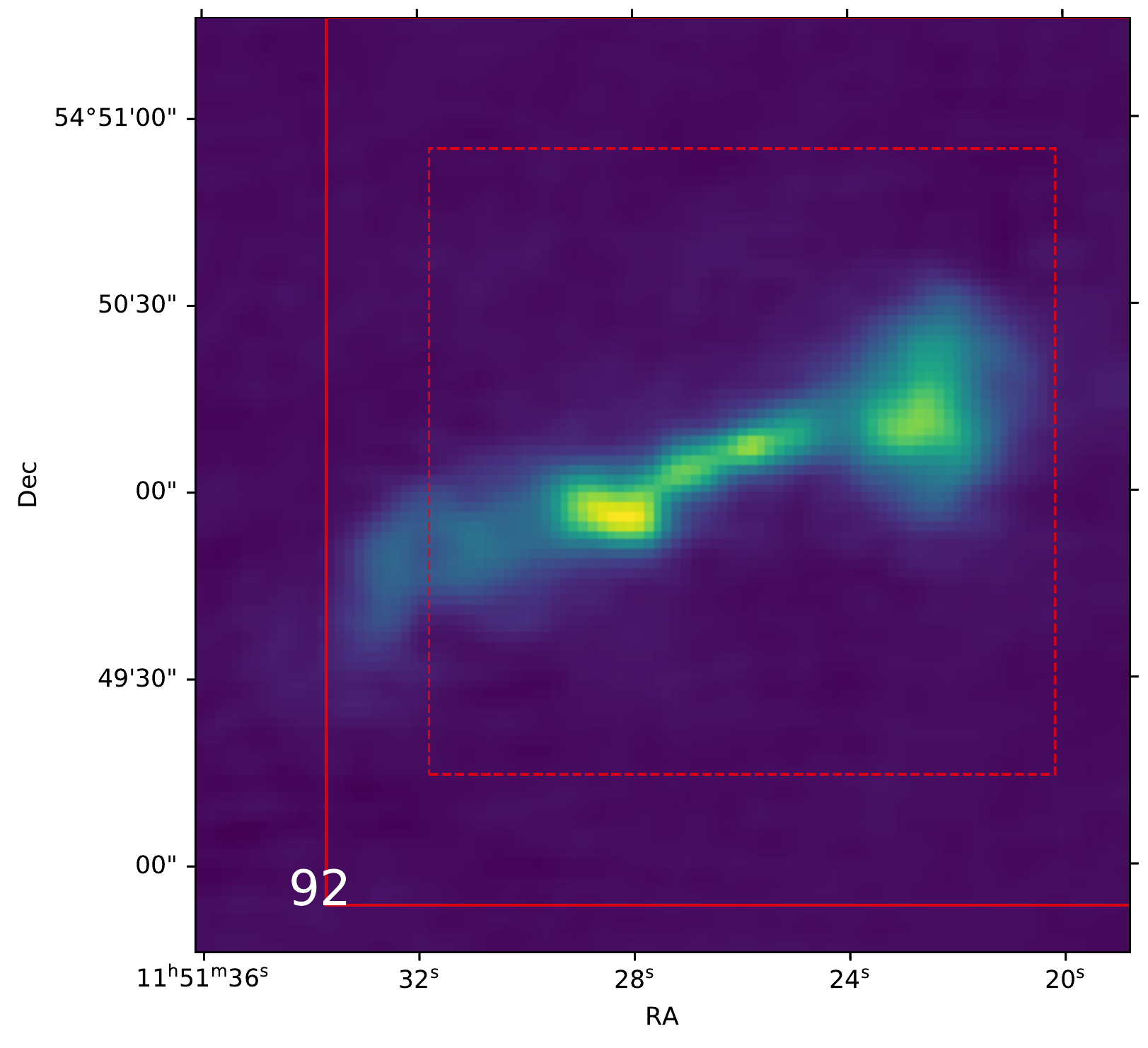}
\includegraphics[width=0.19\textwidth]{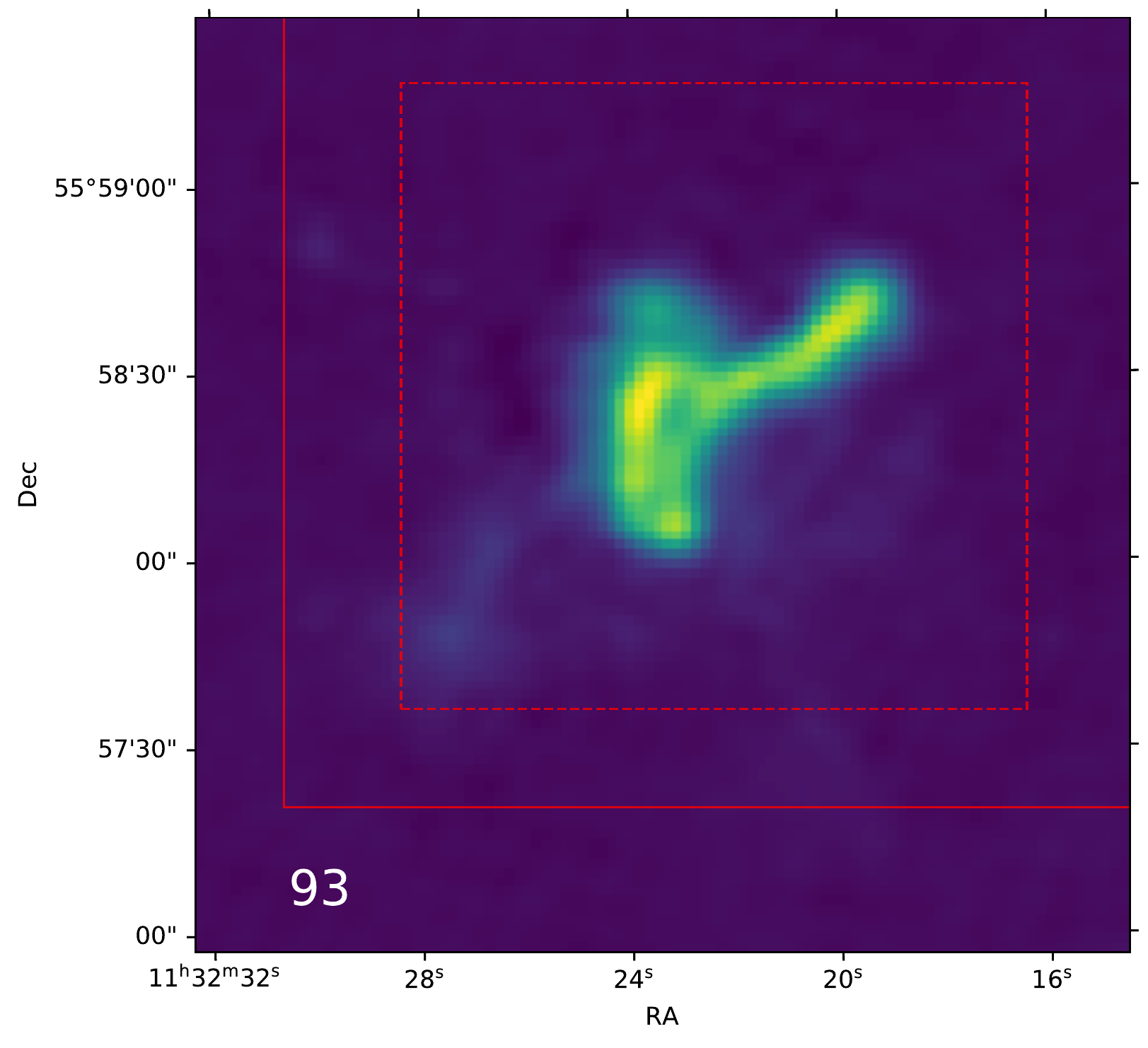}
\includegraphics[width=0.19\textwidth]{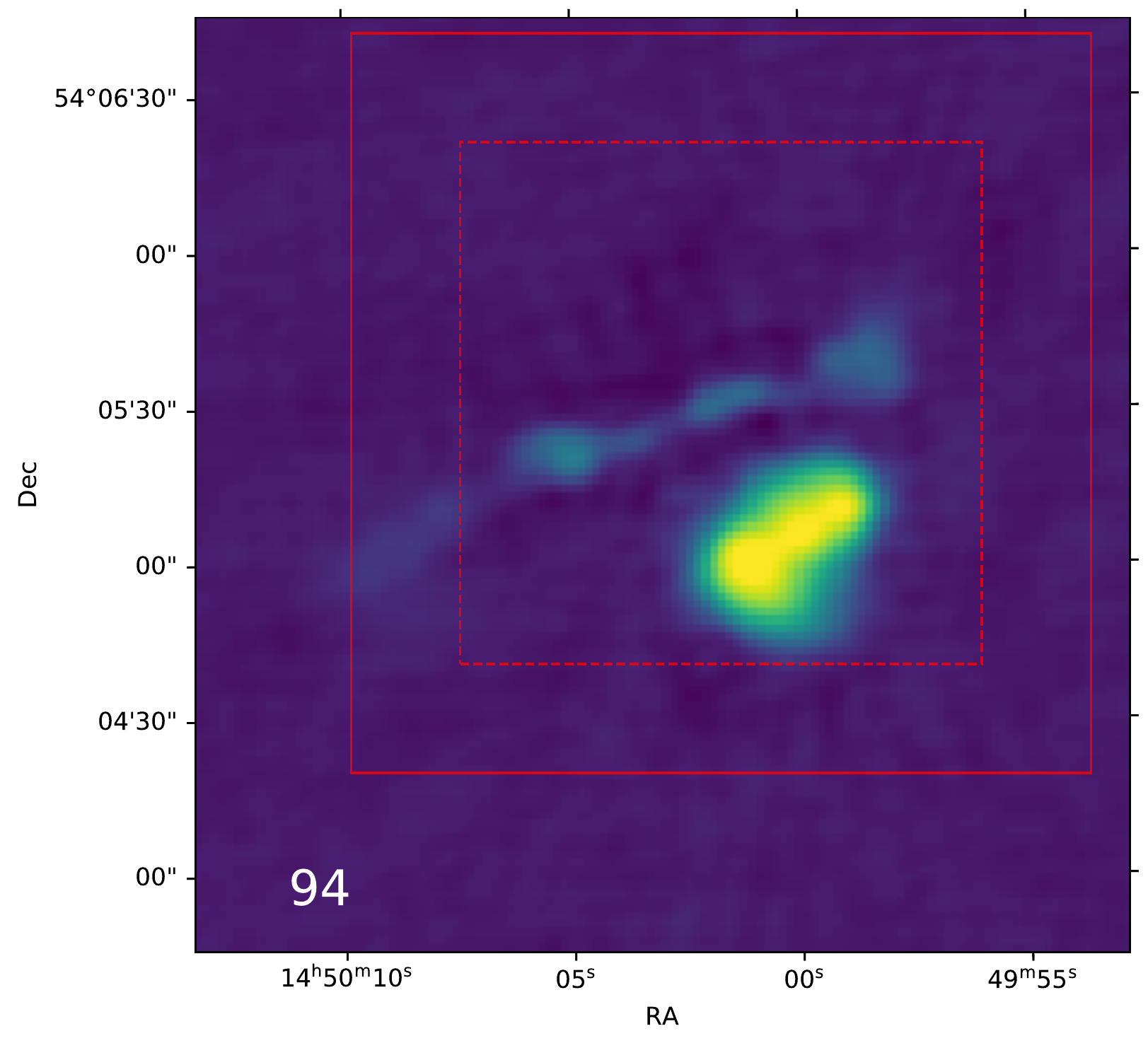}
\includegraphics[width=0.19\textwidth]{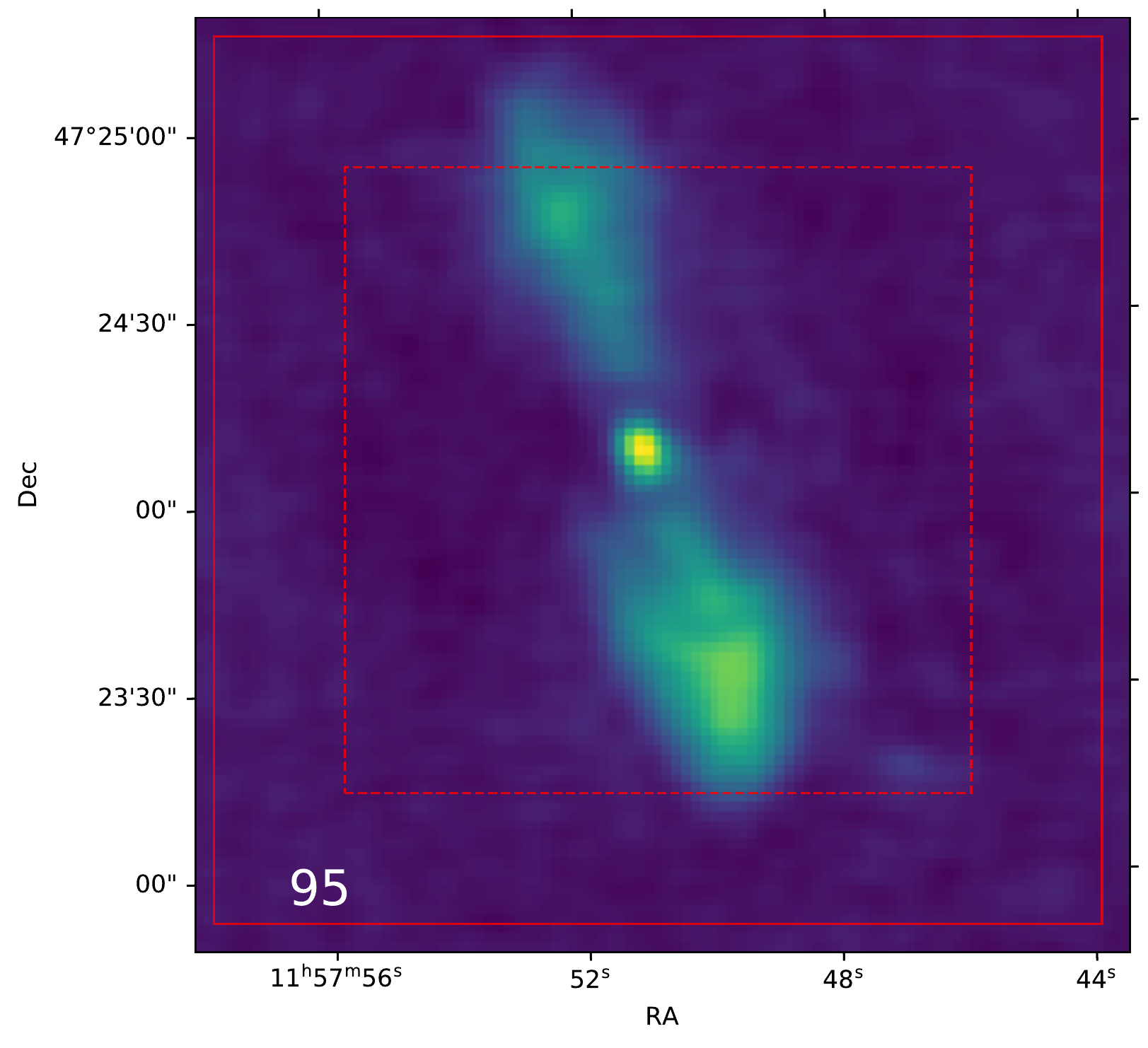}
\includegraphics[width=0.19\textwidth]{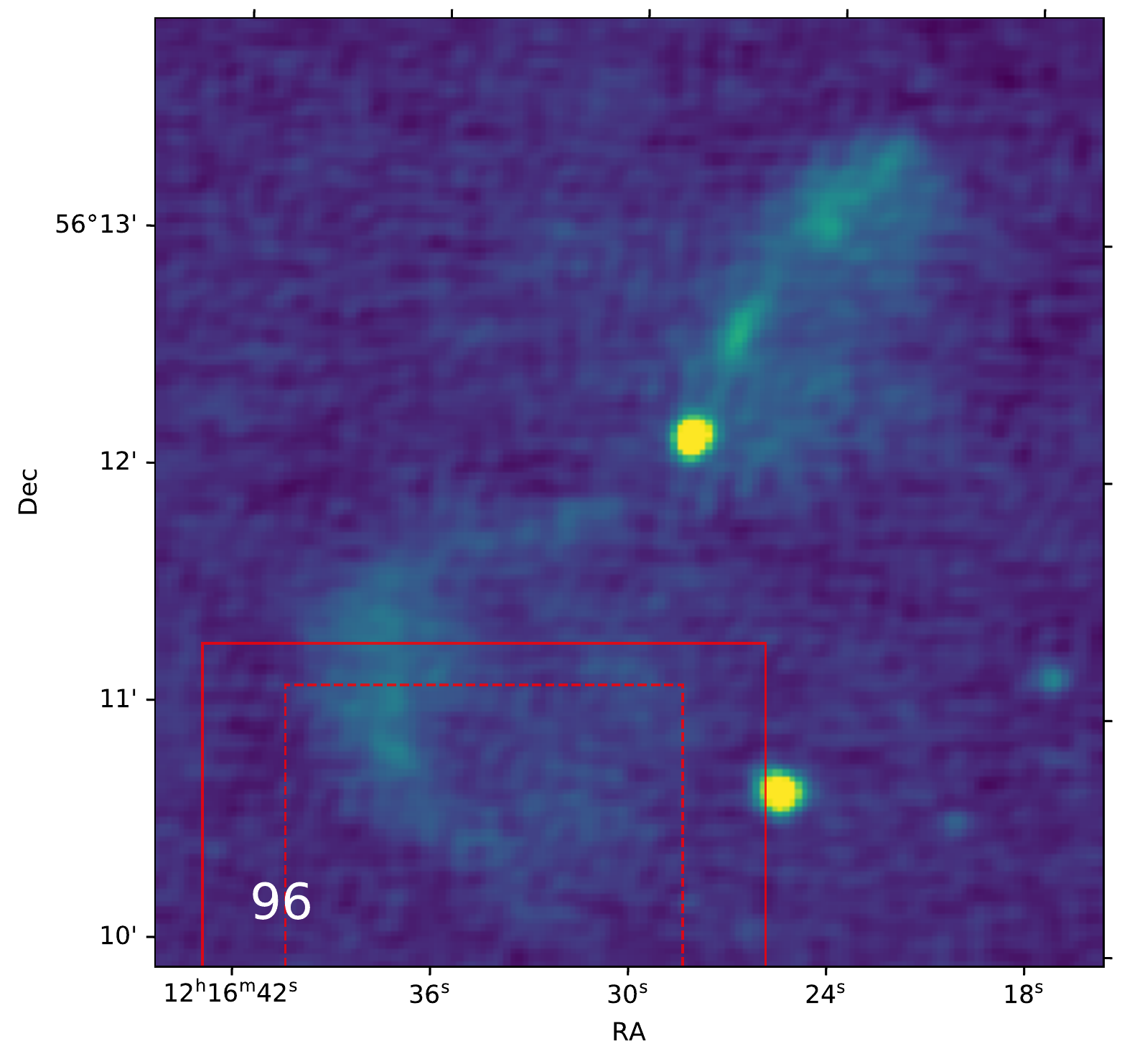}
\includegraphics[width=0.19\textwidth]{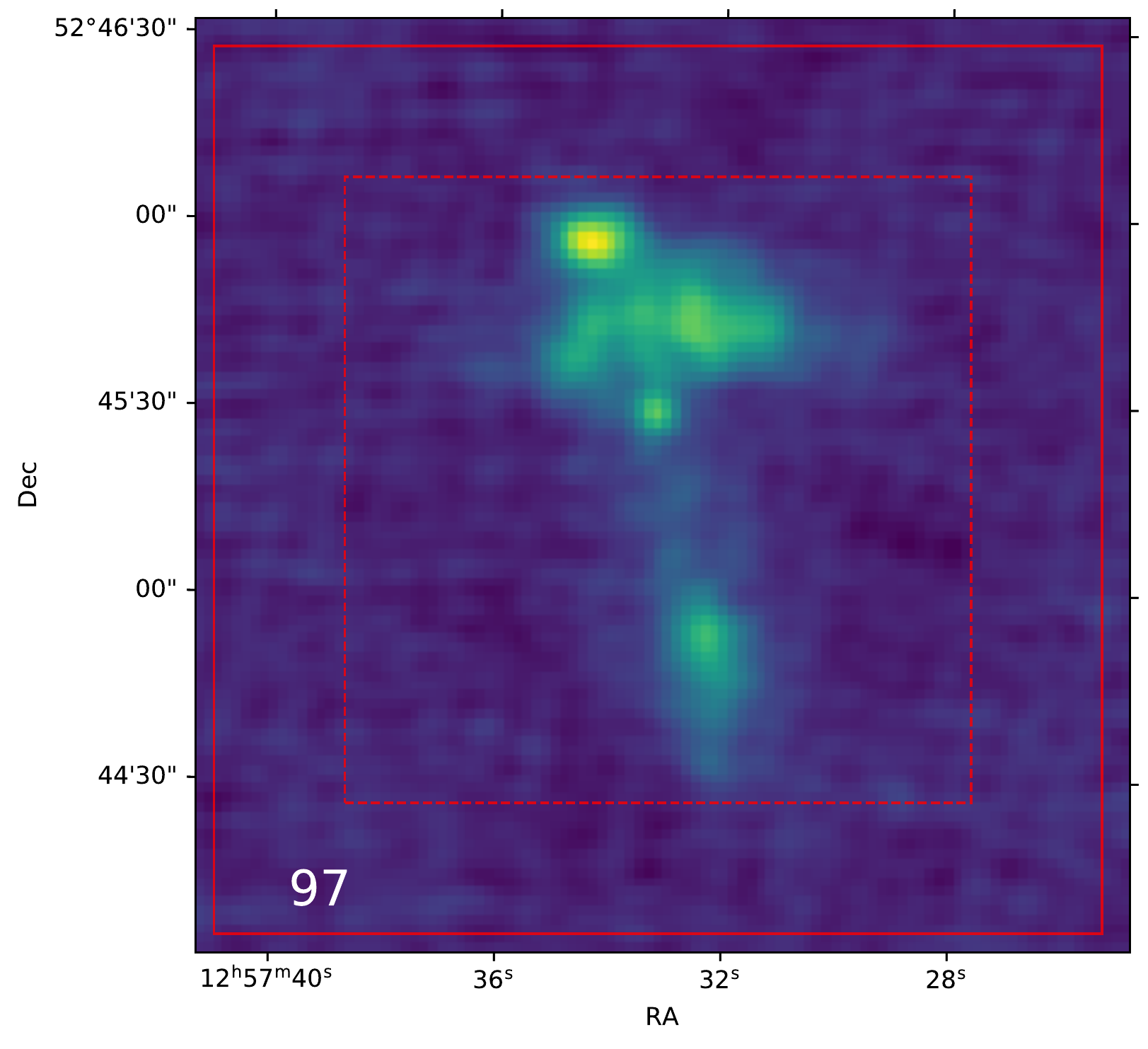}
\includegraphics[width=0.19\textwidth]{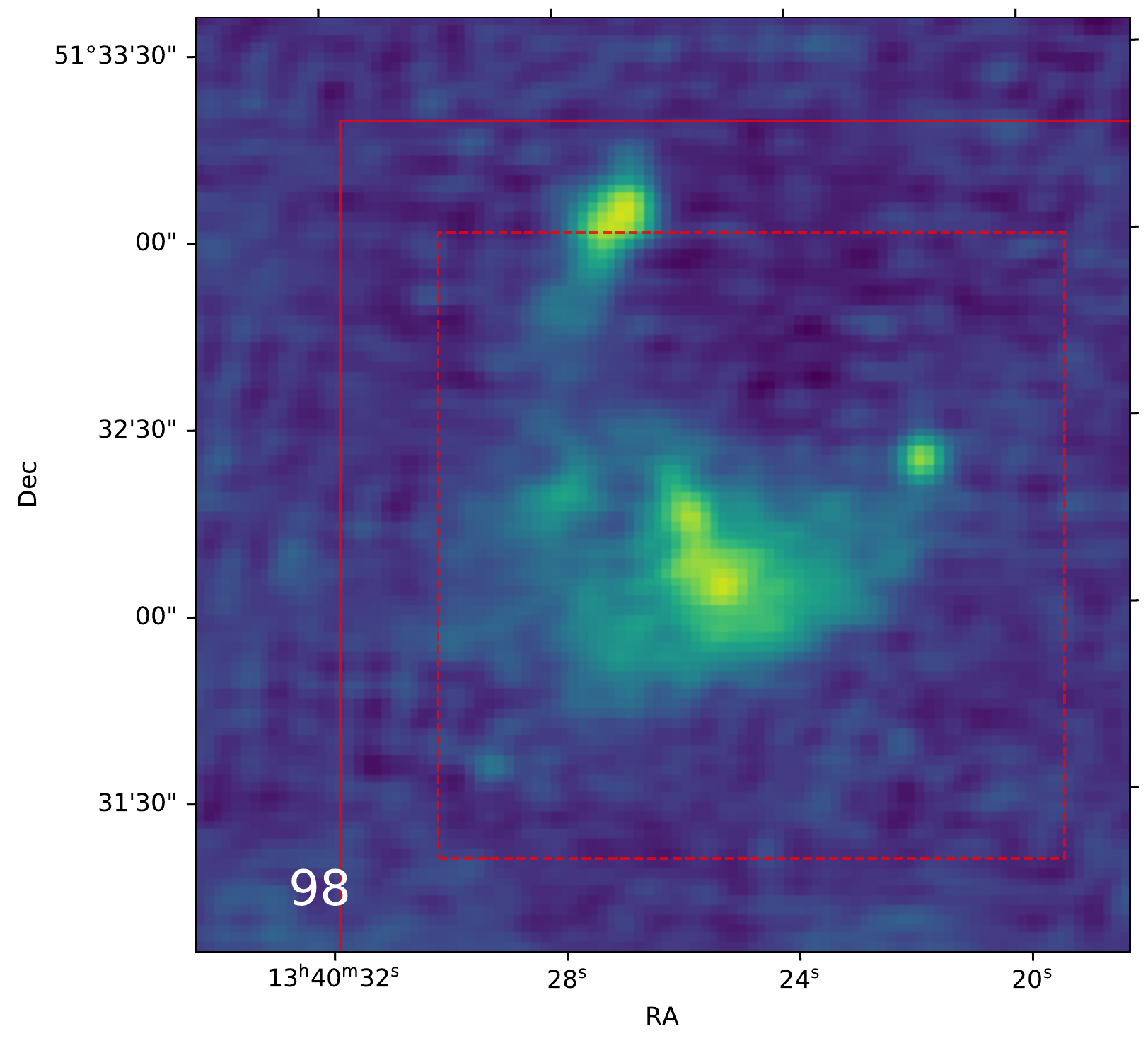}
\includegraphics[width=0.19\textwidth]{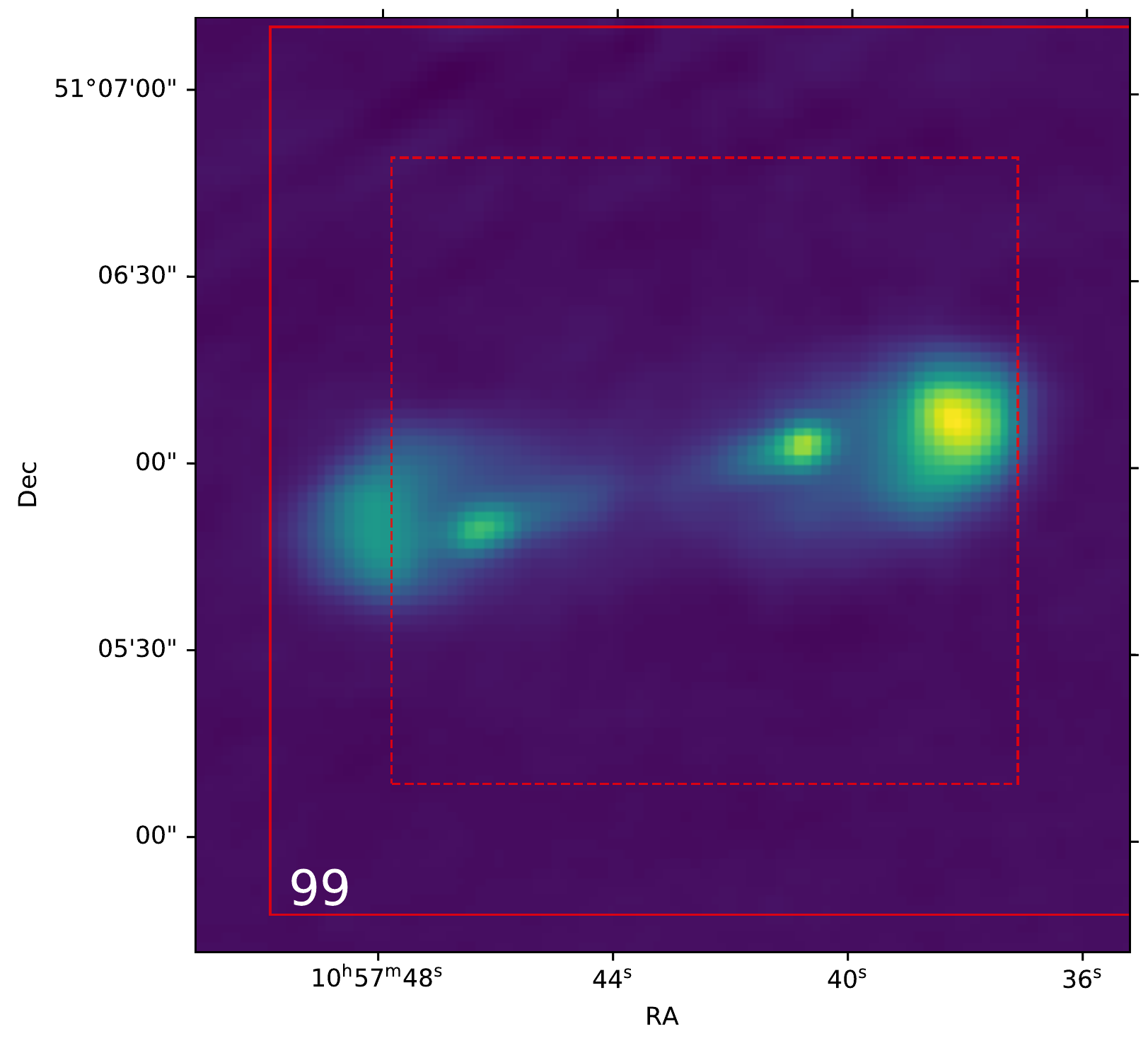}
\includegraphics[width=0.19\textwidth]{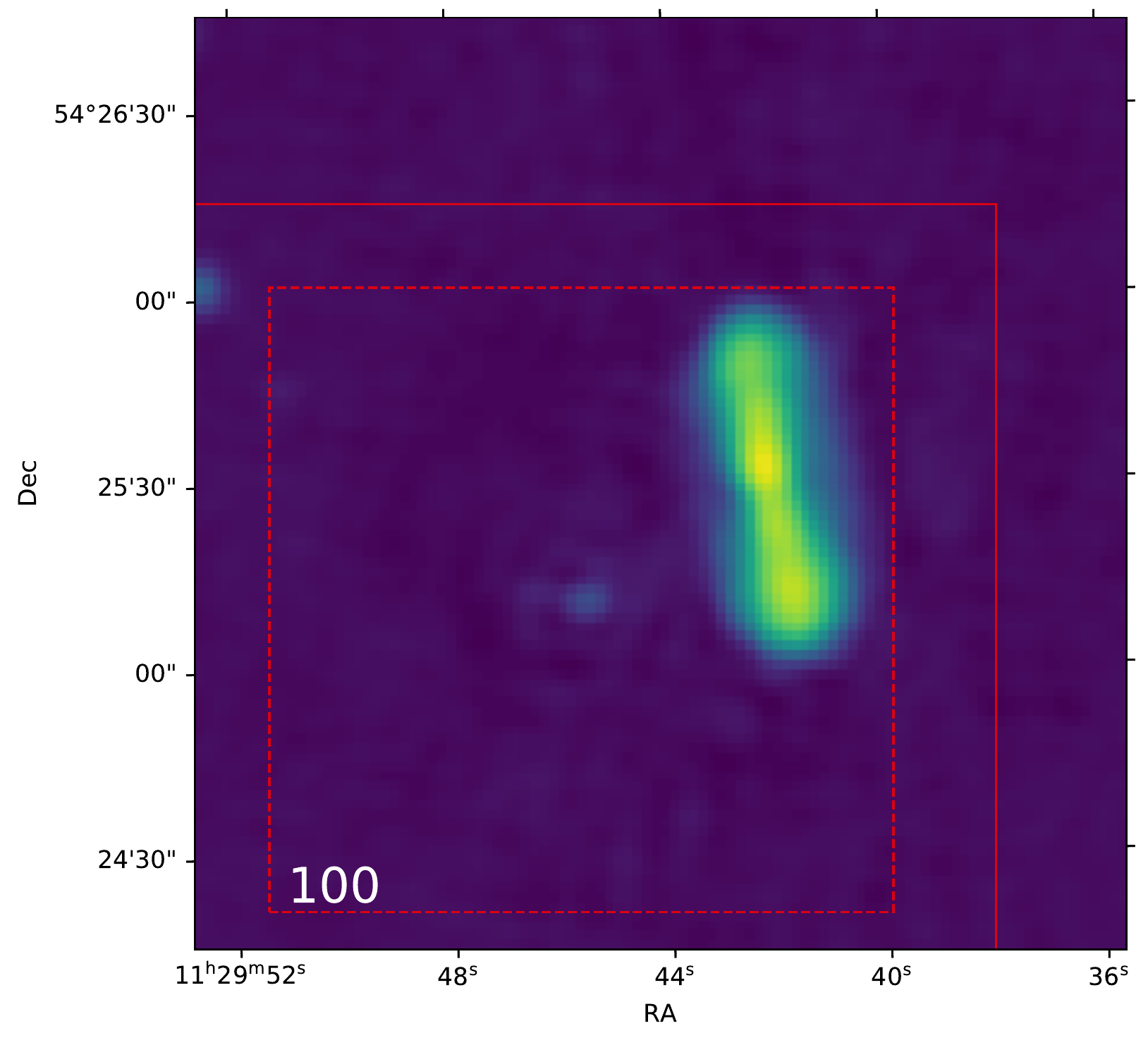}
\caption{\textit{Continued:} See previous page for the figure caption.}
\end{center}\end{figure*} 

\begin{figure}\begin{center}

\includegraphics[width=0.19\textwidth]{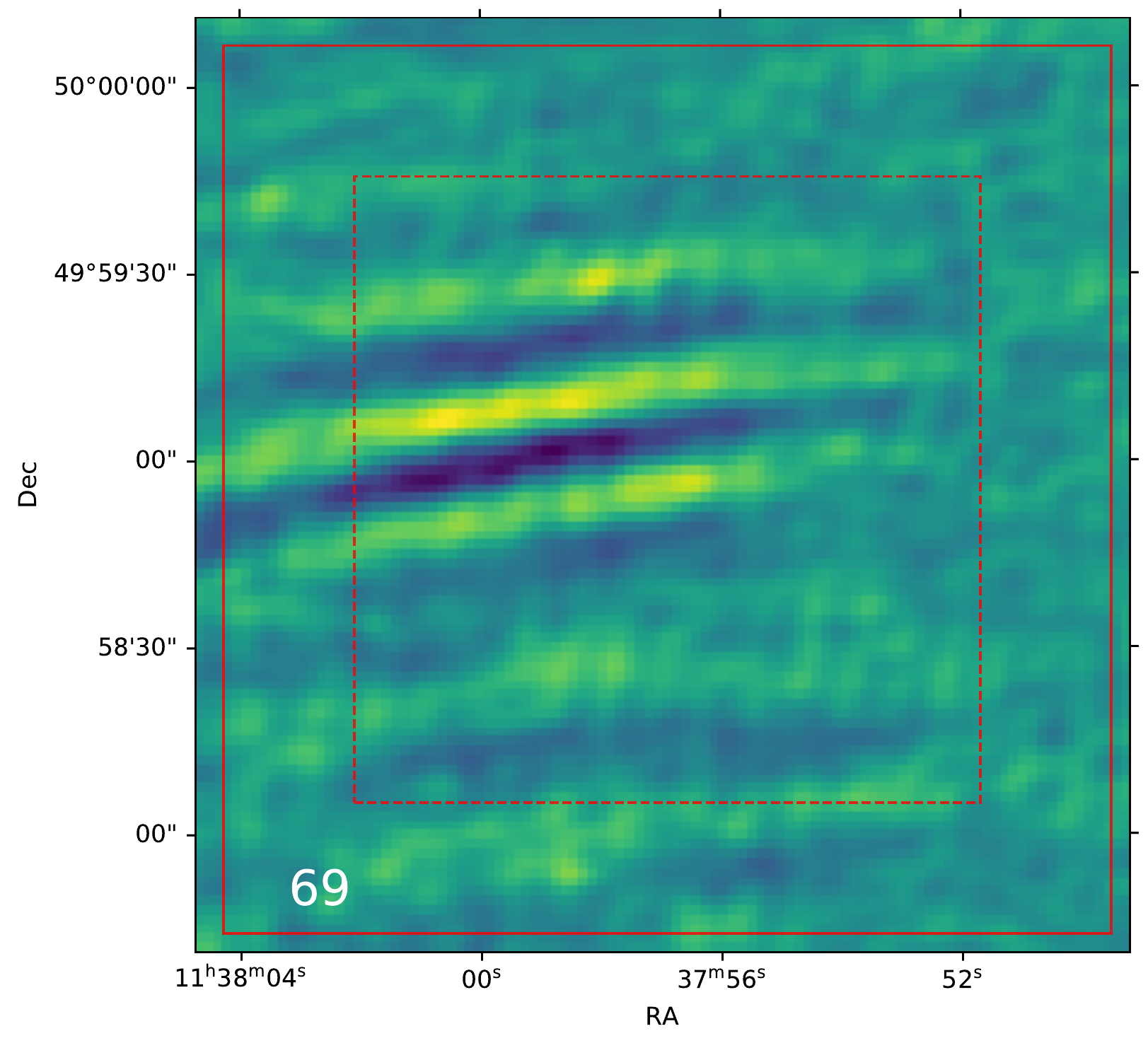}
\caption{Artefact within the 100 morphological outliers.}
\label{fig:100outliers_artefact}
\end{center}\end{figure}

\begin{table*}[t]
  \centering
\caption{Table containing the 100 sources in the DR1 area with the largest Euclidean norm to their best matching representative image (or outlier score) in the SOM shown in Fig. \ref{fig:final_som}. The numbers (\#) match the numbers shown in Fig. \ref{fig:100outliers1}. Each radio description includes a morphology-based FR-class when applicable -- the `?' indicates the FR-class could not be clearly determined.}
\label{tab:out100}
\resizebox{0.9\textwidth}{!}{
\begin{tabular}{lllllll}
\toprule
          \# & Outlier score & Rep. Image & Radio description (FR-class) &                 Radio ID &                  Optical description &                    Optical ID \\
\midrule
      1, 67 &          22.5 &      (1,1) &                         M101 &                        - &                               Galaxy &                         M 101 \\
          2 &          20.6 &      (1,1) &                      complex &   FIRST J142857.6+543627 &                                    - &      SDSS J142857.65+543627.7 \\
          3 &          19.9 &      (1,1) &                    ambiguous &                        - &                               Galaxy &               SDSSCGB 25342.4 \\
          4 &          18.4 &      (1,1) &                nearby galaxy &      NVSS J115341+475132 &          Galaxy in Group of Galaxies &                     NGC 3949 \\
          5 &          18.2 &      (1,1) &                    ambiguous &      NVSS J131358+532822 &  Brightest galaxy in a Cluster (BCG) &       2MASX J13135837+5328092 \\
          6 &          18.0 &      (1,1) &                nearby galaxy &      NVSS J114837+484242 &           Galaxy in Pair of Galaxies &                     NGC  3893 \\
          7 &          17.7 &      (1,1) &                nearby galaxy &      NVSS J121103+502901 &           Galaxy in Pair of Galaxies &                     NGC  4157 \\
          8 &          16.7 &      (1,1) &                    WAT (FRI) &                        - &                  Cluster of Galaxies &                  [YSS2008] 96 \\
          9 &          16.7 &      (1,1) &                nearby galaxy &      NVSS J143246+492730 &          Galaxy in Group of Galaxies &                     NGC  5676 \\
         10 &          14.8 &      (1,1) &  AGN remnant candidate (FRI) &                        - &                                    - &                             - \\
         11 &          14.6 &      (1,1) &                    ambiguous &                        - &  Brightest galaxy in a Cluster (BCG) &      SDSS J122718.54+562529.8 \\
         12 &          14.5 &      (1,1) &                  NEBD (FRII) &                        - &                               Quasar &      SDSS J105817.89+514017.7 \\
     13, 18 &          14.2 &      (1,1) &                  NEBD (FRII) &                        - &                               Galaxy &               SDSSCGB 22520.3 \\
         14 &          13.9 &     (2,10) &                    WAT (FRI) &                        - &        Galaxy in Cluster of Galaxies &       DES J143442.95+495132.2 \\
         15 &          13.9 &      (1,1) &                    ambiguous &      NVSS J112624+475044 &                                    - &                             - \\
         16 &          13.8 &      (1,1) &  AGN remnant candidate (FRI) &                        - &                                    - &                             - \\
         17 &          13.8 &     (2,10) &                    WAT (FRI) &                        - &        Galaxy in Cluster of Galaxies &       DES J125051.17+531309.4 \\
         19 &          13.5 &     (2,10) &                   double (?) &                        - &                                    - &      SDSS J132435.19+504102.3 \\
         20 &          13.5 &      (1,1) &                nearby galaxy &      NVSS J120535+503231 &           Galaxy in Pair of Galaxies &                     NGC  4088 \\
         21 &          13.4 &      (1,1) &                 cluster halo &                        - &                  Cluster of Galaxies &                    ZwCl  3570 \\
         22 &          13.2 &      (1,1) &                cluster relic &                        - &                    Group of Galaxies &                 [LH2011] 3085 \\
         23 &          13.2 &     (2,10) &                nearby galaxy &                        - &          Galaxy in Group of Galaxies &                         M 109 \\
         24 &          12.4 &     (2,10) &     asymmetric double (FRII) &                        - &                                    - &                             - \\
         25 &          12.3 &      (1,1) &                nearby galaxy &   7C 121620.39+473545.00 &                               Galaxy &                         M 106 \\
         26 &          12.0 &      (1,1) &                    ambiguous &                        - &                               Galaxy &               SDSSCGB 22905.1 \\
         27 &          12.0 &      (1,1) &                    ambiguous &                        - &                                    - &                             - \\
         28 &          11.7 &     (2,10) &                cluster relic &                        - &                                    - &                             - \\
         29 &          11.6 &      (1,1) &                  NEBD (FRII) &   7C 134413.90+541648.00 &        Galaxy in Cluster of Galaxies &       DES J134558.40+540344.9 \\
         30 &          11.5 &      (1,1) &                cluster relic &                        - &                               Galaxy &                  LEDA 2333622 \\
         31 &          11.4 &     (2,10) &                nearby galaxy &      NVSS J120601+472846 &          Galaxy in Group of Galaxies &                     NGC  4096 \\
         32 &          11.3 &     (2,10) &  AGN remnant candidate (FRI) &                        - &  Brightest galaxy in a Cluster (BCG) &      SDSS J144704.03+503332.9 \\
         33 &          11.1 &     (2,10) &                   WDS (FRII) &                        - &                                    - &                             - \\
         34 &          10.8 &      (1,1) &                   WDS (FRII) &                        - &                                    - &                             - \\
         35 &          10.7 &     (2,10) &                    WAT (FRI) &      NVSS J114019+535028 &                                    - &                             - \\
         36 &          10.7 &     (2,10) &                    WAT (FRI) &          [OR76] 1200+519 &                                    - &                             - \\
         37 &          10.5 &     (2,10) &                double (FRII) &                        - &                                    - &                             - \\
 38, 48, 54 &          10.5 &     (2,10) &       cluster halo, NAT, WAT &                        - &        Galaxy in Cluster of Galaxies &       2MASX J11343395+4905157 \\
         39 &          10.4 &      (1,1) &        asymmetric double (?) &                        - &                                    - &                             - \\
         40 &          10.4 &     (1,10) &                 X-type (FRI) &                        - &                               Galaxy &                  LEDA 2298950 \\
         41 &          10.1 &      (1,1) &                  NEBD (FRII) &                        - &                                    - &                             - \\
         42 &          10.0 &      (1,1) &                    ambiguous &                        - &                                    - &                             - \\
         43 &           9.9 &      (1,1) &  AGN remnant candidate (FRI) &            MPWK 1151+488 &                                    - &                             - \\
         44 &           9.9 &      (1,1) &                 cluster halo &                        - &                               Galaxy &               SDSSCGB 14124.3 \\
         45 &           9.8 &     (2,10) &                  NEBD (FRII) &  7C 114226.19+535535.00  &                                    - &       2MASX J11450656+5338521 \\
         46 &           9.8 &     (2,10) &                double (FRII) &                        - &                                    - &                             - \\
         47 &           9.7 &     (2,10) &  AGN remnant candidate (FRI) &                        - &                                    - &                             - \\
         49 &           9.7 &     (2,10) &                    NAT (FRI) &                        - &        Galaxy in Cluster of Galaxies &       DES J130257.91+511936.0 \\
         50 &           9.6 &      (1,1) &                    ambiguous &      NVSS J112922+540735 &  Brightest galaxy in a Cluster (BCG) &       2MASX J11291925+5407344 \\
         51 &           9.6 &      (1,1) &                double (FRII) &                        - &                                    - &      SDSS J112213.75+493326.5 \\
         52 &           9.4 &     (2,10) &     asymmetric double (FRII) &  ILT J133135.09+455957.0 &  Brightest galaxy in a Cluster (BCG) &      SDSS J133135.25+455955.4 \\
         53 &           9.4 &     (2,10) &                  NEBD (FRII) &                        - &                                    - &                             - \\
         55 &           9.1 &      (1,1) &                    ambiguous &                        - &                                    - &                             - \\
         56 &           9.0 &     (2,10) &                    ambiguous & [YHW2016] 181.11848+50.17200 &  Brightest galaxy in a Cluster (BCG) &   SDSS J120428.43+501018.9 \\
         57 &           8.9 &     (2,10) &                    WAT (FRI) &                        - &                                    - &                             - \\
         58 &           8.8 &     (2,10) &     asymmetric double (FRII) &                        - &                                    - &                             - \\
         59 &           8.7 &     (2,10) &                double (FRII) &                        - &                                    - &                             - \\
         60 &           8.7 &      (1,1) &                    WAT (FRI) &                        - &                                    - &                             - \\
         61 &           8.7 &     (2,10) &  AGN remnant candidate (FRI) &                        - &                                    - &                             - \\
         62 &           8.7 &     (2,10) &                    WAT (FRI) &                        - &                                    - &      SDSS J141157.99+550515.4 \\
         63 &           8.5 &      (1,1) &                      complex &                        - &                                    - &                             - \\
         64 &           8.4 &      (2,9) &                    WAT (FRI) &                        - &  Brightest galaxy in a Cluster (BCG) &      SDSS J124154.22+483835.4 \\
         65 &           8.3 &      (1,1) &                X-type (FRII) &                  54P 106 &                                    - &                             - \\
         66 &           8.3 &      (1,1) &  AGN remnant candidate (FRI) &                        - &                                    - &                             - \\
         68 &           8.2 &     (2,10) &      asymmetric double (FRI) &                        - &                                    - &                             - \\
         69 &           8.2 &      (3,9) &                     artefact &                        - &                                    - &                             - \\
         70 &           8.2 &      (2,9) &                    ambiguous &                        - &                                    - &                             - \\
         71 &           8.2 &      (1,1) &  AGN remnant candidate (FRI) &                        - &                                    - &                             - \\
         72 &           8.2 &      (1,1) &                    WDS (FRI) &                        - &                                    - &                  LEDA 2515619 \\
         73 &           8.1 &     (2,10) &  AGN remnant candidate (FRI) &      NVSS J123637+491143 &                               Galaxy &                  LEDA 2336215 \\
         74 &           8.1 &     (2,10) &                  NEBD (FRII) &   7C 104315.90+551446.00 &                               Galaxy &               SDSSCGB 20905.1 \\
         75 &           8.0 &      (1,1) &  AGN remnant candidate (FRI) &                        - &     LINER-type Active Galaxy Nucleus &      SDSS J130331.06+535948.5 \\
         76 &           7.9 &      (2,1) &                double (FRII) &                        - &                                    - &                             - \\
         77 &           7.9 &      (1,1) &                      complex &                        - &  Brightest galaxy in a Cluster (BCG) &       2MASX J12044737+4834115 \\
         78 &           7.9 &      (2,9) &                      complex &                        - &                               Quasar &     [VV2006] J133721.0+563329 \\
         79 &           7.9 &     (2,10) &                double (FRII) &                        - &                                    - &                  LEDA 2350076 \\
         80 &           7.8 &      (1,1) &                cluster relic &                        - &        Galaxy in Cluster of Galaxies &       DES J122852.54+473657.7 \\
         81 &           7.8 &      (2,9) &                    ambiguous &                        - &                                    - &                             - \\
         82 &           7.8 &      (2,9) &                    ambiguous &                        - &                                    - &                             - \\
         83 &           7.7 &      (2,1) &                   WDS (FRII) &      NVSS J133334+553949 &                                    - &                             - \\
         84 &           7.7 &      (1,1) &                  NEBD (FRII) &   7C 110931.60+554242.00 &  Brightest galaxy in a Cluster (BCG) &      SDSS J111226.86+552612.8 \\
         85 &           7.7 &     (3,10) &                double (FRII) &                        - &                               Quasar &      SDSS J114837.41+465319.5 \\
         86 &           7.7 &     (2,10) &                      WAT (?) &                        - &                  Cluster of Galaxies &                [SPD2011] 9215 \\
         87 &           7.7 &     (2,10) &                    ambiguous &                        - &                                    - &                             - \\
         88 &           7.7 &      (1,1) &     asymmetric double (FRII) &      NVSS J111218+475603 &                                    - &                             - \\
         89 &           7.7 &     (10,1) &                    WDS (FRI) &                        - &                                    - &                             - \\
         90 &           7.6 &      (2,1) &                      complex &                        - &        Galaxy in Cluster of Galaxies &       DES J131810.69+512724.8 \\
         91 &           7.6 &      (1,1) &                         DDRG &                        - &                     Seyfert 1 Galaxy &       2MASS J14510640+5333537 \\
         92 &           7.6 &     (2,10) &      asymmetric double (FRI) &      NVSS J115126+545005 &  Brightest galaxy in a Cluster (BCG) &                  LEDA 2481883 \\
         93 &           7.6 &      (1,1) &                      complex &                        - &                               Galaxy &                  LEDA 2520943 \\
         94 &           7.6 &     (3,10) &                    ambiguous &      NVSS J145000+540504 &                                    - &       2MASX J14500017+5405042 \\
         95 &           7.6 &     (2,10) &                double (FRII) &                        - &                                    - &                             - \\
         96 &           7.5 &     (2,10) &                    WAT (FRI) &   FIRST J121628.4+561209 &                                    - &      SDSS J121628.37+561209.4 \\
         97 &           7.4 &     (2,10) &                    ambiguous &                        - &                                    - &                             - \\
         98 &           7.4 &      (1,1) &                      complex &                        - &                                    - &                             - \\
         99 &           7.4 &     (2,10) &                         DDRG &  ILT J105742.50+510558.5 &                                    - &                             - \\
        100 &           7.4 &      (3,9) &                 double (FRI) &      NVSS J112942+542528 &  Brightest galaxy in a Cluster (BCG) &    	 SDSS J112942.17+542528.8 \\
\bottomrule
\end{tabular}}
\end{table*}

Figure \ref{fig:100outliers1} shows 100 morphological outliers from the first data release of the LoTSS survey \citep{Shimwell2019}, detected based on the large dissimilarity between the source and the SOM in Fig. \ref{fig:final_som}. The 100 detections are distributed over 96 physical sources as large objects enter the radio catalogue multiple times. We mitigated all but four duplicate outlier entries by imposing a minimum separation distance of 400 arcsec between each outlier. As our outlier metric is continuous, 100 is an arbitrary number of outliers and the procedure could instantly be repeated to find the $n$ sources with the highest outlier score. The white numbers in the cutouts in Fig. \ref{fig:100outliers1} correspond to the numbers in Tab. \ref{tab:out100}, which shows the outlier score for each source, a manual description of its radio morphology and FR class, its best matching representative image and, if present, the radio and optical ID, cross-matched using the SIMBAD astronomical database \citep{wenger2000simbad}. 

The radio IDs refer to the objects published by \citet[FIRST;][]{White1997}; \citet[NVSS;][]{Condon1998}; \citet[7C;][]{7Cfinal}; \citet[OR;][]{OR}; \citet[MPWK;][]{MPWK}; \citet[ILT;][]{ILT};  \citet[YHW2016;][]{YHW2016}; and \citet[P;][]{Green1995}. 
The optical IDs refer to the objects published by \citet[M;][]{M}; \citep[SDSS;][]{Adelman-McCarthy2009,Szabo2011,Ahn2012a,Alam2015,Paris2017,Wake2017}; \citet[NGC;][]{NGC}; \citet[2MASX;][]{jarrett20002mass}; \citet[YSS2008; ZwCl;][]{YSS2008}; \citet[DES;][]{Rykoff2016}; \citet[LH2011;][]{LH2011}; \citet[LEDA;][]{Paturel2003}; \citet[VV2006;][]{VV2006}; \citet[SPD2011;][]{SPD2011}; and \citet[2MASS;][]{skrutskie2006two}.

\clearpage

\section{Visualisation and exploration tool}
\label{app:visualisation}
The \textsc{LOFAR-PINK Visualisation Tool}: an interactive website for SOM exploration is available at \url{https://rafaelmostert.com/lofar/ID51/som.php} and will eventually be hosted at \url{https://www.astron.nl/lofar-som}.

The visualisation tool features: the image of a trained SOM; a heatmap overlay, which shows how many inputs from the dataset best matched each representative image; for each representative image, cutouts of ten radio sources from the training set that best match it; these cutouts are visible on the sky using an Aladin Lite snippet allowing users to see whether the object is part of some larger structure; a histogram of the Euclidean norm of each cutout to its best matching representative image; 100 cutouts that have the largest Euclidean norm and are thus most rare in morphology again coupled to an Aladin Lite snippet.
 
After a new SOM is trained and a corresponding dataset is mapped to it, we prepare the static content of the website, all within the same Jupyter notebook.
The project source code combined with the static content are uploaded to a web hosting server. For personal use, the upload step can be omitted and starting a virtual PHP server in the command prompt suffices. 
 
The webtool is built using a combination of open source (\textsc{Aladin Lite}, \textsc{jQuery}, \textsc{PHP}, \textsc{Python}) and open format code (\textsc{HTML}, \textsc{CSS}).
The source-code and its documentation is published on GitHub\footnote{\url{https://github.com/RafaelMostert/lofar-pink-visualization-tool}}. 

\end{appendix}

\end{document}